\documentclass{jfm}
\usepackage[dvipsnames, svgnames, x11names]{xcolor}
\usepackage{graphicx}
\usepackage{todonotes}
\usepackage{epstopdf, epsfig}
\usepackage{amsmath}
\usepackage{lipsum}  
\usepackage{hyperref}
\usepackage{verbatim}
\usepackage[hang,splitrule,symbol]{footmisc}
\usepackage{amssymb}
\usepackage{multirow}
\usepackage{lineno}
\usepackage{mathrsfs}
\usepackage{graphicx}
\usepackage{adjustbox}
\usepackage{subcaption}
\usepackage{booktabs}
\usepackage{makecell}

\usepackage{bigints}

\usepackage{xspace}

\shorttitle{A wall model for separated flows}
\shortauthor{Z. D. Zhou, X.-L. Zhang, G.-W. He and X. Yang}
\title{A wall model for separated flows: embedded learning to improve \emph{a posteriori} performance} 
\author{Zhideng Zhou\aff{1,2},
  Xin-lei Zhang\aff{1,2},
  Guo-wei He\aff{1,2},
 \and Xiaolei Yang\aff{1,2}\corresp{\email{xyang@imech.ac.cn}}}

\affiliation{\aff{1}The State Key Laboratory of Nonlinear Mechanics, Institute of Mechanics, Chinese Academy of Sciences, Beijing 100190, China
\aff{2}School of Engineering Sciences, University of Chinese Academy of Sciences, Beijing 100049, China}

\begin{document}
%\linenumbers
\maketitle

\begin{abstract}
The development of a wall model using machine learning methods for the large-eddy simulation (LES) of separated flows is still an unsolved problem. Our approach is to leverage the significance of separated flow data, for which existing theories are not applicable, and the existing knowledge of wall-bounded flows (such as the law of the wall) along with embedded learning to address this issue. The proposed so-called features-embedded-learning (FEL) wall model comprises two submodels: one for predicting the wall shear stress and another for calculating the eddy viscosity at the first off-wall grid nodes. We train the former using the wall-resolved LES data of the periodic hill flow and the law of the wall. For the latter, we propose a modified mixing length model, with the model coefficient trained using the ensemble Kalman method. The proposed FEL model is assessed using the separated flows with different {\color{black} flow configurations,} grid resolutions, and Reynolds numbers. Overall good \emph{a posteriori} performance is observed for predicting the statistics of the recirculation bubble, wall stresses, and turbulence characteristics. The statistics of the modelled subgrid-scale (SGS) stresses at the first off-wall grids are compared with those calculated using the wall-resolved LES data. The comparison shows that the amplitude and distribution of the SGS stresses obtained using the proposed model agree better with the reference data when compared with the conventional wall model.
\end{abstract}

%\begin{keywords}
%Authors should not enter keywords on the manuscript, as these must be chosen by the author during the online submission process and will then be added during the typesetting process (see http://journals.cambridge.org/data/\linebreak[3]relatedlink/jfm-\linebreak[3]keywords.pdf for the full list)
%\end{keywords}

\section{Introduction}\label{sec:Introduction}
%
%Large-eddy simulation (LES) is a powerful tool for the mechanism study of
%has shown promise in simulating complex 
%turbulent flows~\citep{Lesieur_ARFM_1996, Georgiadis_etal_AIAA_2010, He_Jin_Yang_ARFM_2017}. 
%, including large-scale flow phenomena like massive separation. 
%
In large-eddy simulation (LES) of high Reynolds number wall-bounded turbulent flows occurring in industrial and environmental applications, a wall model~\citep{Piomelli_Balaras_ARFM_2002, Bose_Park_ARFM_2018, Goc_Moin_Bose_2020, Yang_etal_AE_2021} is necessary even on the state-of-the-art supercomputers, which models the flows in the near-wall region to avoid the need to resolve the small-scale turbulence structures therein~\citep{Choi_Moin_PoF_2012, Yang_Griffin_PoF_2021}. 
%to avoid 
%However, the use of wall-resolved LES (WRLES) in high Reynolds number applications is still unfeasible due to 
%the prohibitively high grid resolution requirements ($N \sim Re^{13/7}$~\citep{Choi_Moin_PoF_2012, Yang_Griffin_PoF_2021}). 
%needed to resolve the viscous scale  
% 
The existence of universality for the near-wall flow is beneficial for such treatments, which, however, is not yet observed for all types of turbulent flows especially for those under non-equilibrium states. 
This makes the traditional wall models based on the equilibrium hypothesis unable to accurately simulate non-equilibrium flows, such as flows with separation~\citep{Breuer_etal_CAF_2007, Duprat_etal_PoF_2011}. 
The machine learning method offers a novel avenue for constructing models for non-equilibrium flows. Nonetheless, the acquired models frequently face challenges related to limited generalizability and suboptimal \emph{a posteriori} performance~\citep{Zhou_He_Yang_PRF_2021, Zhou_etal_PoF_2023}.
In this study, we introduce a features-embedded-learning (FEL) wall model for LES of separated flows. The model enhances generalizability by leveraging high-fidelity data from separated flows and the law of the wall. The \emph{a posteriori} performance is enhanced through embedded learning using the ensemble Kalman method.

%To address this issue and reduce computational costs, wall modeling techniques have been proposed to approximate the effect of the wall on the outer flow. Wall-modeled LES (WMLES) reduces the grid resolution requirement to $N \sim Re$, making LES more applicable to industrial turbulent flows at high Reynolds numbers~\citep{Piomelli_Balaras_ARFM_2002, Bose_Park_ARFM_2018, Goc_Moin_Bose_2020, Yang_etal_AE_2021}.
Lack of generalizability is one of the most critical problems for data-driven turbulence models, although many efforts have been placed for both 
%The application of ML methods in fluid mechanics 
%The machine learning methods have been employed to develop 
%has been extensive~\citep{Duraisamy_etal_ARFM_2019, Brunton_etal_ARFM_2020}, including the development of 
%turbulence models for 
Reynolds-averaged Navier-Stokes (RANS) methods~\citep{Ling_etal_JFM_2016, Wu_Xiao_Paterson_2018, Zhang_etal_JFM_2022} and % subgrid-scale (SGS) models for 
LES methods~\citep{Vollant_etal_JOT_2017, Park_Choi_JFM_2021, Xu_etal_JFM_2023}. 
Employing neural networks to model near-wall flows dates back to the work by~\citet{Milano_Koumoutsakos_JCP_2002}. 
In wall-modelled LES (WMLES), approximate boundary conditions, such as the shear stress at the wall, define how the near-wall unresolved flow structures influence the outer flow. In traditional wall models, the wall shear stress is often computed using the law of the wall~\citep{Deardorff_JFM_1970, Werner_Wengle_1993} or by solving the thin-boundary-layer equation~\citep{Cabot_Moin_FTC_2000, Wang_Moin_PoF_2002, Park_Moin_PoF_2014, Yang_etal_PoF_2015}.
%
%
%No {\it a posteriori cases} were carried out in their work. 
%
%temporal prediction of turbulence~\citep{Lee_You_JFM_2019, Nakamura_etal_PoF_2021, Qu_etal_PoF_2022}, reconstruction of the turbulent flow fields~\citep{Fukami_Fukagata_Taira_JFM_2021, Zhou_etal_CAF_2022, Laima_etal_PoF_2023, Arun_Bae_McKeon_arXiv_2023}, turbulence identification~\citep{Wu_etal_PRF_2019, Li_etal_JFM_2020, Fukami_etal_JFM_2021}, and flow control~\citep{Park_Choi_JFM_2020, Zhou_etal_JFM_2020, Ren_Rabault_Tang_PoF_2021, Guastoni_etal_arXiv_2023}.
In the work by~\citet{Yang_etal_PRF_2019}, a feedforward neural network model was constructed to compute the wall shear stress using the flow quantities at the first off-wall grid node, and successfully applied to WMLES of turbulent channel flows at various Reynolds numbers and spanwise rotating turbulent channel flows~\citep{Huang_etal_PoF_2019}. 
In the model developed by \citet{Lee_etal_AST_2023}, the input features are extracted from the Fukagata-Iwamoto-Kasagi identity~\citep{FIK_identity_PoF_2002} to describe the effects of flow dynamics on the wall shear stress. 
%This model could be applied to the TCF and separated turbulent boundary layer flow with untrained Reynolds numbers. 
%
It was shown that incorporating known physics in the construction of input features improves the model generalizability for different Reynolds numbers ~\citep{Yang_etal_PRF_2019, Lee_etal_AST_2023}. 
Generalization of a data-driven wall model for different flow regimes is even more challenging. One idea is to build the model using data from various flows, which can be in the same flow regime but with different parameters or in different flow regimes. 
To simulate shock-boundary layer interaction, \citet{Bhaskaran_etal_IEEE_2021} trained a wall model using the data from eight wall-resolved LES (WRLES) cases with different curvatures near the blade trailing edge and different shock locations and strengths. 
To simulate supersonic turbulent flows with separation, \citet{Zangeneh_PoF_2021} trained a wall model using data from the zero pressure gradient turbulent flow over a flat plate and supersonic flow around an expansion-compression corner. 
%Although the model showed some advantages over the baseline WMLES model, the improvement was less significant compared to models trained with data from a single flow. 
%
To simulate separated flows, \citet{Dupuy_etal_JCP_2023} trained a wall model using the filtered high-fidelity data from turbulent channel flow, the flow in a three-dimensional diffuser, and the backward-facing step flow. 
%and two turbulent flows with separated regions. 
%
%This model could be applied to the similar flow configuration. These supervised learning wall models were mainly trained using the high-fidelity (DNS or WRLES) flow data and tested with the coarse grid of WMLES. 
%Although the accurate prediction can be achieved in specific flow configurations, ensuring consistency between the learning and prediction environments remains a challenge.
%
Assuming that the flow in the near-wall region can be modelled using a finite set of canonical flows, \citet{Lozano_Bae_JFM_2023} proposed a building-block-flow wall model, tested the model for canonical flows, and applied it to two aircraft configurations. 
The lack of the law of the wall makes that the state of flow at a single off-wall grid node is not enough to fully describe the near-wall flow for separated flows. 
%In the case of flow over periodic hills (PH), 
\citet{Zhou_He_Yang_PRF_2021} employed the velocity and pressure gradient at three off-wall grid nodes as input features to construct a neural network model for computing wall shear stress, which was trained using the wall-resolved data of the periodic hill flow. The predicted instantaneous wall shear stress agree well with the reference data with the correlation coefficients higher than 0.6. 
%
%developed the FNN\_PH model using the WRLES flow data at various streamwise locations as training samples and considering the pressure gradients as input features. The model accurately predicted the wall shear stress in the PH cases at different Reynolds numbers. 
%
To improve the generalizability of the model, the law of the wall was later introduced in the training of the model~\citep{Zhou_etal_PoF_2023}, 
showing good performance for both the periodic hill flow and the turbulent channel flow. 

Suboptimal \emph{a posteriori} performance remains as a challenge for data-driven wall models especially for non-equilibrium flows. One main cause is the inconsistency in the environments for model training and prediction~\citep{Duraisamy_Perspective_PRF_2021}. This mainly includes the numerical discretization error (which is small in wall-resolved simulations while is large in coarse-grid WMLES) and the error from the subgrid-scale (SGS) modelling (which is zero or small in wall-resolved simulations while is significant in WMLES). To address this problem, efforts have been made in the way preparing the training data and using different machine learning methods. 
%
%The challenge can be addressed through a novel strategy known as model-consistent training~\citep{Duraisamy_Perspective_PRF_2021}, which involves the interaction between model training and the prediction environment. 
%
In the work by~\citet{Lozano_Bae_JFM_2023}, the training data were generated from the so-called E-WMLES, in which the ``exact'' SGS model and wall shear stress are employed to match the mean velocity profiles from direct numerical simulation (DNS). Performance better than the model based on filtered DNS data was obtained. 
%
%To guarantee the model consistency, the training data were obtained directly from WMLES optimized for correct mean flow quantities, instead of the filtered DNS data in their previous work~\citep{Lozano_Duran_Bae_2020}. 
%
In the work by~\citet{bae2022scientific}, the multi-agent reinforcement learning was employed to train a wall model in WMLES environment for turbulent channel flows. In the proposed model, the agents are evenly distributed points on the wall with their actions of adjusting the applied wall shear stress and the reward based on the predicted wall shear stress. 
Later development of the model based on reinforcement learning to consider pressure gradient effects was carried out by~\citet{Zhou_etal_AIAA_2023, Zhou_Bae_arXiv_2023}.
Possible sources of errors for wall models based on different machine learning methods were analyzed by~\citet{Vadrot_etal_PRF_2023}. 
Wall shear stress boundary condition is often employed to approximate the effects of the unresolved near-wall flow structures on the outer flow in WMLES. However, it has drawbacks for separated flows {\color{black}because of the lack of a constant shear layer}. Consider the flow near the separation or reattachment point. Although the amplitude of wall shear stress is small in the region, there still exist small-scale near-wall flow structures acting on the outer flow. In WMLES, zero-wall shear stress indicates no effects of the wall on the outer flow, together with the SGS model for the very coarse first-wall grid nodes poorly modelling the effects of the unresolved near-wall flow structures. 

We propose to enable the generalization ability of the neural network wall model by integrating high-fidelity data and knowledge and improve the \emph{a posteriori} performance by accounting for the SGS modelling defect in an embedded-learning environment. The proposed wall model is composed of a model for wall shear stress and a model for the eddy viscosity at the first off-wall grid nodes, in which the latter is trained in WMLES environment, and the former is trained using the separated flow data and the law of the wall. Such a training approach prevents contaminating the wall shear stress model with errors in WMLES.

The rest of the paper is organized as follows: the proposed wall model is introduced in 
%, we introduce the machine learning strategy for the wall model in 
section~\ref{sec:ML_strategy}; the training of the model is described in section~\ref{sec:Training_PH};
% This strategy encompasses the flow solver, the supervised learning of the wall shear stress, and the embedded training of the near-wall eddy viscosity using the ensemble Kalman method. Then 
systematic assessment of the proposed model in the periodic hill flows is presented in section~\ref{sec:Results}; {\color{black}the application of the proposed model to other flow configurations is presented in section~\ref{sec:Application_flow_config};} the conclusions are drawn in section~\ref{sec:Conclusion}.

\section{The features-embedded-learning wall model} \label{sec:ML_strategy}
\begin{figure}
    \centering
    \includegraphics[width=1.0\textwidth]{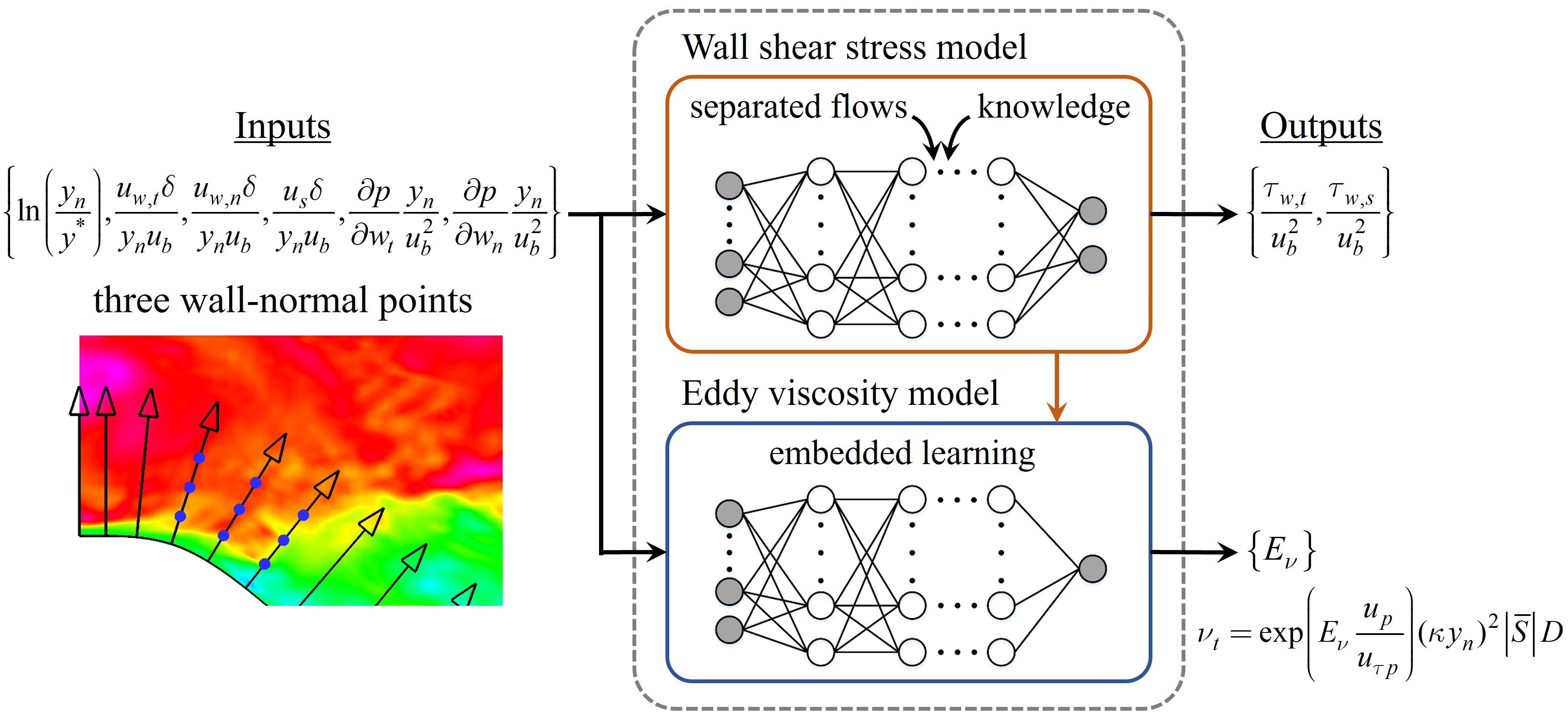}
    \caption{Schematic of the features-embedded-learning wall model.}
    \label{fig:schematic-cDK}
\end{figure}

%
%The framework of the ensemble-based training of the embedded-neural-network wall model (ENWM) for LES of incompressible flows
%
The FEL wall model approximates the effects of near-wall flows on the outer flow using wall shear stress and eddy viscosity at the first off-wall grids. The former is computed using a neural network model trained separately using high-fidelity data and the law of the wall. The model for the latter is trained in an embedded way in the WMLES environment to account for the SGS modelling issue for separated flows. A schematic of the FEL wall model is shown in figure~\ref{fig:schematic-cDK}.
%
%Using a well-validated model for wall shear stress  
%The framework of ensemble-based training of the critical-Data-and-Knowledge-embedded (cDK-embedded) wall model for LES of incompressible flows, which includes the FNN\_PH-LoW model for the wall shear stress and an embedded neural network for the eddy-viscosity coefficient, 
%
\subsection{Wall shear stress model}\label{subsec:tau_w}
%As we known, there are various influential factors that determine the quality of WMLES, such as the wall model, SGS model, grid resolution and numerical scheme. In this study, a ML based model is employed to couple the wall shear stress and the near-wall eddy viscosity, thereby representing the effect of the wall on the outer flow.
%
We empower the capability of the neural network model in estimating wall shear stress for different flow regimes using the data from separated flows and the law of the wall. Specifically, the data from the periodic hill flow and the logarithmic law for the mean streamwise velocity profile, {\color{black}which is introduced in appendix~\ref{appendix:data},} are employed for training the wall shear stress model. The periodic hill case, even though its geometry is relatively simple, contains several flow regimes, i.e., flows with pressure gradients, flow separation, and flow reattachment. 
%However, the boundary layer flow in the equilibrium state does not exist in the periodic hill flow. With 
The inclusion of the logarithmic law in model training then enables the trained model to 
%, the neural network model is then trained to learn how the wall shear stress 
react properly to flows in the equilibrium state. 
%

%The neural network model developed
%the wall shear stress, the FNN\_PH-LoW model, which was trained 
%in our previous work~\citep{Zhou_etal_PoF_2023} is adopted for estimating the wall shear stress. To construct this model, the data from the WRLES of flow over periodic hills and the law of the wall are both prepared, and a FNN is employed for the training. The FNN architecture 
A feedforward neural network (FNN) is employed for building the connection between the near-wall flow and the wall shear stress. The neural network comprises an input layer, six hidden layers with 15 neurons in each layer, and an output layer. {\color{black}And the hyperbolic tangent (tanh) function serves as the activation function.} The input features consist of six flow quantities at three wall-normal points, with the distance between two adjacent points denoted as $\Delta y_n/\delta_0 = 0.03$,
\begin{equation}
X_I = \left\{ \ln\left(\frac{y_n}{y^{*}}\right), \frac{u_{w,t}}{y_n} \frac{\delta_0}{u_b}, \frac{u_{w,n}}{y_n} \frac{\delta_0}{u_b}, \frac{u_s}{y_n} \frac{\delta_0}{u_b}, \frac{\partial p}{\partial w_t} \frac{y_n}{\delta_0} \frac{\delta_0}{u_b^2}, \frac{\partial p}{\partial w_n} \frac{y_n}{\delta_0} \frac{\delta_0}{u_b^2} \right\},
\label{eq_input}
\end{equation}
where $y_n$ is the wall-normal distance, $u_{w,t}$, $u_{w,n}$ and $u_s$ are the velocity components in the wall-tangential, wall-normal and spanwise directions, $\frac{\partial p}{\partial w_t}$, $\frac{\partial p}{\partial w_n}$ are the pressure gradients in the wall-tangential and wall-normal directions, $u_b$ is the bulk velocity, $\delta_0$ is the global length scale. 
{\color{black}
It is noted that the length scale $\delta_0$ represents the scale of the outer flow. It is set as the hill height $h$ in periodic hill flows, and the half channel width $\delta$ in turbulent channel flows. For flows in a complex geometry with several geometric length scales, a systematic way for defining the length scale is yet to be developed. A case-by-case approach is probably needed.
}
The wall-normal distance is normalized by a near-wall length scale $y^{*} = \nu/u_{\tau p}$~\citep{Duprat_etal_PoF_2011}, where $\nu = \mu/\rho$ is the kinematic viscosity, $u_{\tau p} = \sqrt{u_{v}^2+u_p^2}$, $u_{v} = \sqrt{\left| \frac{\nu u_{w, t}}{y_n} \right|}$, $u_p = \left| \frac{\nu}{\rho} \frac{\partial p}{\partial w_t} \right|^{1/3}$.
The output labels are the wall-tangential and spanwise wall shear stresses,
\begin{equation}
Y_O^{(w)} = \left\{ \frac{\tau_{w, t}}{u_b^2}, \frac{\tau_{w, s}}{u_b^2} \right\}.
\label{eq_Tauw_out}
\end{equation}
{\color{black}The cost function is defined as the mean square error between the predicted output and the real output. The error backpropagation (BP) scheme~\citep{Rumelhart_etal_Nature_1986} implemented with TensorFlow~\citep{Google_Brain_2016} is employed to train the FNN model by optimizing the weight and bias coefficients to minimize the cost function. A detailed description of the training procedures can be found in~\citet{Zhou_He_Yang_PRF_2021}.}
%The procedure for calculating the output based on the input in the FNN involves linear matrix manipulation of the weight and bias coefficients, as well as nonlinear mapping using the activation function. A detailed description of this procedure can be found in Appendix C of Ref.~\citep{Zhou_He_Yang_PRF_2021}.

\subsection{Eddy viscosity model}\label{subsec:nu_t}
SGS modelling error and discretization error are the two major causes for the suboptimal \emph{a posteriori} performance of a data-driven wall model in WMLES environment. 
%In this work, we attempt to address the SGS modelling error for separated flows using an eddy viscosity approach. 
In separated flows, a near-wall layer of constant shear stress does not exist. In WMLES, the near-wall layer, which is in non-equilibrium state for separated flows, is not resolved by the grid. The wall shear stress boundary condition alone is not enough to model the way the unresolved near-wall flow affecting the outer flow. In this work, we postulate that a proper eddy viscosity can approximate the effects of the near-flow not captured by the wall shear stress.

There are several options to establish an eddy viscosity model, e.g., an analytical approach, a data-driven approach, or a hybrid approach. To compromise between generalizability and predictive capability of the model, a hybrid approach is employed. Specifically, we employ a modified version of the mixing length model with its coefficients learned in an embedded way, in the following form,
%the damped mixing length model~\citep{Cabot_Moin_FTC_2000} to replace the SGS stress in the first wall-normal grid, as shown below:
%
\begin{equation}
%\nu_t = \exp(E_\nu) \kappa y{u_\tau}D, \quad D = \left[ 1 - \exp(-y^+ / A^+) \right]^2,
    \nu_t = \exp\left( E_\nu \frac{u_p}{u_{\tau p}} \right) (\kappa y_n)^2 \left| \overline{S} \right| D \text{ at the first off-wall grid nodes,}
\label{eq_nut}
\end{equation}
where the von K{\'a}rm{\'a}n constant $\kappa \approx 0.4$, $D = \left[ 1 - \exp(-(y^+ / A^+)^3) \right]$, $A^+=25$, $y^+ = y_n u_\tau/\nu$, $u_\tau = \sqrt{\tau_w/\rho}$ is the friction velocity given by the wall shear stress model {\color{black}in section~\ref{subsec:tau_w}}. {\color{black}With eq.~(\ref{eq_nut}), the eddy viscosity at the first off-wall grid node is modelled using the the modified mixing length model instead of the dynamic Smagorinsky model (DSM) employed in the outer flows.}
The exponential form coefficient, i.e., $\exp\left( E_\nu \frac{u_p}{u_{\tau p}} \right)$ is proposed to prevent negative eddy viscosity. While a clipping procedure for negative $\nu_t$ can be employed as well~\citep{Zhou_etal_CAF_2019, Park_Choi_JFM_2021}, it may affect the predictive capability of the model.

To approximate the modified mixing length model using a neural network, {\color{black}the model coefficient $E_\nu$ is used as the output label.} The neural network is composed of an input layer, six hidden layers with 15 neurons in each layer, and an output layer. In the input layer, the input features are the same as those for the wall shear stress model, i.e., eq. (\ref{eq_input}). {\color{black}The hyperbolic tangent (tanh) function is also employed as the activation function for this new neural network.} Using the ensemble Kalman method, the neural network is trained in the WMLES environment in an embedded way.
%The eddy-viscosity coefficient, denoted as $E_\nu$, serves as the output of a new neural network,
%\begin{equation}
%Y_O^{(n)} = \left\{ E_\nu \right\}.
%\label{eq_nut_out}
%\end{equation}
%
%for $Y_O^{(n)}$ are also assigned as $X_I$ in Eq. \ref{eq_input}. The neural network is embedded into the WMLES solver and trained using the Ensemble Kalman method in the subsequent section.

\subsection{Ensemble Kalman method}\label{subsec:Kalman}
The ensemble Kalman method is a statistical inference method based on Monte Carlo sampling, and has been widely used in various applications~\citep{zhang2020evaluation,schneider2022ensemble,zhang2022acoustic,Liu_Zhang_He_AIAA_2023}. 
In this work, the method is employed to learn the weights in the neural network model for eddy viscosity.
%in Section~\ref{subsec:nu_t}, and
It utilizes the statistics of the weights and model predictions to compute the gradient and Hessian of the cost function to update the model. The cost function is given as,
\begin{equation}
    J= \| w_m^{n+1} - w_m^n  \|_\mathsf{P}+ \| \mathcal{H}[w_m^{n+1}] - \mathsf{y} \|_\mathsf{R},    
\end{equation}
where $w_m$ is the weight of neural network, $m$ is the sample index, $n$ is the iteration index, $\mathsf{P}$ is the error covariance of neural network model, $\mathsf{R}$ is the error covariance of observation data, $\mathsf{y}$ is the observation data that obey the normal distribution with zero mean and variance of $\mathsf{R}$, and $\mathcal{H}$ is the model operator that maps the neural network weights to model predictions. %In this work, the observation data and model predictions denote the various vertical profiles of mean velocity obtained from the WRLES and WMLES, respectively.

The method uses the ensemble of the neural network samples $W$ ($=[w_1, w_2, \cdots, w_M]$) to estimate the sample mean~$\bar{W}$ and covariance~$\mathsf{P}$ as
\begin{equation}
\left\{
\begin{aligned}
	\bar{W} &= \frac{1}{M} \sum_{m=1}^M w_{m} \text{,} \\
	\mathsf{P} &= \frac{1}{M-1} (W - \bar{W})(W-\bar{W})^\top \text{,}
\end{aligned}
\right.
\end{equation}
where $M$ is the sample size.
Based on the Gauss-Newton method, the first- and second-order derivatives of the cost function are required to update the weights.
The ensemble Kalman method uses the statistics of these samples to estimate the derivative information~\citep{luo2015iterative}.
At the $n^{th}$ iteration, each sample~$w_m$ is updated based on
\begin{equation}
    w_m^{n+1} = w_m^n + \mathsf{PH}^\top (\mathsf{HPH}^\top + \mathsf{R})^{-1} (\mathsf{y}_m^n - \mathsf{H}w_m^n) \text{,}
    \label{eq:enkf}
\end{equation}
where $\mathsf{H}$ is the tangent linear model operator.
The readers are referred to Ref.~\cite{Zhang_etal_JFM_2022} for details of the employed ensemble Kalman method.

\subsection{Procedure for embedded training}
The procedure for embedded training of the wall model is divided into four steps as depicted in figure~\ref{fig:schematic}, including 1) Pre-train the model; 2) Obtain the predictions for an ensemble of neural network models; 3) Compute the error of the model predictions; 4) Update the neural network using the ensemble Kalman method; 5) Iterate steps 2 to 4 until a certain threshold or the maximum number of iterations is reached. Specifics for steps 2-4 are listed as follows: 
\begin{figure}
    \centering
    \includegraphics[width=1.0\textwidth]{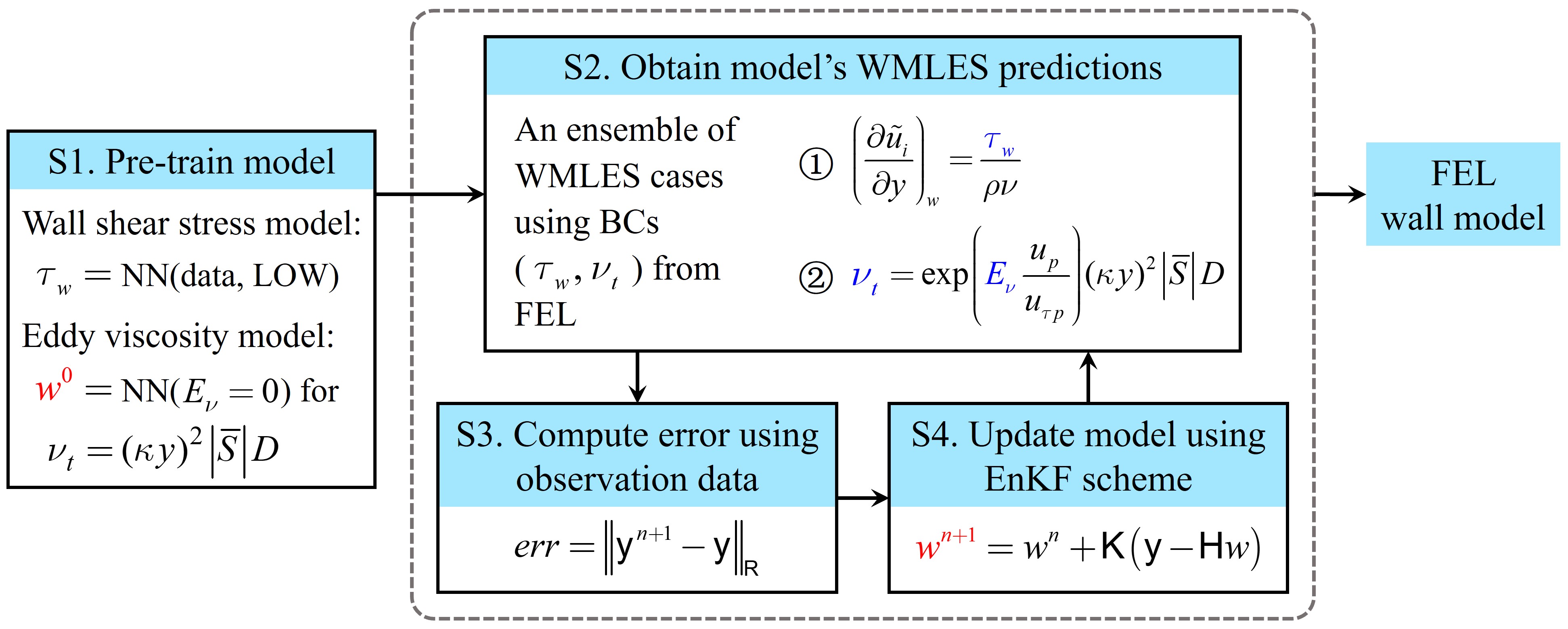}
    \caption{Embedded training of the FEL wall model.}
    \label{fig:schematic}
\end{figure}

%The detailed training procedure is formulated as follows:
\begin{enumerate}[1)]
    \item Pre-train the model: 
    %Before the initial step of figure~\ref{fig:schematic}, 
    the neural network model for the eddy viscosity is trained using %firstly pre-trained as 
    the mixing length model with $E_{\nu}=0.0$ ($\exp(E_\nu) = 1.0$) to obtain the initial guess on the neural network weights. This is important for accelerating the embedded training of the model. The wall shear stress model is trained using the high-fidelity data and the law of the wall, and remains unchanged during the training of the eddy viscosity model. 
    %
    %Using the pre-trained weights $w^0$ and the initial variance, we can generate $M$ samples of neural network weights for calculating $\nu_t$. This is done because the conventional initialization of neural network weight may lead to an excessively large value of eddy viscosity, which may further cause the flow prediction far from the true solution and increase the number of iteration steps required for embedded training.
    %Samples of neural network weights are then drawn around this pre-trained weight, ensuring that the resulting eddy viscosity remains close to the baseline.
    \item Obtain the WMLES predictions for an ensemble of neural network models: this is done by applying the FEL model to WMLES. In this step, it is important to consider the balance between the computational cost and training efficiency when determining the number of samples. 
    %
    %Feature extraction for the cDK-embedded model: The second step is to prepare the input data for the cDK-embedded model. Based on the instantaneous flow field of the WMLES, we can extract the input features, including the wall-normal distance ($y_n$), velocity ($\mathbf{\tilde u}$), and pressure gradient ($\nabla p$). These input features are subsequently normalized by the global length and time scales, as described in Eq.~\ref{eq_input}.
    %\item Coupling of the cDK-embedded model and WMLES solver: The input features are propagated into the wall shear stress model and the eddy viscosity model. The outputs $\nu_t$ and $\tau_w$ are used as the boundary conditions for the WMLES solver. Subsequently, the $M$ samples of flow prediction can be obtained from the realizations of WMLES.
    %
    \item Compute the errors of the model predictions: the WMLES predictions and the observation data are employed to compute the error. The quantities and their locations essentially determine the capability of the learned model, which has to be selected with care.  
    \item Update the neural networks using the ensemble Kalman method: the errors obtained in the last step are employed to update the weights of the neural network model for $\nu_t$ based on the ensemble Kalman method. 
\end{enumerate}

%For the WMLES of separated flows, the VFS-Wind code discussed in section~\ref{subsec:Flow solver} is employed. The ensemble Kalman method is implemented using the publicly available DAFI code~\citep{strofer2021dafi}.

%In this section, the proposed framework in learning embedded-neural-network wall models is applied to the separated flows with different curved surfaces, shown in figure. The geometry, computational domain, the employed curvilinear mesh and time-averaged streamwise velocity contour from a typical WRLES case of the flow over periodic hills and backward-facing hill are plotted in the figure. Table~\ref{tab:table1} lists the parameters of the WRLES cases, in which ``HS'', ``HB'' and ``HL'' respectively denote the periodic hill with $\alpha=0.5$, 1.0 and 1.5, ``BH'' denotes the backward-facing hill with $\alpha=1.0$. The WRLES data is used for the training and testing of the embedded-neural-network wall models.
%

\section{Embedded training in WMLES of the periodic hill flow}\label{sec:Training_PH}
In this section, we first present the embedded training of the FEL wall model in the WMLES of the periodic hill flow. With the learned model, the \emph{a posteriori} performance is systematically examined. 
\subsection{WMLES environment}\label{subsec:Flow solver}
%In this section, we describe the numerical method and the case setup for generating the data employed for developing a data-driven wall model, which can take into account the nonequilibrium effects, e.g., flow separation and reattachment, for turbulent flows over periodic hills.
%
The Virtual Flow Simulator (VFS-Wind) \citep{Yang_etal_WE_2015, Yang_Sotiropoulos_WE_2018} code is employed for WRLES and WMLES.
%for the WMLES of separated flows. The VFS-Wind code has demonstrated successful applications in various turbulent flows within industrial and environmental contexts~\citep{Kang_Sotiropoulos_AWR_2012, Kang_etal_JFM_2014, Khosronejad_Sotiropoulos_JFM_2014, Yang_etal_RE_2017, Yang_Sotiropoulos_BE_2019, Yang_Sotiropoulos_PRF_2019, Foti_etal_JFM_2019, Khosronejad_etal_JHE_2020, Zhou_Wu_Yang_PoF_2021}. In the code, 
The governing equations are the spatially filtered incompressible Navier-Stokes equations in non-orthogonal, generalized curvilinear coordinates as follows,
\begin{equation}
\left\{
  \begin{aligned}
    J \frac{ \partial U^{j} }{\partial \xi^{j}} &= 0, \\
    \frac{1}{J} \frac{\partial U^{i}}{\partial t} &= \frac{\xi^{i}_{l}}{J} \left( -\frac{\partial}{\partial \xi^{j}} (U^{j}u_{l}) - \frac{1}{\rho} \frac{\partial}{\partial \xi^{j}} ( \frac{\xi^{j}_{l} p}{J} ) + \frac{\mu}{\rho} \frac{\partial}{\partial \xi^{j}}( \frac{g^{jk}}{J} \frac{\partial u_{l}}{\partial \xi^{k}} ) - \frac{1}{\rho} \frac{\partial \tau_{lj}}{\partial \xi^{j}} + f_l \right),
  \end{aligned}
\right.
\label{eq_1}
\end{equation}
where $x_{i}$ and $\xi^{i}$ are the Cartesian and curvilinear coordinates, respectively, $\xi^{i}_{l} = \partial \xi^{i} / \partial x_{l}$ are the transformation metrics, $J$ is the Jacobian of the geometric transformation, $u_{i}$ is the $i$-th component of the velocity vector in Cartesian coordinates, $U^{i} = (\xi^{i}_{m} / J) u_{m}$ is the contravariant volume flux, $g^{jk} = \xi^{j}_{l} \xi^{k}_{l}$ are the components of the contravariant metric tensor, $\rho$ is the fluid density, $\mu$ is the dynamic viscosity, and $p$ is the pressure. In the momentum equation, $\tau_{ij}$ represents the anisotropic part of the SGS stress tensor, which is modeled by the Smagorinsky model,
\begin{equation}
\tau_{ij} - \frac{1}{3} \tau_{kk} \delta_{ij} = - 2 \nu_{t} \overline{S}_{ij},
\label{eq_2}
\end{equation}
where $\overline{S}_{ij} = \frac{1}{2} \left( \frac{\partial U_i}{\partial x_j} + \frac{\partial U_j}{\partial x_i} \right)$ is the filtered strain-rate tensor and $\nu_t$ is the eddy viscosity calculated by
\begin{equation}
\nu_{t} = C \Delta^{2} |\overline{S}|,
\label{eq_3}
\end{equation}
where $C$ is the model coefficient calculated dynamically using the procedure of Germano et al. \citep{Germano_etal_PoF_1991}, $|\overline{S}| = \sqrt{2\overline{S}_{ij}\overline{S}_{ij}}$ and $\Delta = J^{-1/3}$ is the filter size, where $J^{-1}$ is the cell volume.

The governing equations are spatially discretized using a second-order accurate central difference scheme, and integrated in time using the fractional step method. An algebraic multigrid acceleration along with generalized minimal residual method (GMRES) solver is used to solve the pressure Poisson equation. A matrix-free Newton-Krylov method is used for solving the discretized momentum equation.
%More details about the flow solver can be found in the literatures\citep{Ge_Sotiropoulos_JCP_2007, Kang_etal_AWR_2011, Yang_etal_WE_2015}.

\subsection{Embedded training}
The ensemble Kalman method is an efficient method for embedded learning. 
%It is a second-order optimization with approximated gradient and Hessian information. 
To learn the model, an ensemble of samples are required to estimate the gradient and Hessian based on the statistics derived from these samples~\citep{luo2015iterative}. 
A sample corresponds to a WMLES case with the FEL wall model, which runs over a time period long enough to obtain the statistics to be compared with the reference data (WRLES results in this work). 
%
%is incorporated with the WMLES solver to predict the fully developed turbulent flow fields. 
%
The ensemble of WMLES samples are executed in parallel without communication during running the cases. 
%, ensuring that the wall time does not experience a significant increase.
%
With the error computed based on the discrepancy between the WMLES predictions and reference data, the FEL model is updated for the next iteration. 
%
%In the subsequent sections, we provide a comprehensive description of the case setup, as well as the training and testing outcomes for the cDK-embedded wall model.

\begin{figure}
\centering{\includegraphics[width=0.75\textwidth]{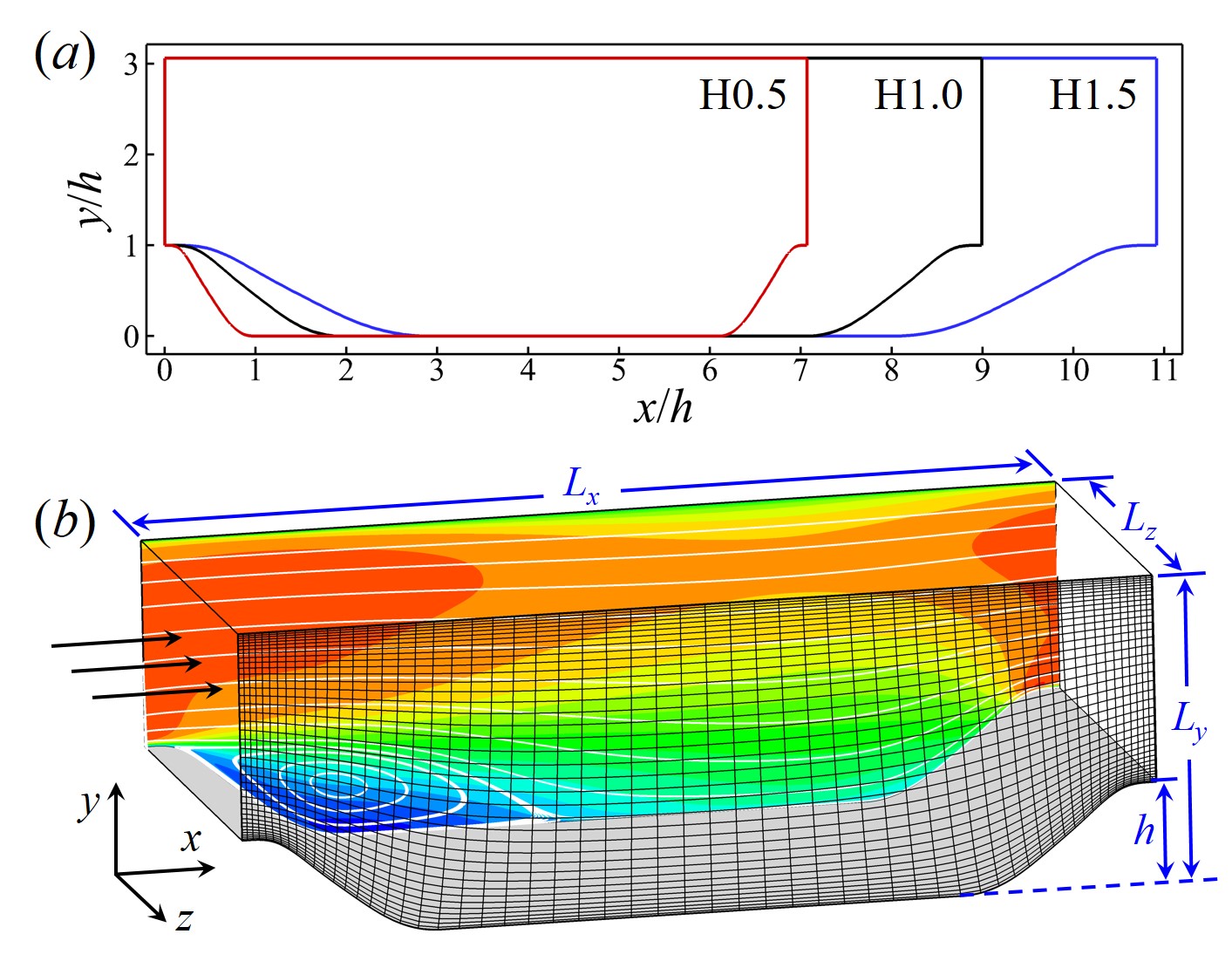}}
  \caption{Schematic of (a) the periodic hill geometry with various slopes (parameterized by multiplying a factor $\alpha$ to the hill width), and (b) the computational domain ($L_x=9.0h$, $L_y=3.036h$, $L_z=4.5h$), the employed curvilinear mesh on an $x-y$ plane (every fifth grid), and the contour of time-averaged streamwise velocity with streamlines for the baseline case ($\alpha = 1.0$) at $Re_h=10595$.}
\label{fig:geo_PH}
\end{figure}

%The proposed framework in learning the cDK-embedded wall model is applied to 
The periodic hill flow case employed for embedded learning of the FEL wall model is illustrated in figure~\ref{fig:geo_PH}. The hill geometry is given by analytical expressions. The different slopes shown in figure~\ref{fig:geo_PH}(a) 
%displays the periodic hill geometry with various slopes, which are parameterized 
are obtained by multiplying a factor $\alpha$ to the hill width~\citep{Xiao_etal_CAF_2020, Zhou_He_Yang_PRF_2021}. 
The computational domain ($L_x=9.0h$, $L_y=3.036h$, $L_z=4.5h$) and the employed curvilinear mesh on an $x-y$ plane are illustrated in figure~\ref{fig:geo_PH}(b) for the case with $\alpha = 1.0$, with the contour of time-averaged streamwise velocity with streamlines from the simulation at $Re_h=\rho u_b h/\mu=10595$ (where $h$ represents the hill height, $u_b = Q/(\rho L_z (L_y-h))$ denotes the bulk velocity, and $Q$ is the mass flux). 
%For the baseline hill geometry ($\alpha = 1.0$), figure~\ref{fig:geo_PH}(b) depicts . The Reynolds number is defined as $Re_h = \rho U_b h/\mu$, where $h$ represents the hill height, $U_b = Q/(\rho L_z (L_y-h))$ denotes the bulk velocity, and $Q$ is the mass flux. 
%
Periodic boundary condition is applied in the streamwise and spanwise directions.
A uniform pressure gradient maintaining a constant mass flux is applied over the entire domain to drive the flow.
On the top wall and the surface of the hills, no-slip boundary condition is applied in WRLES, while in WMLES the FEL wall model is employed.  
\begin{table}
  \begin{center}
\def~{\hphantom{0}}
  \begin{tabular}{c|c|c|c|cc|cc}
  \cline{1-8}
  Case     &  $Re_h$  &  $\alpha$  &  $N_x \times N_y \times N_z$  &  $\Delta x_f/h$  &  $\Delta y_f/h$  &  $\Delta y_c^+$  &  $\Delta y_{c,\text{max}}^+$ \\
  \cline{1-8}\rule{0pt}{8pt}
  H0.5-WR      & \multirow{15}{*}{10595} & \multirow{4}{*}{0.5} & $266 \times 192 \times 186$ & 0.01 & 0.003 & 0.5 & 2.1 \\ \cline{0-0}\cline{4-8}
  H0.5-WM-0.01 &         &  &  $162 \times 64  \times 64 $                      & 0.03 &  0.01  & 2   & 7 \\ \cline{0-0}\cline{4-8}
  H0.5-WM-0.03 &         &  &  \multirow{3}{*}{$102 \times 32  \times 32 $}     & 0.06 &  0.03  & 5   & 21 \\ 
  H0.5-WM-0.06 &         &  &                                                   & 0.08 &  0.06  & 10  & 42 \\
  H0.5-WM-0.09 &         &  &                                                   & 0.10 &  0.09  & 15  & 63 \\ \cline{0-0}\cline{3-8}
  H1.0-WR      &         & \multirow{5}{*}{1.0} &  $296 \times 192 \times 186$  & 0.01 &  0.003 & 0.5 & 2  \\ \cline{0-0}\cline{4-8}
  H1.0-WM-0.01 &         &  &  $172 \times 64  \times 64 $                      & 0.03 &  0.01  & 2   & 7  \\ \cline{0-0}\cline{4-8}
  H1.0-WM-0.03 &         &  &  \multirow{3}{*}{$112 \times 32  \times 32 $}     & 0.06 &  0.03  & 5   & 20 \\ 
  H1.0-WM-0.06 &         &  &                                                   & 0.09 &  0.06  & 10  & 40 \\
  H1.0-WM-0.09 &         &  &                                                   & 0.10 &  0.09  & 15  & 60 \\ \cline{0-0}\cline{3-8}
  H1.5-WR      &         & \multirow{5}{*}{1.5} &  $326 \times 192 \times 186$  & 0.01 &  0.003 & 0.5 & 1.7 \\ \cline{0-0}\cline{4-8}
  H1.5-WM-0.01 &         &  &  $182 \times 64  \times 64 $                      & 0.03 &  0.01  & 2   & 6 \\ \cline{0-0}\cline{4-8}
  H1.5-WM-0.03 &         &  &  \multirow{3}{*}{$122 \times 32  \times 32 $}     & 0.06 &  0.03  & 5   & 17 \\ 
  H1.5-WM-0.06 &         &  &                                                   & 0.09 &  0.06  & 10  & 34 \\
  H1.5-WM-0.09 &         &  &                                                   & 0.06 &  0.09  & 15  & 51 \\
  \cline{1-8}
  H1.0-WR      & \multirow{5}{*}{37000} & \multirow{5}{*}{1.0} & $758 \times 492 \times 476$ & 0.0036 & 0.001 & 0.5 & 1.9 \\
  \cline{0-0}\cline{4-8}
  H1.0-WM-0.01 &         &  &  $172 \times 64  \times 64 $                      & 0.03 &  0.01  & 5   & 19  \\
  \cline{0-0}\cline{4-8}
  H1.0-WM-0.03 &         &  &  \multirow{3}{*}{$112 \times 32  \times 32 $}     & 0.06 &  0.03  & 15  & 57  \\ 
  H1.0-WM-0.06 &         &  &                                                   & 0.09 &  0.06  & 30  & 114 \\
  H1.0-WM-0.09 &         &  &                                                   & 0.10 &  0.09  & 45  & 171 \\
  \cline{1-8}
  \end{tabular}
  \caption{{\color{black}Parameters for the WRLES and WMLES periodic hill with different slopes and Reynolds numbers.}}
  \label{tab:PH_case}
  \end{center}
\end{table}
\begin{figure}
\centering
	\begin{subfigure}[b]{0.56\textwidth}
	\centering
	\includegraphics[width = 1.0\textwidth]{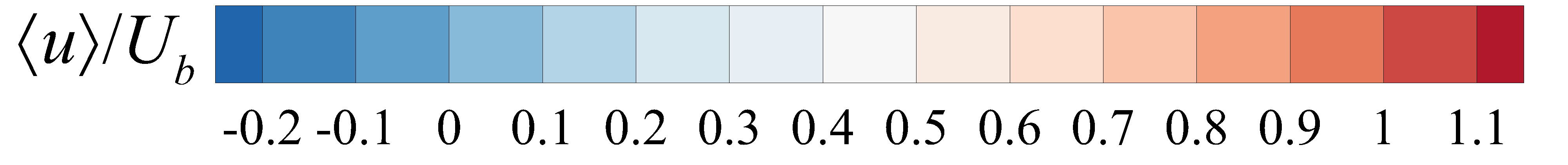}
	\end{subfigure} \\
	\begin{subfigure}[b]{0.4\textwidth}
	\centering
	\includegraphics[width = 1.0\textwidth]{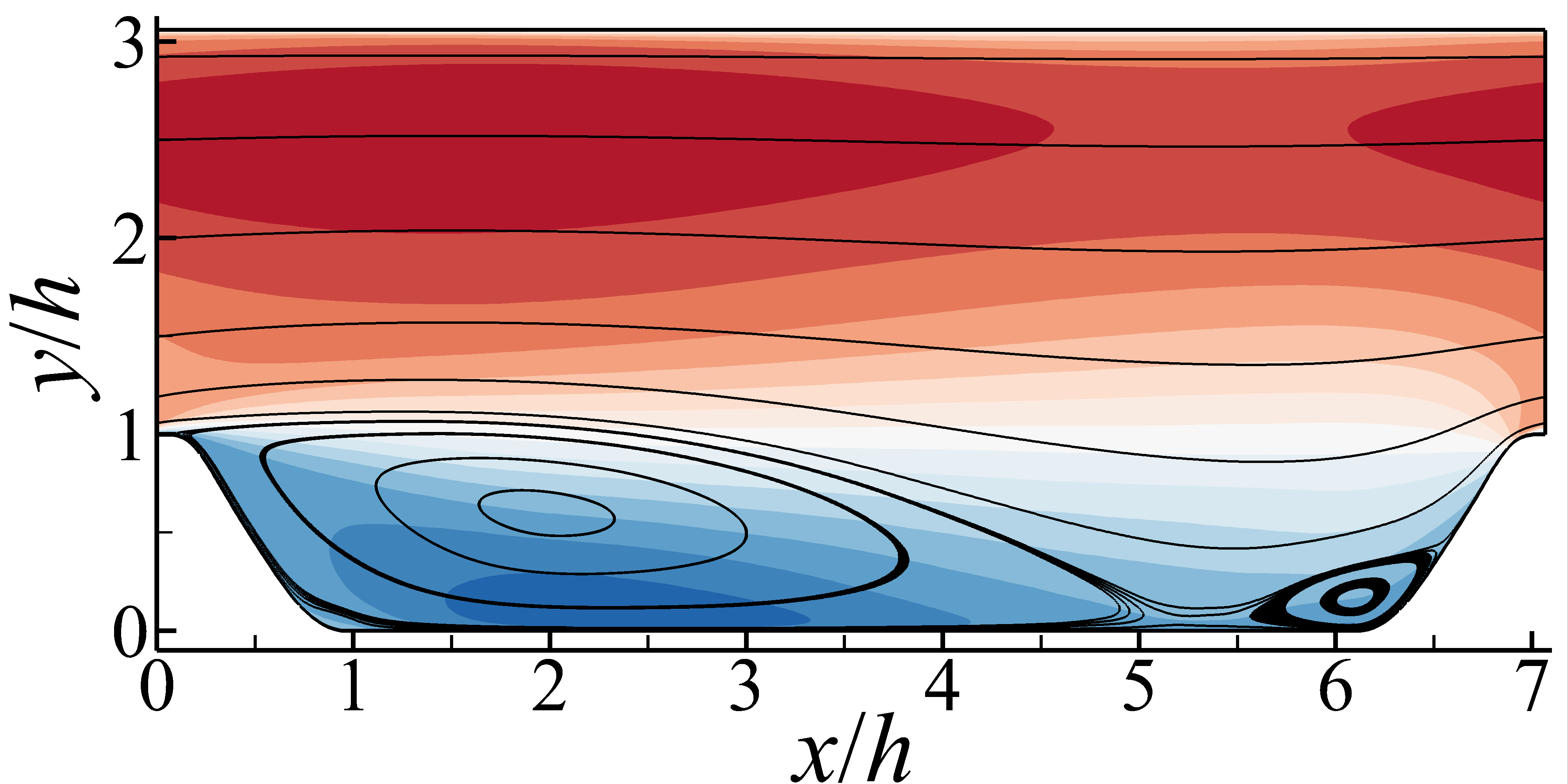}
	\subcaption{H0.5-WR: $\Delta y_f/h=0.003$}
	\end{subfigure}
	\begin{subfigure}[b]{0.59\textwidth}
	\centering
	\includegraphics[width = 1.0\textwidth]{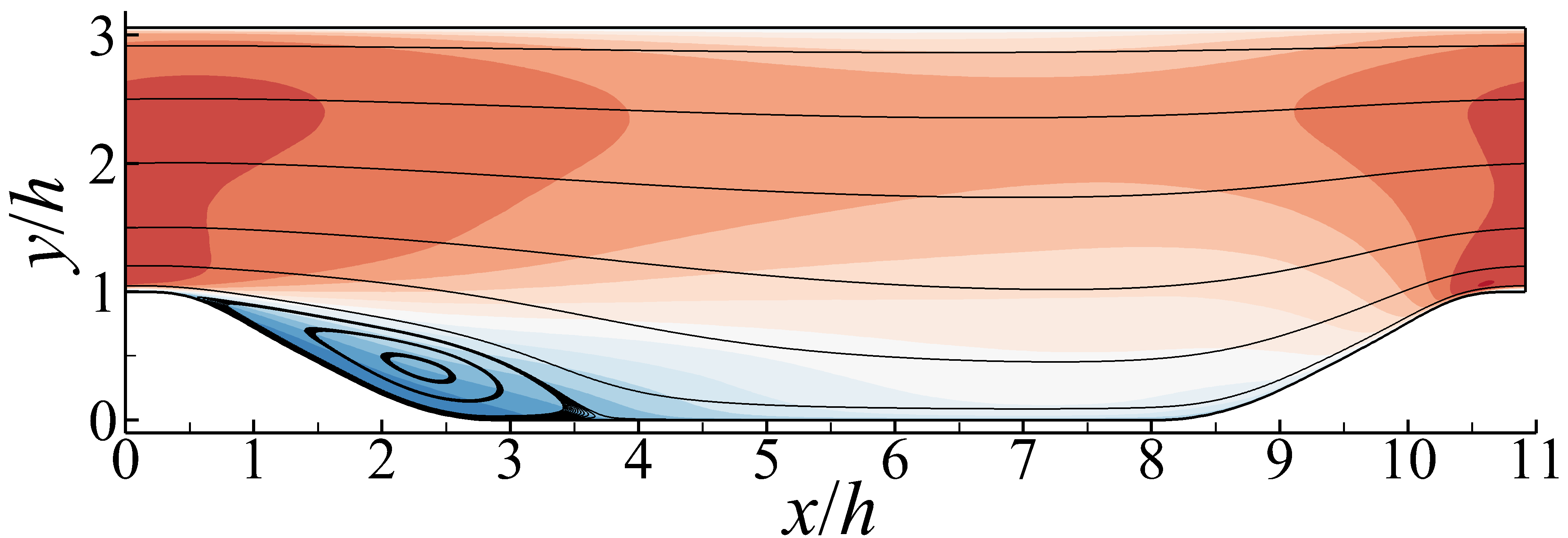}
	\subcaption{H1.5-WR: $\Delta y_f/h=0.003$}
	\end{subfigure}
	\begin{subfigure}[b]{0.4\textwidth}
	\centering
	\includegraphics[width = 1.0\textwidth]{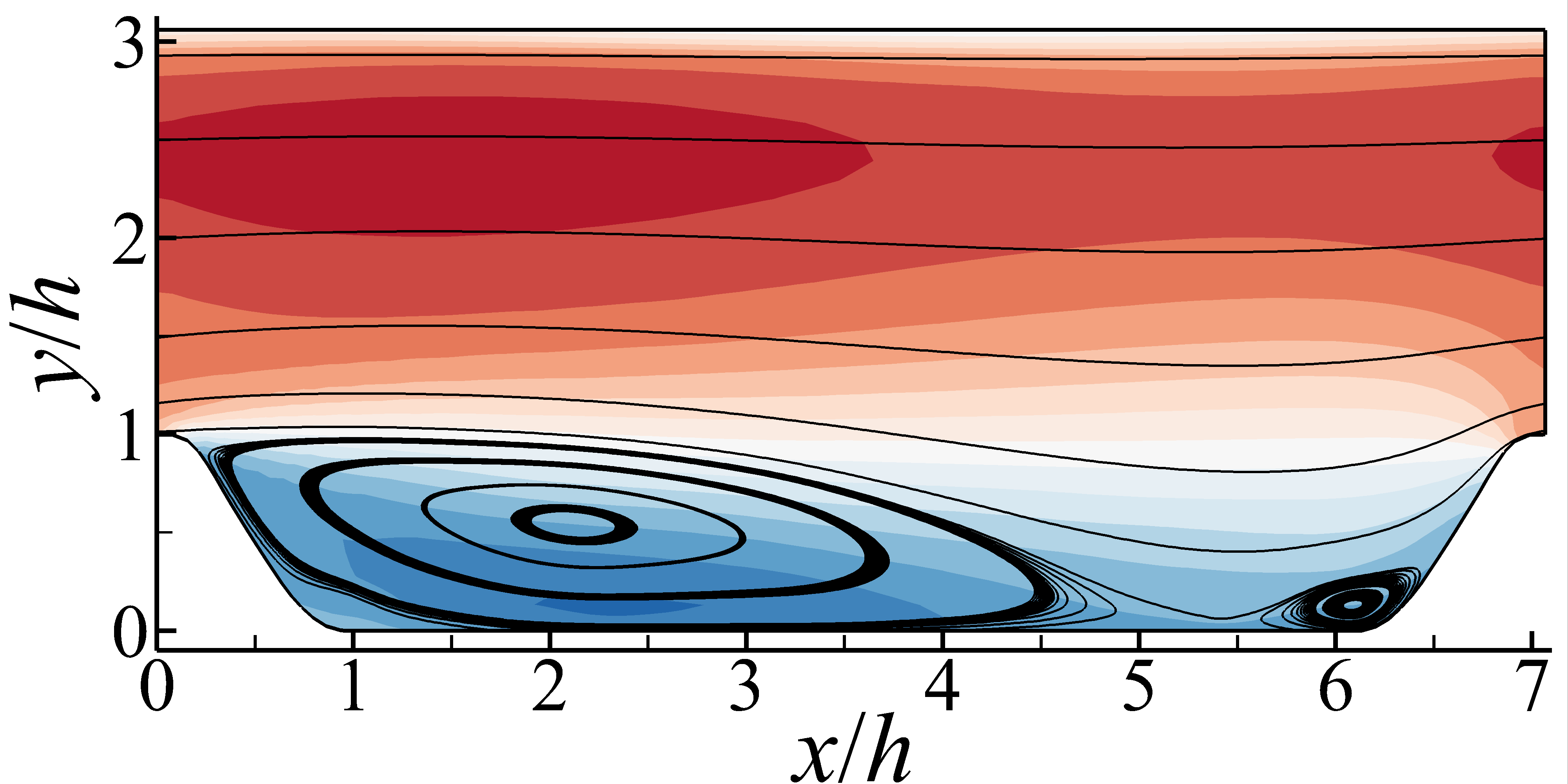}
	\subcaption{FEL: $\Delta y_f/h=0.06$}
	\end{subfigure}
	\begin{subfigure}[b]{0.59\textwidth}
	\centering
	\includegraphics[width = 1.0\textwidth]{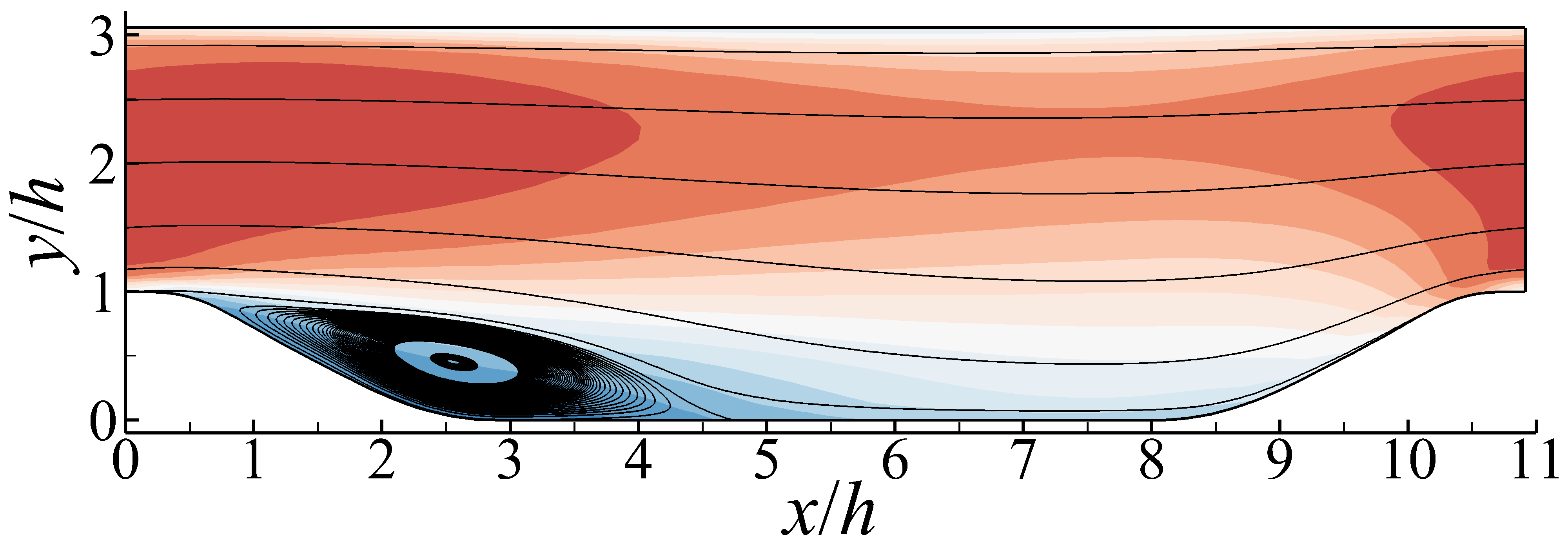}
	\subcaption{FEL: $\Delta y_f/h=0.06$}
	\end{subfigure}
	\begin{subfigure}[b]{0.4\textwidth}
	\centering
	\includegraphics[width = 1.0\textwidth]{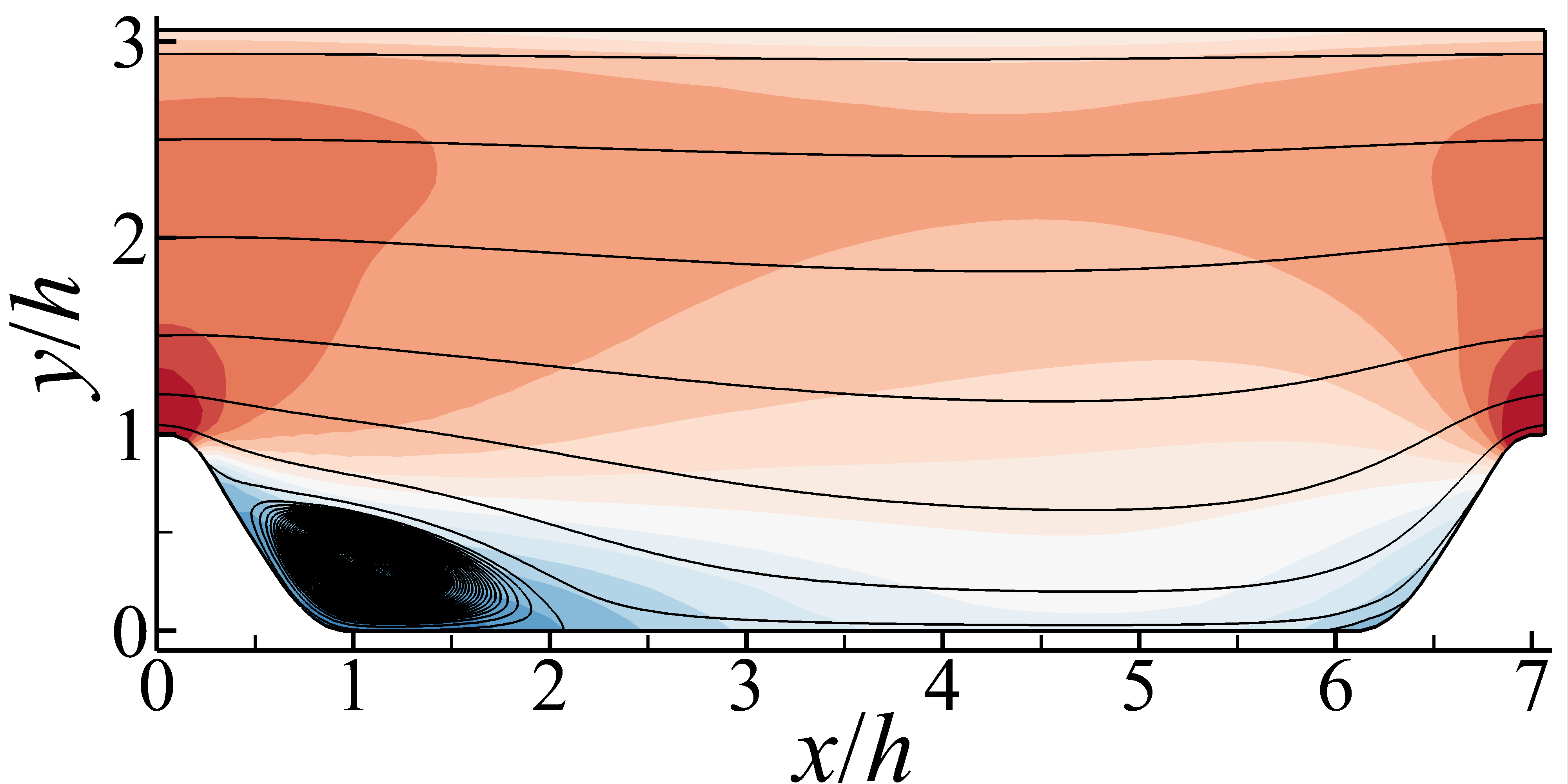}
	\subcaption{WW: $\Delta y_f/h=0.06$}
	\end{subfigure}
	\begin{subfigure}[b]{0.59\textwidth}
	\centering
	\includegraphics[width = 1.0\textwidth]{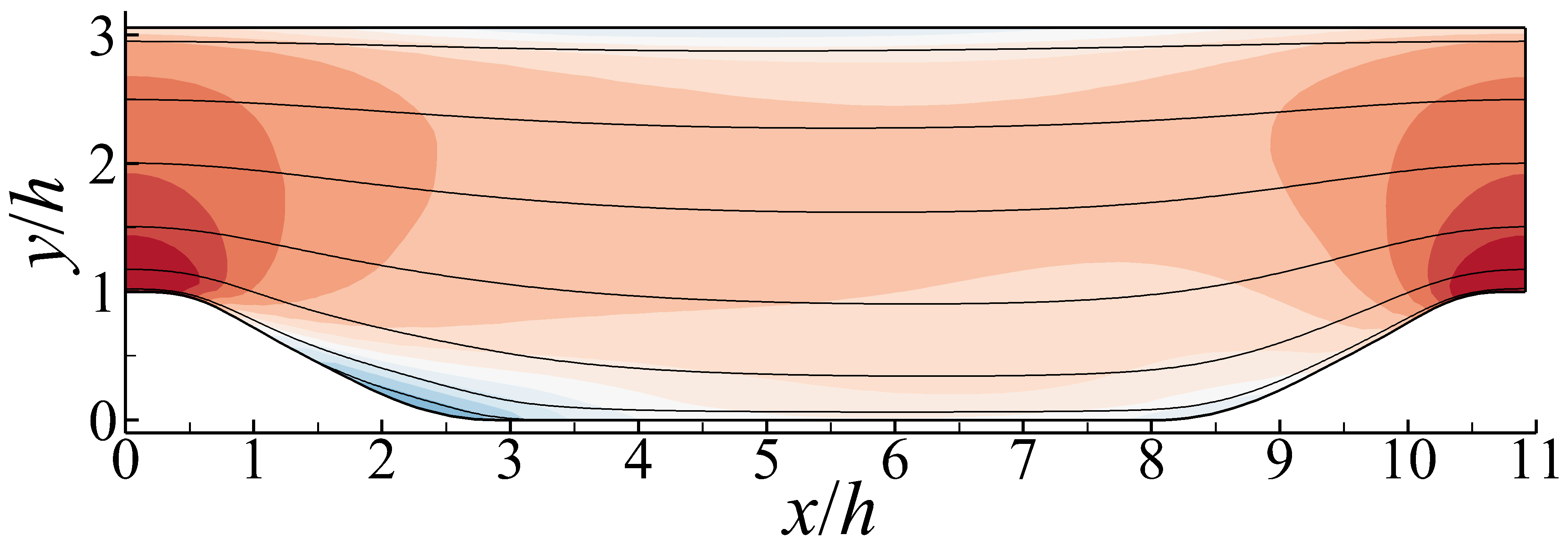}
	\subcaption{WW: $\Delta y_f/h=0.06$}
	\end{subfigure}
\caption{Contours of time-averaged streamwise velocity with streamlines obtained from the two training cases at $Re_h = 10595$.}
\label{fig:HSHL_0p06}
\end{figure}
\begin{figure}
\centering{\includegraphics[width=0.72\textwidth]{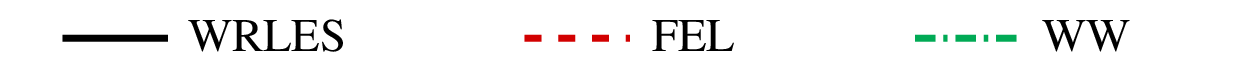}}
\centering{\includegraphics[width=0.40\textwidth]{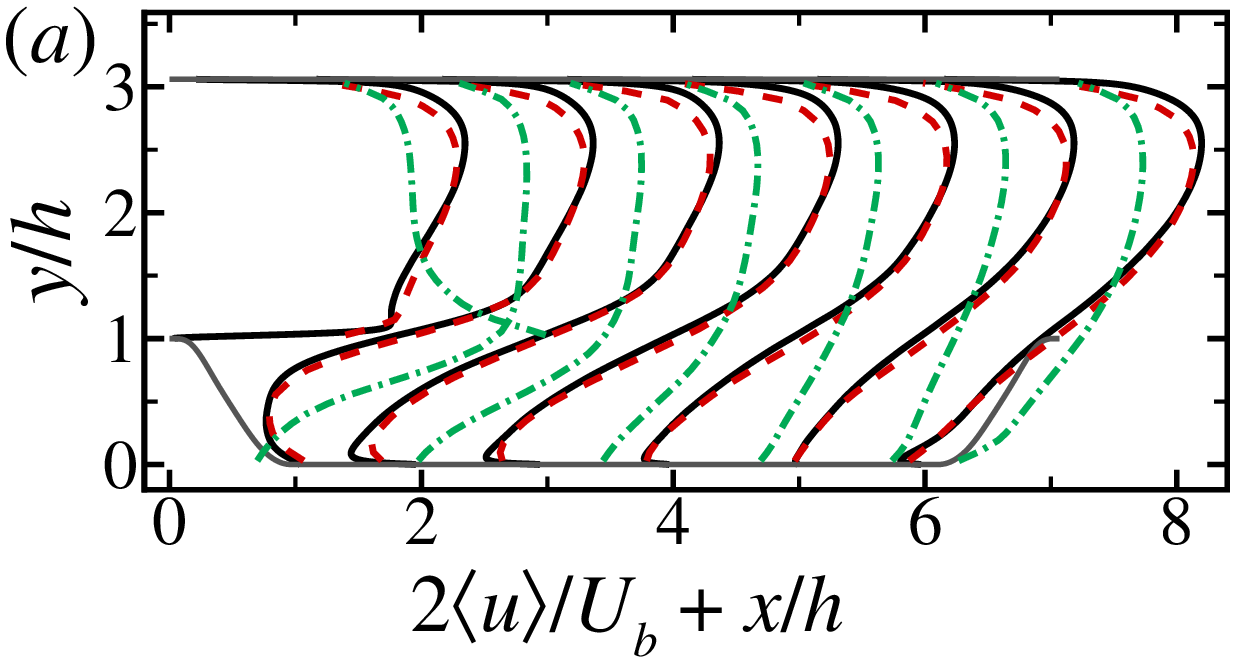}}
\centering{\includegraphics[width=0.59\textwidth]{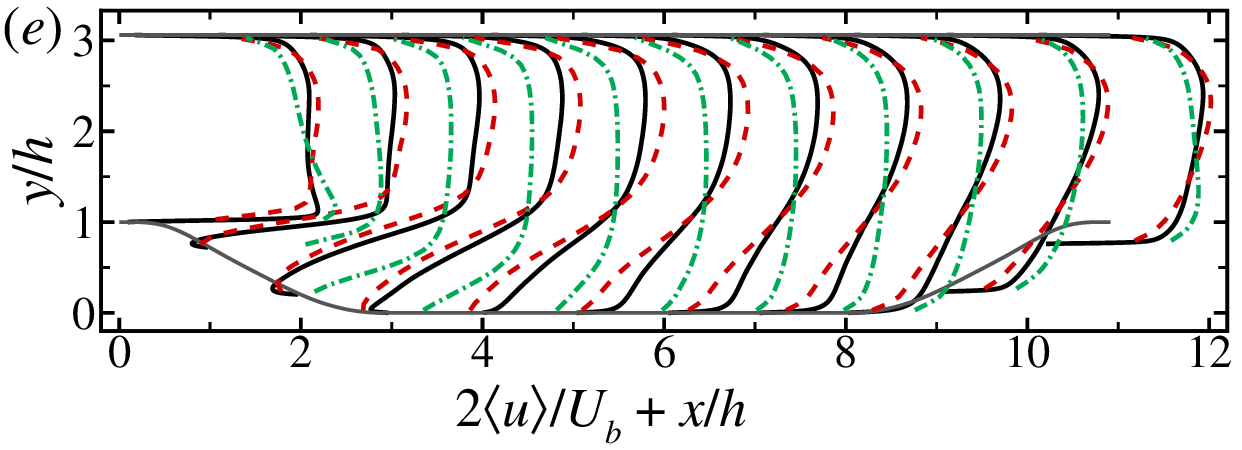}}
\centering{\includegraphics[width=0.40\textwidth]{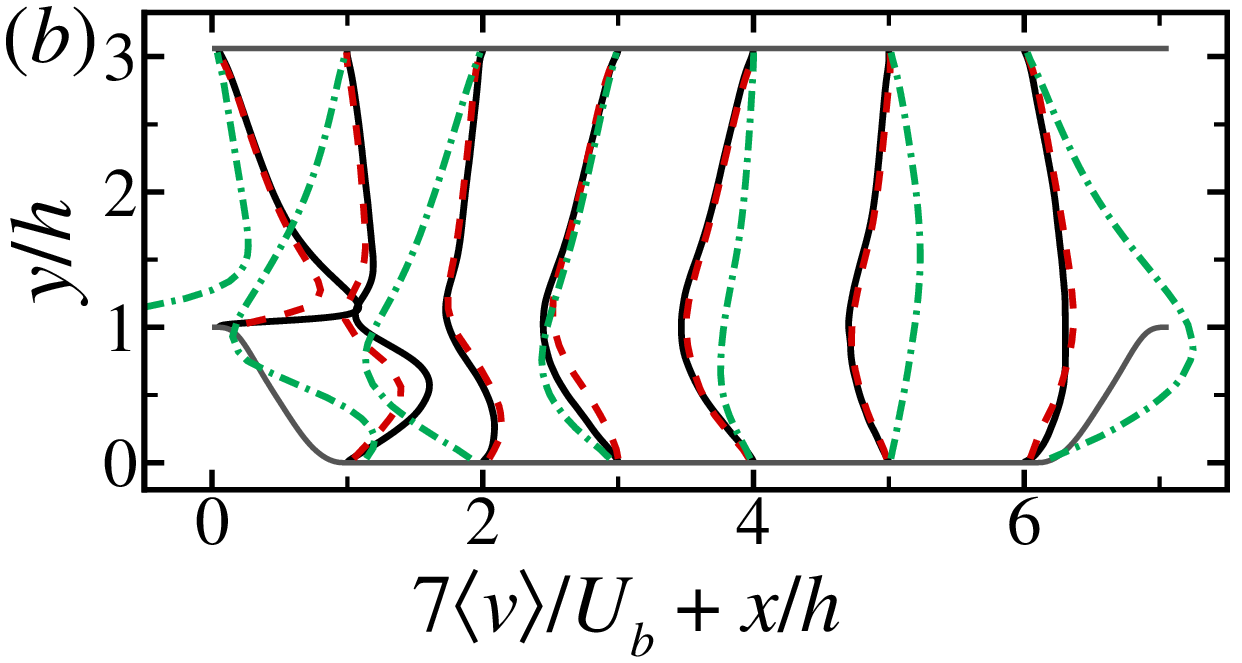}}
\centering{\includegraphics[width=0.59\textwidth]{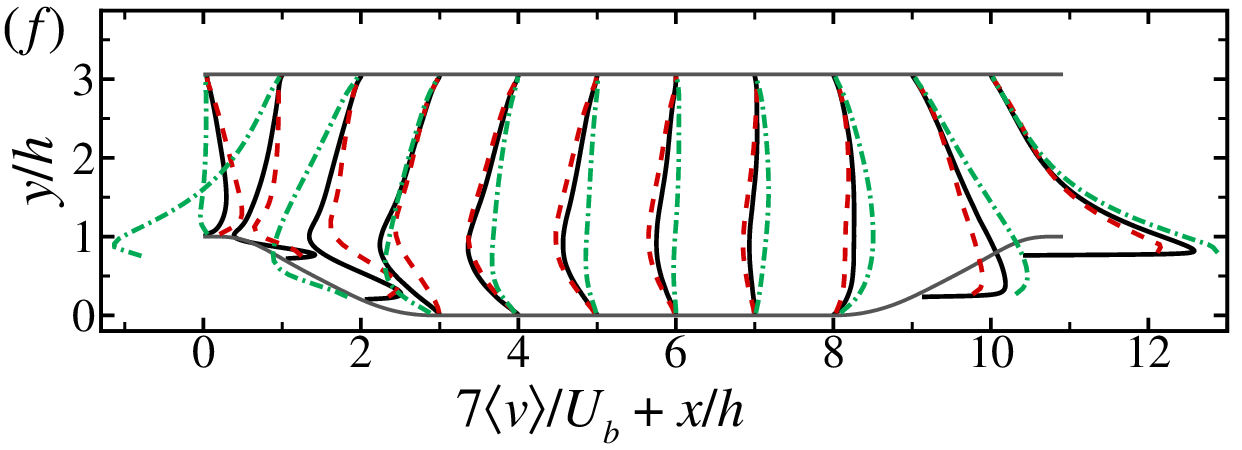}}
\centering{\includegraphics[width=0.40\textwidth]{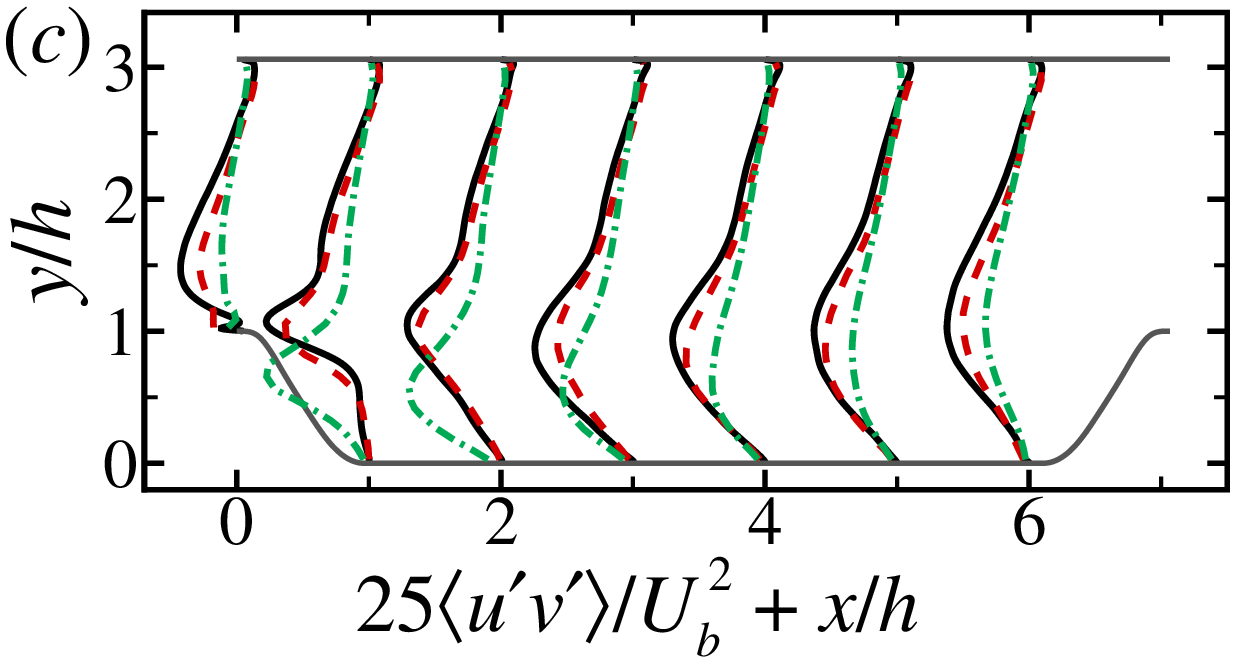}}
\centering{\includegraphics[width=0.59\textwidth]{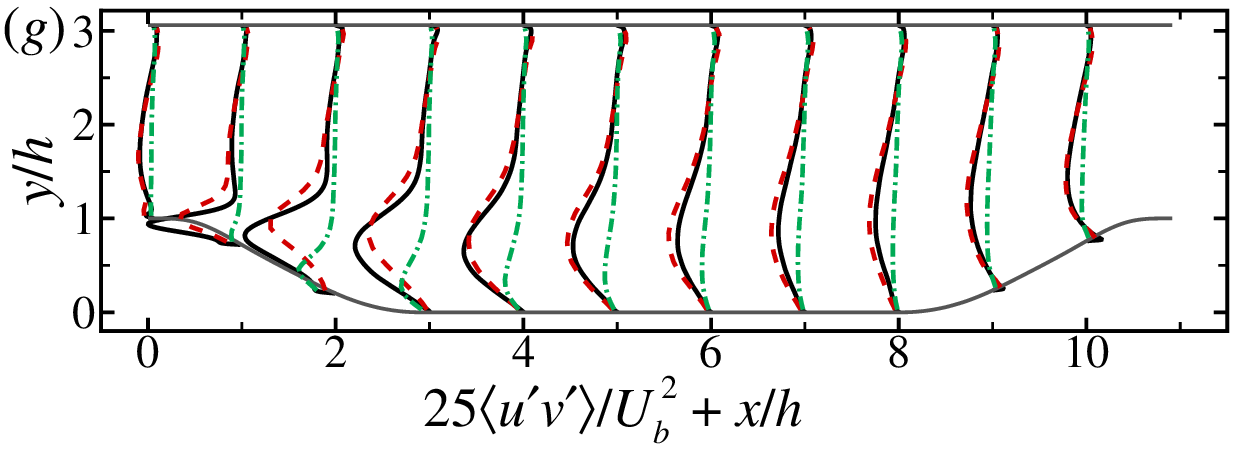}}
\centering{\includegraphics[width=0.40\textwidth]{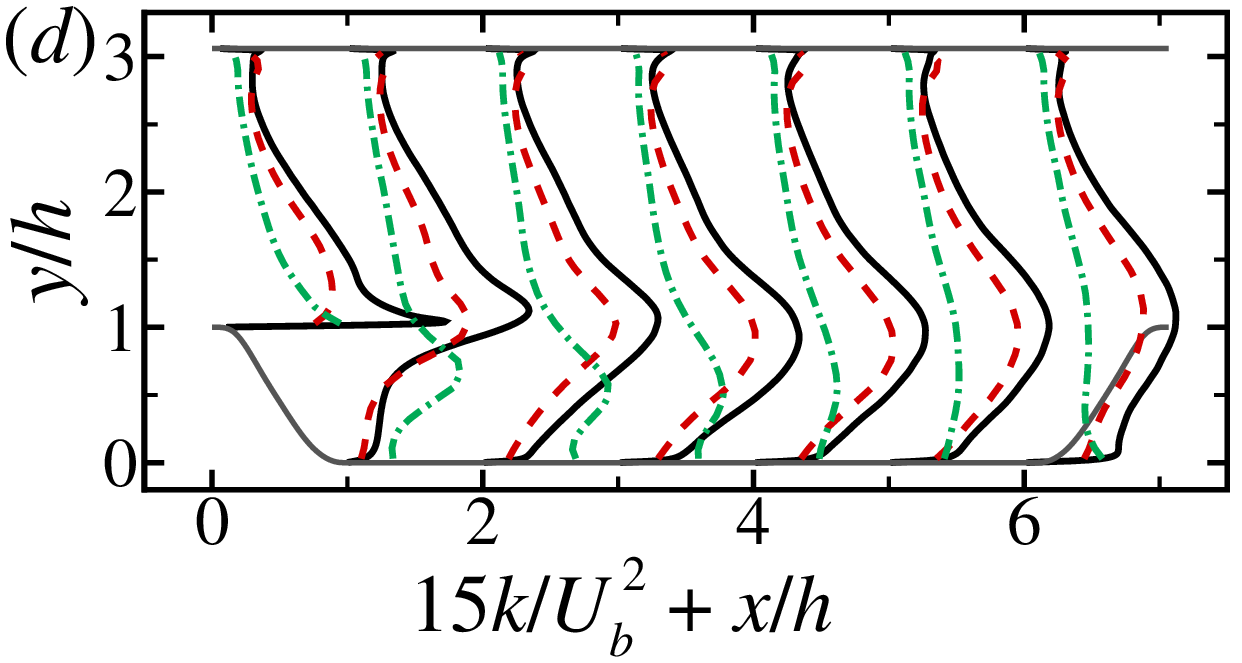}}
\centering{\includegraphics[width=0.59\textwidth]{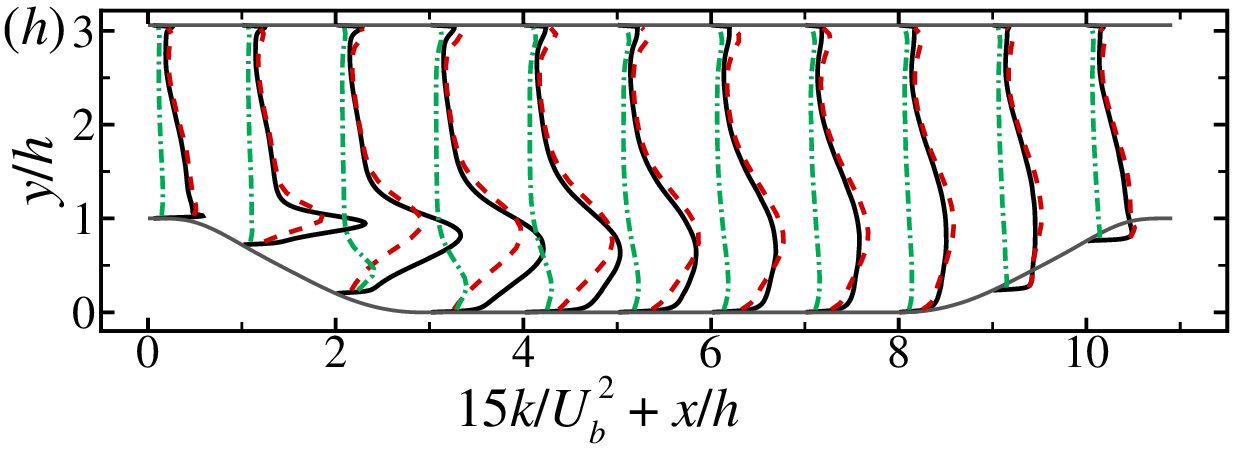}}
  \caption{Vertical profiles of (a, e) the time-averaged streamwise velocity $\left\langle u \right\rangle$ and (b, f) vertical velocity $\left\langle v \right\rangle$, (c, g) primary Reynolds shear stress $\left\langle u'v' \right\rangle$, and (d, h) turbulence kinetic energy $k = \frac{1}{2}\left\langle u'u'+v'v'+w'w' \right\rangle$ from the WRLES and WMLES with the FEL and WW models for the H0.5 case (a$\sim$d) and H1.5 case (e$\sim$h) with $\Delta y_f/h=0.06$ at $Re_h = 10595$.}
\label{fig:profile_HSHL_0p06}
\end{figure}
\begin{figure}
\centering{\includegraphics[width=0.4\textwidth]{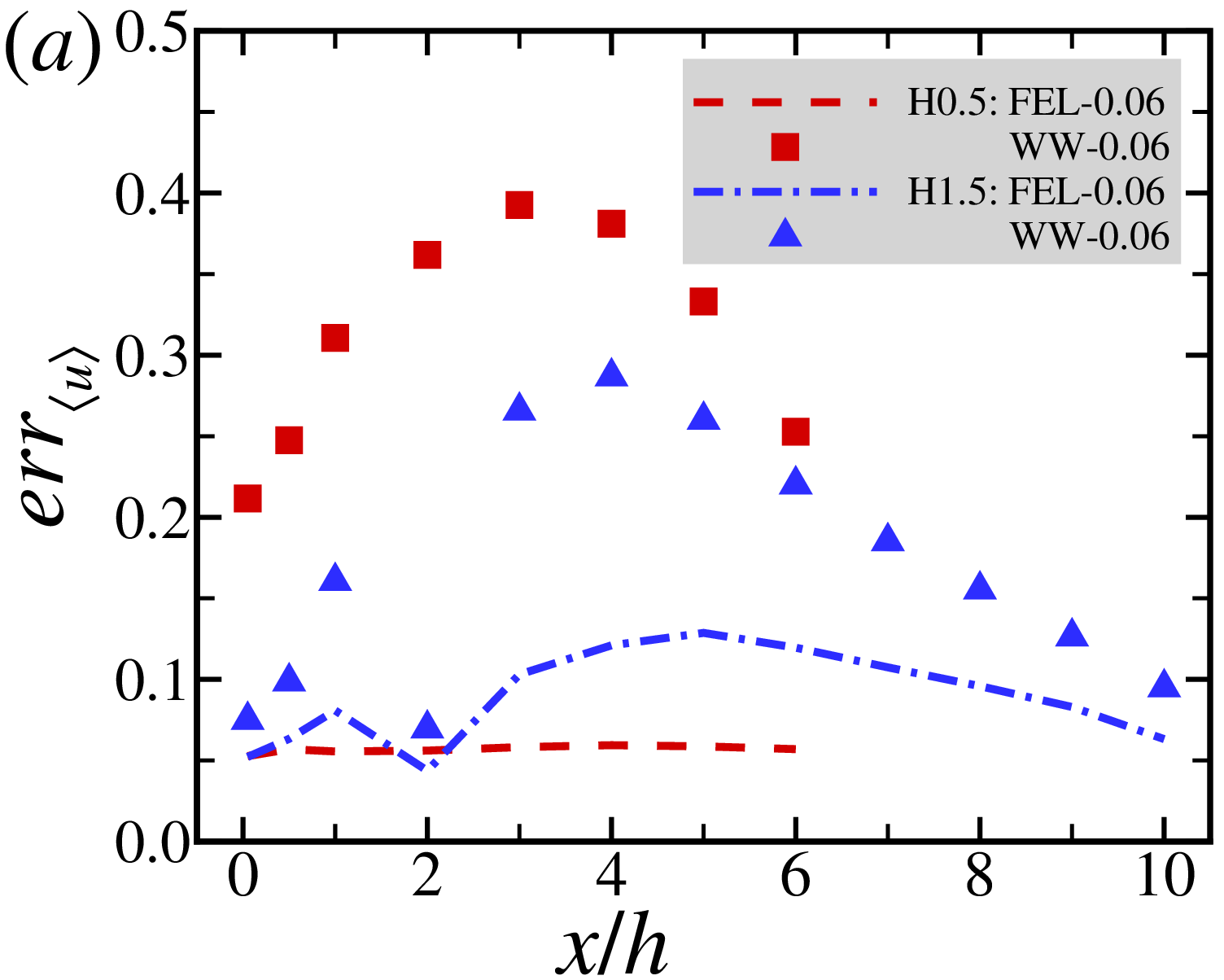}}\quad\quad
\centering{\includegraphics[width=0.4\textwidth]{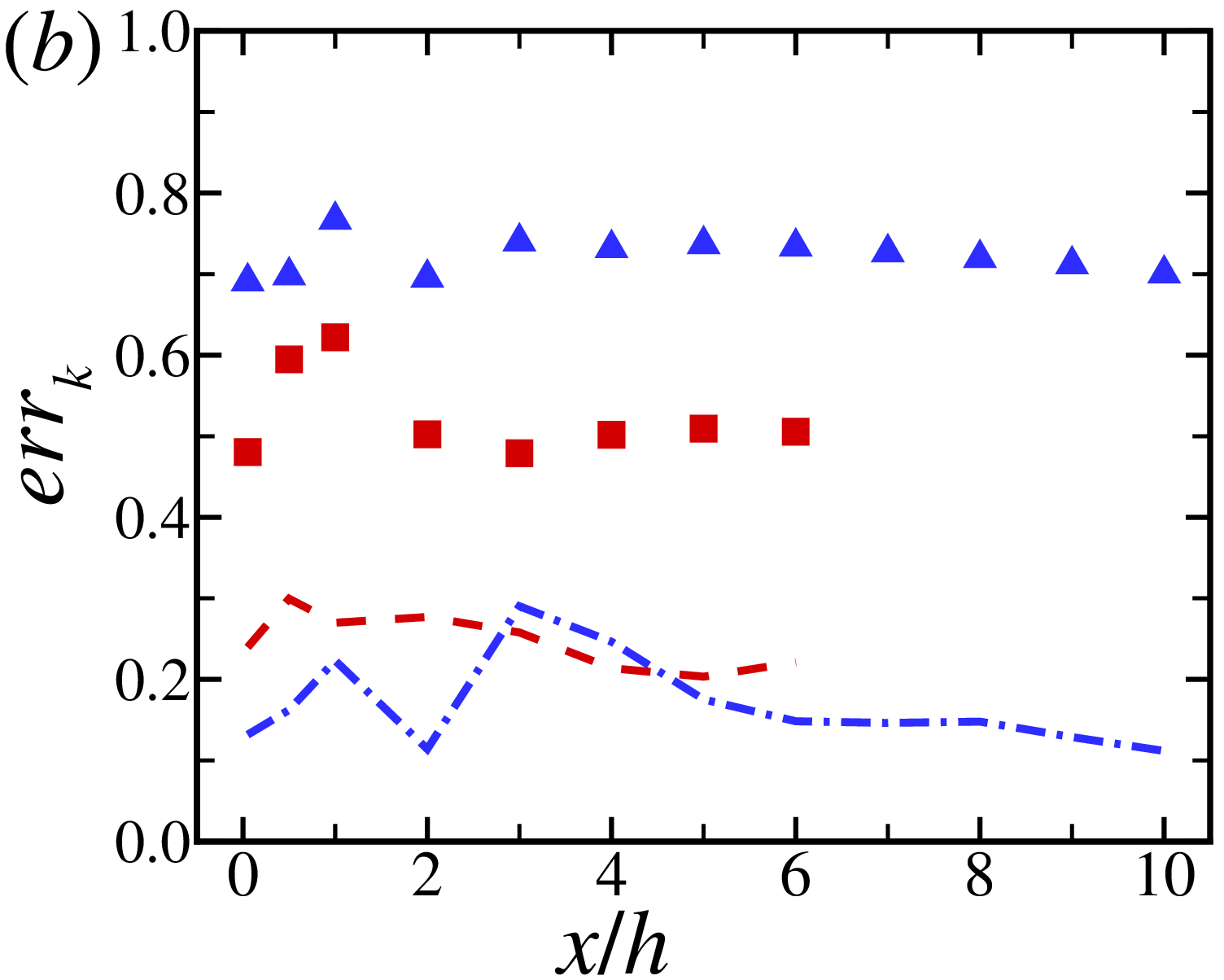}}
  \caption{{\color{black}The relative errors of (a) time-averaged streamwise velocity $\left\langle u \right\rangle$ and (b) turbulence kinetic energy $k$ between the WMLES and WRLES for the H0.5 and H1.5 cases with $\Delta y_f/h=0.06$ at $Re_h = 10595$.}}
\label{fig:profile_err_HSHL_0p06}
\end{figure}

Table~\ref{tab:PH_case} lists the parameters of the WRLES (``WR'') and WMLES (``WM'') cases for the flow over periodic hill with different slopes, grid resolutions and Reynolds numbers. In the table, ``H0.5'', ``H1.0'' and ``H1.5'' represent the periodic hill configurations with $\alpha=0.5$, 1.0 and 1.5, respectively. The parameters $N_x$, $N_y$ and $N_z$ denote the grid number in the streamwise, vertical and spanwise directions, respectively. $\Delta x_f$ corresponds to the grid size in the streamwise direction at the crest of the hill, $\Delta y_f$ represents the height of the first off-wall grid cell, and $\Delta y_c^+ = (\Delta y_f/2) u_\tau/\nu$ is the dimensionless wall-normal distance of the volume center of the first off-wall grid. {\color{black}As the friction velocity $u_\tau$ varies with the streamwise locations, the $\Delta y_c^+$ and $\Delta y_{c,\text{max}}^+$ listed in table~\ref{tab:PH_case} are respectively the approximate mean value and the maximum value over various streamwise locations.
For the grids with $\Delta y_f/h = 0.03$, 0.06 and 0.09, the grid numbers are the same. The grid nodes are non-uniformly distributed in the vertical direction to adjust the sizes of the first off-wall grid cells. The maximum grid spacings $\Delta y/h$ are approximately 0.142, 0.107, and 0.105 for $\Delta y_f/h = 0.03$, 0.06 and 0.09, respectively. 
It is noted that the resolution of the finest WMLES grid employed for the periodic hill case with $Re_h=10595$ is close to that for a wall-resolved LES. For the case at $Re_h=37000$, on the other hand, the grid resolutions with $\delta y_f =$ 0.06, 0.09 are typical for WMLES. The objective is to examine whether the proposed model can work when varying the grid resolution from wall-modelled to that close to wall-resolved. 
}
%, as shown in figure 2 of our previous work~\citep{Zhou_Wu_Yang_PoF_2021}. As for the WMLES cases, $\Delta y_c^+$ is calculated by multiplying with the ratio of first off-wall grid size between the WMLES and WRLES cases.

The FEL model is trained using the H0.5-WM-0.06 and H1.5-WM-0.06 cases at $Re_h=10595$. The vertical profiles of the mean streamwise velocity at $x/h=[0.05, 0.5, 1, 2, \cdots, 6]$ for the H0.5-WR case and $x/h=[0.05, 0.5, 1, 2, \cdots, 10]$ for the H1.5-WR case are employed as the observation data to learn the weights of the embedded neural network.
%The training process utilizes various vertical profiles of mean streamwise velocity obtained from the HS-WR and HL-WR cases as observation data.
%Since the H0.5 and H1.5 cases exhibit different lengths of separation bubble, the trained model can potentially be applied to the flow over periodic hills with varying geometries. 
Here we set the sample size $M=20$, and the maximum iteration index $N=20$. The training employs 40 samples, including 20 samples of H0.5-WM cases and 20 samples H1.5-WM cases, respectively. Each case requires approximately 27 hours to run on 64 CPU cores for the totally 20 iterations. Thus, the computational cost for training the model amounts to 68266 core hours, which is primarily attributed to the WMLES solver. {\color{black}In comparison, the off-line FNN wall models in our previous work~\citep{Zhou_etal_PoF_2023} were trained using a Graphic Processing Unit (GPU) of NVIDIA RTX2080 for about 3 hours, or the Intel-i7 CPU for about 11 hours. The embedded training exhibits a higher demand on the computing environment. As for the large-scale problems which require a much finer grid resolution for WMLES, it is a great challenge to train the wall model using the embedded learning approach because of the enormous computational resources, especially the large number of CPU cores in parallel.}

The obtained FEL wall model is then applied to the cases shown in table~\ref{tab:PH_case}. The computational cost using the proposed model is found comparable to the algebraic wall model, such as the Werner-Wengle (WW) model~\citep{Werner_Wengle_1993}.
%coupled by an embedded-neural-network model for the near-wall eddy-viscosity coefficient and the FNN\_PH-LoW model for the wall shear stress, is extracted for evaluation on
%for a testing case is about 64 core hours, which 
%Firstly, the cDK-embedded model is tested in the training cases. 
Results from the FEL model are shown in the below and compared with those from the WW model for the H0.5-WM-0.06 and H1.5-WM-0.06 cases employed for training. Figure~\ref{fig:HSHL_0p06} displays the contours of time-averaged streamwise velocity with streamlines.
%obtained from the H0.5-WM-0.06 and H1.5-WM-0.06 using the cDK-embedded model. The results are compared with the WW model and the WRLES. 
How the hill slope affects the separation and reattachment points, and the shape of the separation bubble is demonstrated in the wall-resolved results as shown in figures~\ref{fig:HSHL_0p06}(a) and (b).
%, the H0.5-WR case with a steeper hill slope compared to the baseline H1.0-WR case exhibits an earlier flow separation and a later flow reattachment. For the H1.5-WR case in figure~\ref{fig:HSHL_0p06}(b), where the hill becomes flatter, the flow separation occurs later and the reattachment happens sooner compared to the H1.0-WR case. 
%
The FEL model captures such variations of the separation bubble with the hill slope as shown in figures~\ref{fig:HSHL_0p06}(c-d). The WW model, on the other hand, underestimates the size of the separation bubble (figure~\ref{fig:HSHL_0p06}e-f), which was also demonstrated in the literature~\citep{Temmerman_etal_IJHF_2003, Breuer_etal_CAF_2007, Duprat_etal_PoF_2011}.
%, the ability to correctly reproduce the length of the separated region serves as a good indicator of model quality. A poor model tends to delay the flow separation and advance the flow reattachment. Therefore, the trained cDK-embedded model performs better than the WW model at the grid of $\Delta y_f/h=0.06$.
Quantitative evaluations of the FEL model are shown in figure~\ref{fig:profile_HSHL_0p06} for the training cases. For both cases, an overall good agreement with the wall-resolved data is observed for the time-averaged streamwise velocity ($\left\langle u \right\rangle$), vertical velocity ($\left\langle v \right\rangle$), and the primary Reynolds shear stress ($\left\langle u'v' \right\rangle$) for the FEL model.
%the vertical profiles of turbulence statistics at different streamwise locations
%These predictions are much closer to the WRLES results compared to the WW model. 
%Although there are some deviations in the 
The turbulence kinetic energy (TKE, $k = \frac{1}{2}\left\langle u'u'+v'v'+w'w' \right\rangle$), on the other hand, is somewhat underpredicted by the FEL model. This is reasonable because of the limited range of scale resolved by the employed coarse grid. We will show later in appendix~\ref{appendix:filtering} that the TKE predicted by the FEL model agrees well with the filtered WRLES results.
%and the WRLES, they exhibit similar variation tendencies at almost all the streamwise locations.
%The underestimation of the turbulence kinetic energy in the cDK-embedded model can be attributed to the spatial filtering of the coarse grid. 

To further quantify the prediction accuracy of the FEL model, we introduce the relative error for the flow statistics as follows,
\begin{equation}
err_f = \frac{\sum |f_{\text{WR}}-f_{\text{WM}}| \Delta y}{\sum |f_{\text{WR}}| \Delta y},
\label{eq_err}
\end{equation}
where $f$ represents the flow statistics, such as $\left\langle u \right\rangle$ and $k$, $\sum$ denotes the integral along the vertical direction. {\color{black}The obtained relative errors are shown in figure~\ref{fig:profile_err_HSHL_0p06} for the time-averaged streamwise velocity $\left\langle u \right\rangle$ and TKE $k$ for the training cases. For the H0.5-WM-0.06 case, the errors of the FEL model are approximately 5\% for $\left\langle u \right\rangle$ and 25\% for $k$. As for the H1.5-WM-0.06 case, the errors of the FEL model are approximately 10\% for $\left\langle u \right\rangle$ and 20\% for $k$. The errors of the WW model predictions are much larger than those from the FEL model.}

\section{Evaluation of the FEL model {\color{black}using the periodic hill cases}}\label{sec:Results}
In this section, we evaluate the performance of the FEL model for PH cases with different grids, different hill slopes, and different Reynolds numbers in sections~\ref{subsec:Application_grid}, \ref{subsec:Application_geom}, and \ref{subsec:Application_Re}, respectively.

\begin{table}
  \begin{center}
\def~{\hphantom{0}}
  \begin{tabular}{c|c|c|cc|c|cc}
  \cline{1-8}
  \multirow{2}{*}{Case} & \multirow{2}{*}{$Re_h$} & \multicolumn{3}{c|}{FEL} & \multicolumn{3}{c}{WW} \\
  \cline{3-8}\rule{0pt}{8pt}
               &  & $x_{\text{sep}}/h$ & $x_{\text{ret}}/h$ & $E_{\text{ret}}$ (\%) & $x_{\text{sep}}/h$ & $x_{\text{ret}}/h$ & $E_{\text{ret}}$ (\%) \\
  \cline{1-8}\rule{0pt}{8pt}
  H0.5-WR      & \multirow{15}{*}{10595}  & 0.0  & 5.28 & -- & 0.0 & 5.28 & -- \\
  \cline{0-0}\cline{3-8}
  H0.5-WM-0.01 &         &  0.0   &  5.24 & -0.76  & 0.22 & 3.53 & -33.14 \\
  H0.5-WM-0.03 &         &  0.12  &  5.43 & 2.84   & 0.35 & 2.51 & -52.46 \\
  H0.5-WM-0.06 &         &  0.21  &  5.34 & 1.14   & 0.40 & 2.08 & -60.61 \\
  H0.5-WM-0.09 &         &  0.27  &  4.47 & -15.34 & 0.48 & 1.62 & -69.32 \\
  \cline{0-0}\cline{3-8}
  H1.5-WR      &         &  0.45  &  3.89 & --     & 0.45 & 3.89 & --     \\
  \cline{0-0}\cline{3-8}
  H1.5-WM-0.01 &         &  0.53  &  3.54 & -9.0   & 0.52 & 3.64 & -6.43  \\
  H1.5-WM-0.03 &         &  0.58  &  3.89 & 0.0    & 1.40 & 2.72 & -30.08 \\
  H1.5-WM-0.06 &         &  0.66  &  4.60 & 18.25  & --   & --   & -100.0 \\
  H1.5-WM-0.09 &         &  0.86  &  3.93 & 1.03   & --   & --   & -100.0 \\
  \cline{0-0}\cline{3-8}
  H1.0-WR      &         &  0.22  &  4.35 & --     & 0.22 & 4.35 & --     \\
  \cline{0-0}\cline{3-8}
  H1.0-WM-0.01 &         &  0.29  &  4.18 & -3.91  & 0.31 & 4.26 & -2.07  \\
  H1.0-WM-0.03 &         &  0.29  &  4.31 & -0.92  & 0.86 & 2.35 & -45.98 \\
  H1.0-WM-0.06 &         &  0.36  &  5.07 & 16.55  & 1.27 & 1.98 & -54.48 \\
  H1.0-WM-0.09 &         &  0.54  &  4.15 & -4.60  & --   & --   & -100.0 \\
  \cline{1-8}
  H1.0-WR      & \multirow{5}{*}{37000} &  0.29    &  4.01 & -- & 0.29 & 4.01 & -- \\
  \cline{0-0}\cline{3-8}
  H1.0-WM-0.01 &         &  0.30  &  4.21 & 4.99   & 0.57 & 2.78 & -30.67 \\
  H1.0-WM-0.03 &         &  0.30  &  4.50 & 12.22  & 1.06 & 2.06 & -48.63 \\
  H1.0-WM-0.06 &         &  0.37  &  5.38 & 34.16  & --   & --   & -100.0 \\
  H1.0-WM-0.09 &         &  0.54  &  4.18 & 4.24   & --   & --   & -100.0 \\
  \cline{1-8}
  \end{tabular}
  \caption{{\color{black}The streamwise locations of the separation point and reattachment point predicted by the WRLES and WMLES using the FEL and WW models.}}
  \label{tab:sep_ret}
  \end{center}
\end{table}

Before a systematic evaluation of the proposed model, the obtained streamwise locations of the separation point ($x_{\text{sep}}$) and reattachment point ($x_{\text{ret}}$) are summarized in table~\ref{tab:sep_ret} for the WMLES and WRLES cases. Noted that, in the H0.5-WR case, the steep hill flow has two separation bubbles and the flow does not fully reattach at the bottom plane wall, as plotted in figure~\ref{fig:HSHL_0p06}(a). The reattachment point is defined as the location closest to the wall with zero streamwise velocity between the primary and secondary separation bubbles.
{\color{black}Compared with the WRLES results, both the FEL and WW models accurately capture the separation and reattachment locations for the finest grid $\Delta y_f/h=0.01$, demonstrating the grid convergence of the wall models. When the grid is coarsened, the separation location predicted by the FEL model somewhat moves downstream, while the reattachment location is barely affected for most cases.
With the relative error $E_{\text{ret}}$ of the reattachment location defined as
\begin{equation}
E_{\text{ret}} = \frac{x_{\text{ret}}^{\text{WM}}-x_{\text{ret}}^{\text{WR}}}{x_{\text{ret}}^{\text{WR}}}\times 100\%,
\label{eq_err_xret}
\end{equation}
the prediction accuracy of $x_{\text{ret}}$ is measured quantitatively, that $E_{\text{ret}}$ is less than 5\% for the most cases for the proposed FEL model, being significantly lower than that for the WW model.}

\subsection{Cases with different grid resolutions}\label{subsec:Application_grid}
\begin{figure}
\centering
	\begin{subfigure}[b]{0.56\textwidth}
	\centering
	\includegraphics[width = 1.0\textwidth]{Fig_contour_legend.png}
	\end{subfigure} \\
	\begin{subfigure}[b]{0.4\textwidth}
	\centering
	\includegraphics[width = 1.0\textwidth]{Fig_contour_HS_WR.png}
	\subcaption{H0.5-WR: $\Delta y_f/h=0.003$}
	\end{subfigure} \\
	\begin{subfigure}[b]{0.4\textwidth}
	\centering
	\includegraphics[width = 1.0\textwidth]{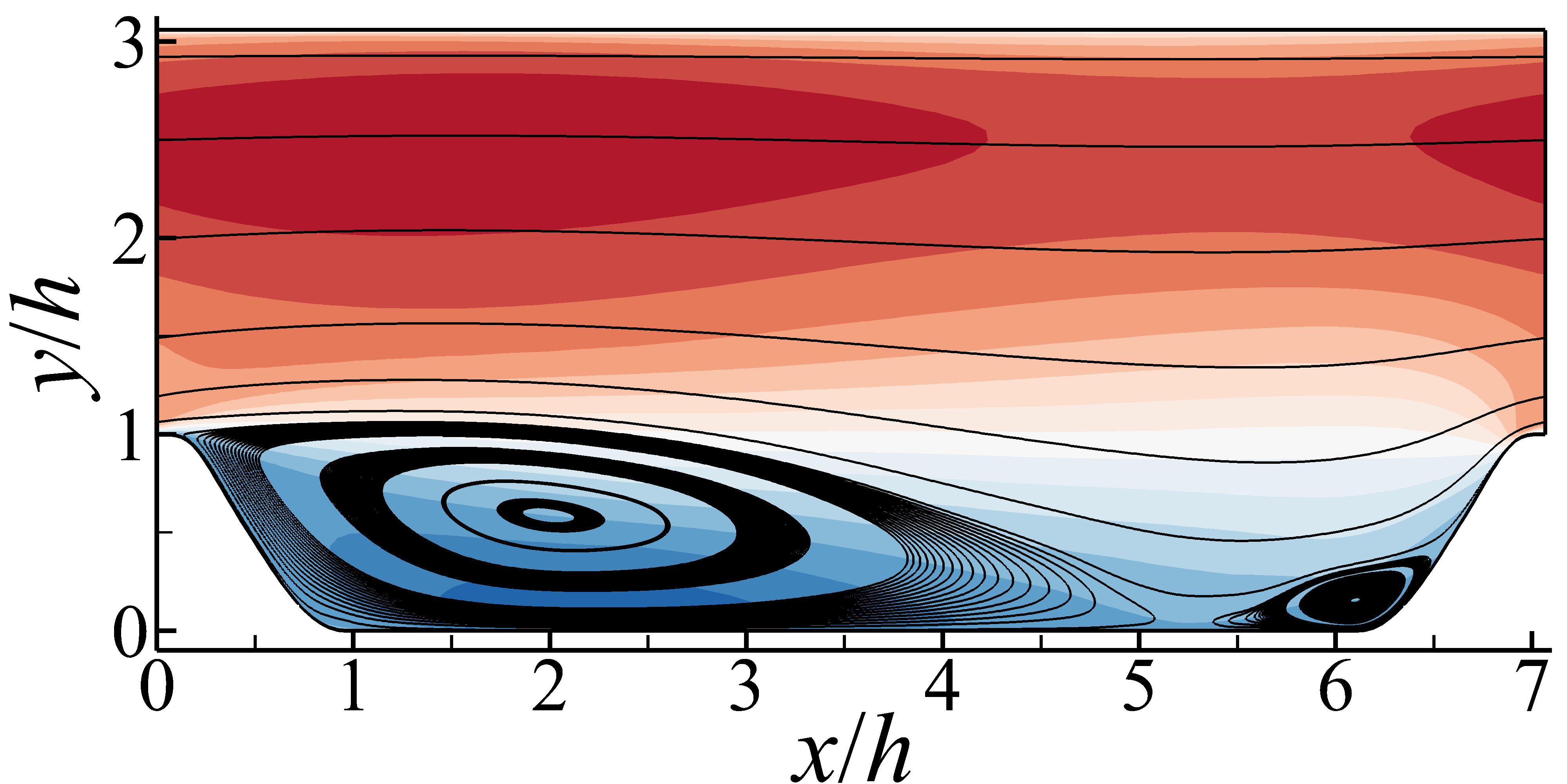}
	\subcaption{FEL: $\Delta y_f/h=0.01$}
	\end{subfigure}\quad\quad
	\begin{subfigure}[b]{0.4\textwidth}
	\centering
	\includegraphics[width = 1.0\textwidth]{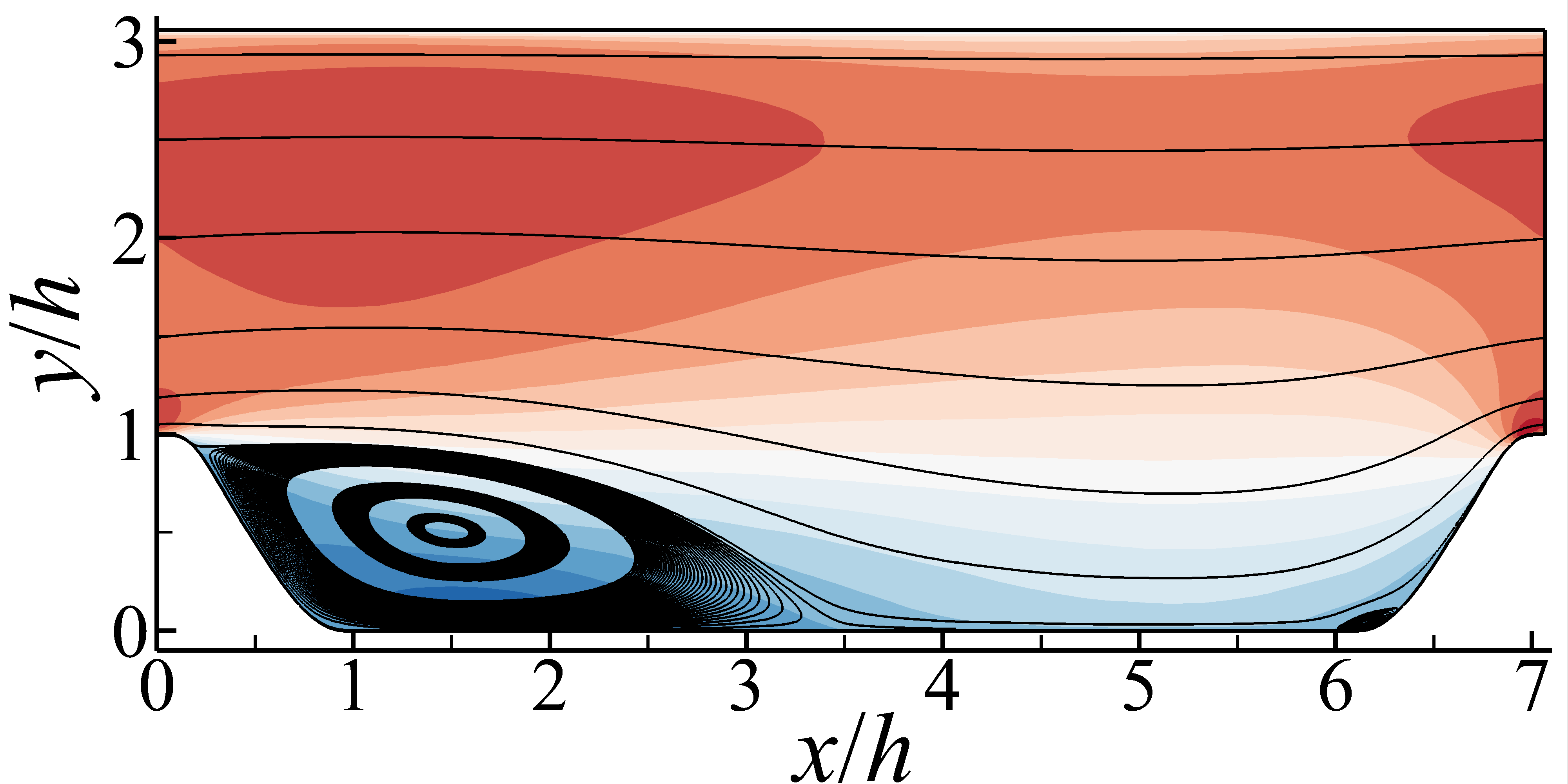}
	\subcaption{WW: $\Delta y_f/h=0.01$}
	\end{subfigure}
	\begin{subfigure}[b]{0.4\textwidth}
	\centering
	\includegraphics[width = 1.0\textwidth]{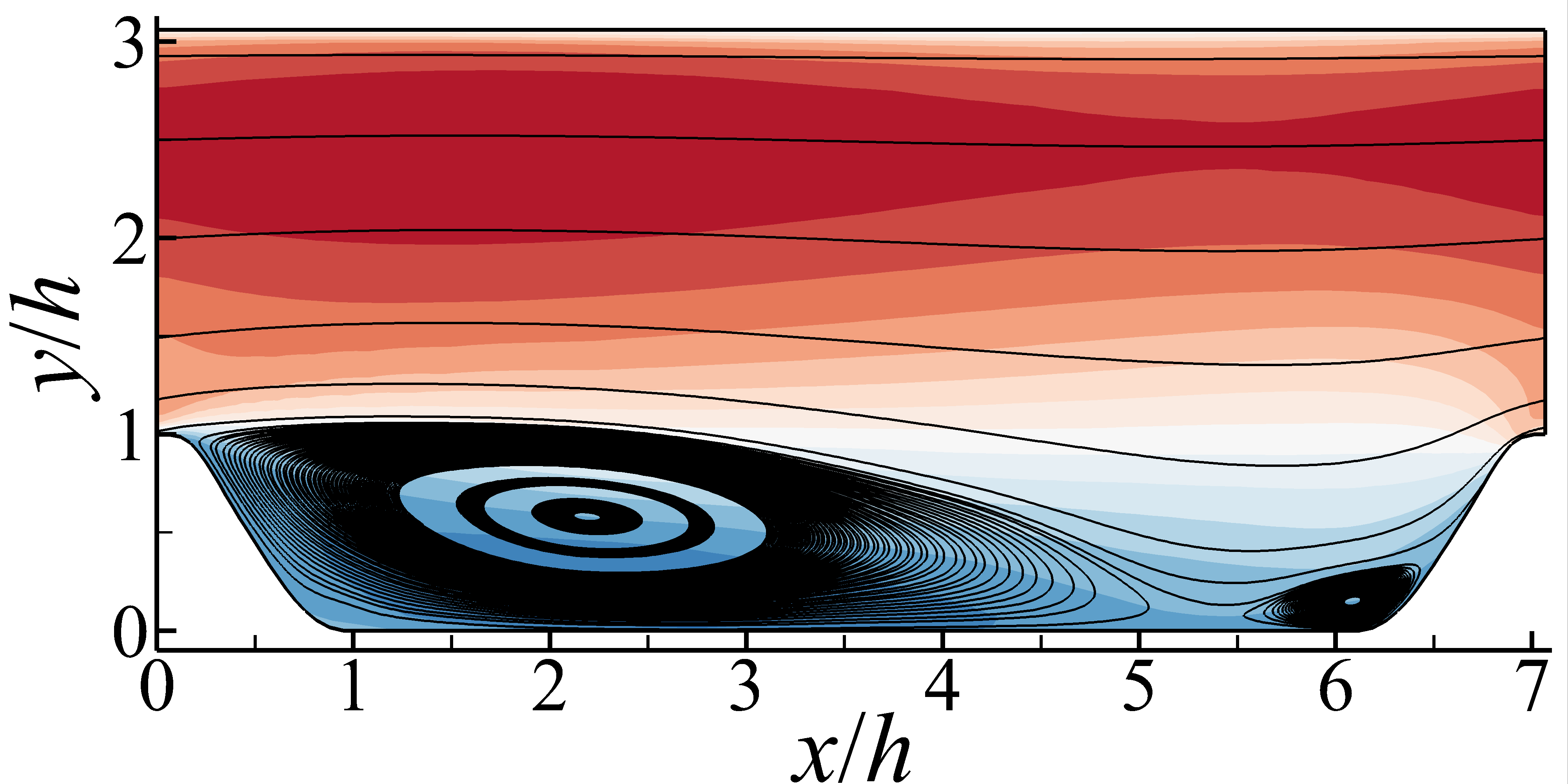}
	\subcaption{FEL: $\Delta y_f/h=0.03$}
	\end{subfigure}\quad\quad
	\begin{subfigure}[b]{0.4\textwidth}
	\centering
	\includegraphics[width = 1.0\textwidth]{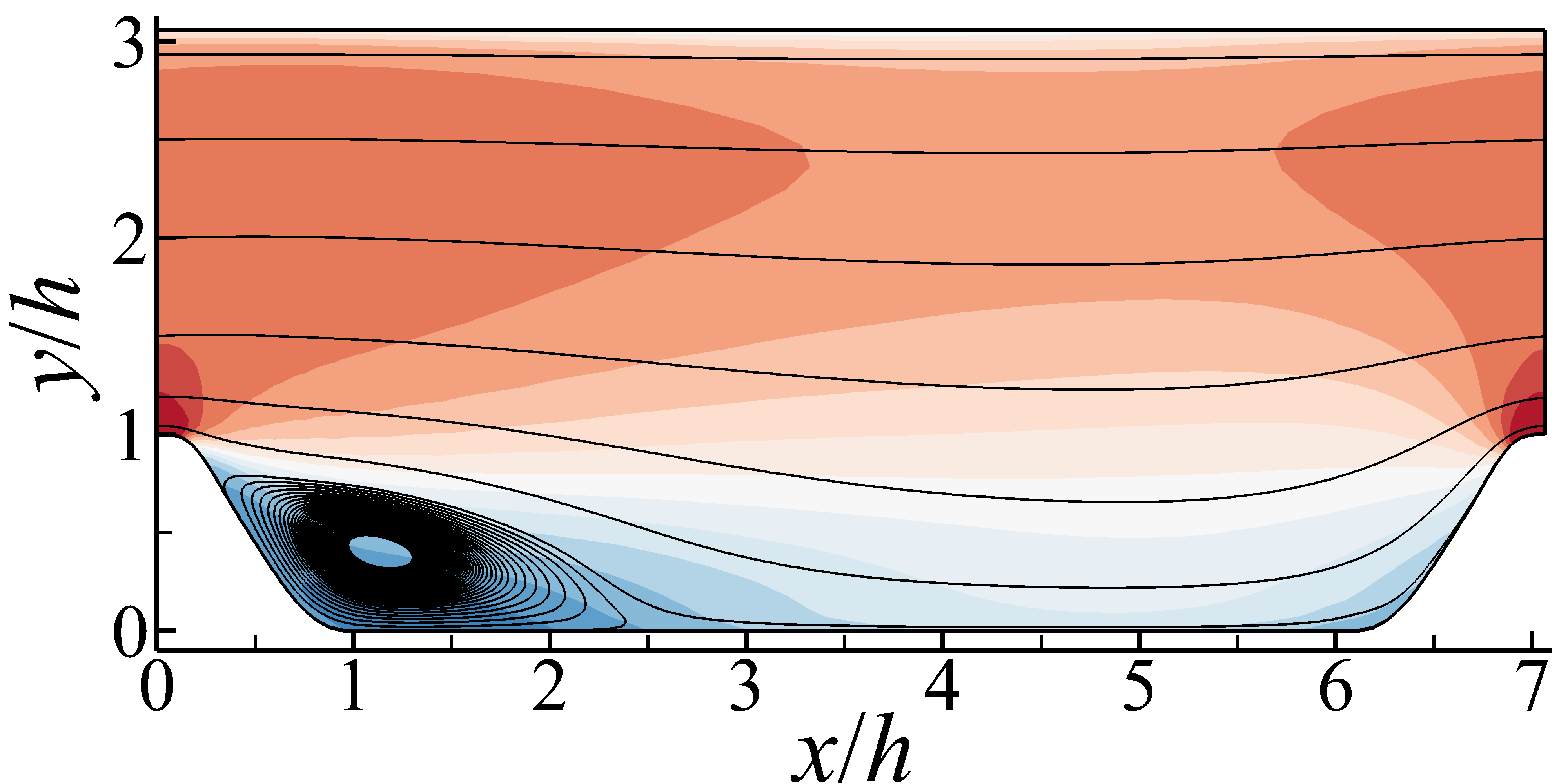}
	\subcaption{WW: $\Delta y_f/h=0.03$}
	\end{subfigure}
	\begin{subfigure}[b]{0.4\textwidth}
	\centering
	\includegraphics[width = 1.0\textwidth]{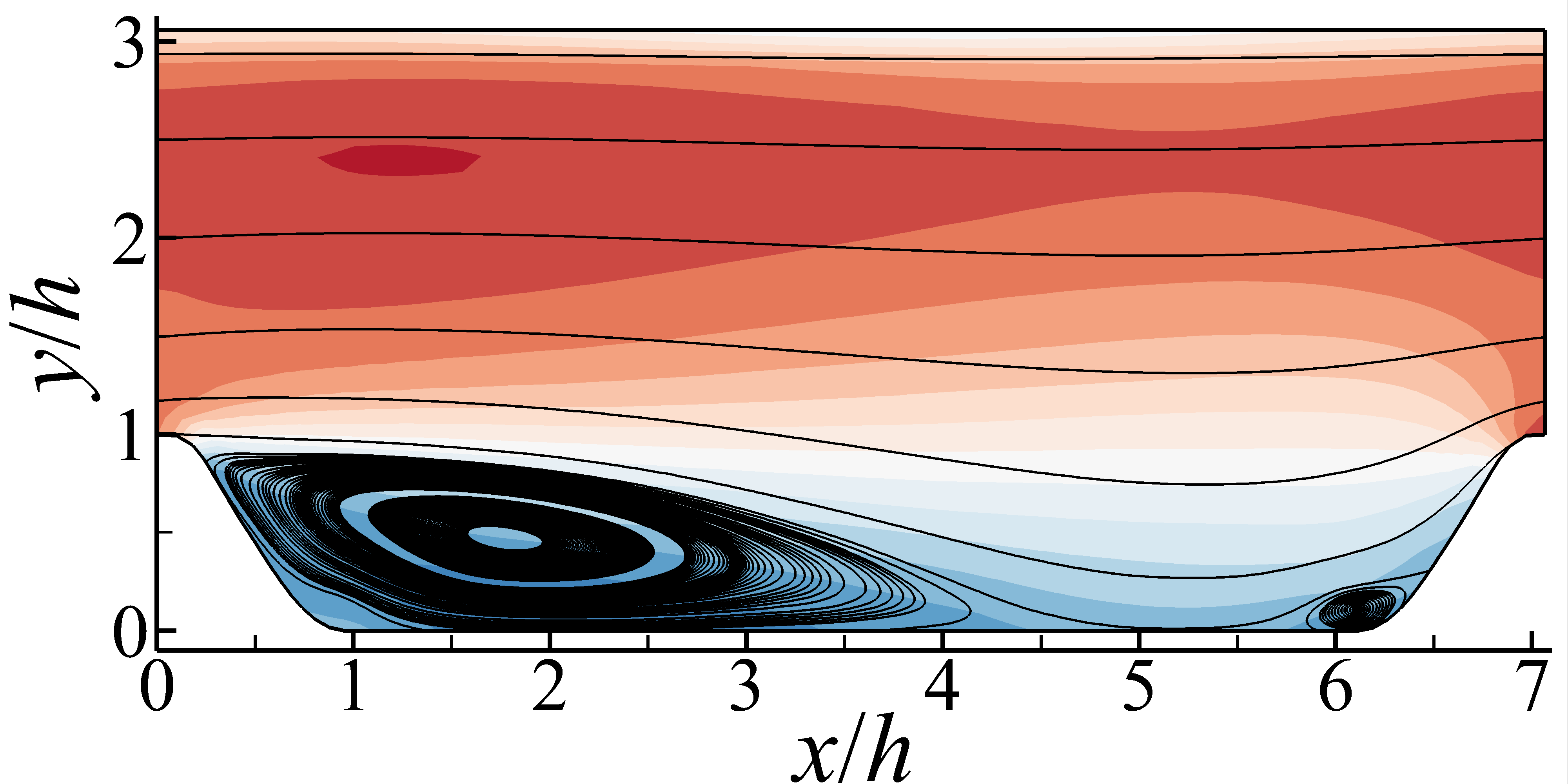}
	\subcaption{FEL: $\Delta y_f/h=0.09$}
	\end{subfigure}\quad\quad
	\begin{subfigure}[b]{0.4\textwidth}
	\centering
	\includegraphics[width = 1.0\textwidth]{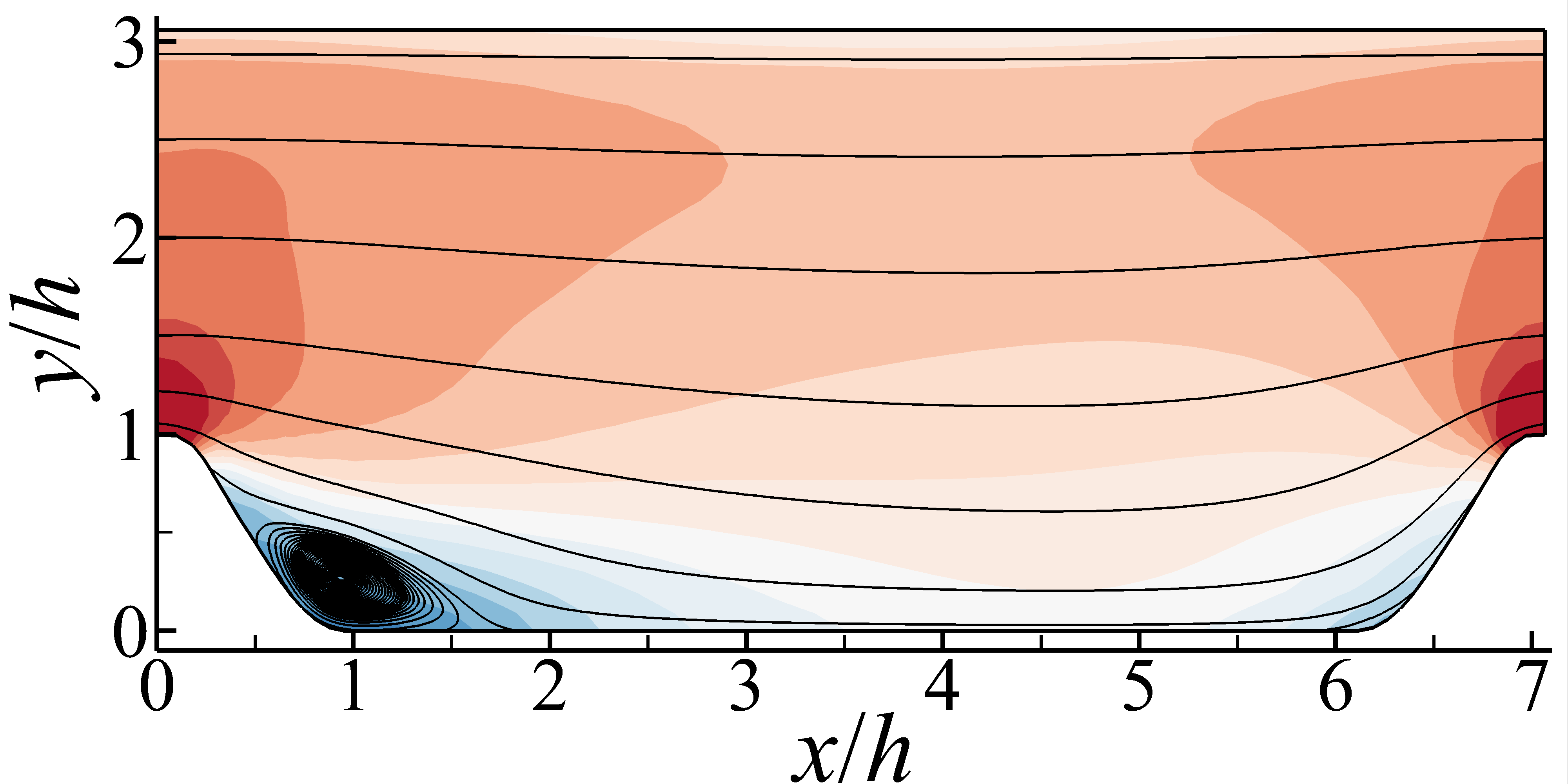}
	\subcaption{WW: $\Delta y_f/h=0.09$}
	\end{subfigure}
\caption{Contours of time-averaged streamwise velocity with streamlines obtained from the H0.5-WR and H0.5-WM cases with the FEL and WW models at $Re_h = 10595$.}
\label{fig:HS_grid}
\end{figure}
\begin{figure}
\centering
	\begin{subfigure}[b]{0.56\textwidth}
	\centering
	\includegraphics[width = 1.0\textwidth]{Fig_contour_legend.png}
	\end{subfigure}
	\begin{subfigure}[b]{0.495\textwidth}
	\centering
	\includegraphics[width = 1.0\textwidth]{Fig_contour_HL_WR.png}
	\subcaption{H1.5-WR: $\Delta y_f/h=0.003$}
	\end{subfigure} \quad\quad
	\begin{subfigure}[b]{0.495\textwidth}
	\centering
	\includegraphics[width = 1.0\textwidth]{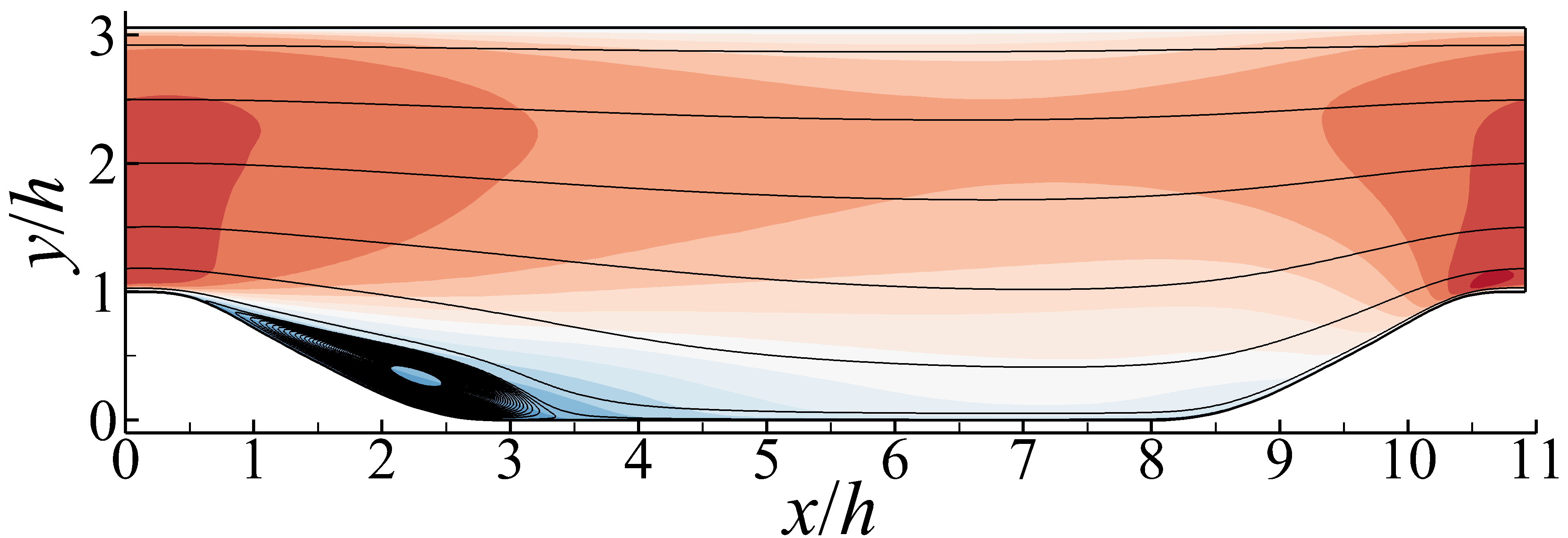}
	\subcaption{FEL: $\Delta y_f/h=0.01$}
	\end{subfigure}
	\begin{subfigure}[b]{0.495\textwidth}
	\centering
	\includegraphics[width = 1.0\textwidth]{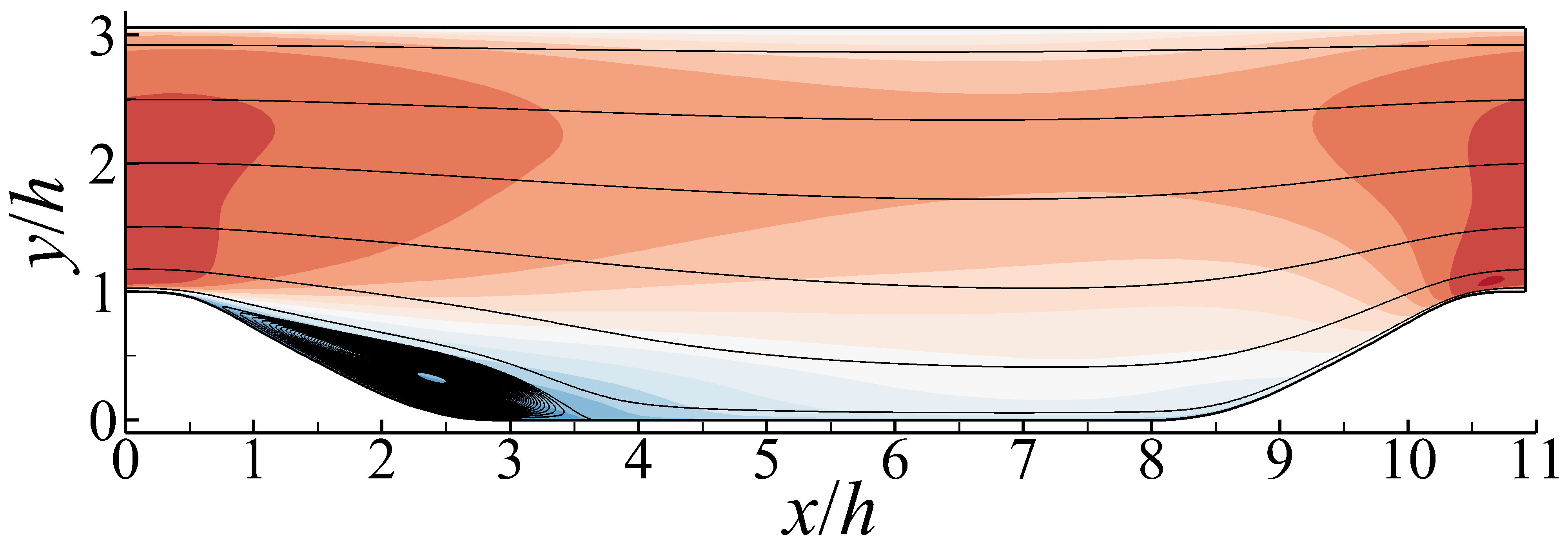}
	\subcaption{WW: $\Delta y_f/h=0.01$}
	\end{subfigure}
	\begin{subfigure}[b]{0.495\textwidth}
	\centering
	\includegraphics[width = 1.0\textwidth]{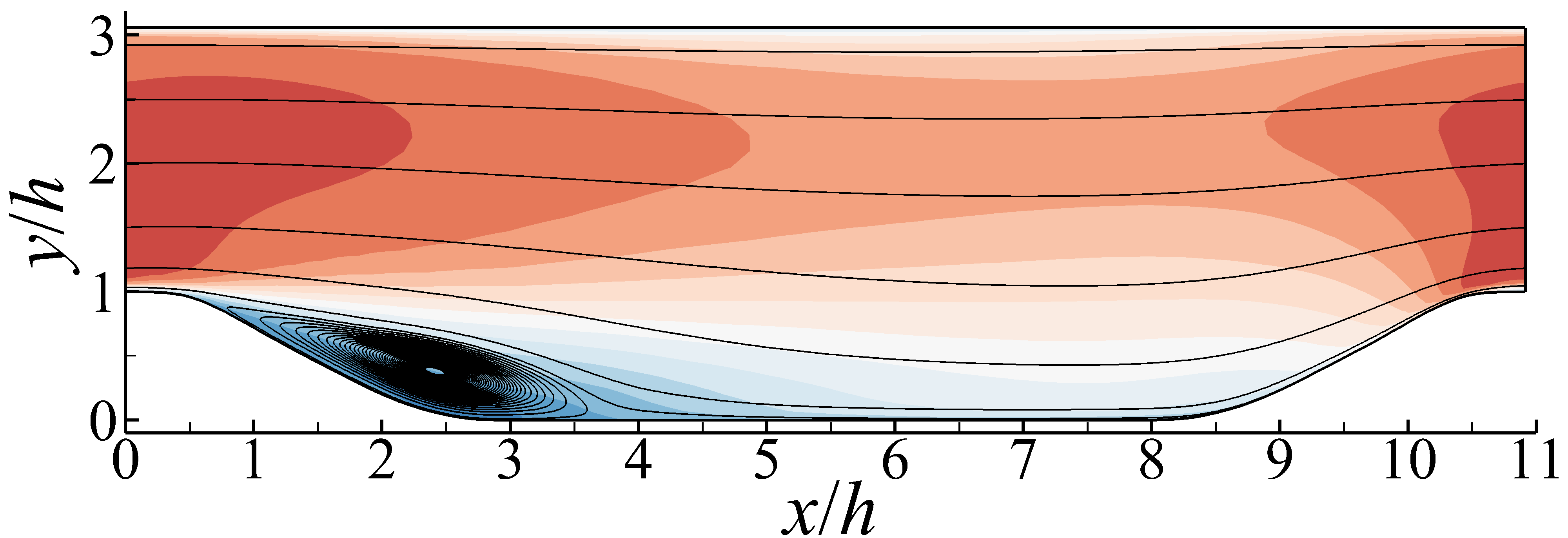}
	\subcaption{FEL: $\Delta y_f/h=0.03$}
	\end{subfigure}
	\begin{subfigure}[b]{0.495\textwidth}
	\centering
	\includegraphics[width = 1.0\textwidth]{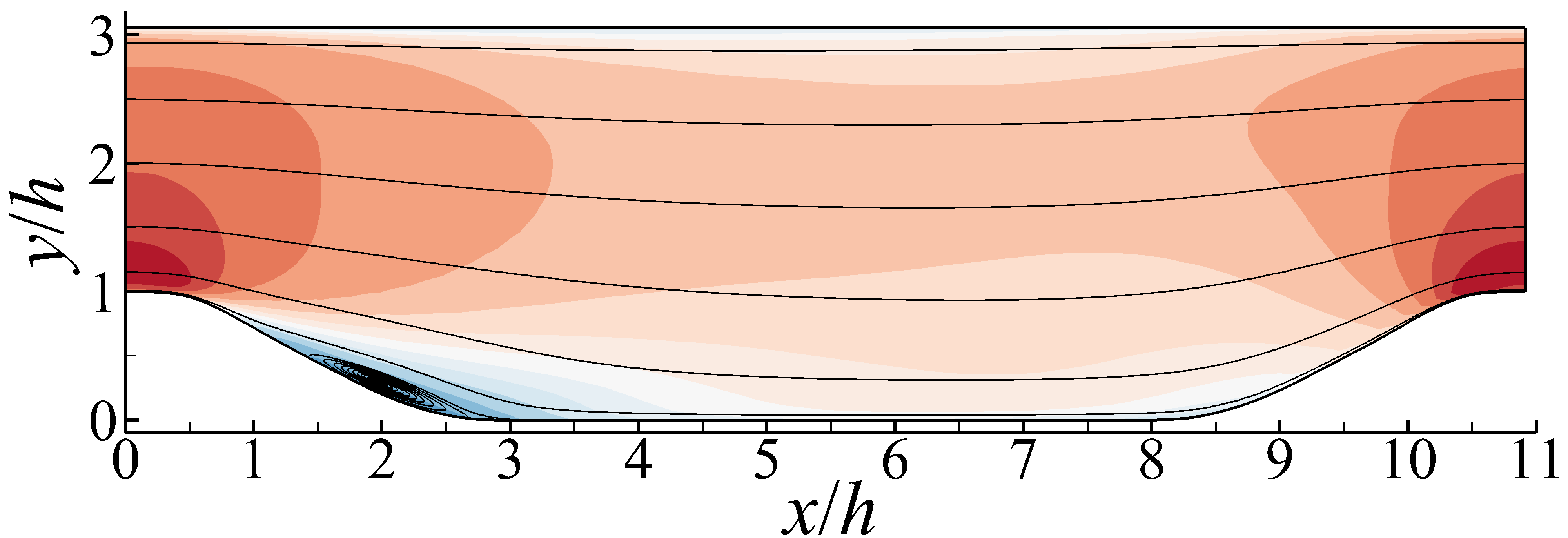}
	\subcaption{WW: $\Delta y_f/h=0.03$}
	\end{subfigure}
	\begin{subfigure}[b]{0.495\textwidth}
	\centering
	\includegraphics[width = 1.0\textwidth]{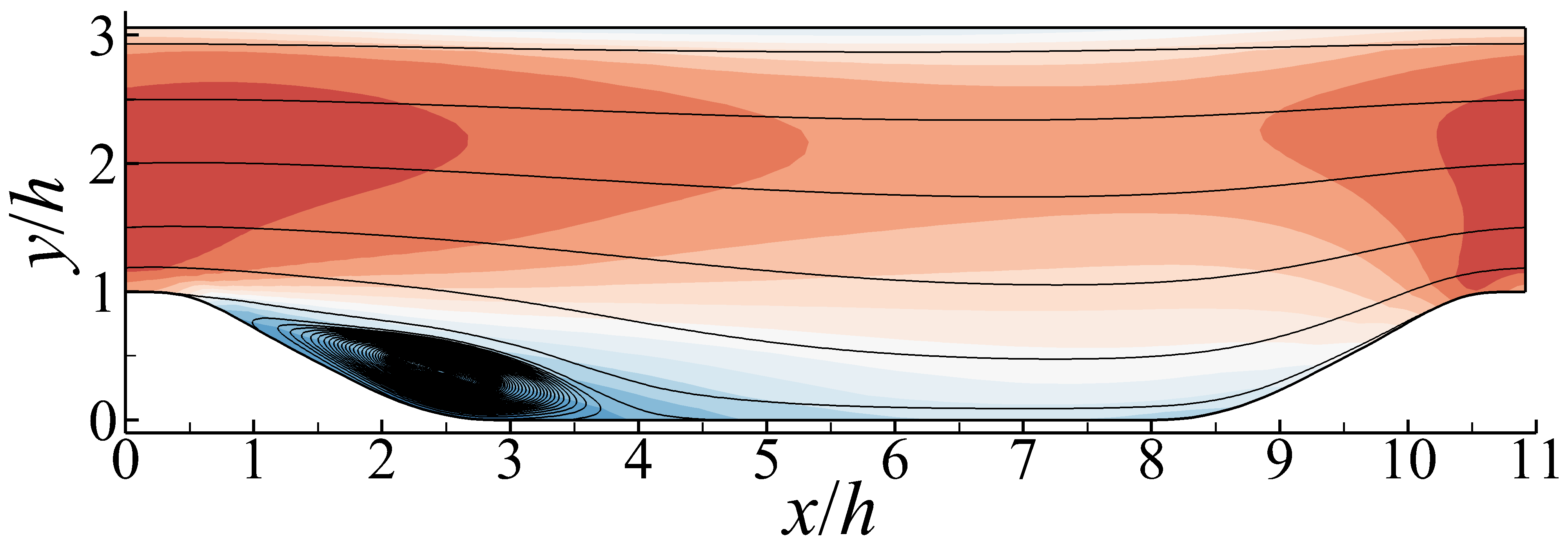}
	\subcaption{FEL: $\Delta y_f/h=0.09$}
	\end{subfigure}
	\begin{subfigure}[b]{0.495\textwidth}
	\centering
	\includegraphics[width = 1.0\textwidth]{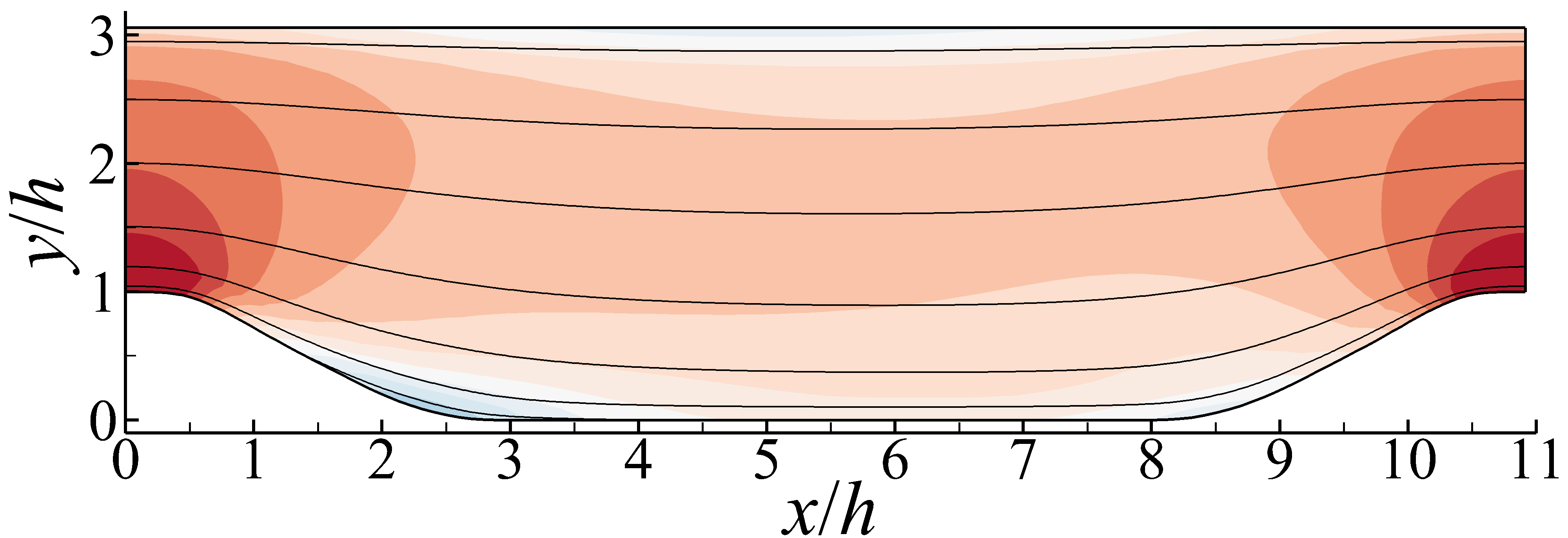}
	\subcaption{WW: $\Delta y_f/h=0.09$}
	\end{subfigure}
\caption{Contours of time-averaged streamwise velocity with streamlines obtained from the H1.5-WR and H1.5-WM cases with the FEL and WW models at $Re_h = 10595$.}
\label{fig:HL_grid}
\end{figure}
\begin{figure}
\centering{\includegraphics[width=0.9\textwidth]{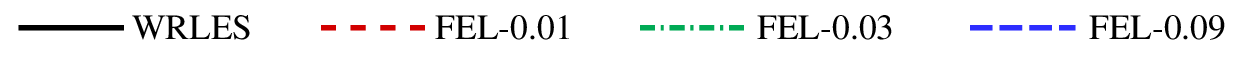}}
\centering{\includegraphics[width=0.40\textwidth]{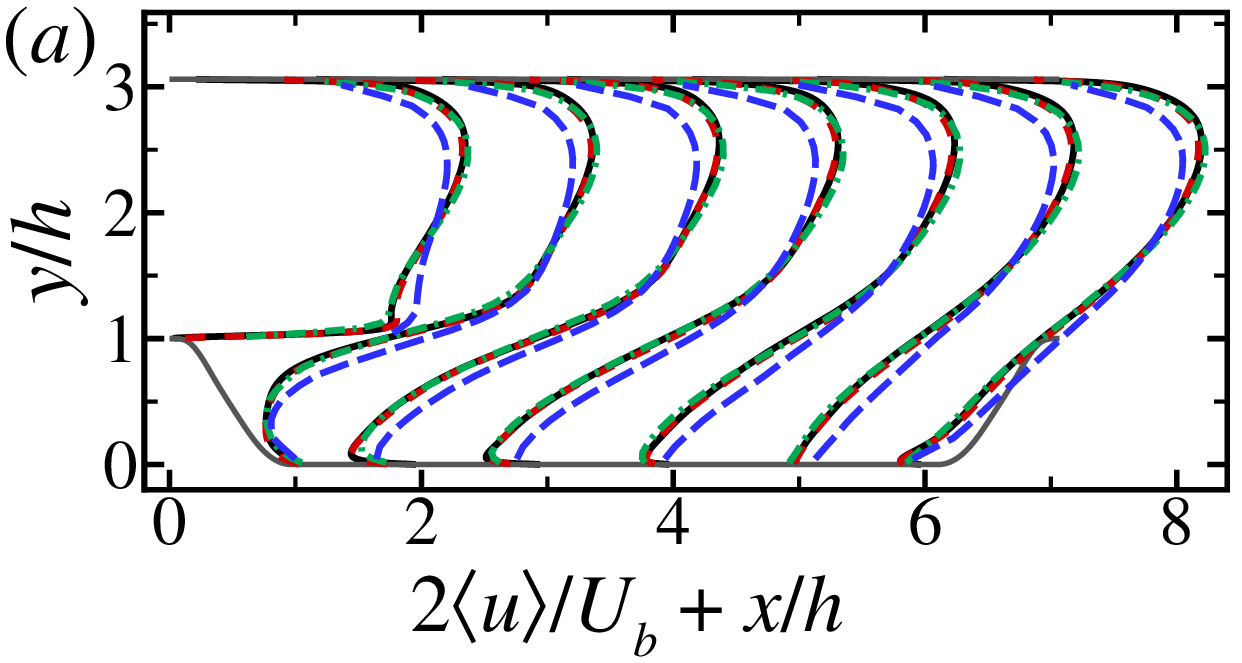}}
\centering{\includegraphics[width=0.59\textwidth]{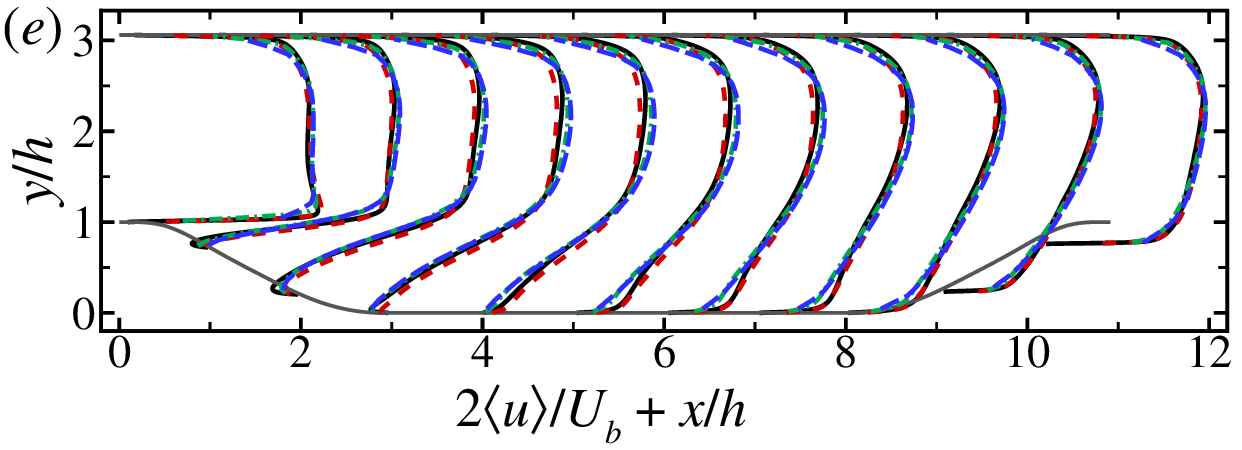}}
\centering{\includegraphics[width=0.40\textwidth]{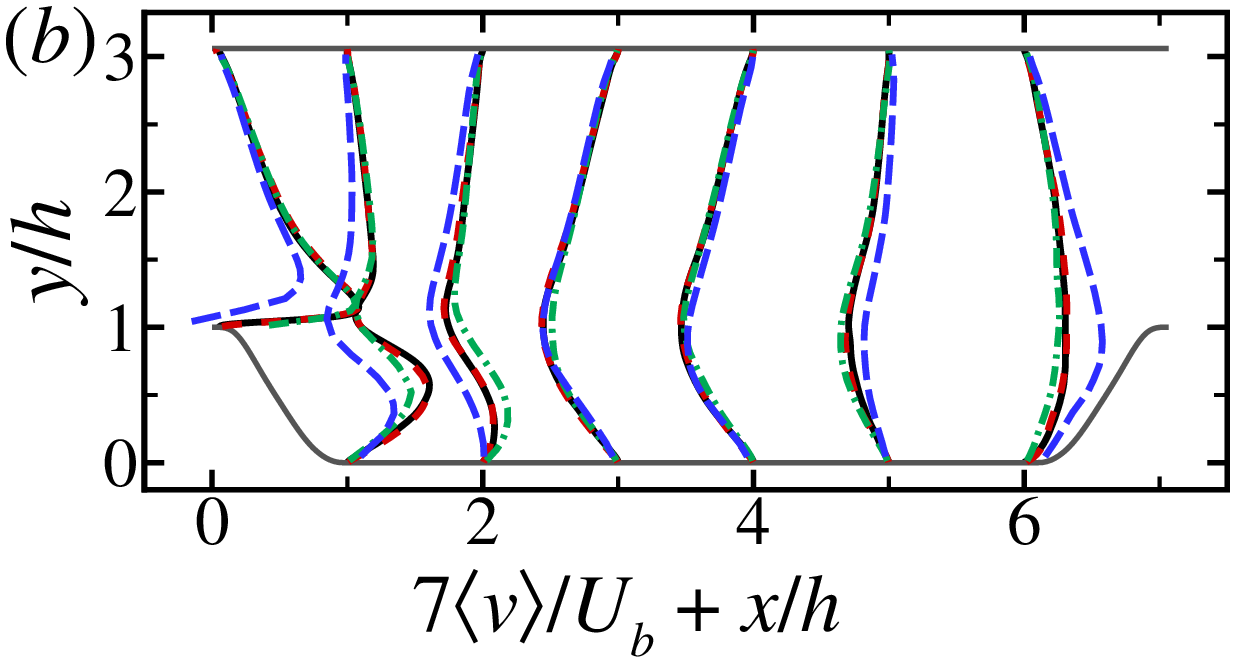}}
\centering{\includegraphics[width=0.59\textwidth]{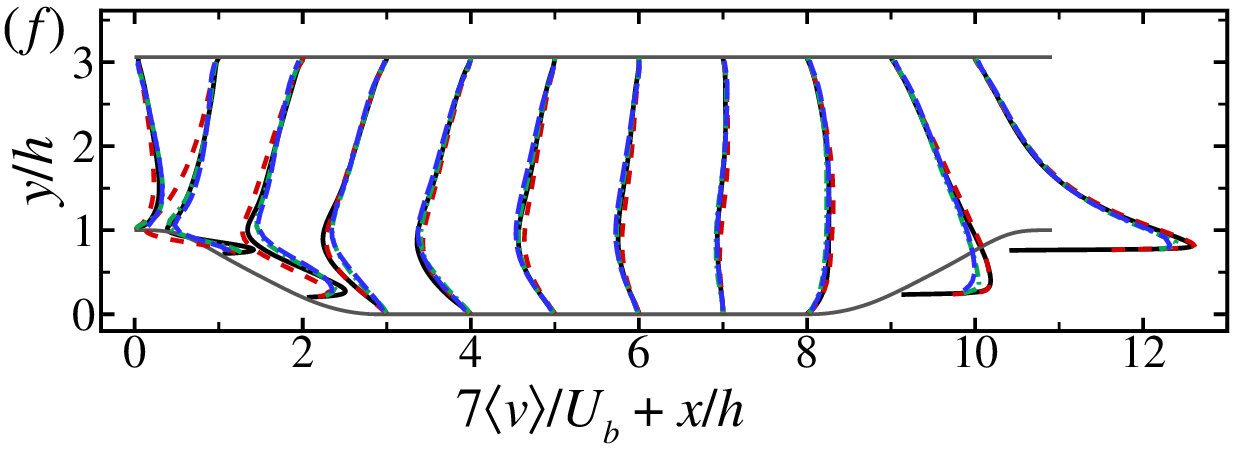}}
\centering{\includegraphics[width=0.40\textwidth]{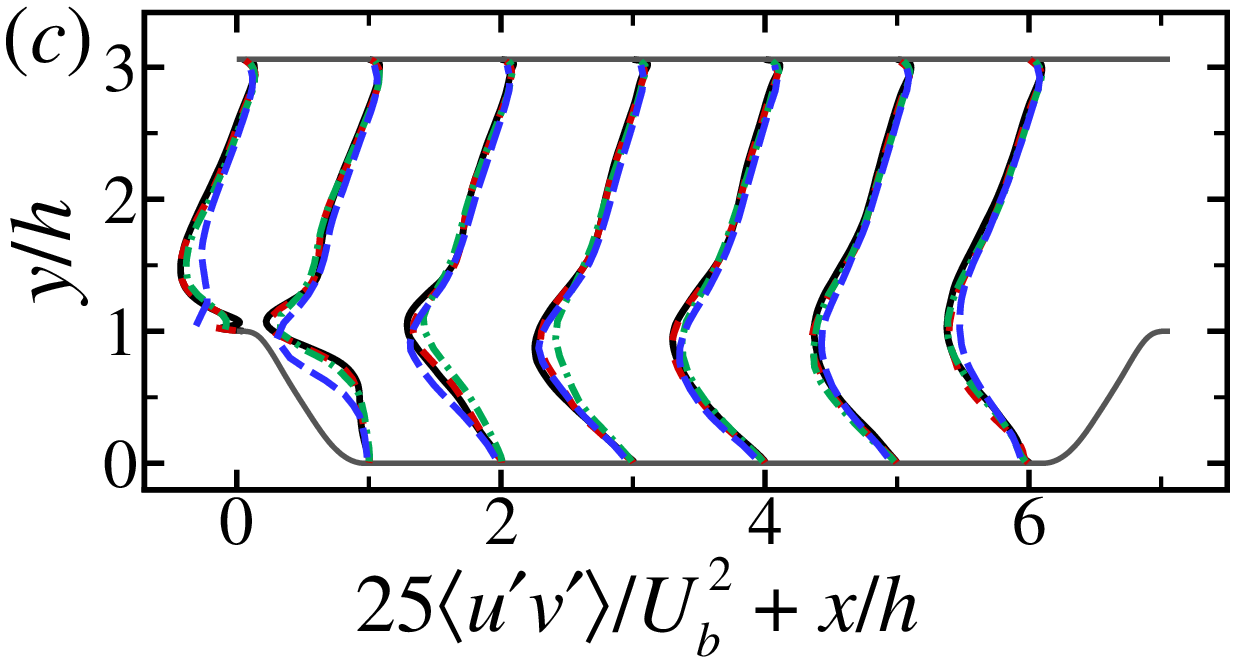}}
\centering{\includegraphics[width=0.59\textwidth]{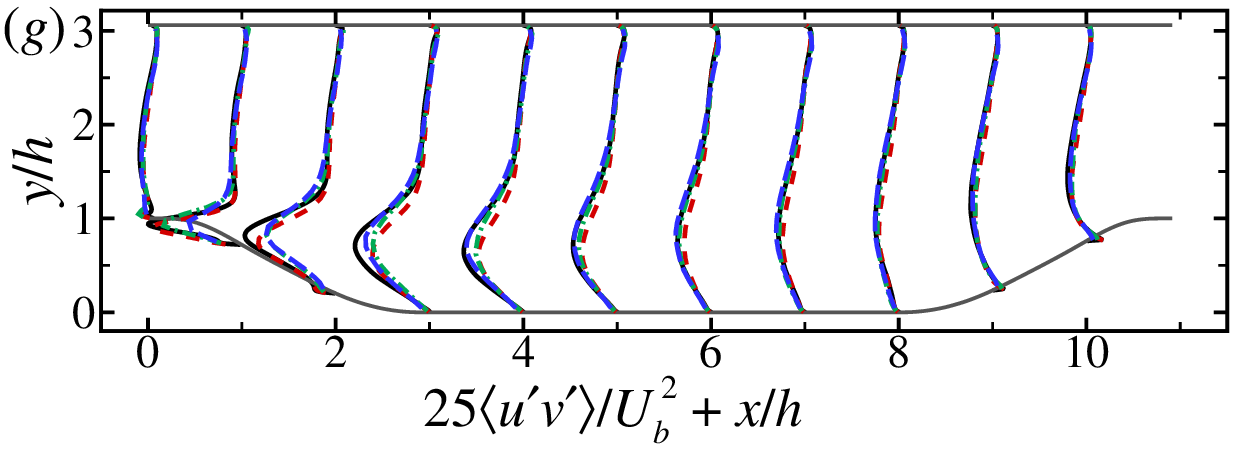}}
\centering{\includegraphics[width=0.40\textwidth]{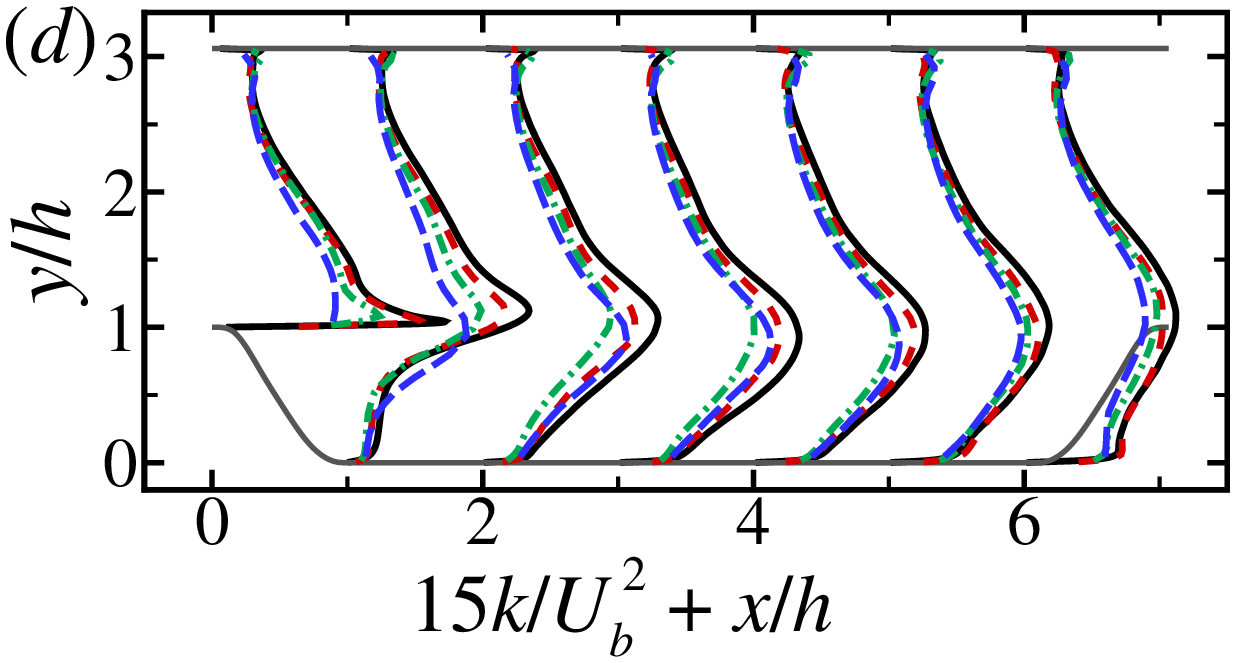}}
\centering{\includegraphics[width=0.59\textwidth]{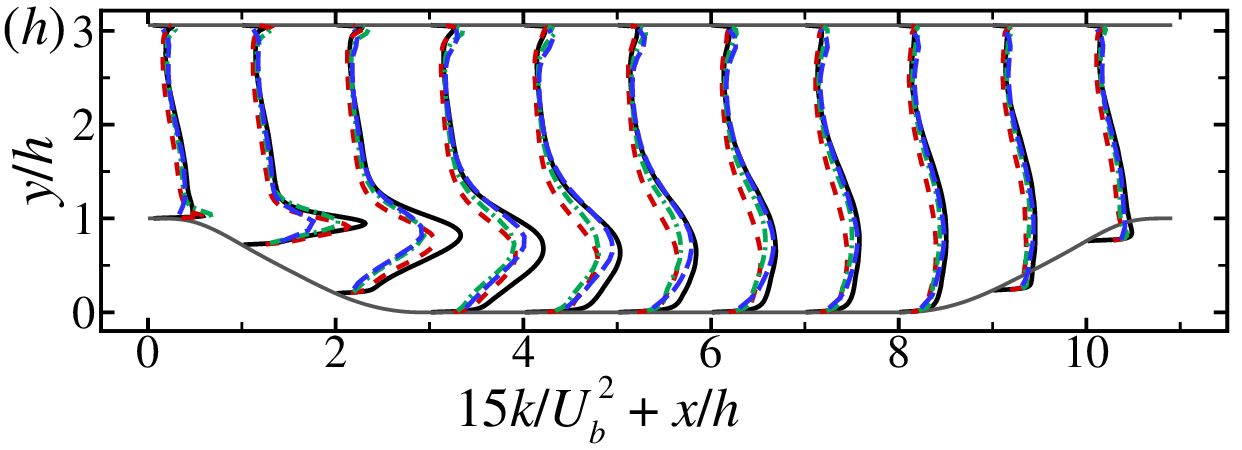}}
  \caption{Vertical profiles of (a, e) time-averaged streamwise velocity $\left\langle u \right\rangle$ and (b, f) vertical velocity $\left\langle v \right\rangle$, (c, g) primary Reynolds shear stress $\left\langle u'v' \right\rangle$, and (d, h) turbulence kinetic energy $k$ from the WRLES and WMLES with the FEL model for the H0.5 case (a$\sim$d) and H1.5 case (e$\sim$h) at $Re_h = 10595$.}
\label{fig:profile_HSHL_grid}
\end{figure}
\begin{figure}
\centering{\includegraphics[width=0.9\textwidth]{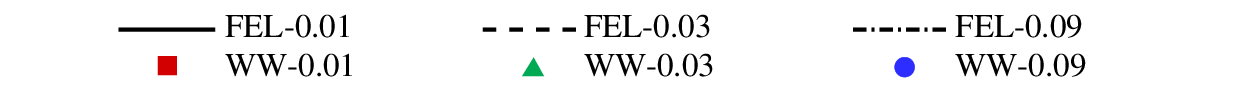}}
\centering{\includegraphics[width=0.4\textwidth]{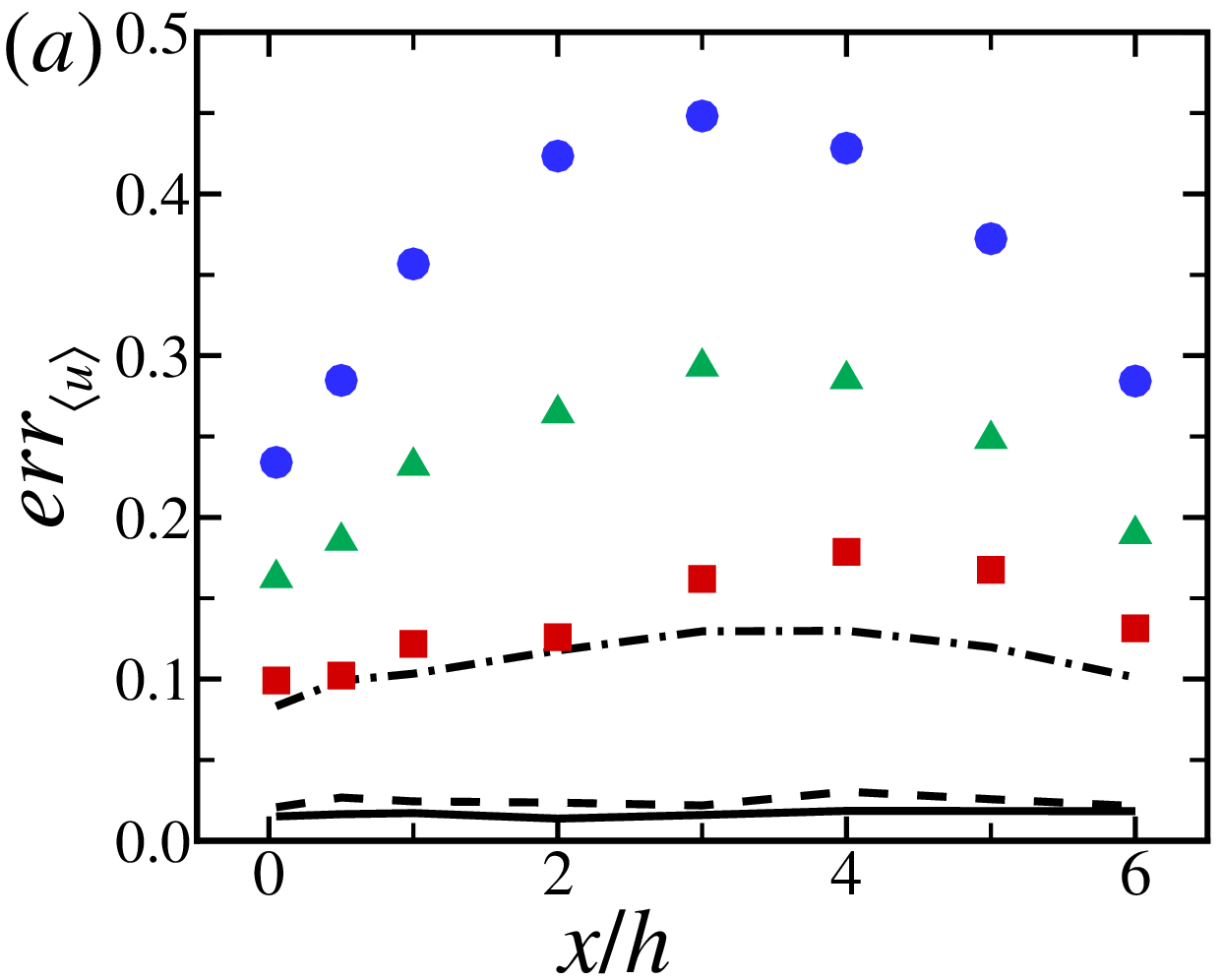}}\quad\quad
\centering{\includegraphics[width=0.4\textwidth]{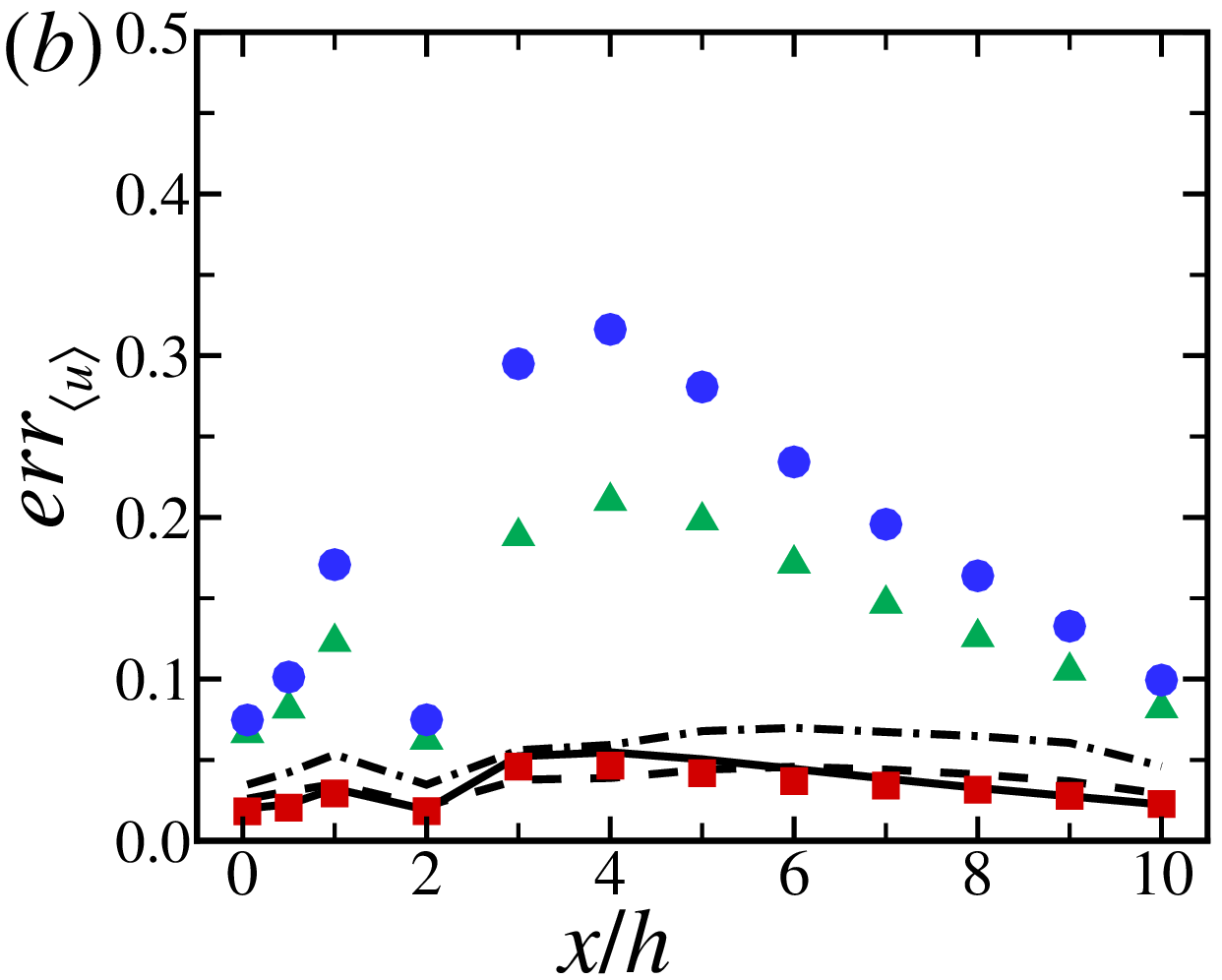}}
\centering{\includegraphics[width=0.4\textwidth]{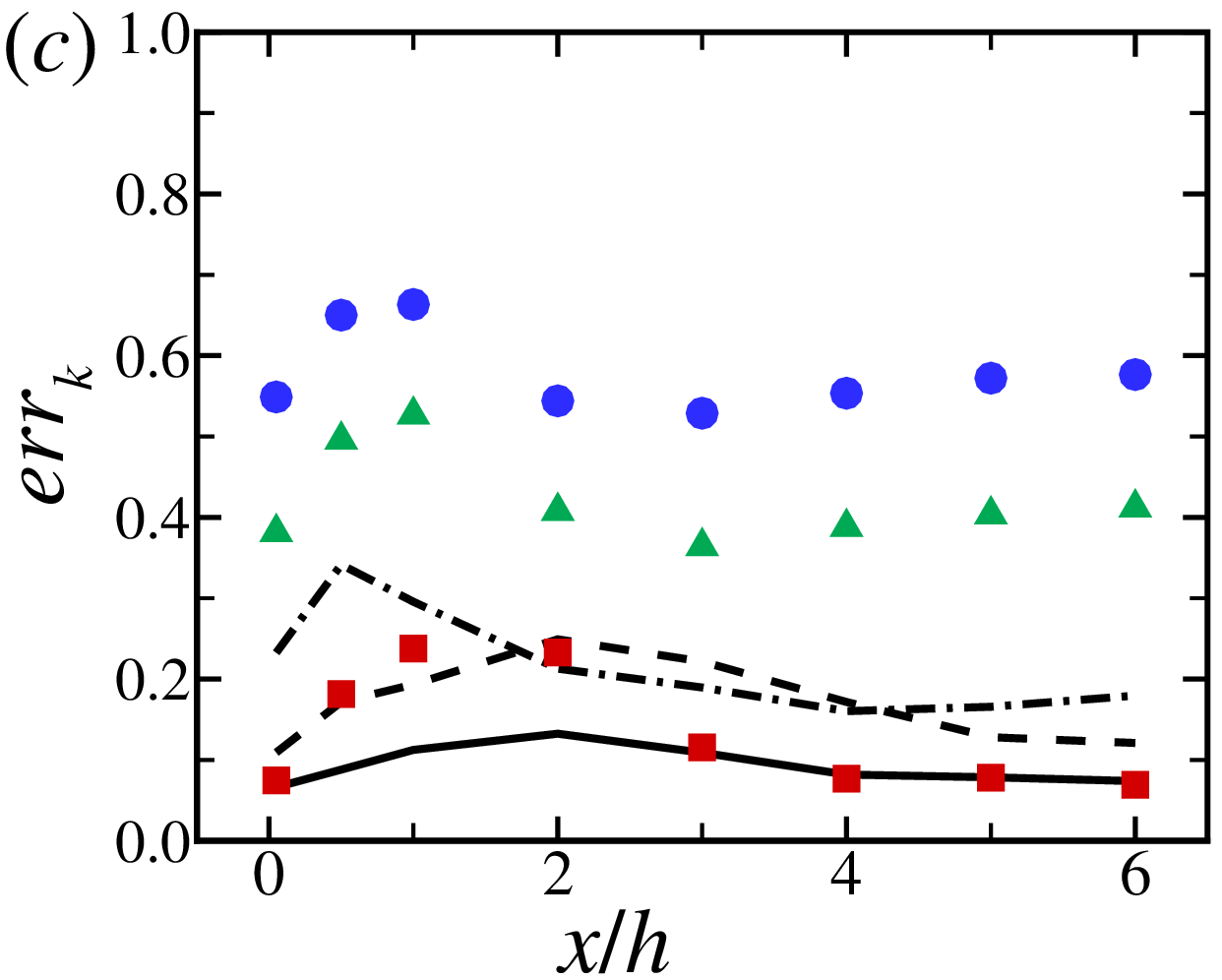}}\quad\quad
\centering{\includegraphics[width=0.4\textwidth]{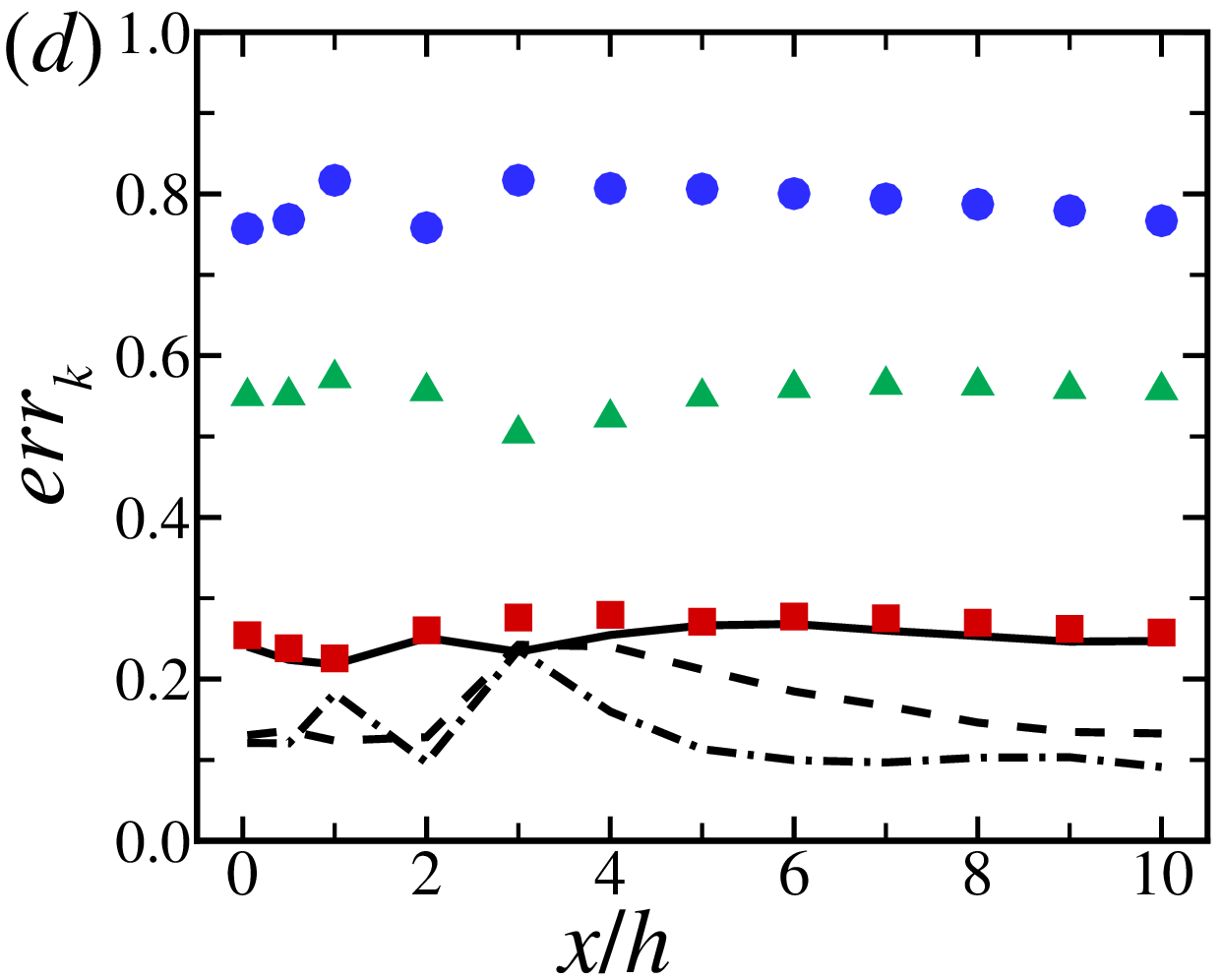}}
  \caption{The relative errors of (a$\sim$b) time-averaged streamwise velocity $\left\langle u \right\rangle$ and (c$\sim$d) turbulence kinetic energy $k$ between the WMLES and WRLES for the H0.5 case (a, c) and H1.5 case (b, d) at $Re_h = 10595$.}
\label{fig:profile_err_HSHL}
\end{figure}
\begin{figure}
\centering{\includegraphics[width=0.48\textwidth]{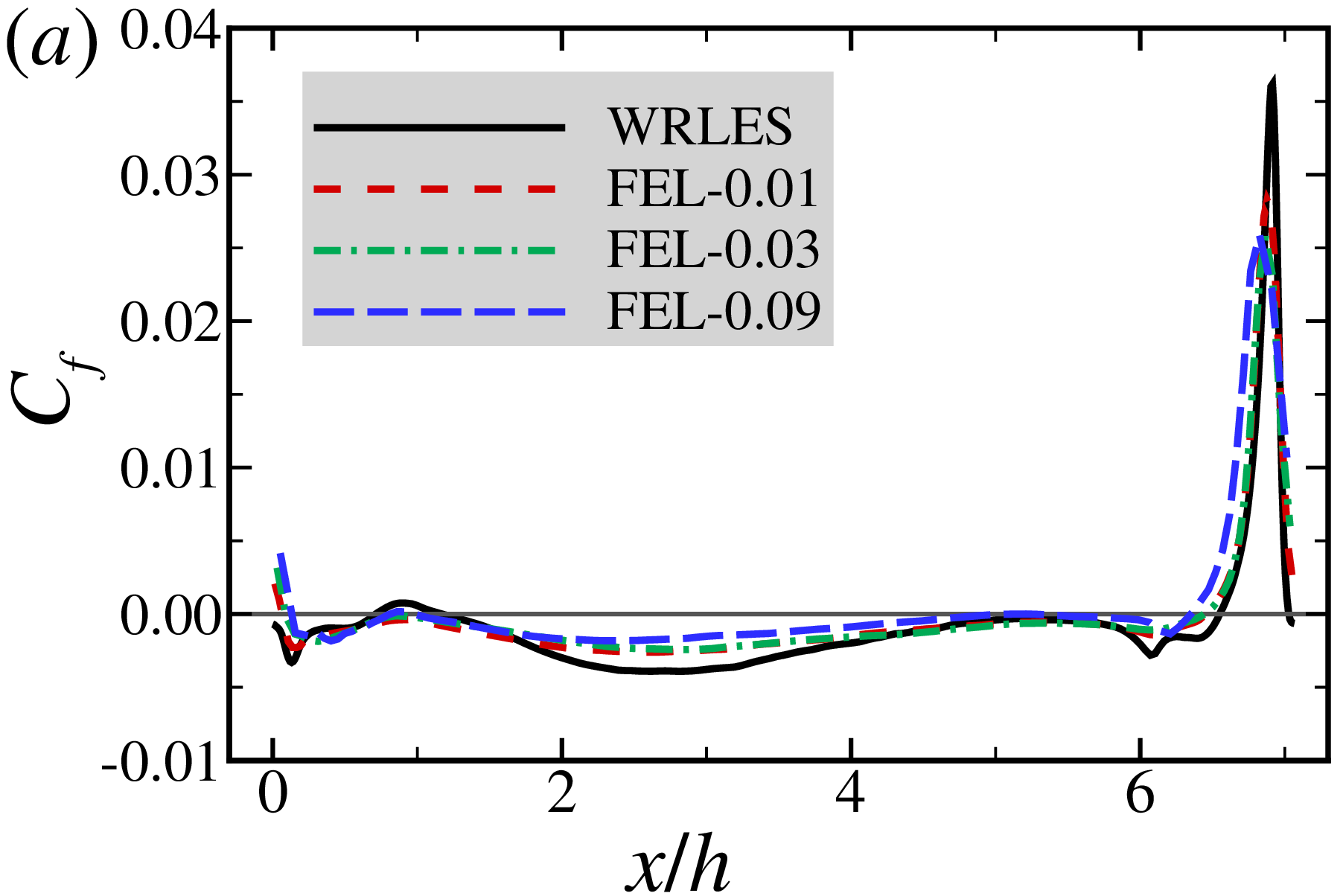}}\;
\centering{\includegraphics[width=0.48\textwidth]{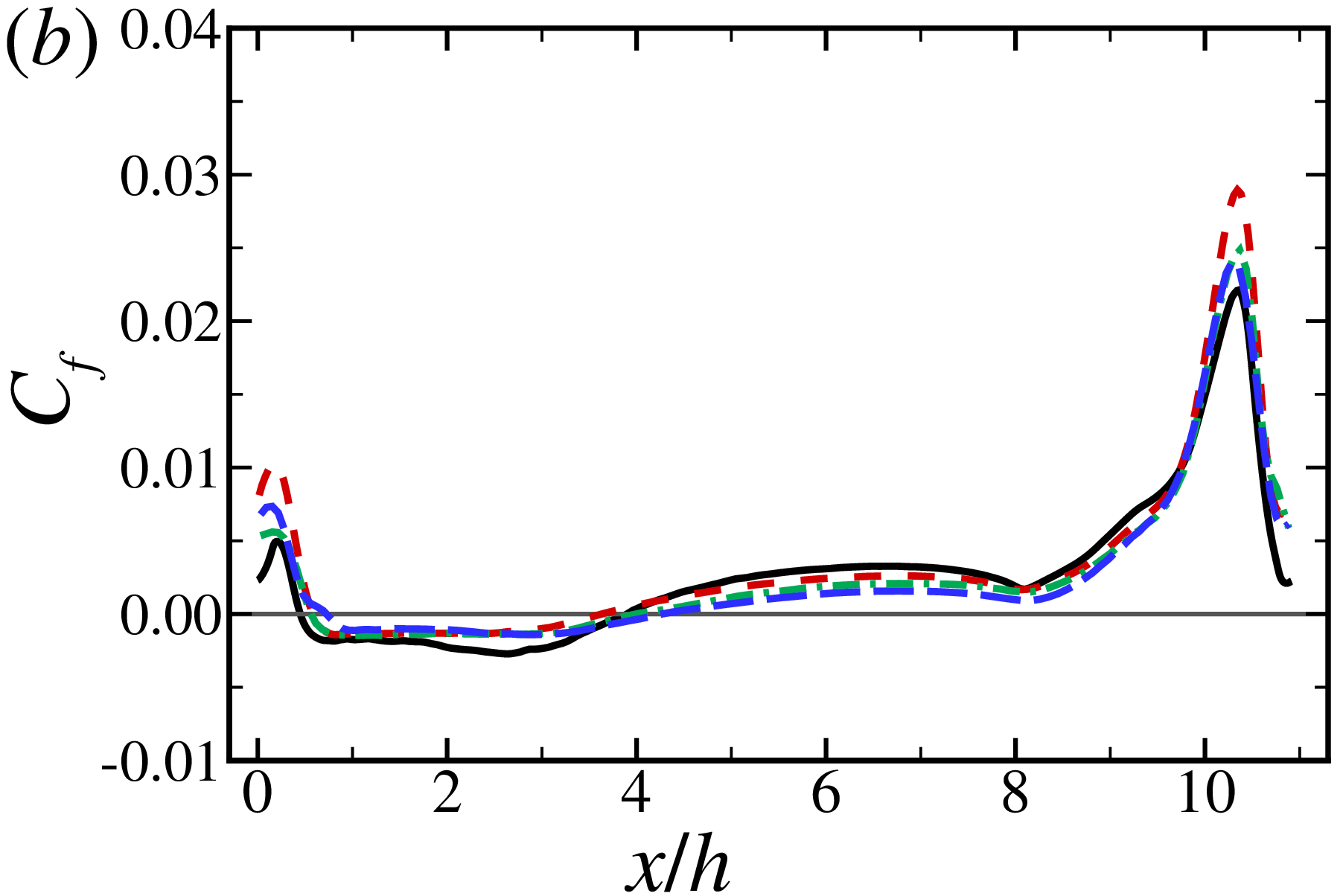}}
  \caption{Comparison of the time-averaged skin friction coefficient from the WMLES with the FEL model and the WRLES of (a) H0.5 and (b) H1.5 case at $Re_h = 10595$.}
\label{fig:Cf_HSHL}
\end{figure}

The FEL model was trained on a specific grid resolution. In this section, we examine the performance of the proposed model for different grid resolutions with $\Delta y_f/h=0.01$, $0.03$ and $0.09$. The contours of the time-averaged streamwise velocity with streamlines obtained from the three grid resolutions different from the training cases are presented in figures~\ref{fig:HS_grid} and \ref{fig:HL_grid}. It is seen that with the refining or coarsening of the grids, consistent results are obtained for both grids. Surprisingly, with the coarsest grid (with only around 11 grid cells over the hill height), not bad recirculation bubbles are still predicted using the FEL model for both cases, which are significantly underpredicted or even not captured using the WW model.

Quantitative assessments of the FEL model for different grid resolutions are demonstrated in figure~\ref{fig:profile_HSHL_grid}(a)$\sim$(d), it is seen that the FEL model performs well in predicting the turbulence statistics at grid resolutions of $\Delta y_f/h=0.01$ and 0.03 for the H0.5 case, but shows some discrepancies at the coarsest grid with $\Delta y_f/h=0.09$. 
Regarding the assessment for the H1.5 case shown in figure~\ref{fig:profile_HSHL_grid}(e)$\sim$(g), the FEL model generally predicts the vertical profiles of the turbulence statistics at the coarser grid resolutions with $\Delta y_f/h=0.03$ and 0.09. A certain degree of discrepancy is observed at the finest grid with $\Delta y_f/h=0.01$ for the second-order turbulence statistics (i.e., $\left\langle u'v' \right\rangle$ and $k$).
%This demonstrates that grid resolution is an important factor influencing the application of WMLES.

The relative errors of vertical profiles between the WMLES and WRLES are shown in figure~\ref{fig:profile_err_HSHL} for the time-averaged streamwise velocity $\left\langle u \right\rangle$ and TKE $k$ for the H0.5 and H1.5 cases at $Re_h = 10595$. As shown in figures~\ref{fig:profile_err_HSHL} (a, c) for the H0.5 case, the errors for the FEL model are only 2\% for $\left\langle u \right\rangle$ and 10\% for $k$ at grid resolution with $\Delta y_f/h=0.01$, and increase to approximately 3\% for $\left\langle u \right\rangle$ and 17\% for $k$ at grid resolution with $\Delta y_f/h=0.03$. For the coarsest grid resolution with $\Delta y_f/h=0.09$, the errors further increase to approximately 10\% and 20\% for $\left\langle u \right\rangle$ and $k$, respectively.
The errors of the WW model predictions are much larger than those from the FEL model. For the assessment using the H1.5 case as shown in figures~\ref{fig:profile_err_HSHL} (b, d), an overall better performance is observed for the proposed model as well in comparison with the WW model. 
For the two coarser grids, the errors for $\left\langle u \right\rangle$ and $k$ are approximately 5\% and 15\%, respectively, for the FEL model. {\color{black}For the fine grid with $\Delta y_f/h=0.01$, the errors for $\left\langle u \right\rangle$ are less than 5\%, but the errors for $k$ increase anomalously to 24\%.}

The mean skin friction coefficients ($C_f = \tau_w / \frac{1}{2} \rho u_b^2$) predicted by the FEL model are compared with WRLES results for different grid resolutions in figure~\ref{fig:Cf_HSHL}. It can be observed that the skin friction coefficients predicted by the FEL model are in good agreement with the WRLES results at most streamwise locations for the H0.5 case with $\Delta y_f/h=0.01$ and 0.03 and the H1.5 case with $\Delta y_f/h=0.03$ and 0.09, respectively. For the H0.5 case with $\Delta y_f/h=0.09$ and the H1.5 case with $\Delta y_f/h=0.01$, the peak values of the skin friction coefficient near the hill crest are somewhat overestimated.
%This demonstrates that the FNN\_PH-LoW model, trained \emph{a priori} in our previous work~\citep{Zhou_etal_PoF_2023}, is capable of capturing the wall boundary condition in the \emph{a posteriori} test.
%However, the overall variation tendencies of the skin friction coefficient along with the streamwise locations are captured reasonably well by the cDK-embedded model. It can be anticipated that the FNN\_PH-LoW model would provide more accurate wall shear stress predictions if the WMLES yields better prediction of the general flow field.

{\color{black}It is noted that the WMLES prediction on a coarser grid is more accurate for the H1.5 case (with the error comparison shown in figure~\ref{fig:profile_err_HSHL}(d)). As several factors can affect the prediction accuracy of WMLES, such as grid resolution, discretization errors, errors from the subgrid scale models, and the employed wall models, it is hard to identify the cause for the observed grid inconsistency. Such inconsistency was also reported in the literature.
%This is mainly because the WMLES prediction of separated flows is sensitive to the grid resolution. The consistency of results convergence with grid refinement is significantly influenced by the SGS model
For instance, \citet{Zhou_Bae_JCP_2024} observed non-monotonic convergence in capturing the separation bubble of a two-dimensional Gaussian-shaped bump for the anisotropic minimum dissipation model and Vreman model.
It is acceptable considering that no similarity property was incorporated into the design of the input and output quantities of the model to ensure its applicability to different grid resolutions, and cases with different grid resolutions were not included in the embedded model training, either.}

\subsection{Cases with different hill slopes}\label{subsec:Application_geom}
\begin{figure}
\centering
	\begin{subfigure}[b]{0.48\textwidth}
	\centering
	\includegraphics[width = 1.0\textwidth]{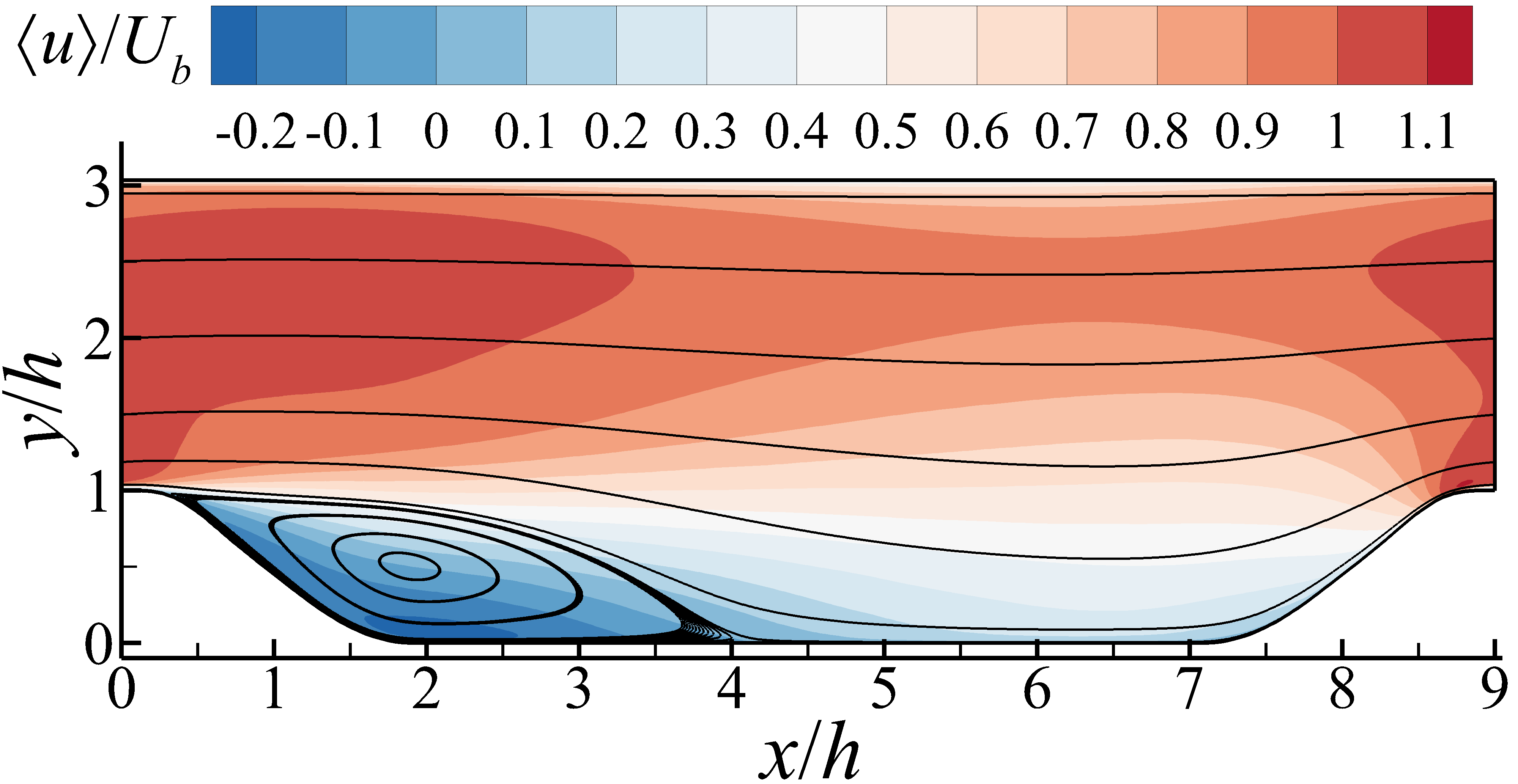}
	\subcaption{H1.0-WR: $\Delta y_f/h=0.003$}
	\end{subfigure} \\
    \begin{subfigure}[b]{0.48\textwidth}
	\centering
	\includegraphics[width = 1.0\textwidth]{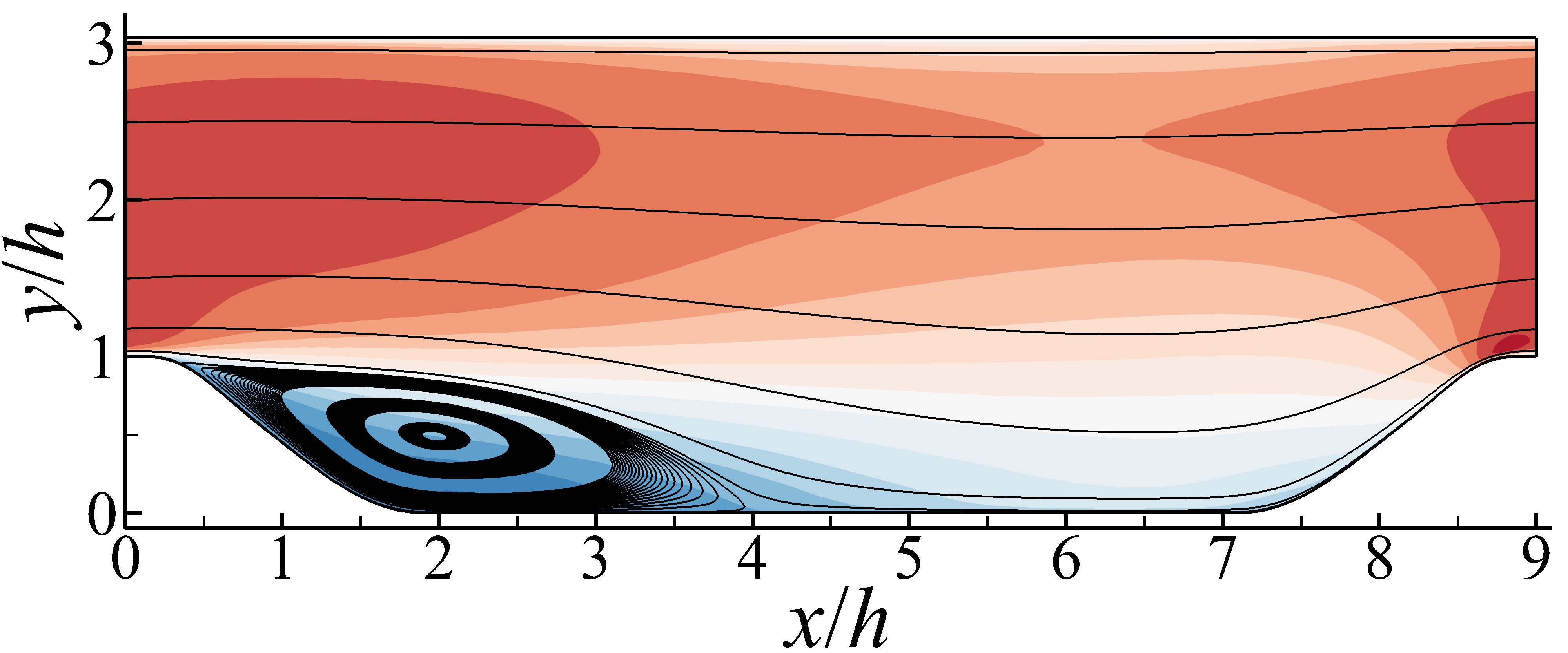}
	\subcaption{FEL: $\Delta y_f/h=0.01$}
	\end{subfigure}
	\begin{subfigure}[b]{0.48\textwidth}
	\centering
	\includegraphics[width = 1.0\textwidth]{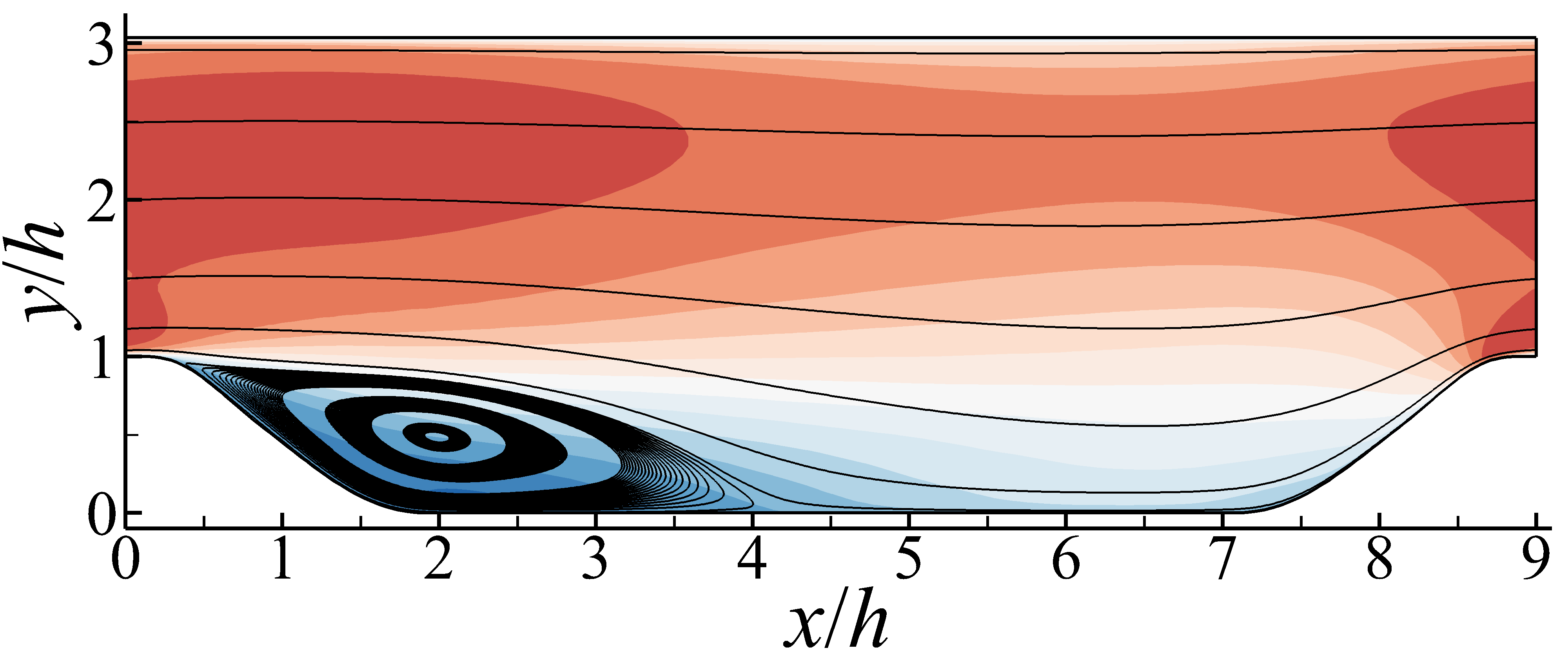}
	\subcaption{WW: $\Delta y_f/h=0.01$}
	\end{subfigure}
    \begin{subfigure}[b]{0.48\textwidth}
	\centering
	\includegraphics[width = 1.0\textwidth]{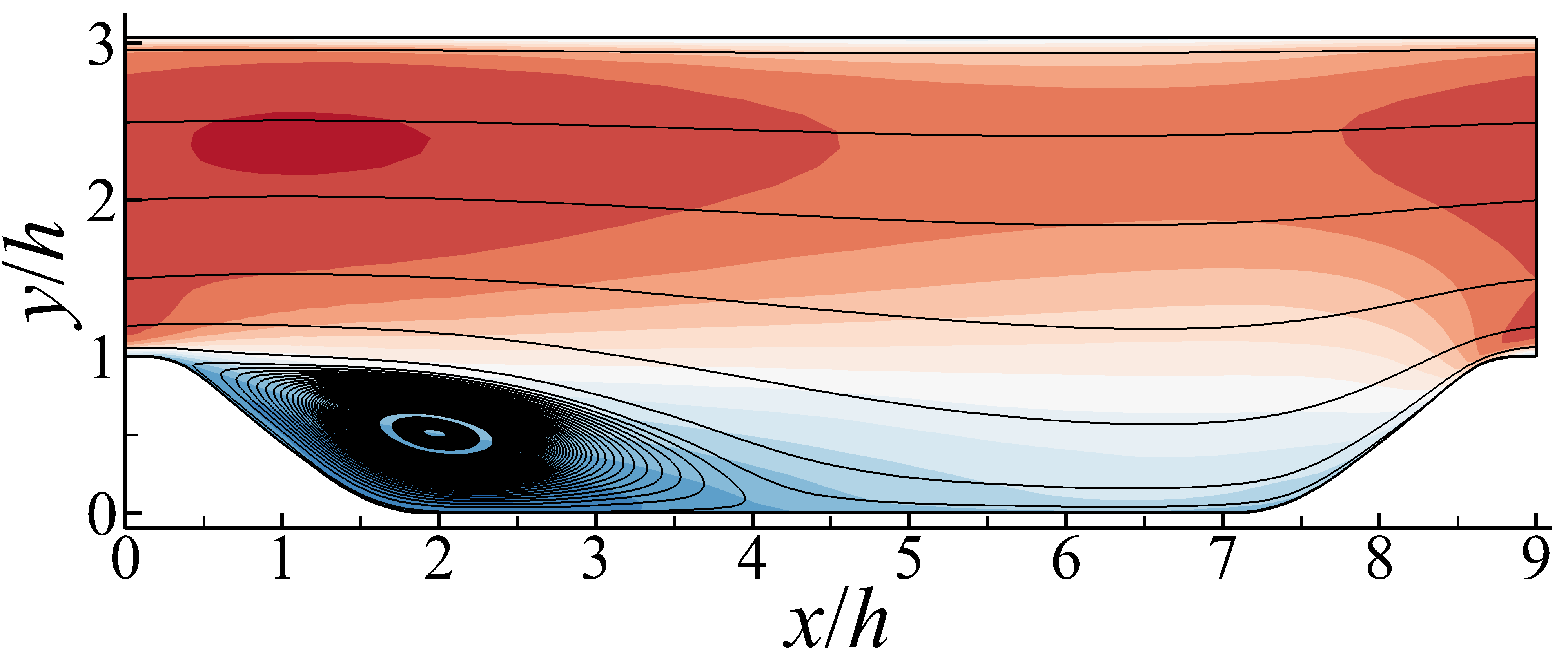}
	\subcaption{FEL: $\Delta y_f/h=0.03$}
	\end{subfigure}
	\begin{subfigure}[b]{0.48\textwidth}
	\centering
	\includegraphics[width = 1.0\textwidth]{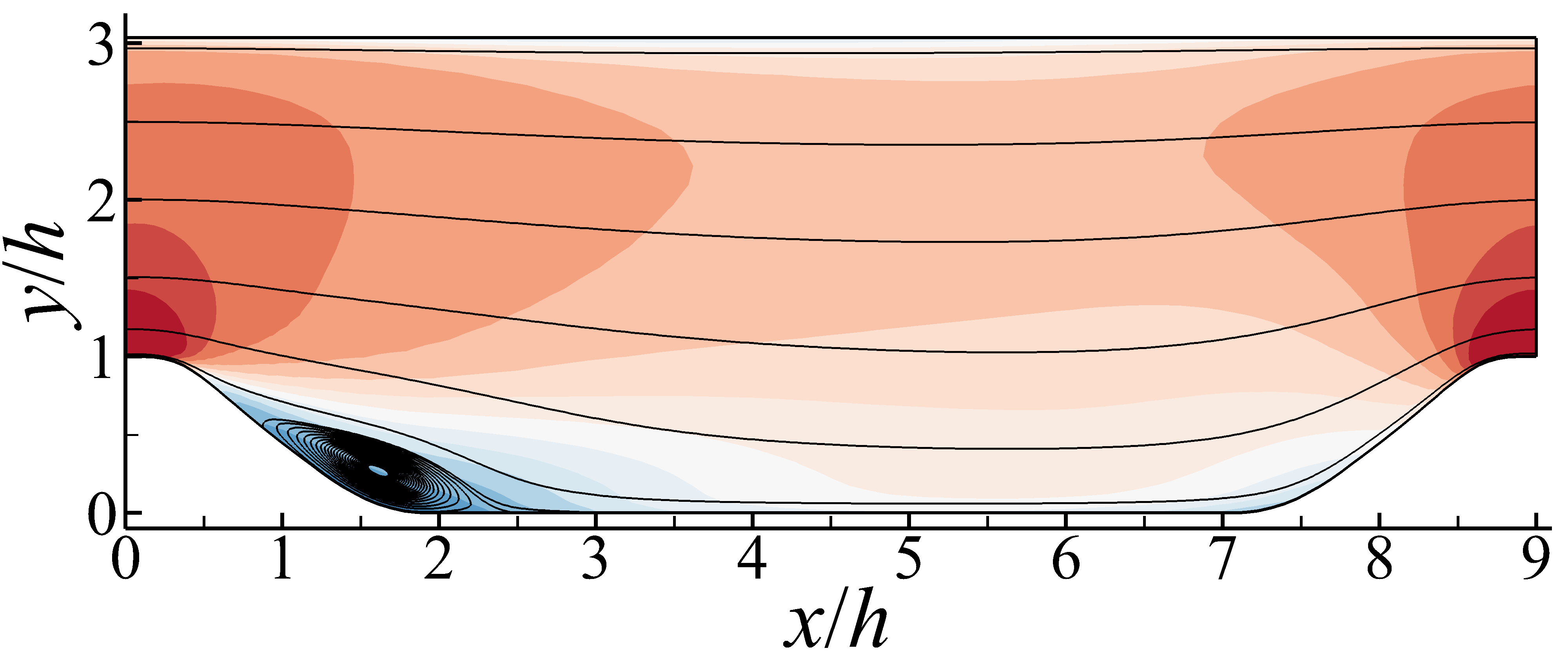}
	\subcaption{WW: $\Delta y_f/h=0.03$}
	\end{subfigure}
    \begin{subfigure}[b]{0.48\textwidth}
	\centering
	\includegraphics[width = 1.0\textwidth]{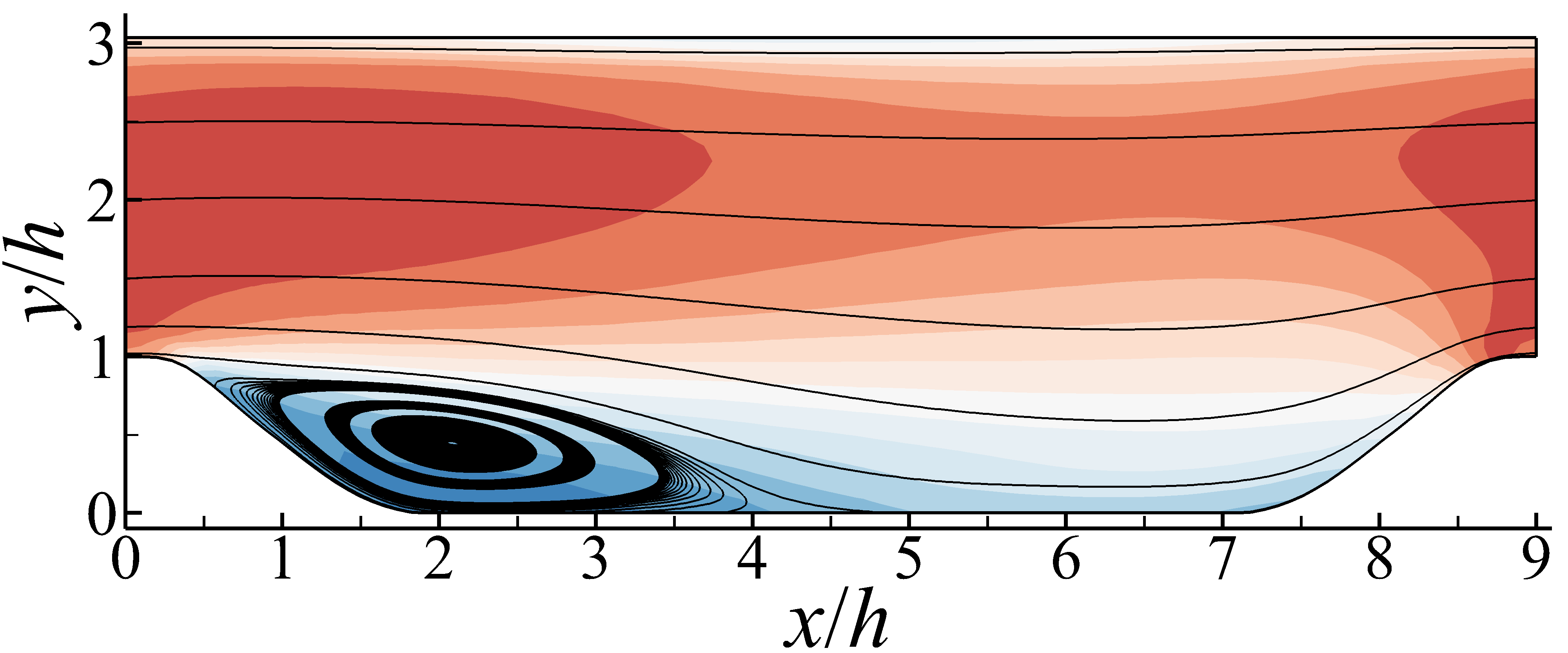}
	\subcaption{FEL: $\Delta y_f/h=0.09$}
	\end{subfigure}
	\begin{subfigure}[b]{0.48\textwidth}
	\centering
	\includegraphics[width = 1.0\textwidth]{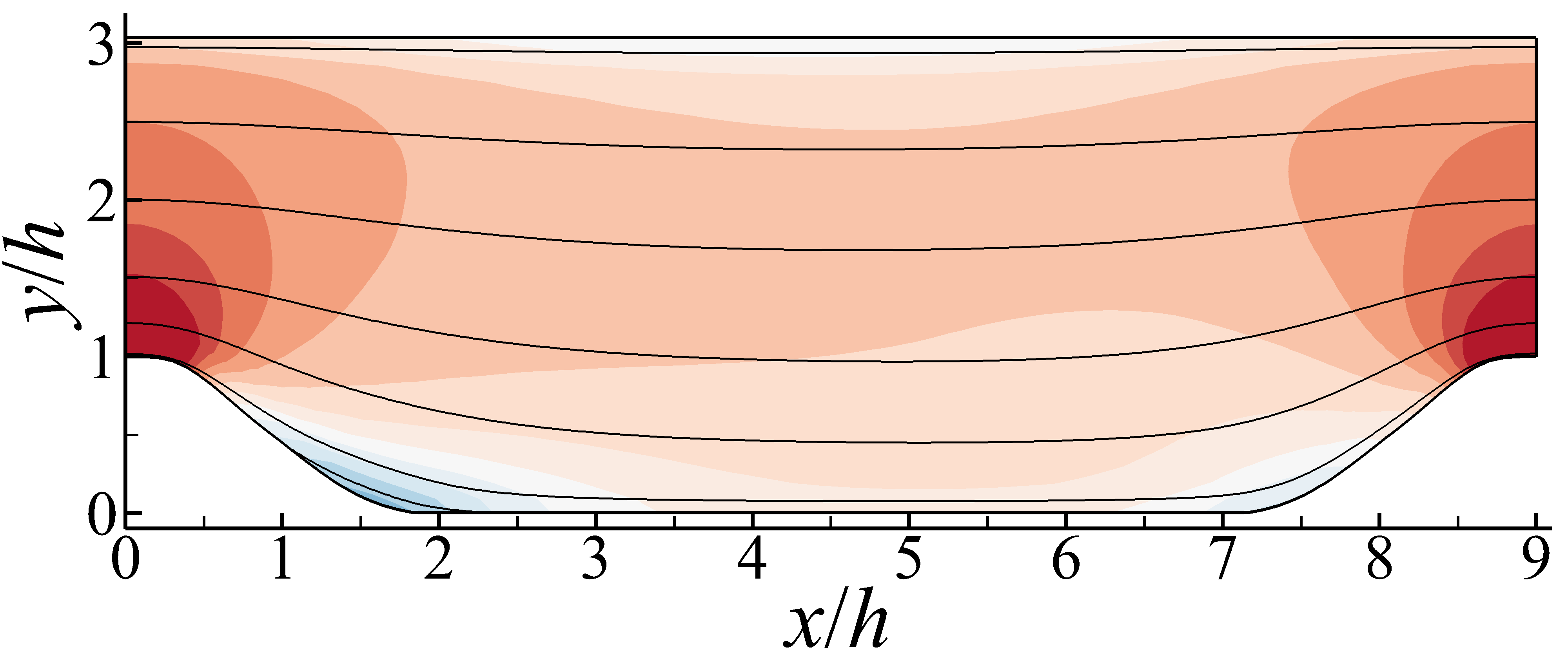}
	\subcaption{WW: $\Delta y_f/h=0.09$}
	\end{subfigure}
\caption{Contours of time-averaged streamwise velocity with streamlines obtained from the H1.0-WR and H1.0-WM cases with the FEL and WW models at $Re_h = 10595$.}
\label{fig:HB_Re10595}
\end{figure}
\begin{figure}
\centering{\includegraphics[width=0.9\textwidth]{Fig_profile_HSHL_legend.eps}}
\centering{\includegraphics[width=0.495\textwidth]{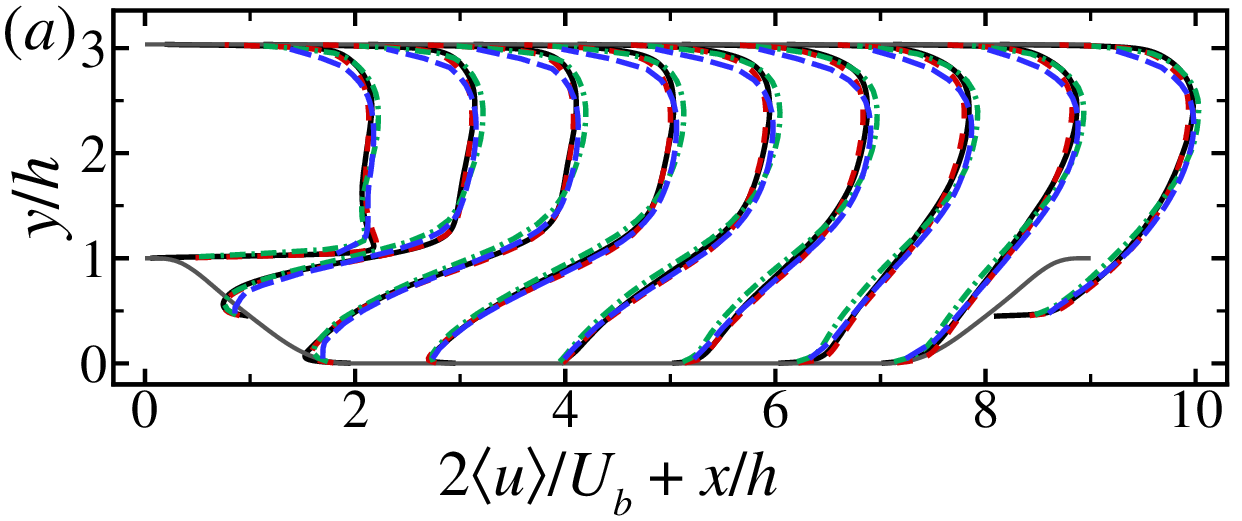}}
\centering{\includegraphics[width=0.495\textwidth]{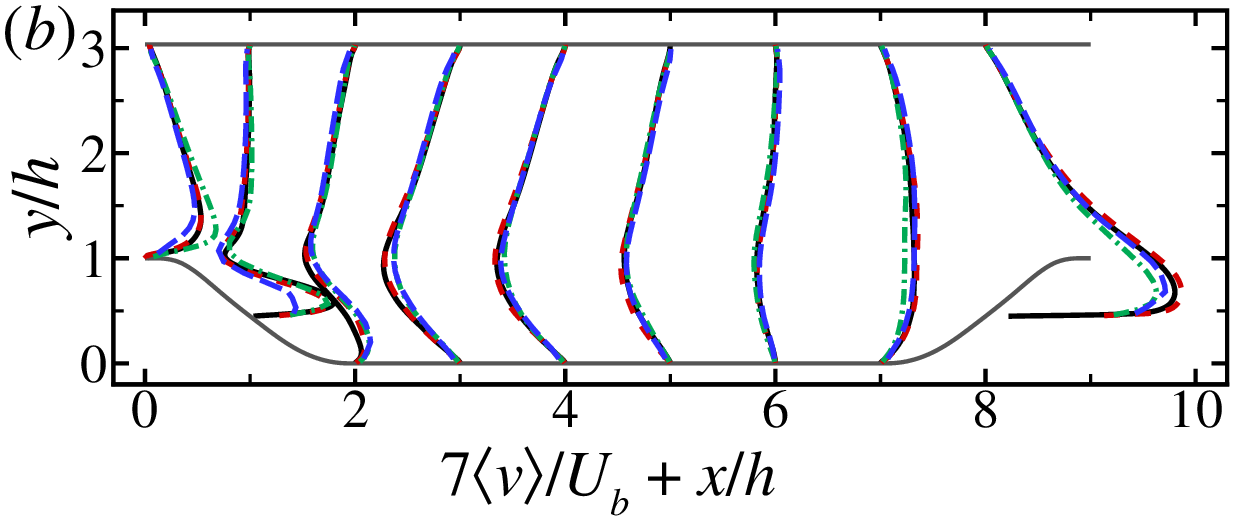}}
\centering{\includegraphics[width=0.495\textwidth]{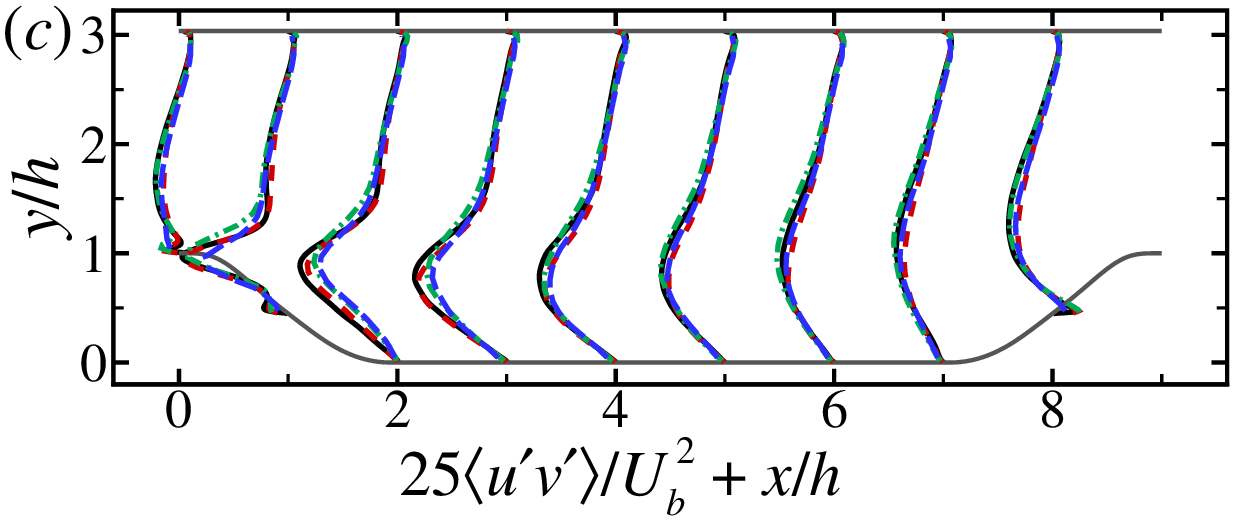}}
\centering{\includegraphics[width=0.495\textwidth]{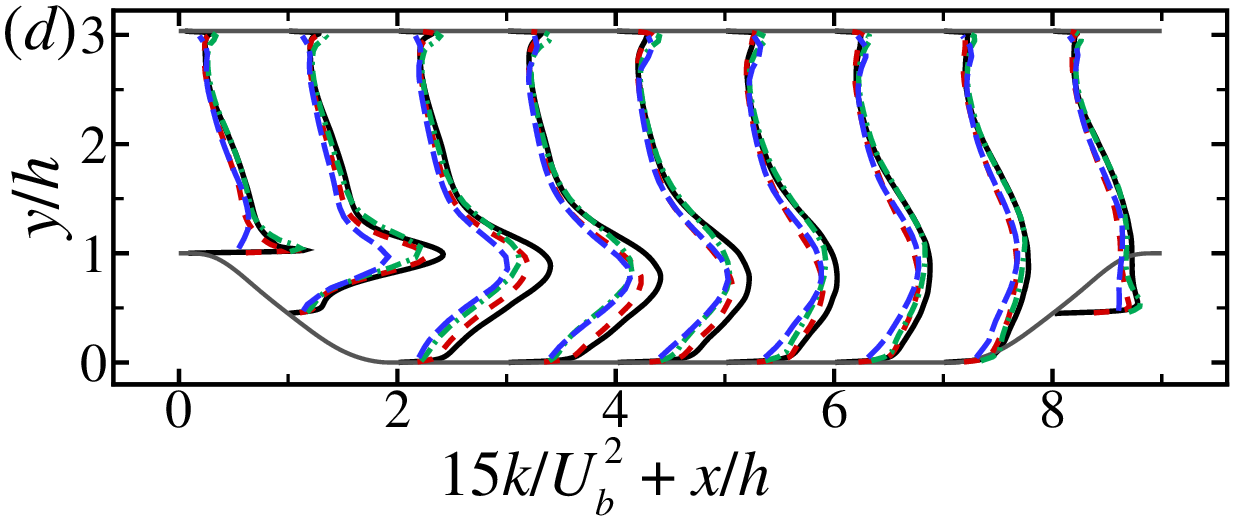}}
  \caption{Vertical profiles of (a) time-averaged streamwise velocity $\left\langle u \right\rangle$ and (b) vertical velocity $\left\langle v \right\rangle$, (c) primary Reynolds shear stress $\left\langle u'v' \right\rangle$, and (d) turbulence kinetic energy $k$ from the H1.0-WR case, the H1.0-WM cases with the FEL model at $Re_h = 10595$.}
\label{fig:profile_HB_10595}
\end{figure}
\begin{figure}
\centering{\includegraphics[width=0.9\textwidth]{Fig_profile_err_legend.eps}}
\centering{\includegraphics[width=0.4\textwidth]{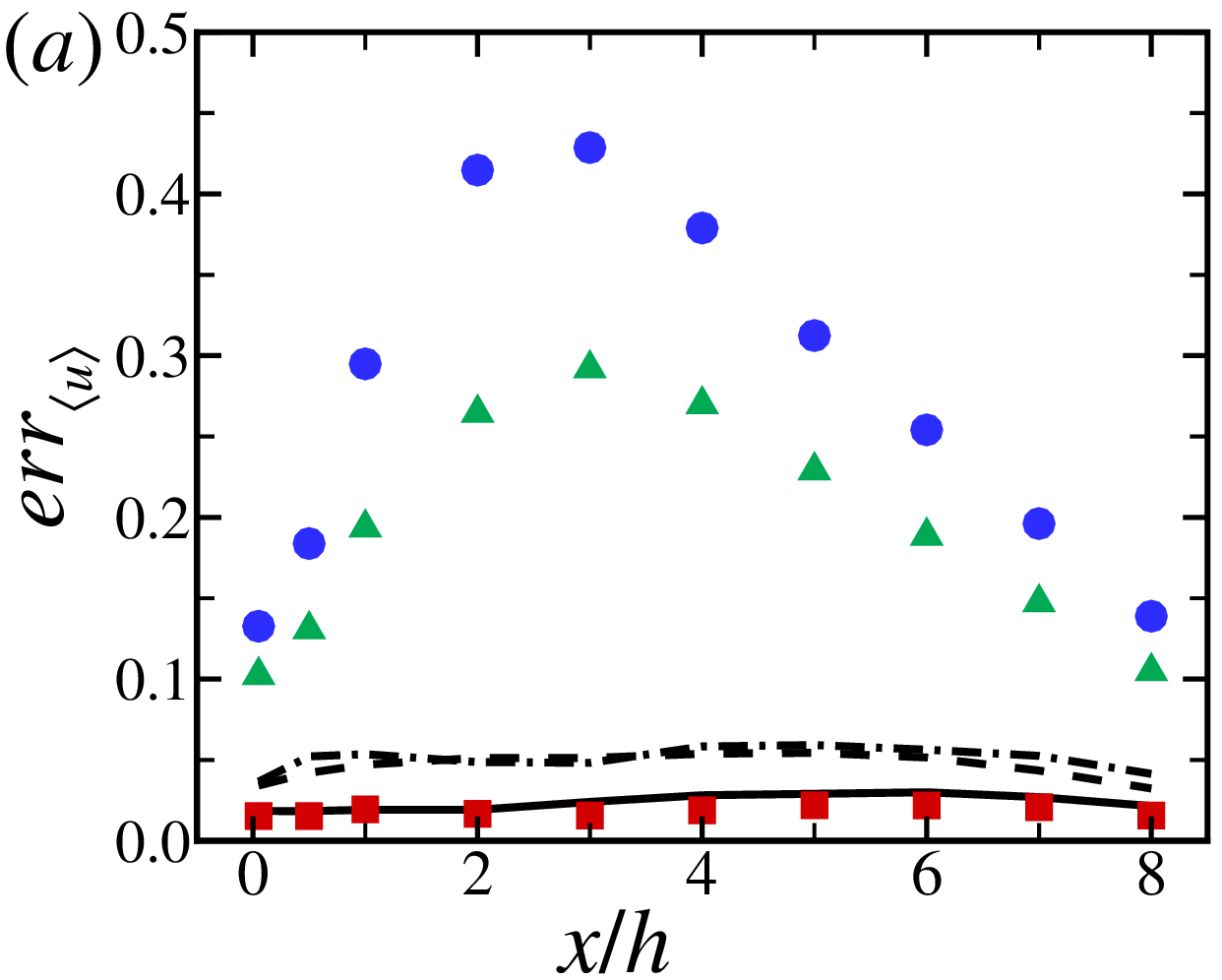}} \quad\quad
\centering{\includegraphics[width=0.4\textwidth]{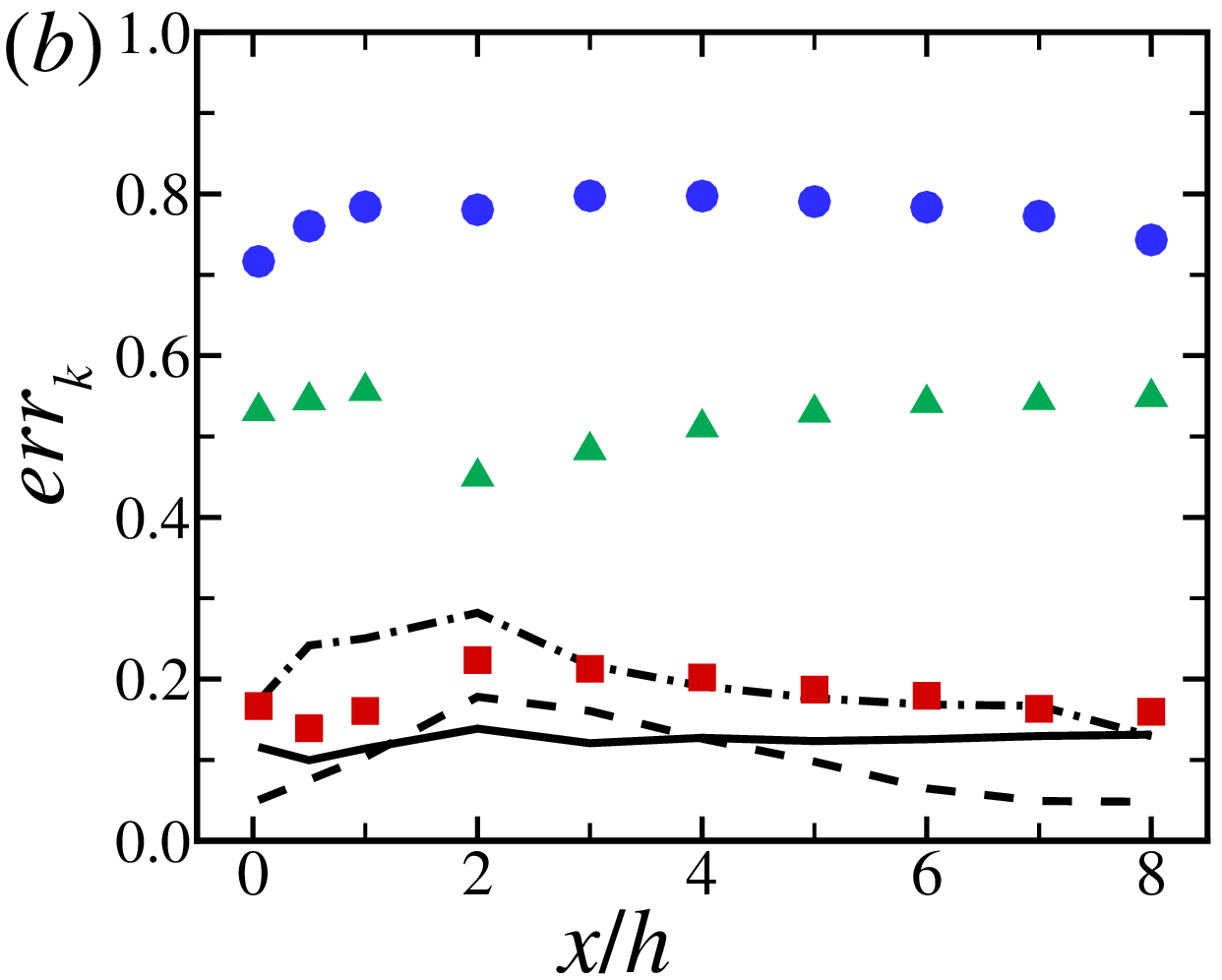}}
  \caption{Relative errors of (a) time-averaged streamwise velocity $\left\langle u \right\rangle$ and (b) turbulence kinetic energy $k$ between the WMLES and WRLES for H1.0 case at $Re_h = 10595$.}
\label{fig:profile_err_HB_10595}
\end{figure}
\begin{figure}
\centering{\includegraphics[width=0.48\textwidth]{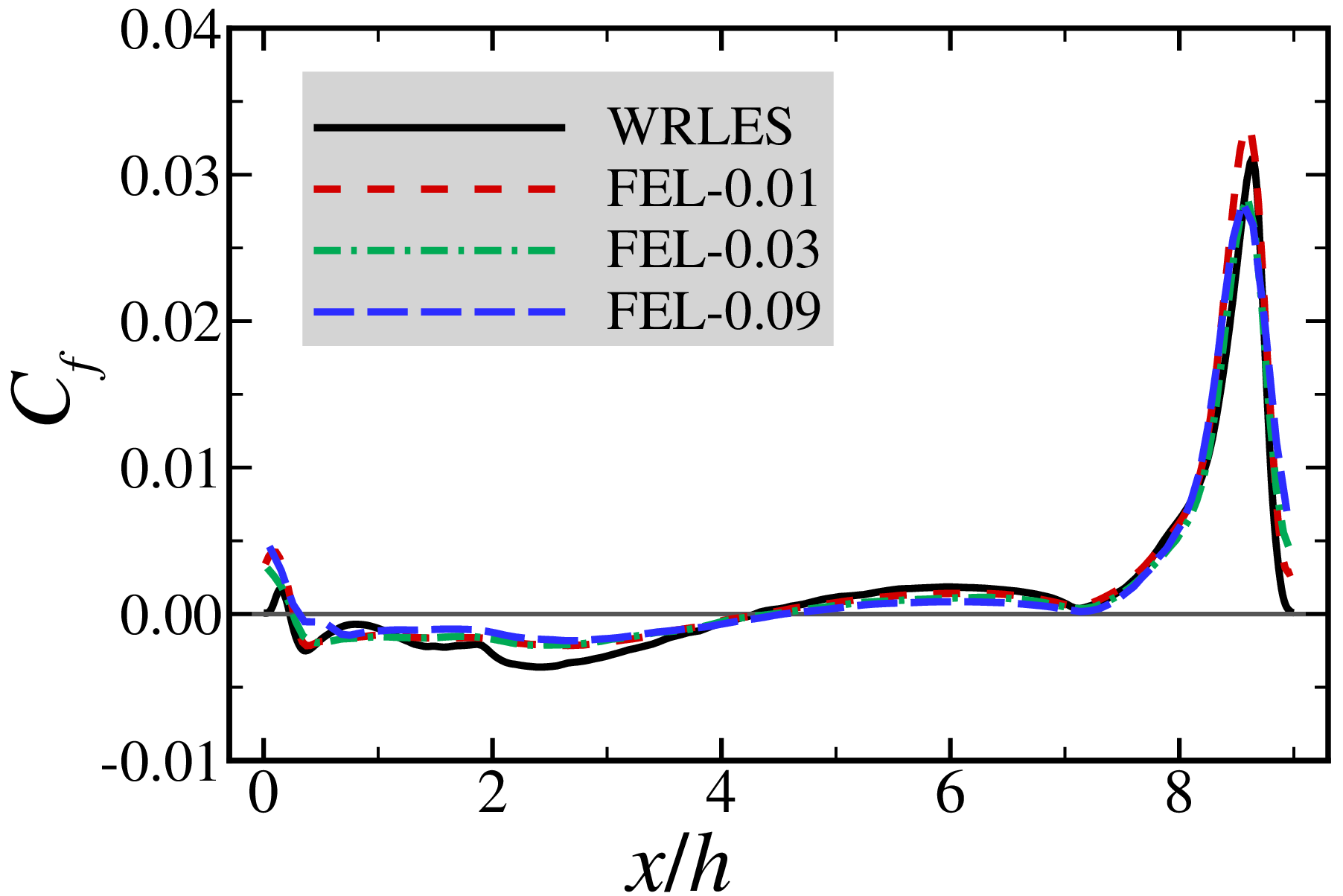}}
  \caption{Comparison of the time-averaged skin friction coefficient from the WMLES with the FEL model and the WRLES of H1.0 case at $Re_h = 10595$.}
\label{fig:Cf_HB}
\end{figure}

In this section, the performance of the FEL model is evaluated using the H1.0 case, which has a slope different from the training cases.

Figure~\ref{fig:HB_Re10595} shows the contours of time-averaged streamwise velocity with streamlines obtained from the FEL model for the H1.0 case with different grid resolutions. It is seen that the FEL and WW models have similar predictions of the separation bubble for the grid with $\Delta y_f/h=0.01$. For the coarser grids with $\Delta y_f/h=0.03$ and 0.09, the separation bubble predicted by the FEL model is close to those from the WRLES, with large discrepancies observed for the WW model.
%the separation bubble predicted by 
%quickly diminishes and eventually disappears.
%However, the predictions of the cDK-embedded model remain close to those from the WRLES. 
%The streamwise locations of the separation and reattachment points can be examined in the $7^{th} \sim 10^{th}$ cases of table~\ref{tab:PH_case}. 
%It can be observed that the separation point predicted by the cDK-embedded model gradually approaches that from the WRLES as the value of $\Delta y_c^+$ decreases. 
%This indicates that the cDK-embedded model is able to capture the location of separation more accurately with finer grid resolutions.

Figure~\ref{fig:profile_HB_10595} compares the vertical profiles of the flow statistics obtained from the FEL model with the WRLES results for the H1.0-WM-0.01/0.03/0.09 cases with $Re_h = 10595$ for various streamwise locations. Good agreements with the WRLES results are observed for all the three grid resolutions, although somewhat discrepancies are observed for TKE $k$ in the lower half of the vertical profiles.

The relative errors in $\left\langle u \right\rangle$ and $k$ predicted by the FEL model and the WW model are shown in figure~\ref{fig:profile_err_HB_10595}. As observed, the error is less than 5\% (sometimes even 2\%) for $\left\langle u \right\rangle$, and the maximum error is around 20\% for $k$, respectively, for the cases with different grid resolutions for the FEL model. The errors for the WW model, on the other hand, are significantly higher in comparison with the proposed model.
%
%exhibits errors in $\left\langle u \right\rangle$ and $k$ that closely match those of the cDK-embedded model only at the grid resolution of $\Delta y_f/h=0.01$. However, the errors rapidly escalate to (25\%, 50\%) and (40\%, 80\%) at $\Delta y_f/h=0.03$ and $0.09$, respectively. 
%Consequently, the cDK-embedded model demonstrates robust performance in simulating the flow over periodic hills with out-of-distribution geometry.

The mean skin friction coefficients $C_f$ predicted by the FEL model are compared with WRLES predictions in figure~\ref{fig:Cf_HB} for the three grid resolutions. It is evident that the predictions from the FEL model closely align with the WRLES results across the three grid resolutions.
%This agreement demonstrates that the FNN\_PH-LoW model is capable of accurately capturing the wall boundary condition based on the precise prediction of flow fields.

%{\color{black}After evaluating the prediction accuracy of the FEL model, the relative significance of the wall shear stress submodel and eddy viscosity submodel is also assessed in the H1.0 case at $Re_h = 10595$, as shown in appendix~\ref{appendix:viscosity}.}

\subsection{Cases with different Reynolds numbers}\label{subsec:Application_Re}

\begin{figure}
\centering
	\begin{subfigure}[b]{0.48\textwidth}
	\centering
	\includegraphics[width = 1.0\textwidth]{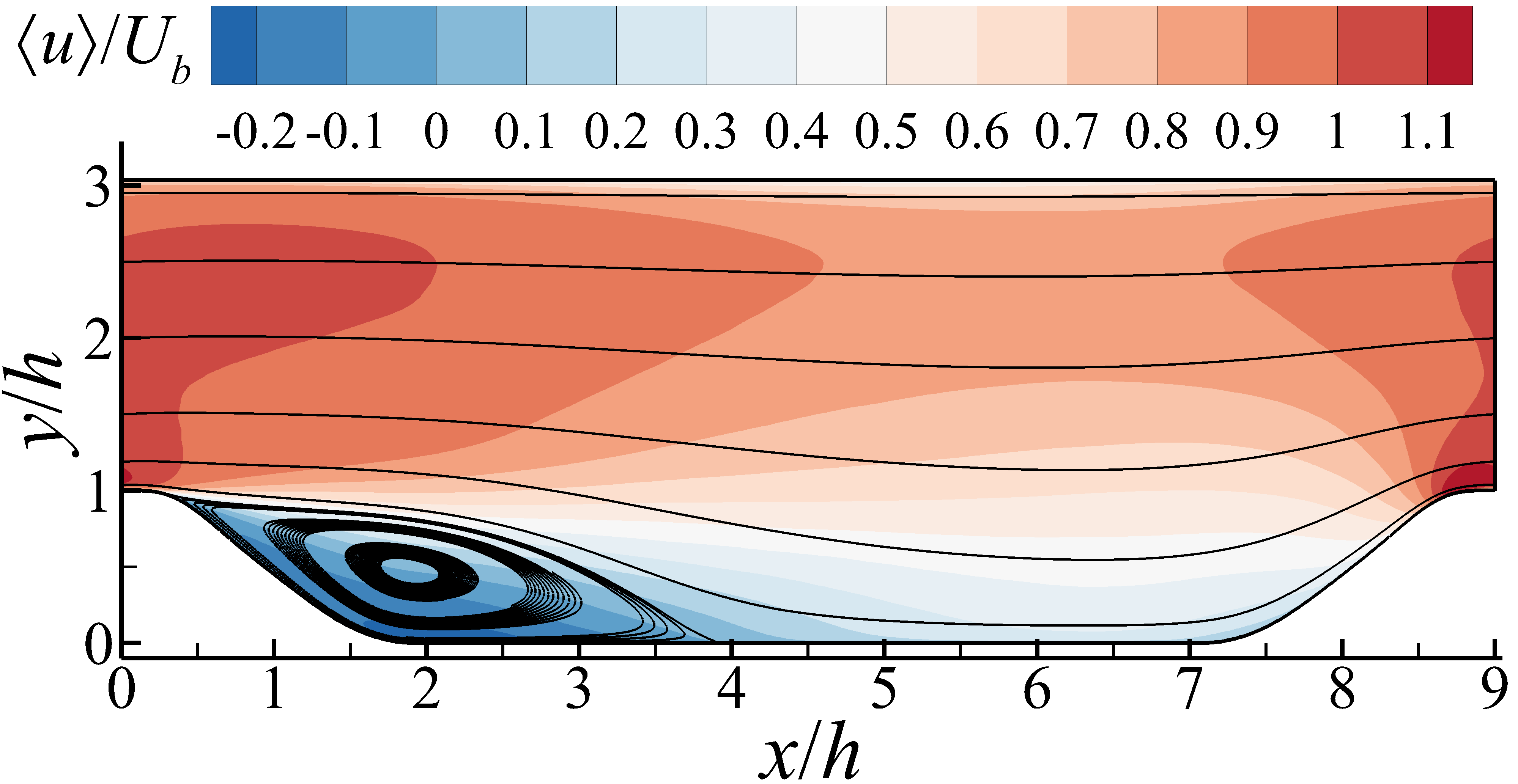}
	\subcaption{H1.0-WR: $\Delta y_f/h=0.003$}
	\end{subfigure} \\
    \begin{subfigure}[b]{0.48\textwidth}
	\centering
	\includegraphics[width = 1.0\textwidth]{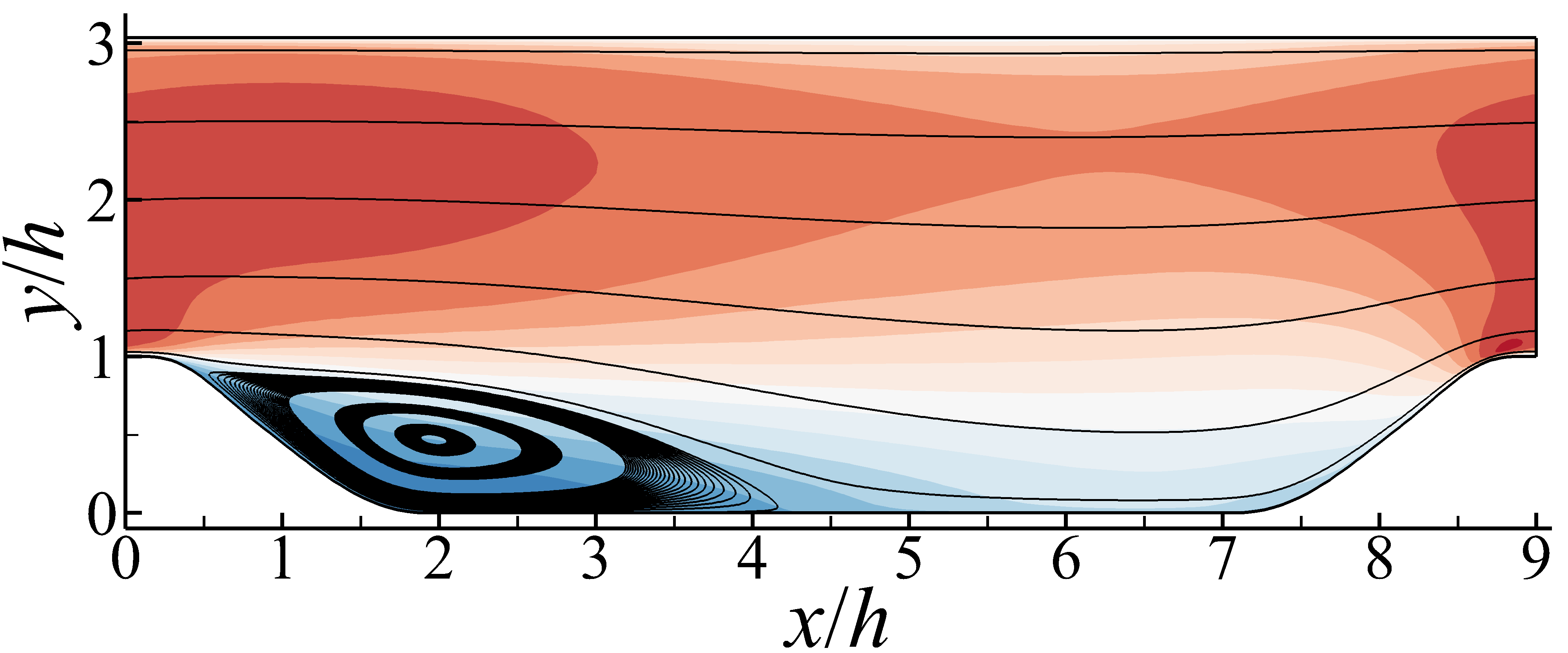}
	\subcaption{FEL: $\Delta y_f/h=0.01$}
	\end{subfigure}
	\begin{subfigure}[b]{0.48\textwidth}
	\centering
	\includegraphics[width = 1.0\textwidth]{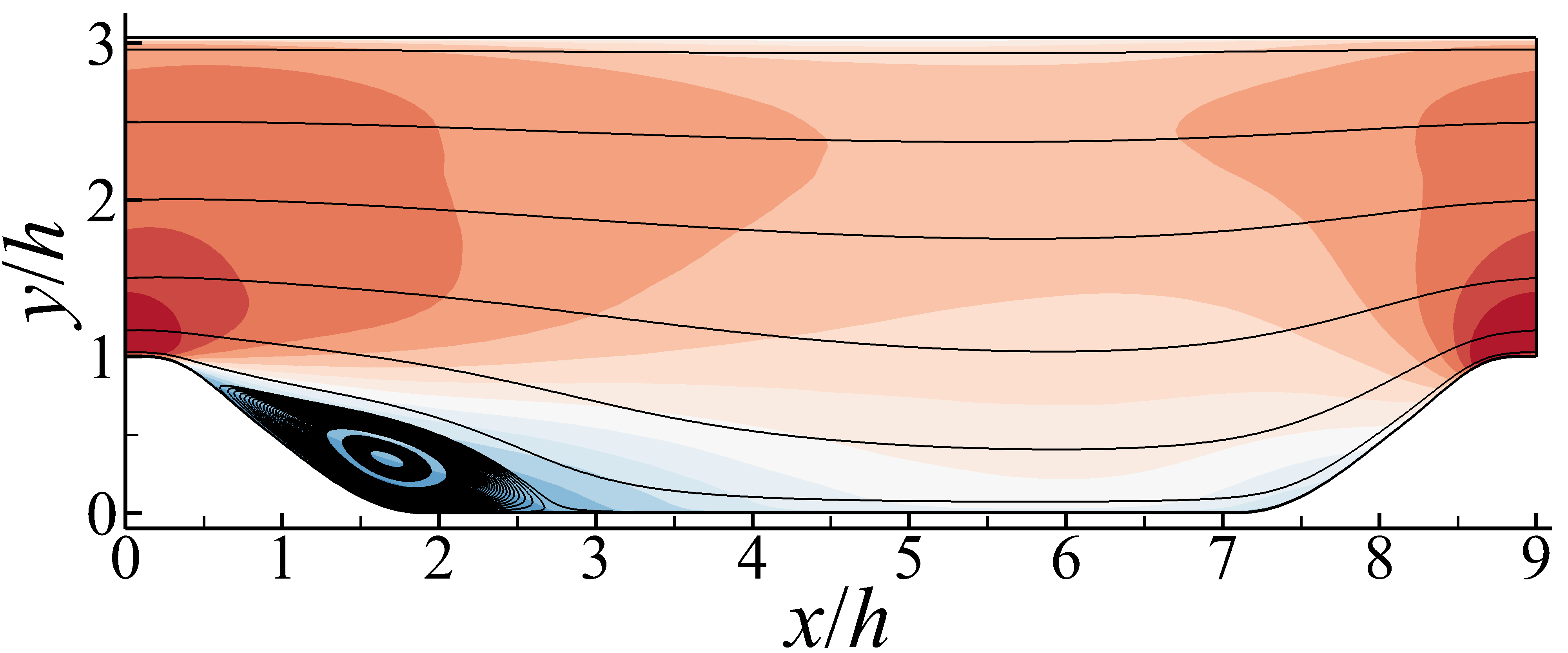}
	\subcaption{WW: $\Delta y_f/h=0.01$}
	\end{subfigure}
    \begin{subfigure}[b]{0.48\textwidth}
	\centering
	\includegraphics[width = 1.0\textwidth]{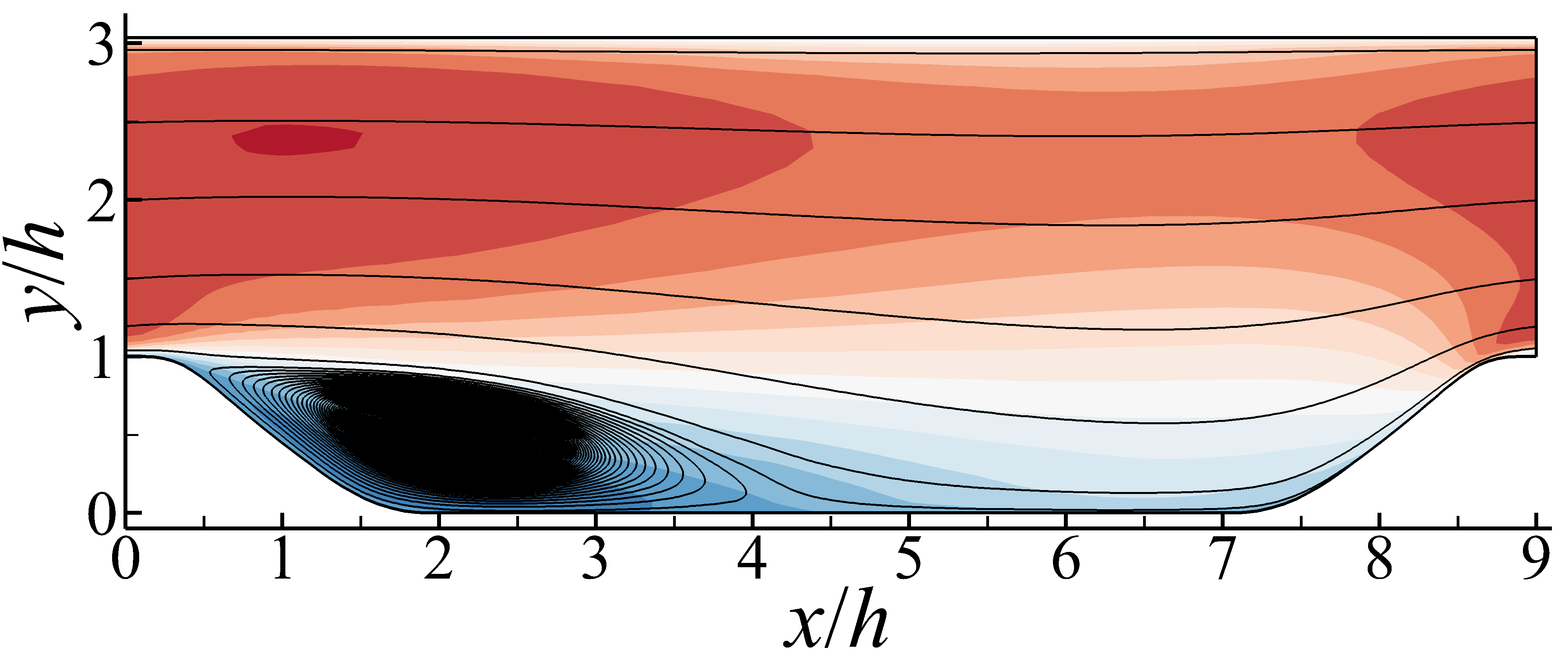}
	\subcaption{FEL: $\Delta y_f/h=0.03$}
	\end{subfigure}
	\begin{subfigure}[b]{0.48\textwidth}
	\centering
	\includegraphics[width = 1.0\textwidth]{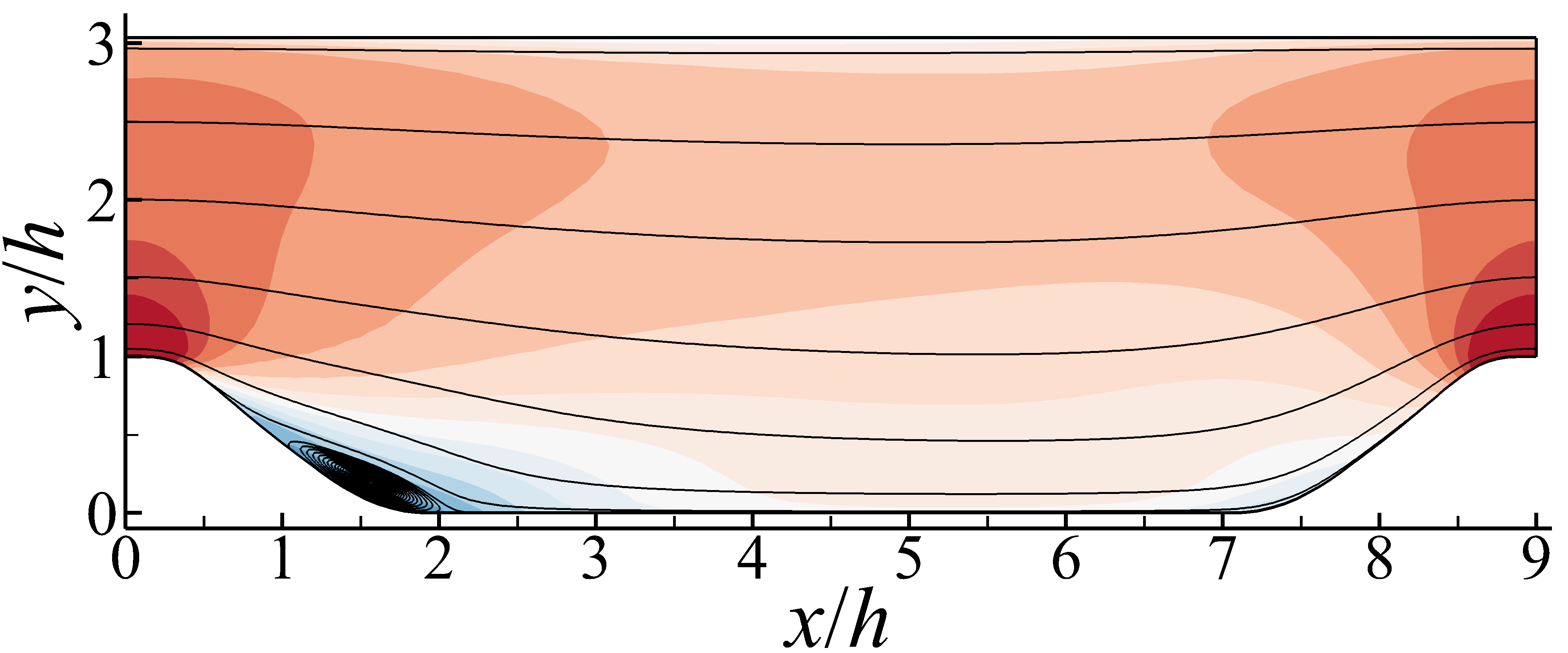}
	\subcaption{WW: $\Delta y_f/h=0.03$}
	\end{subfigure}
    \begin{subfigure}[b]{0.48\textwidth}
	\centering
	\includegraphics[width = 1.0\textwidth]{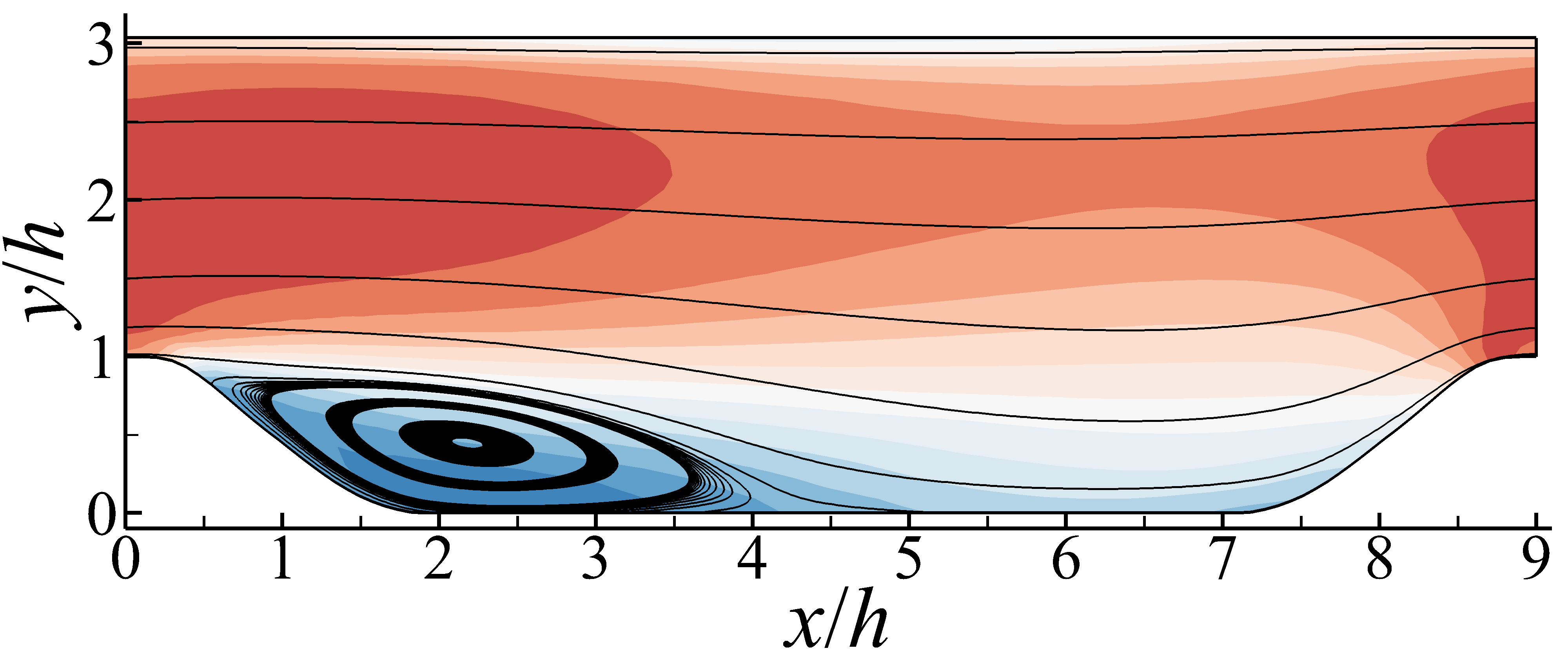}
	\subcaption{FEL: $\Delta y_f/h=0.09$}
	\end{subfigure}
	\begin{subfigure}[b]{0.48\textwidth}
	\centering
	\includegraphics[width = 1.0\textwidth]{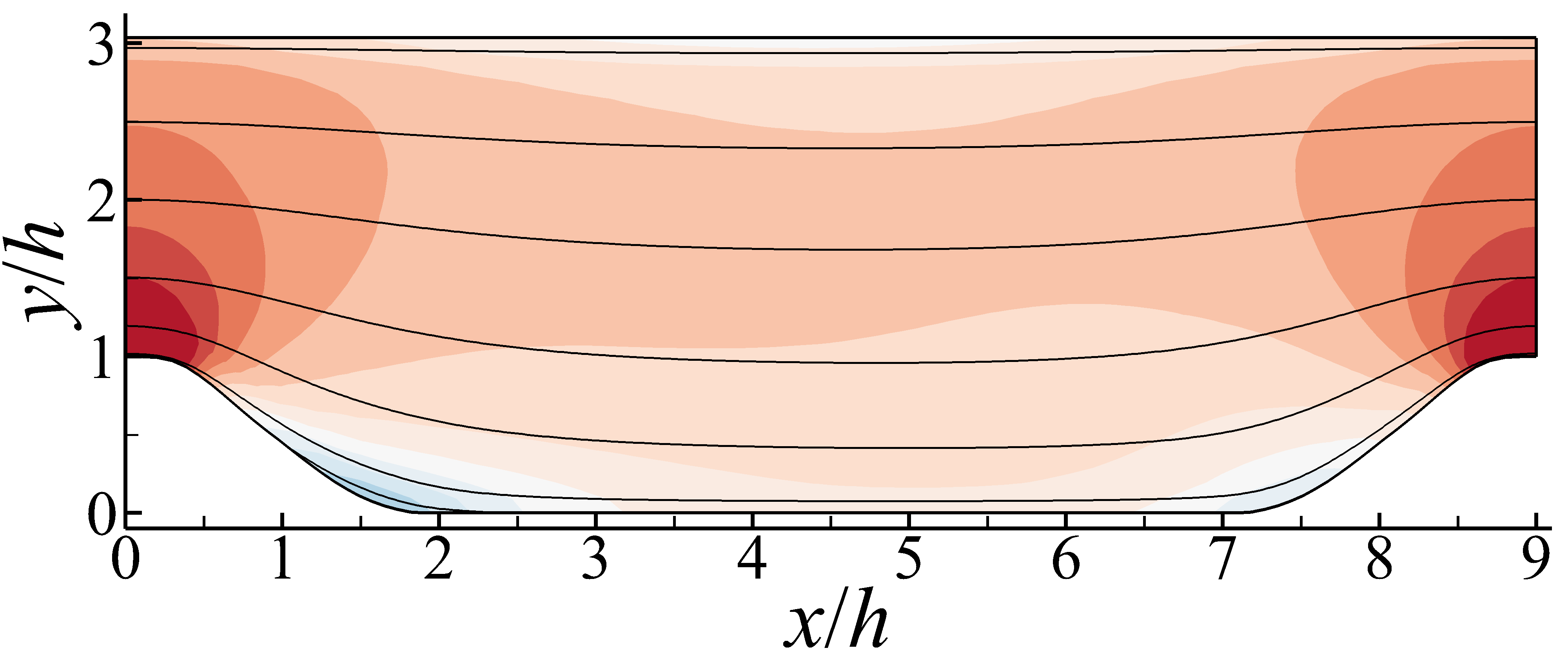}
	\subcaption{WW: $\Delta y_f/h=0.09$}
	\end{subfigure}
\caption{Contours of time-averaged streamwise velocity with streamlines obtained from the H1.0-WR and H1.0-WM cases with the FEL and WW models at $Re_h = 37000$.}
\label{fig:HB_Re37000}
\end{figure}
\begin{figure}
\centering{\includegraphics[width=0.9\textwidth]{Fig_profile_HSHL_legend.eps}}
\centering{\includegraphics[width=0.495\textwidth]{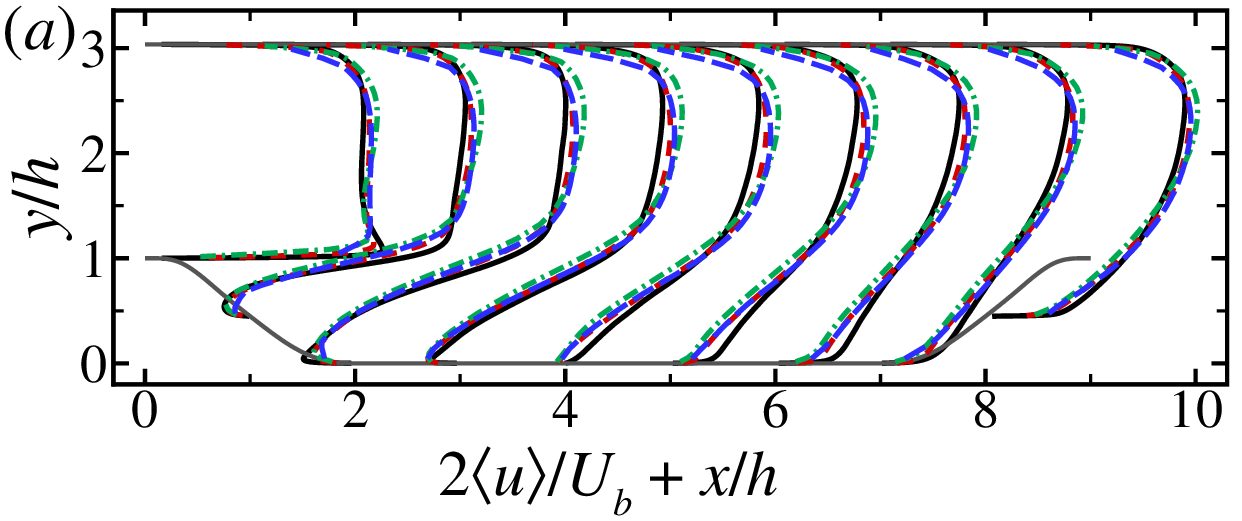}}
\centering{\includegraphics[width=0.495\textwidth]{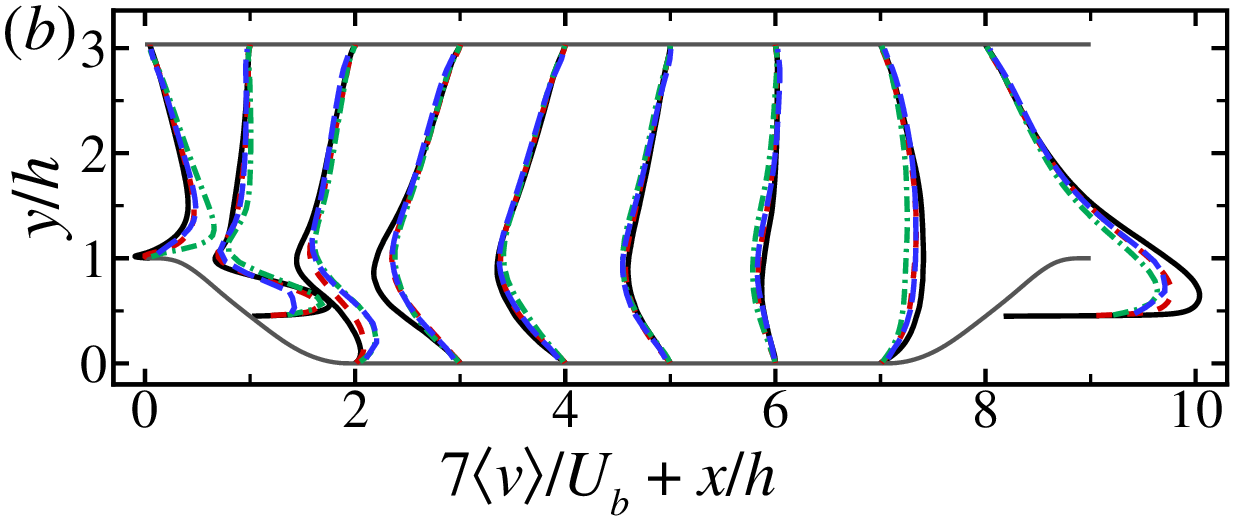}}
\centering{\includegraphics[width=0.495\textwidth]{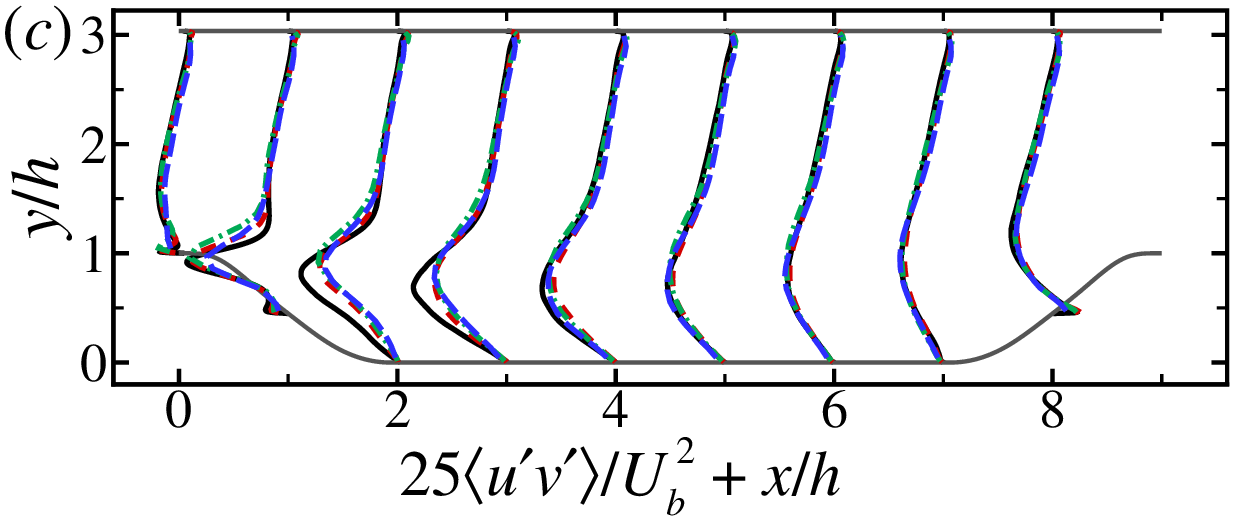}}
\centering{\includegraphics[width=0.495\textwidth]{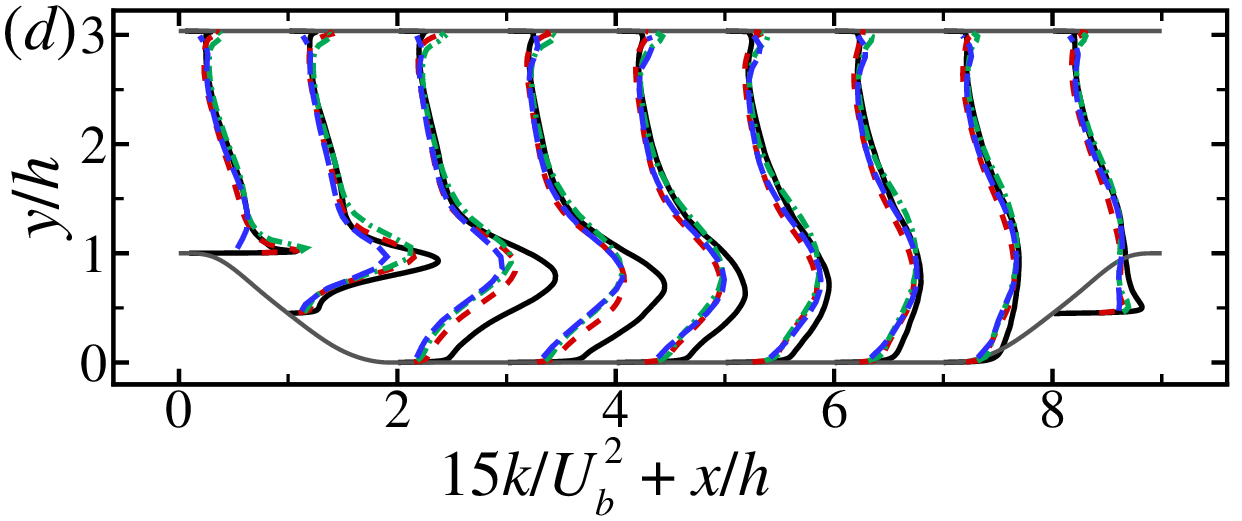}}
  \caption{Vertical profiles of the (a) time-averaged streamwise velocity $\left\langle u \right\rangle$ and (b) vertical velocity $\left\langle v \right\rangle$, (c) primary Reynolds shear stress $\left\langle u'v' \right\rangle$, and (d) turbulence kinetic energy $k$ from the H1.0-WR and H1.0-WM cases with the FEL model at $Re_h = 37000$.}
\label{fig:profile_HB_Re37000}
\end{figure}
\begin{figure}
\centering{\includegraphics[width=0.9\textwidth]{Fig_profile_err_legend.eps}}
\centering{\includegraphics[width=0.4\textwidth]{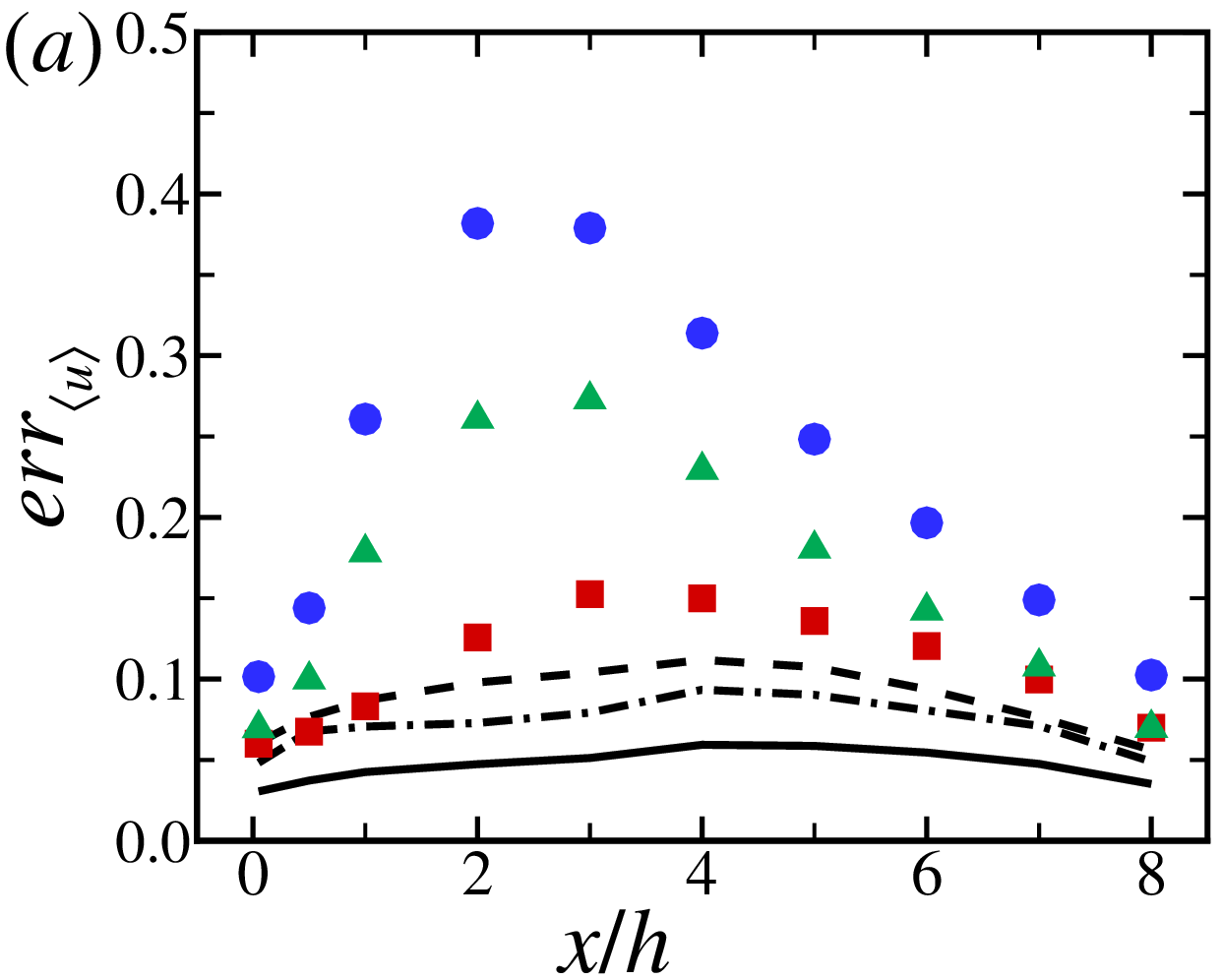}}\quad\quad
\centering{\includegraphics[width=0.4\textwidth]{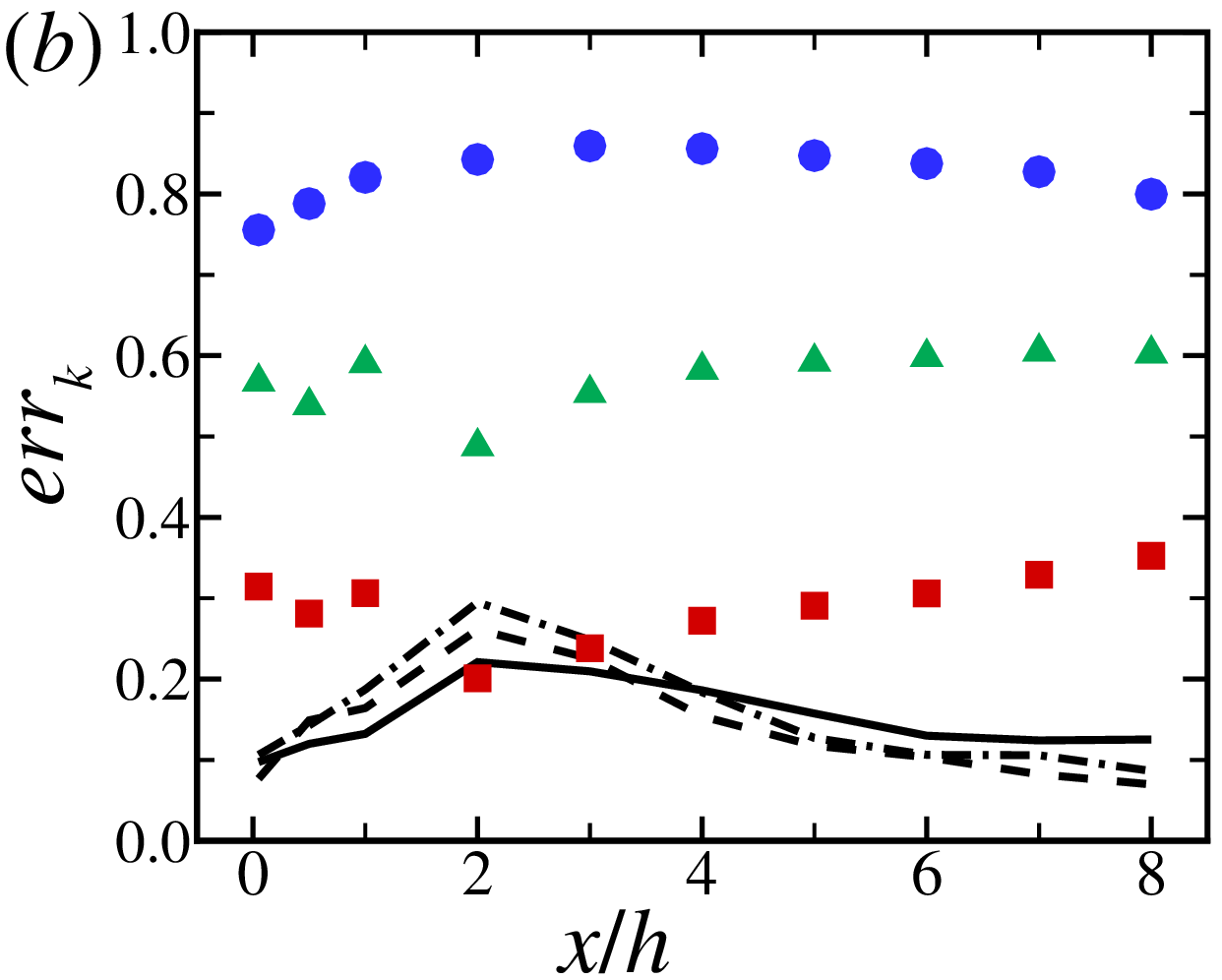}}
  \caption{Relative errors of (a) time-averaged streamwise velocity $\left\langle u \right\rangle$ and (b) turbulence kinetic energy $k$ between the WMLES and WRLES for H1.0 case at $Re_h = 37000$.}
\label{fig:profile_err_HB_Re37000}
\end{figure}
\begin{figure}
\centering{\includegraphics[width=0.48\textwidth]{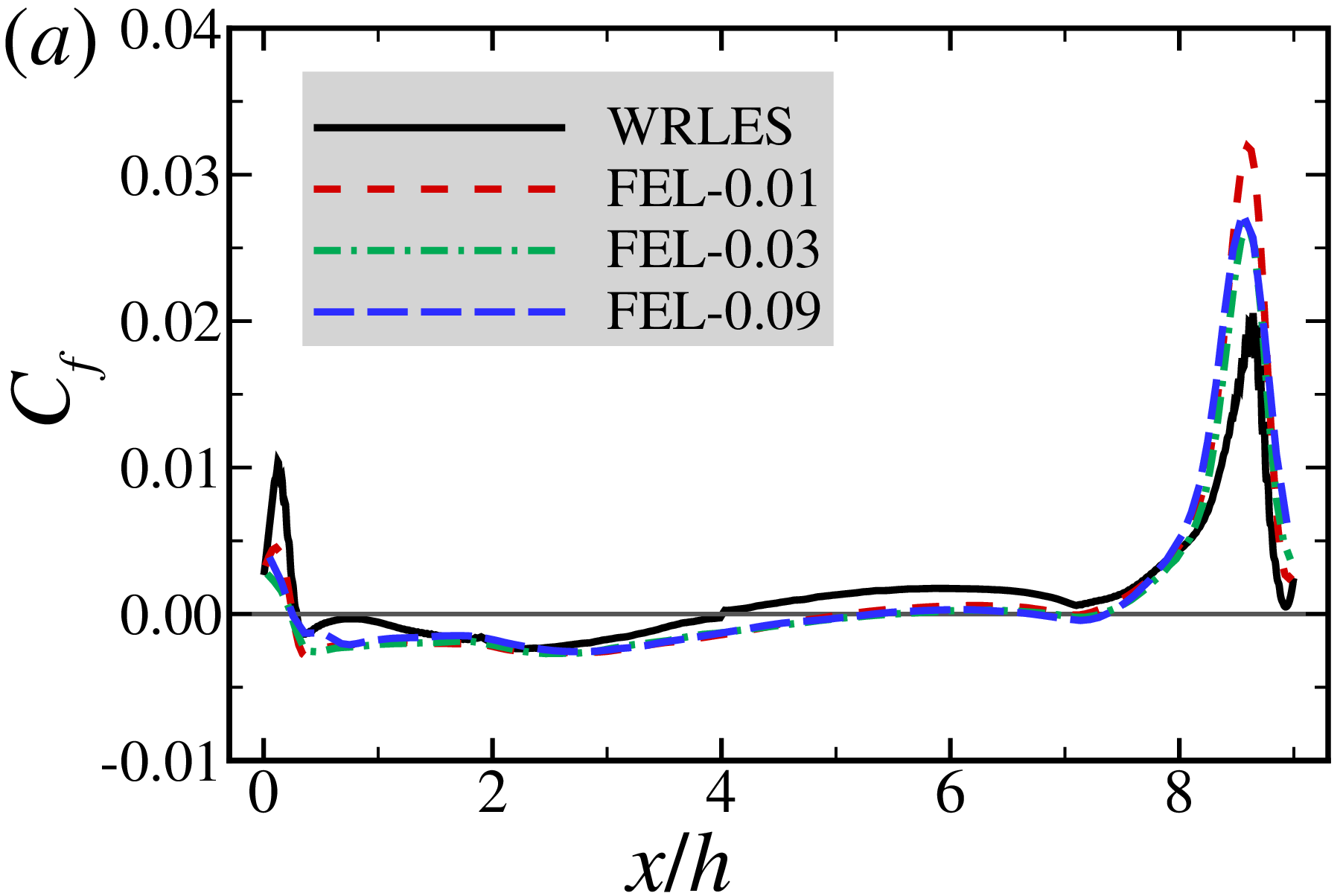}} \;
\centering{\includegraphics[width=0.48\textwidth]{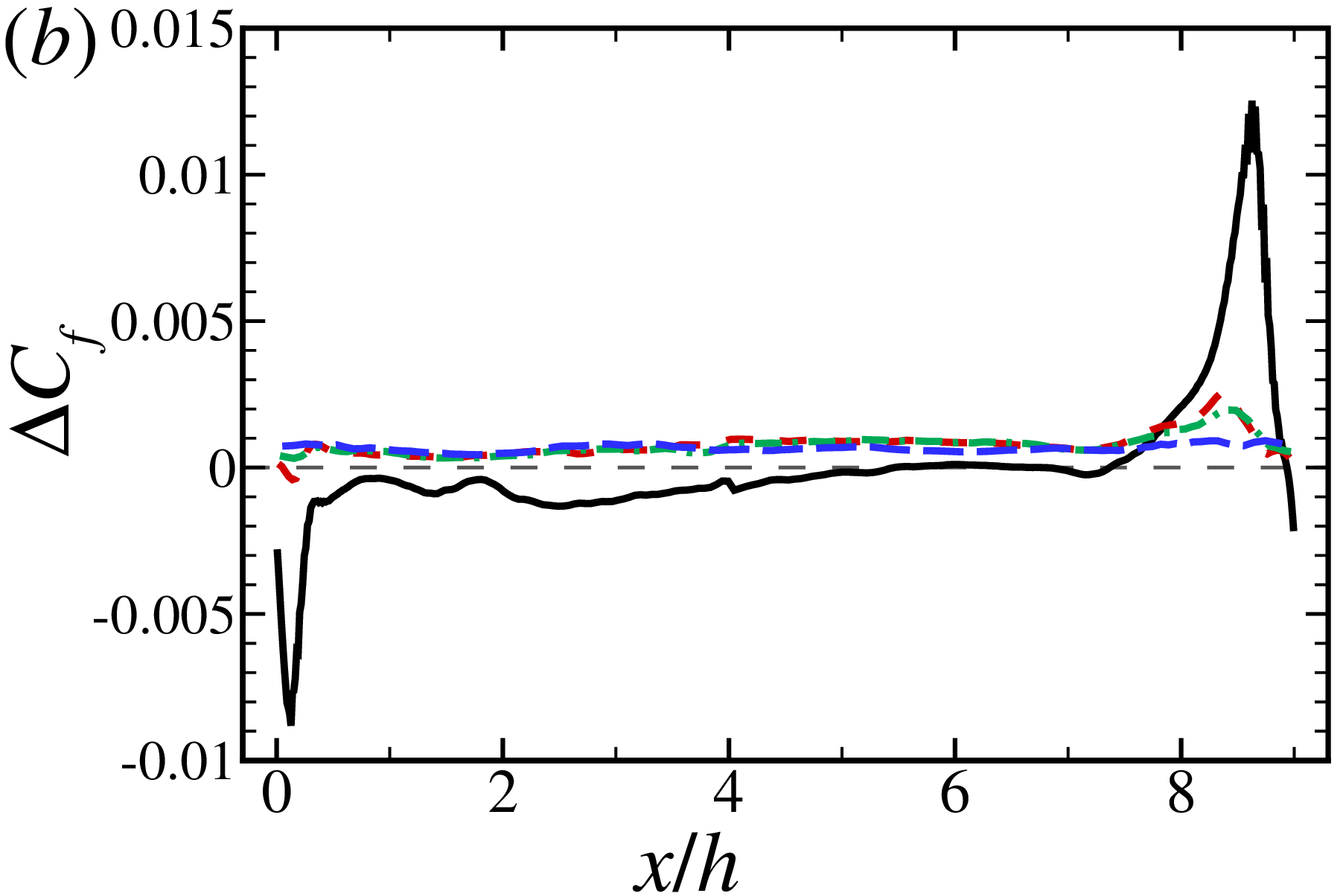}}
  \caption{{\color{black}(a) The time-averaged skin friction coefficient from the WMLES with the FEL model and the WRLES of H1.0 case at $Re_h = 37000$ and (b) the difference $\Delta C_f = C_f (Re_h=10595) - C_f (Re_h=37000)$.}}
\label{fig:Cf_HB_err}
\end{figure}

In this section, the generalization ability of the FEL model is tested for a higher Reynolds number (i.e., $Re_h = 37000$) using the H1.0 case. 

The contours of time-averaged streamwise velocity obtained from the FEL model are compared with WRLES in figure~\ref{fig:HB_Re37000}. It is seen that the FEL model successfully predicts the separation bubble for the three grid resolutions for this higher Reynolds number case. On the other hand, the WW model significantly underpredicts the size of the separation bubble. 
%gradually decreases as $\Delta y_c^+$ increases and even disappears at $\Delta y_c^+ = 45$ (figure~\ref{fig:HB_Re37000}g). 
%Importantly, the cDK-embedded model accurately predicts the separation point at grid resolutions of $\Delta y_c^+=5$ and $15$, as demonstrated in the $14^{th}$ and $15^{th}$ cases of table~\ref{tab:PH_case}.

The vertical profiles of the flow statistics predicted by the FEL model are shown in figure~\ref{fig:profile_HB_Re37000} for the high Reynolds number case. Good agreement with the WRLES results is observed for all the three grid resolutions. The relative errors in $\left\langle u \right\rangle$ and $k$ are shown in figure~\ref{fig:profile_err_HB_Re37000} for the FEL model and the WW model. As seen, the errors in $\left\langle u \right\rangle$ are less than 10\%, and the errors in $k$ are less than 20\%, respectively, for the three grid resolutions for the FEL model. In comparison, the errors of the WW model predictions are relatively larger.

{\color{black}The mean skin friction coefficients $C_f$ predicted by the FEL model are compared with the WRLES predictions in figure~\ref{fig:Cf_HB_err}, with the difference $\Delta C_f = C_f (Re_h=10595) - C_f (Re_h=37000)$ calculated to show the Reynolds number effect. It is seen that the magnitudes of both the maximum $C_f$ at the windward of the hill ($x/h \approx 8.5$) and the minimum $C_f$ in the recirculation zone ($x/h \approx 2\sim5$) decrease while the magnitude of the second maximum $C_f$ close to the separation point increases with the increase of Reynolds number. The FEL model succeeds in predicting the variation of the maximum $C_f$ with the Reynolds number, but does not predict the variations with Reynolds numbers at other locations.
%basically predicts the variation trend of $C_f$ along the streamwise location, but the Reynolds number effect on the peak values cannot be well captured.
}

\subsection{Assessment of the proposed eddy viscosity model}\label{subsec:viscosity}
\begin{figure}
\centering{\includegraphics[width=0.75\textwidth]{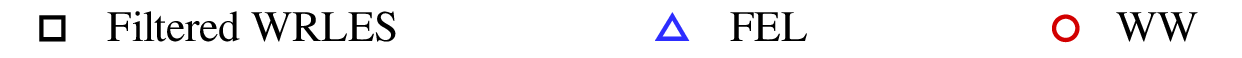}}
\centering{\includegraphics[width=0.48\textwidth]{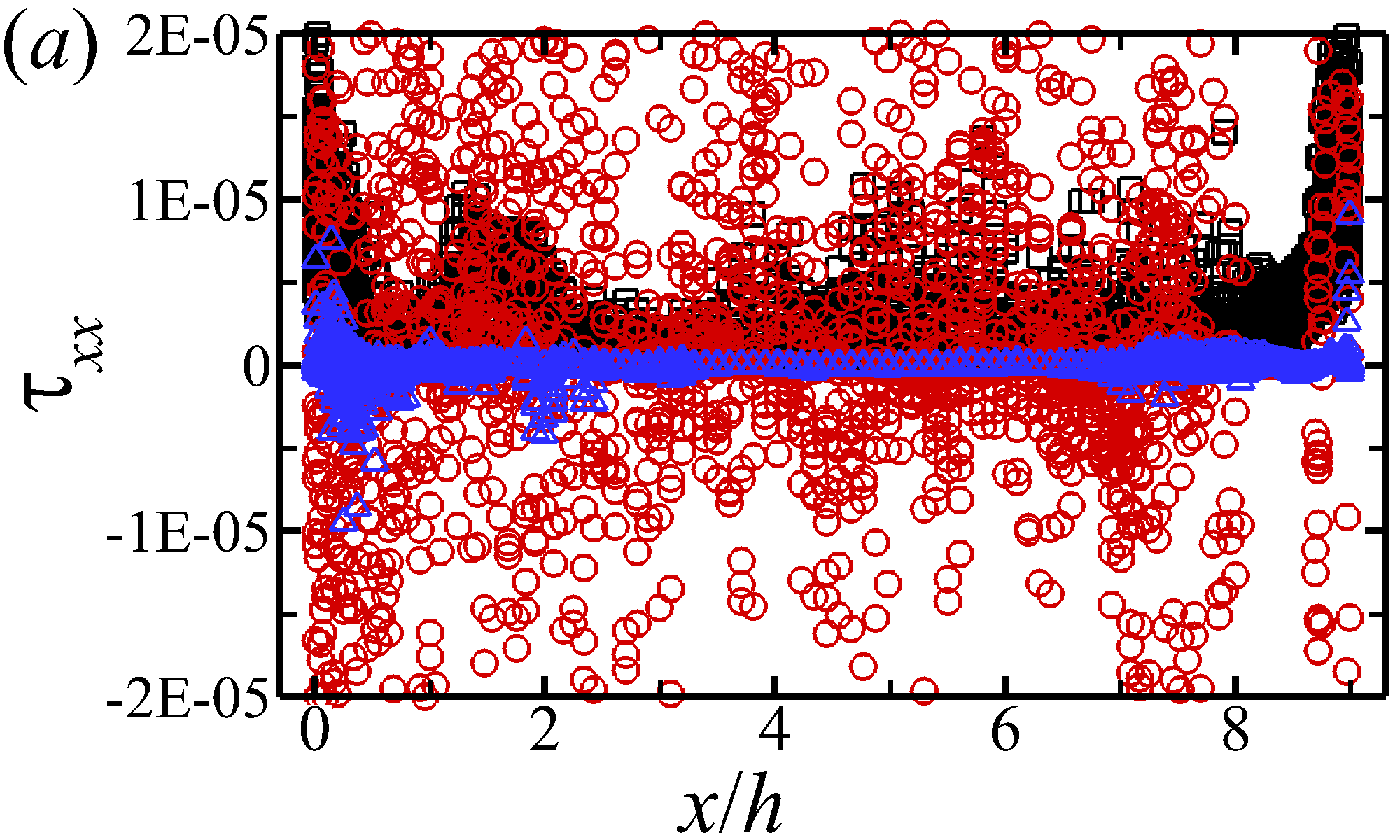}}\quad
\centering{\includegraphics[width=0.48\textwidth]{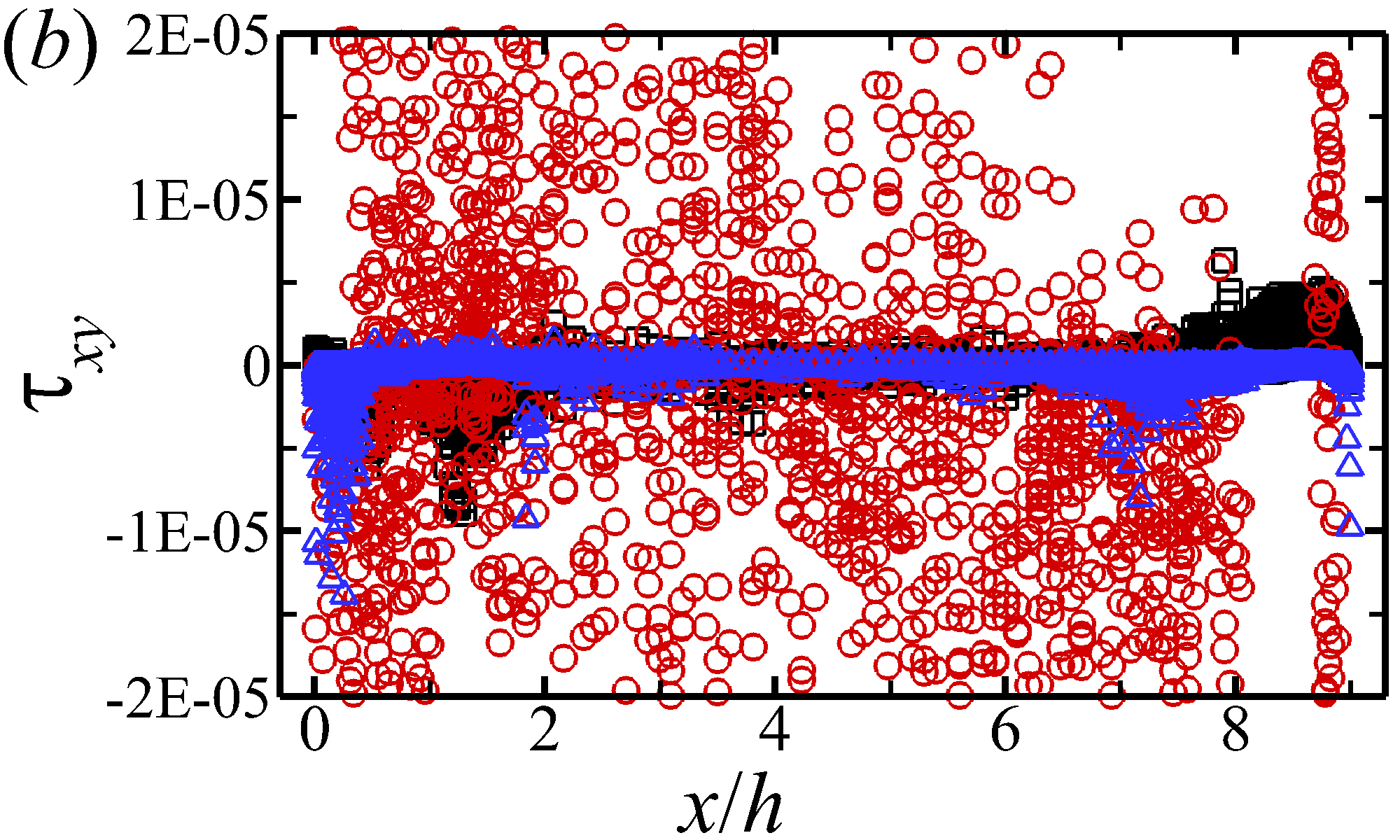}}
\centering{\includegraphics[width=0.48\textwidth]{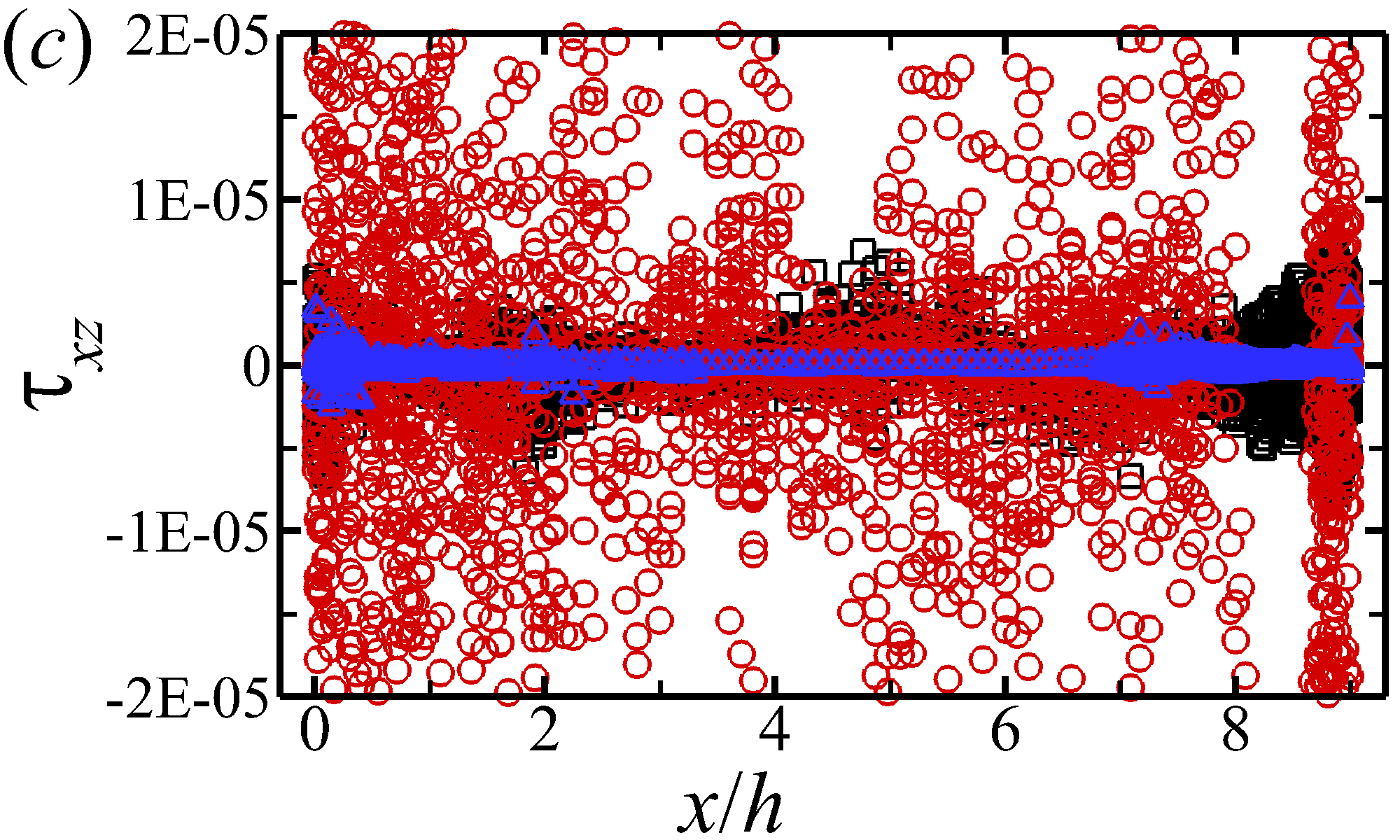}}\quad
\centering{\includegraphics[width=0.48\textwidth]{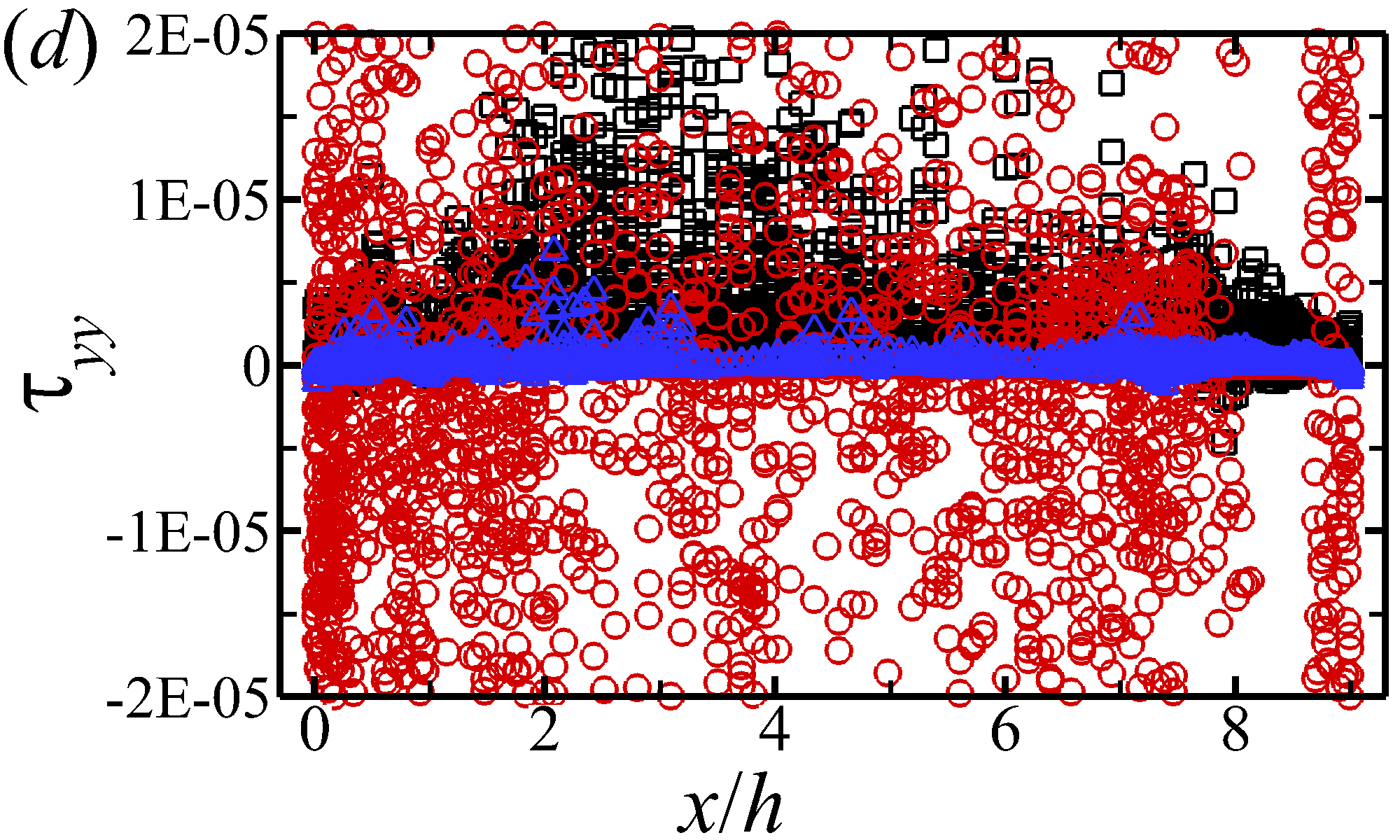}}
\centering{\includegraphics[width=0.48\textwidth]{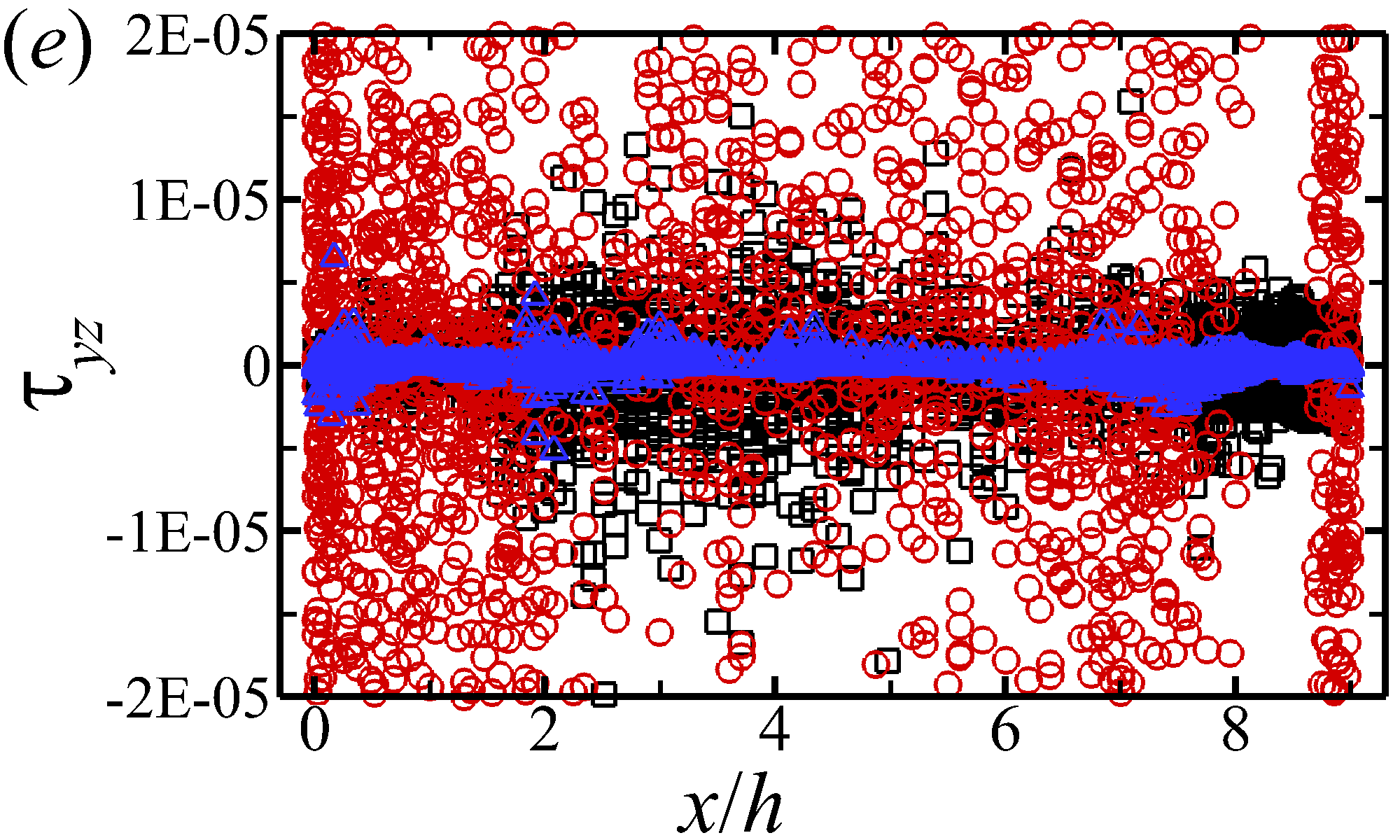}}\quad
\centering{\includegraphics[width=0.48\textwidth]{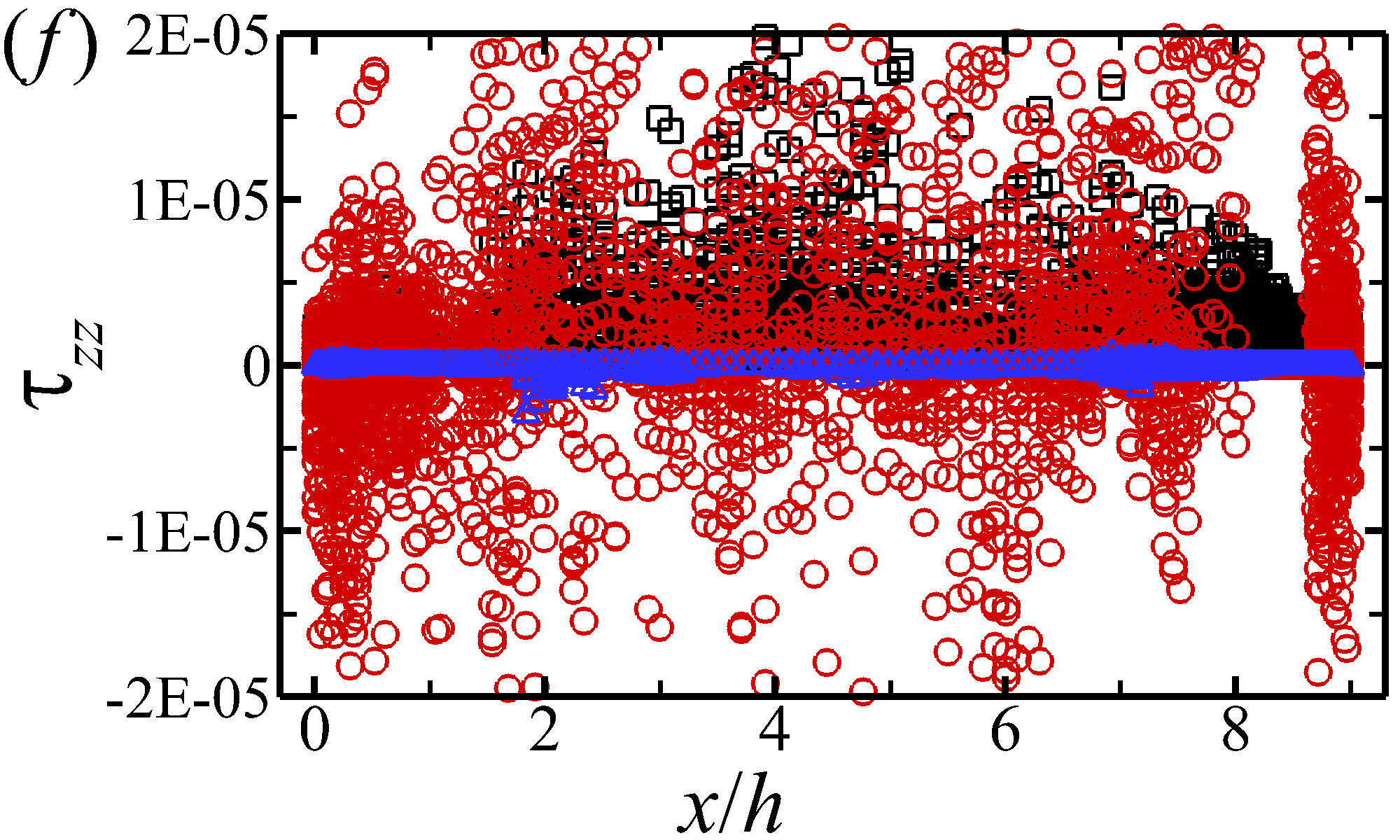}}
  \caption{SGS stresses at the first grid in the vertical direction from an instantaneous snapshot of flow field: (a) $\tau_{xx}$, (b) $\tau_{xy}$, (c) $\tau_{xz}$, (d) $\tau_{yy}$, (e) $\tau_{yz}$, (f) $\tau_{zz}$. {\color{black}Here, $U_b^2$ is used for non-dimensionalization.}}
\label{fig:filter_tau_ij}
\end{figure}
\begin{figure}
\centering{\includegraphics[width=0.75\textwidth]{Fig_Filter_scatter_legend.eps}}
\centering{\includegraphics[width=0.5\textwidth]{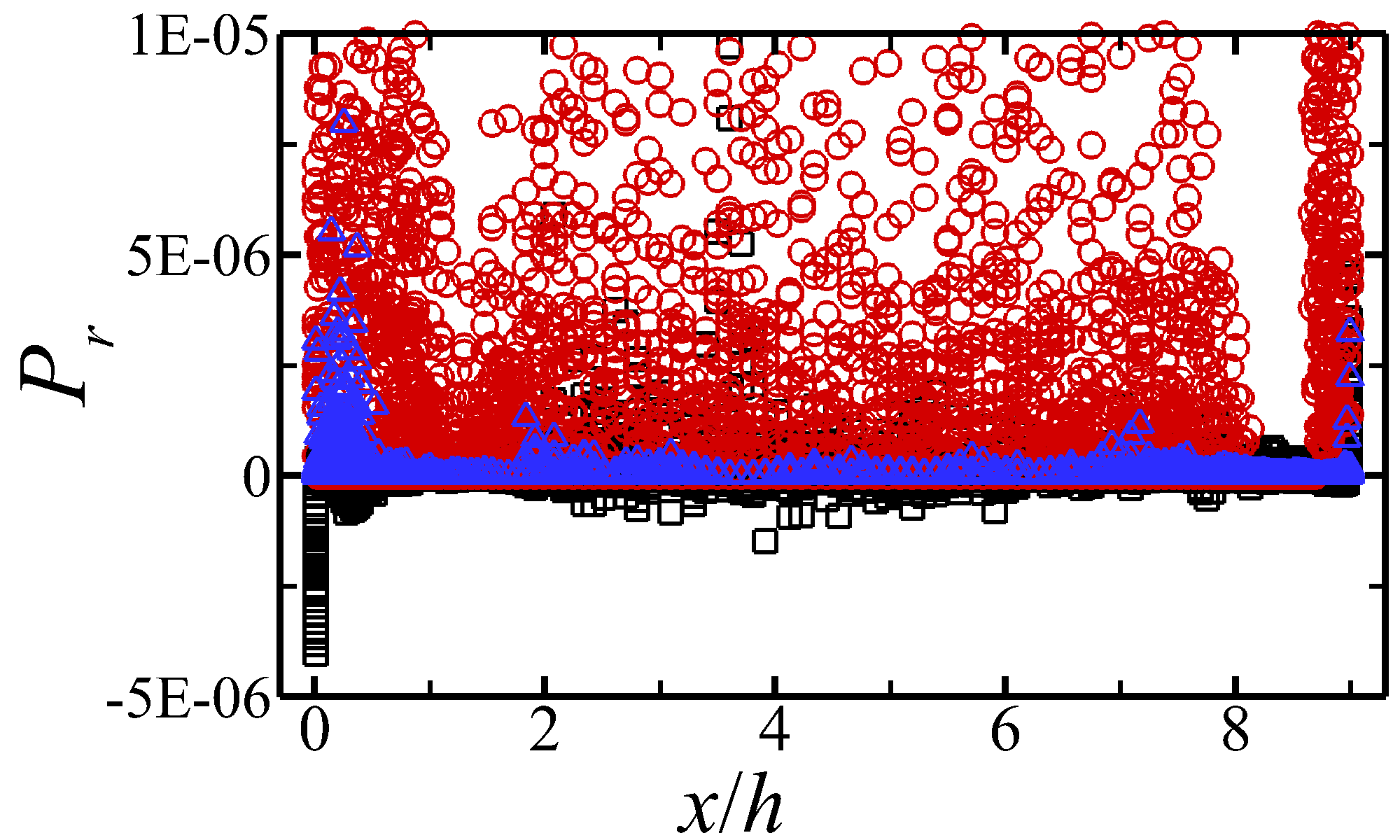}}
  \caption{Rate of energy transfer $P_r$ at the first grid in the vertical direction from an instantaneous snapshot of flow field, {\color{black}where $U_b^3/h$ is used for non-dimensionalization.}}
\label{fig:filter_Pr}
\end{figure}
\begin{figure}
\centering{\includegraphics[width=0.82\textwidth]{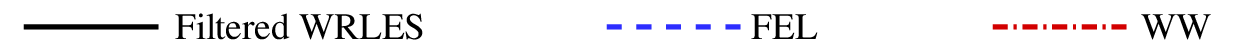}}
\centering{\includegraphics[width=0.45\textwidth]{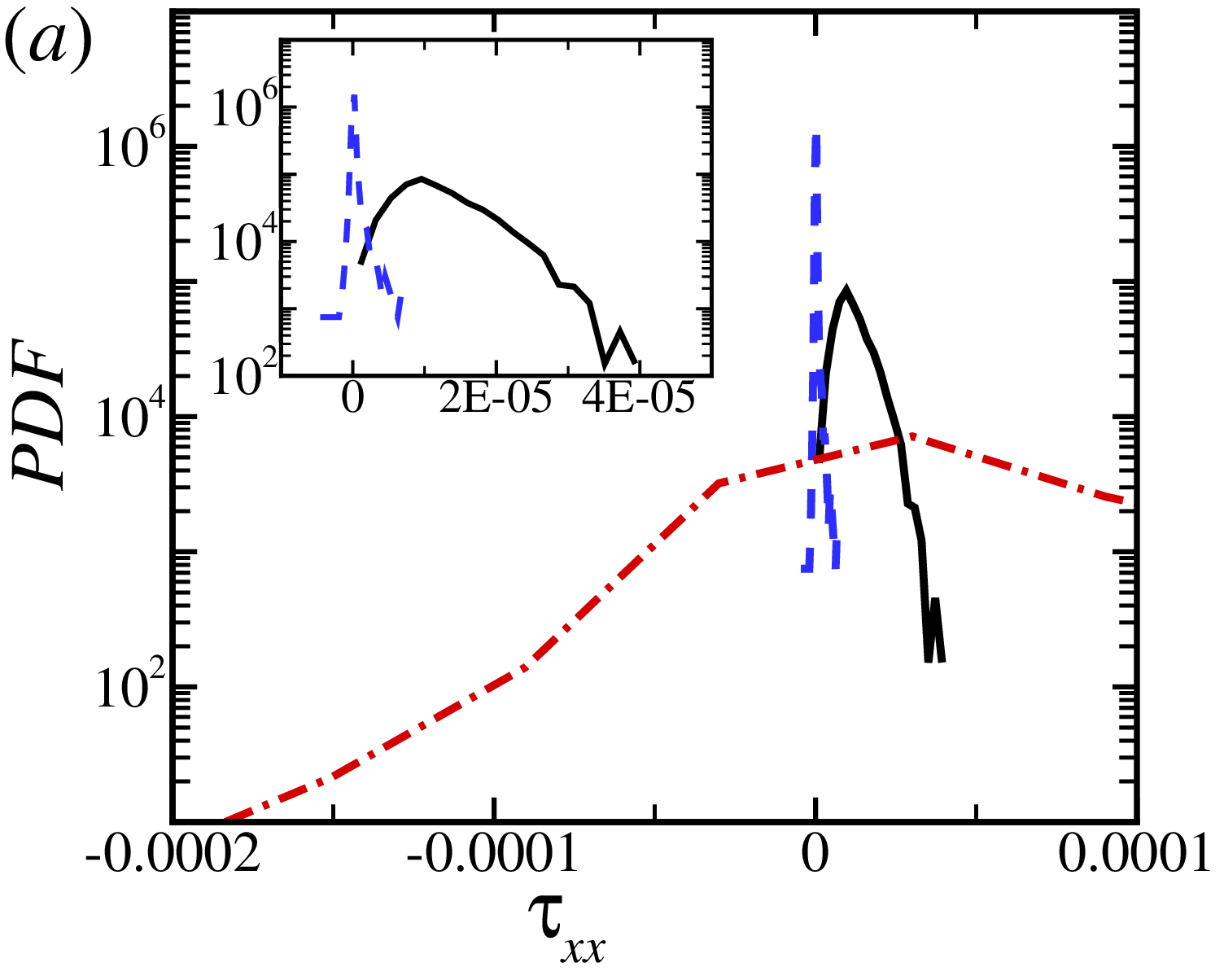}}
\centering{\includegraphics[width=0.45\textwidth]{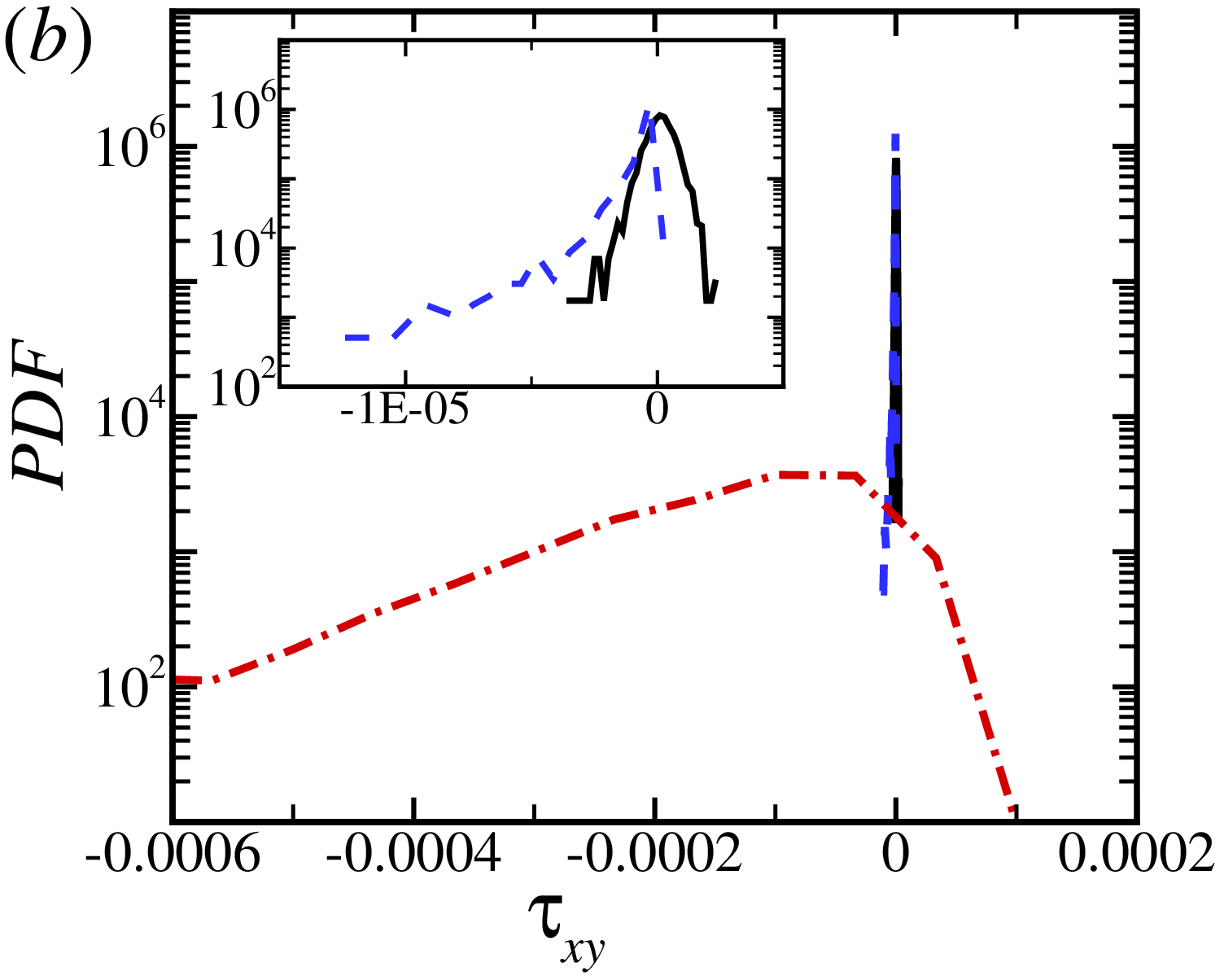}}
\centering{\includegraphics[width=0.45\textwidth]{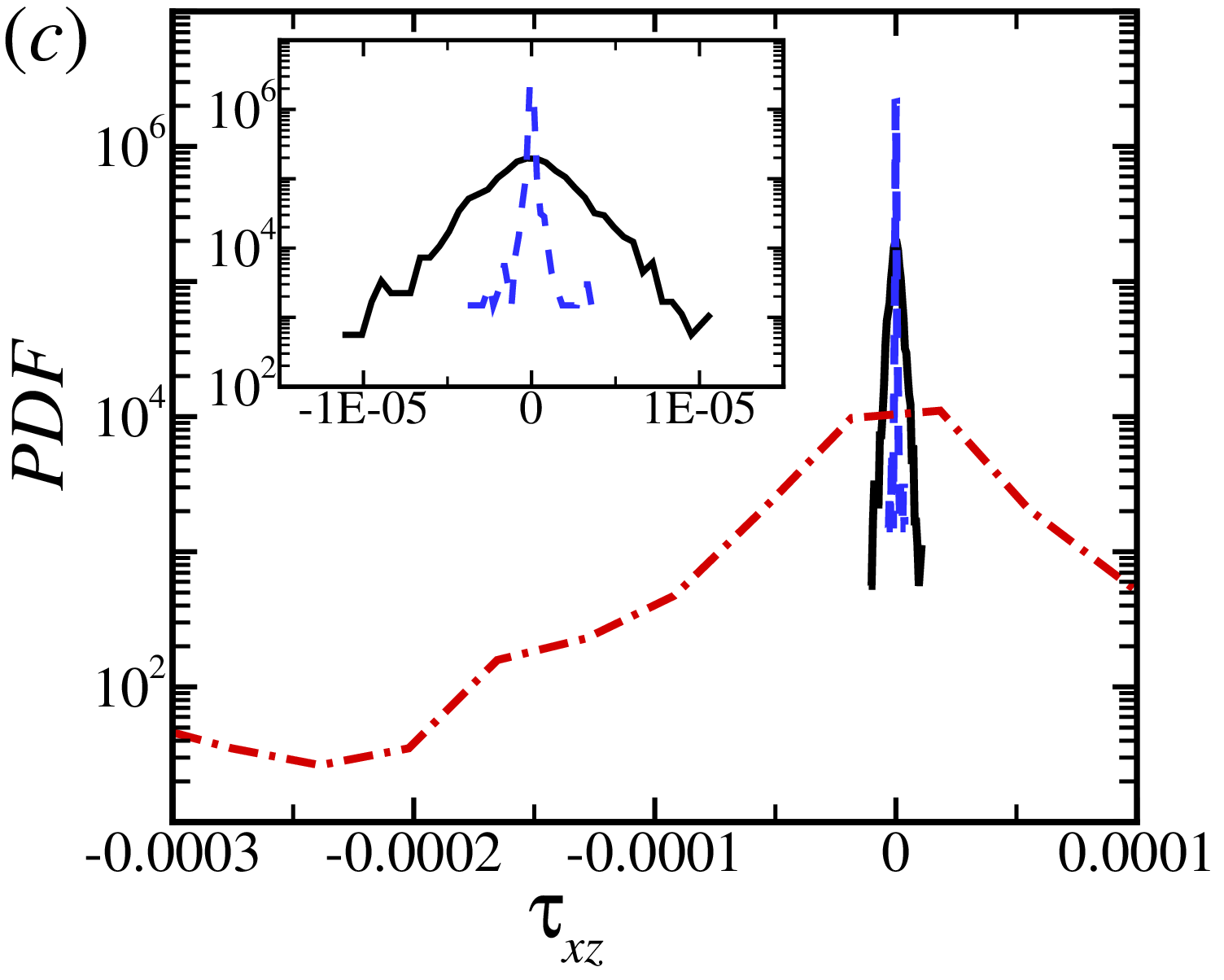}}
\centering{\includegraphics[width=0.45\textwidth]{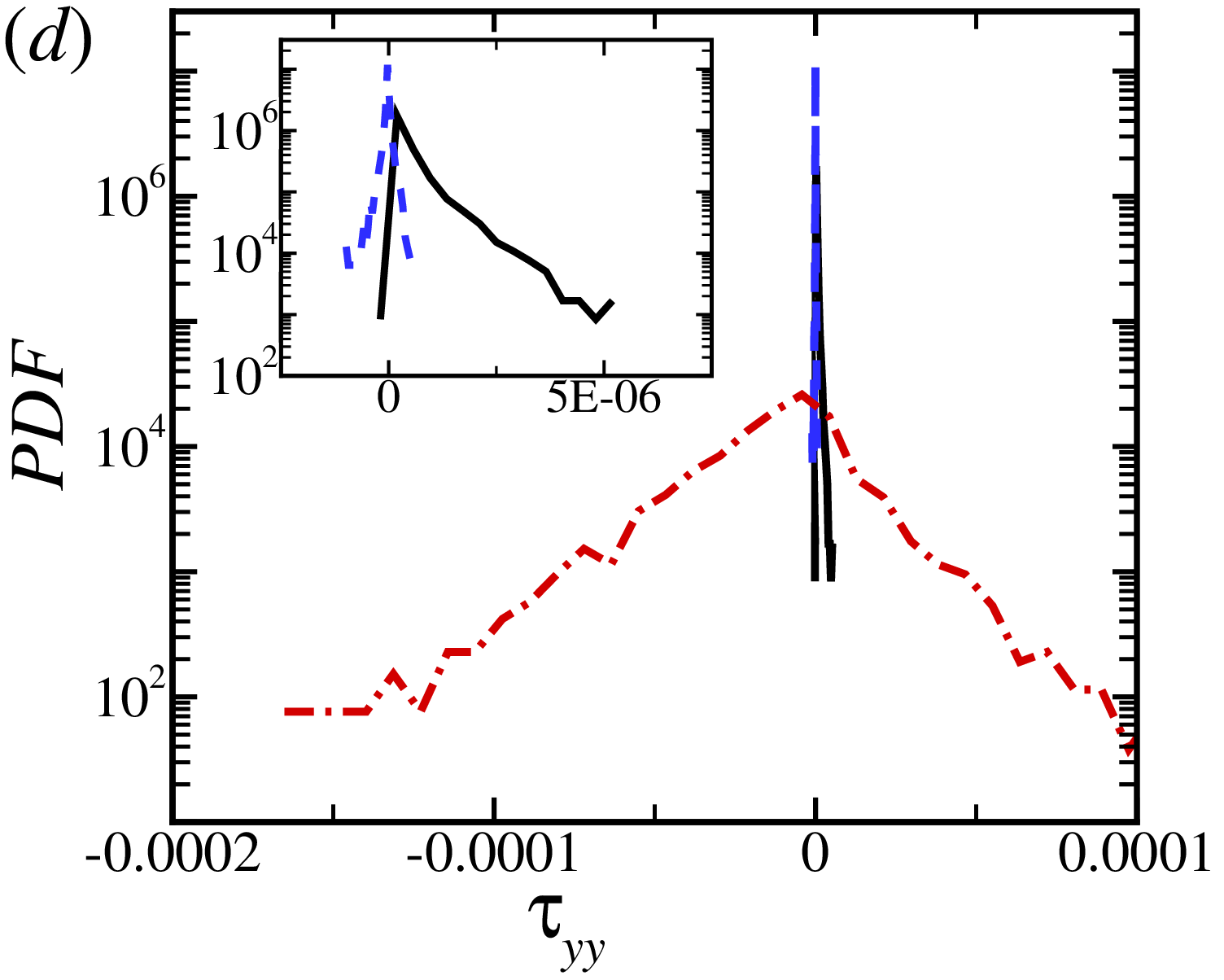}}
\centering{\includegraphics[width=0.45\textwidth]{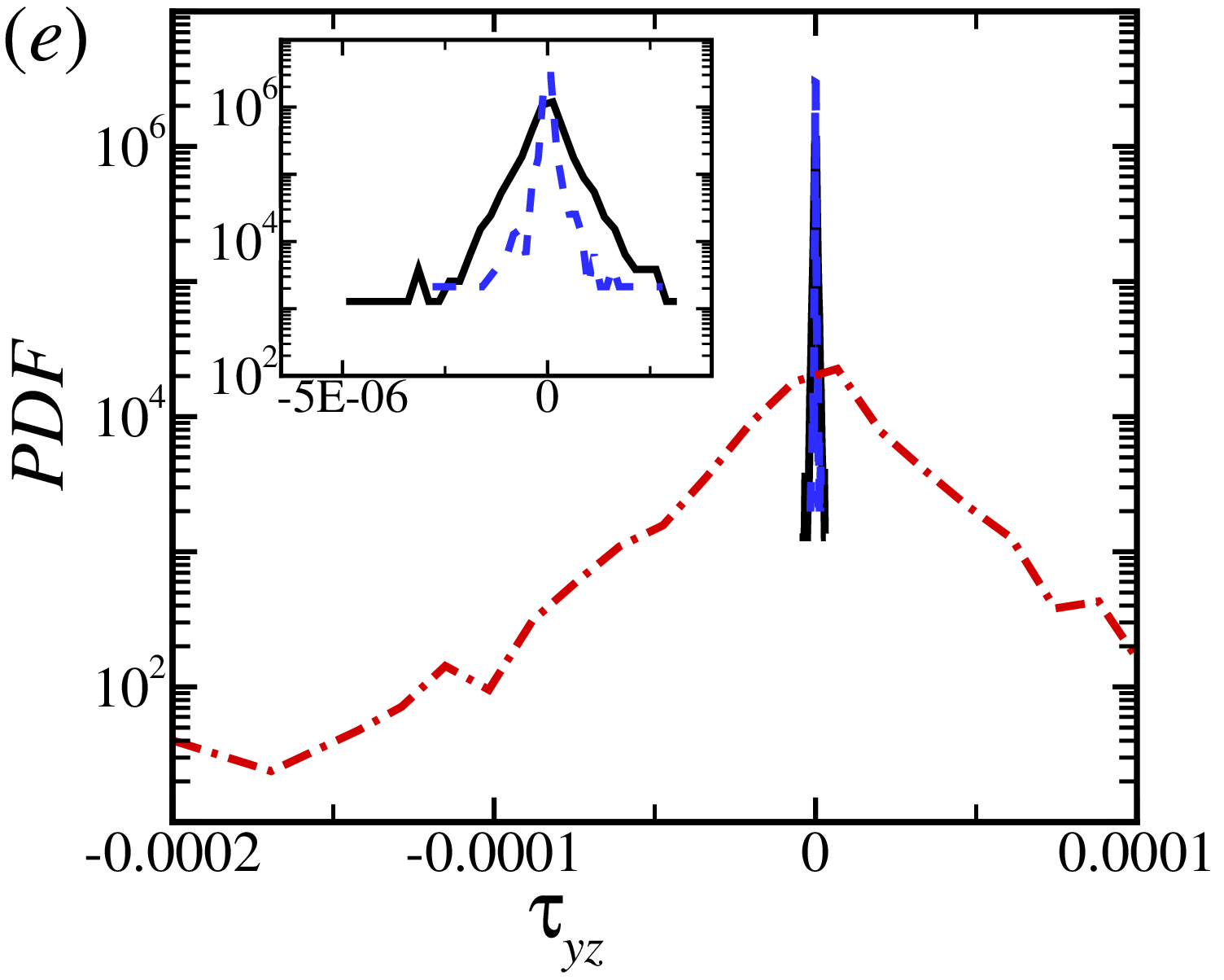}}
\centering{\includegraphics[width=0.45\textwidth]{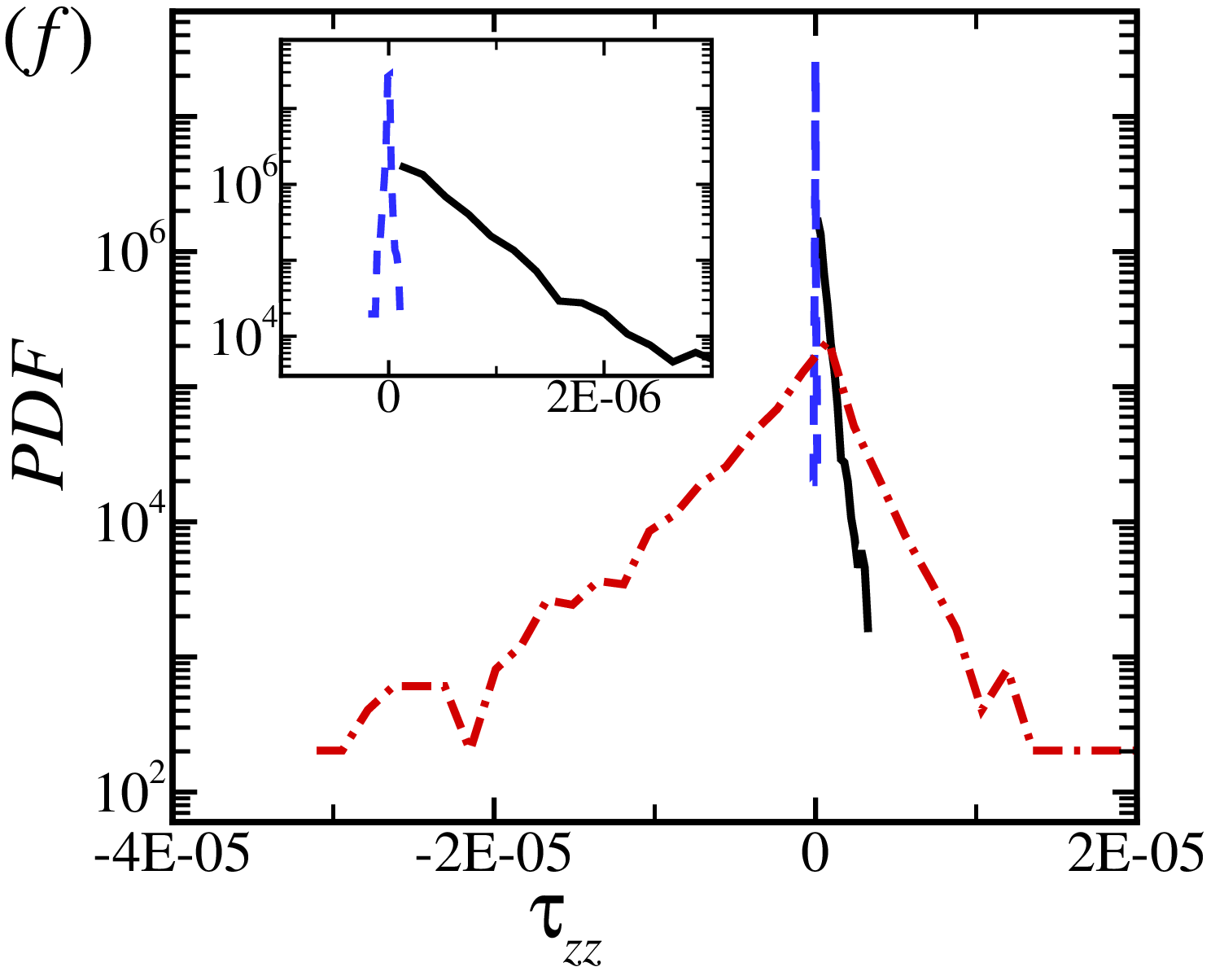}}
  \caption{Normalized PDF of the SGS stresses at the first grid in both the streamwise and vertical directions for the spatial filtering, the FEL model and the dynamic approach: (a) $\tau_{xx}$, (b) $\tau_{xy}$, (c) $\tau_{xz}$, (d) $\tau_{yy}$, (e) $\tau_{yz}$, (f) $\tau_{zz}$. The results are calculated from 200 snapshots of flow fields that cover a total simulation time $22T$.}
\label{fig:filter_tau_pdf}
\end{figure}
\begin{figure}
\centering{\includegraphics[width=0.82\textwidth]{Fig_Filter_pdf_legend.eps}}
\centering{\includegraphics[width=0.46\textwidth]{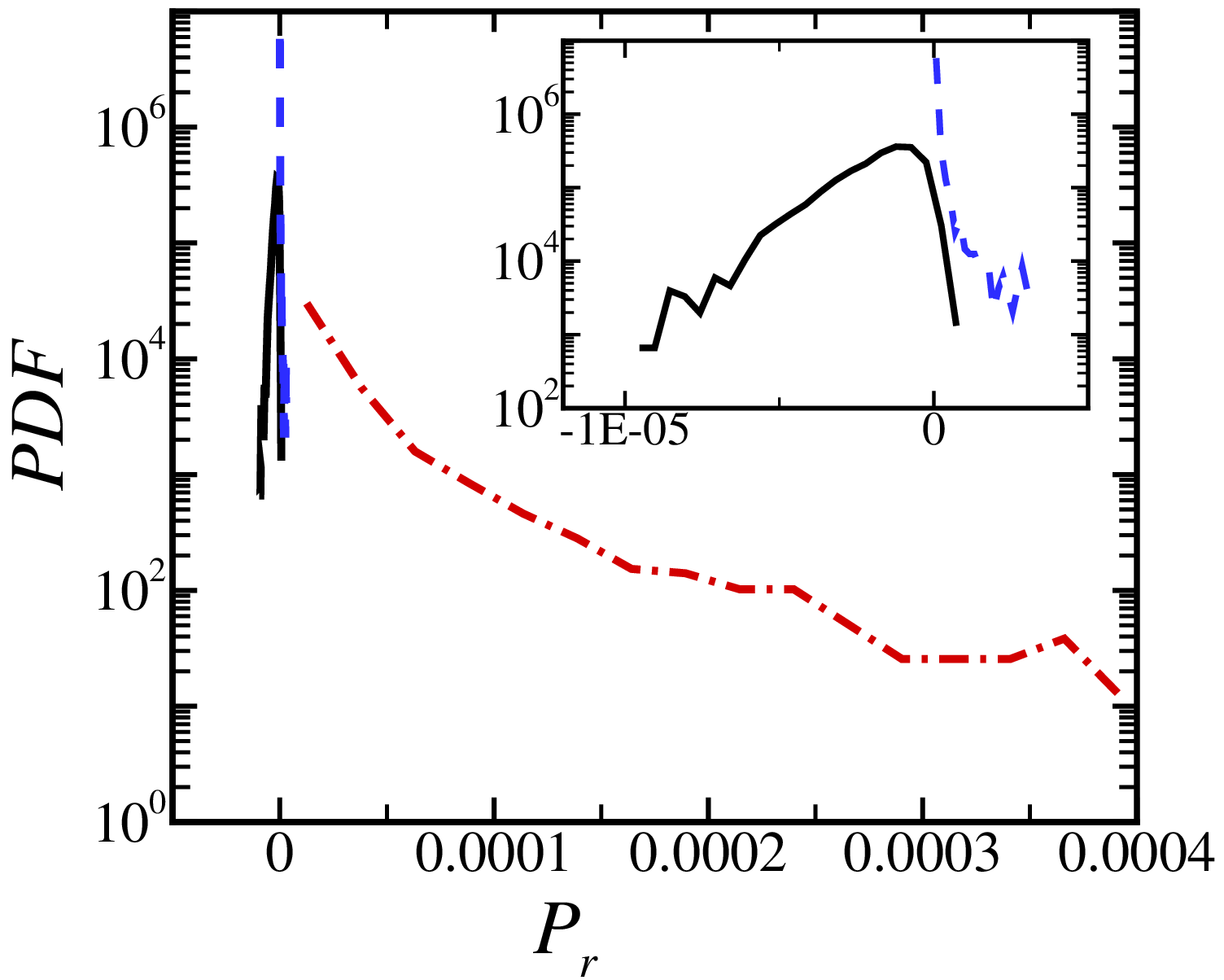}}
  \caption{Normalized PDF of the energy transfer rate $P_r$ at the first grid in both the streamwise and vertical directions for the spatial filtering, the FEL model and the dynamic approach.}
\label{fig:filter_Pr_pdf}
\end{figure}

{\color{black} The proposed eddy viscosity model, i.e., the $\text{FEL}_{\nu_t}$ submodel is assessed in two different ways: 1) comparison of the statistics of the SGS stresses with those from the WRLES results, and 2) comparison of the results from various combinations of the wall shear stress model and the eddy viscosity model for the first off-wall grid nodes.}
Eddy viscosity in the first off-wall grid cell models how the subgrid-scale flow structures there influence the outer flow. By filtering the WRLES flow field, the SGS stresses can be obtained. In this section, the flow fields obtained from the H1.0-WR case (with grid number $296 \times 192 \times 186$) are spatially filtered onto a grid with the resolution of $148 \times 16 \times 31$ in the streamwise, spanwise and vertical directions. {\color{black}Based on this coarse grid with $\Delta y_f/h \approx 0.06$, the WMLES cases with the FEL and WW models are carried out and compared with the filtered WRLES results, as shown in appendix~\ref{appendix:filtering}.}

For a curvilinear grid, the contravariant counterpart of the SGS stress tensor is in the following form~\citep{Armenio_Piomelli_FTC_2000},
%
%According to the spatial filtering of Eq.~\ref{eq_1}, the effect of the unresolved scales appears through 
%the residual stress tensor~\citep{Armenio_Piomelli_FTC_2000},
\begin{equation}
\sigma_i^k = \overline{J^{-1} \xi_j^k u_j u_i} - \overline{J^{-1} \xi_j^k u_j} \overline{u_i} = \overline{U_k u_i} - \overline{U_k} \overline{u_i}.
\label{eq:filter_tau_ij}
\end{equation}
%
%which represents the contravariant counterpart of the SGS stress tensor $\tau_{ij} = \overline{u_i u_j} - \overline{u_i} \ \overline{u_j}$.
In figure~\ref{fig:filter_tau_ij}, the scatter distribution of the SGS stresses at the first off-wall grid computed from the WMLES are compared with those from the filtered WRLES. The scatter points are extracted from an instantaneous flow field at all spanwise positions. In the FEL model, the eddy viscosity is approximated using an embedded NN model. In the WW model, on the other hand, it is computed using the dynamic approach (i.e., the DSM)~\citep{Germano_etal_PoF_1991}.
%
%in the vertical direction. The symbolds ``$\Box$'', ``$\triangle$'' and ``$\bigcirc$'' respectively denote the SGS stress generated by the spatial filtering of the WRLES flow field, the eddy viscosity of the cDK-embedded model (eq.~(\ref{eq_nut})) and the DSM (eq.~(\ref{eq_3})) in the WMLES flow field.
It is seen that the magnitudes of the different component SGS stresses are confined in a narrow region for the FEL model, which exhibits over a much larger range if computed using the dynamic approach.

The energy transfer rate to the residual motions is another important quantity to examine, which is given by
\begin{equation}
P_r = - \tau_{ij} \overline{S_{ij}},
\label{eq:filter_dissp}
\end{equation}
where the residual stress is computed by
\begin{equation}
\tau_{ij} = -2 \nu_t \overline{S_{ij}}.
\label{eq:filter_tau_ij_2}
\end{equation}
Figure~\ref{fig:filter_Pr} shows the scatter distribution of the rate of energy transfer $P_r$. For the filtered flow field, the energy transfer can be positive or negative, but it is always positive (sometimes zero) for the FEL model and the dynamic approach. The energy dissipation (positive value of $P_r$) predicted by the FEL model is closer to that from the filtered flow field when compared with the dynamic approach, particularly at the streamwise locations near the hill crest ($x/h<1.0$ and $x/h>8.0$).

To further analyze the distribution of near-wall energy dissipation, the probability distribution functions (PDFs) of the SGS stresses and energy transfer rate from the filtered flow field, the FEL model and the dynamic approach are plotted using 200 snapshots. The time interval between two adjacent snapshots is $2T/9$. Figures~\ref{fig:filter_tau_pdf}$\sim$\ref{fig:filter_Pr_pdf} show the normalized PDFs of $\tau_{ij}$ and $P_r$ at the first off-wall grid located at $x/h \approx 0.01$ and $y/h \approx 1.021$. It is seen that the PDFs of the SGS stresses predicted by the FEL model are similar to the the spatially filtered results, especially for the shear stresses $\tau_{xy}$, $\tau_{xz}$, and $\tau_{yz}$. As for the energy transfer rate, the PDF of the FEL model predictions is also in a better agreement with the filtered WRLES results when compared with WW model results.

{\color{black}
%\section{Tests of various combinations of the wall shear stress model and the eddy viscosity model}\label{appendix:viscosity}
%
\begin{figure}
\centering{\includegraphics[width=1.0\textwidth]{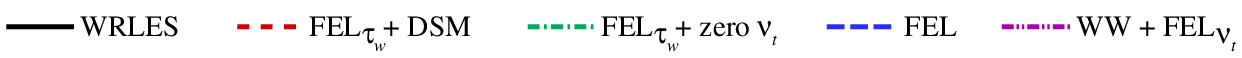}}
\centering{\includegraphics[width=0.495\textwidth]{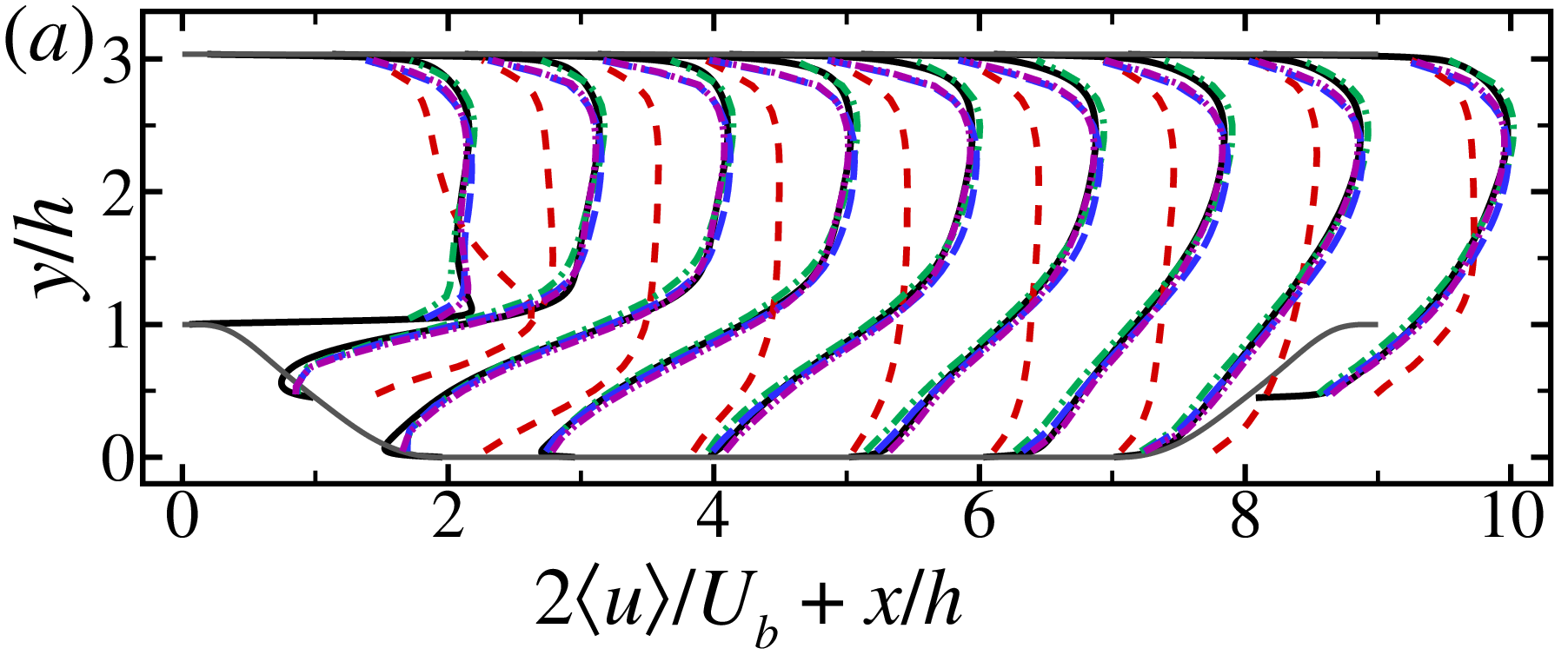}}
\centering{\includegraphics[width=0.495\textwidth]{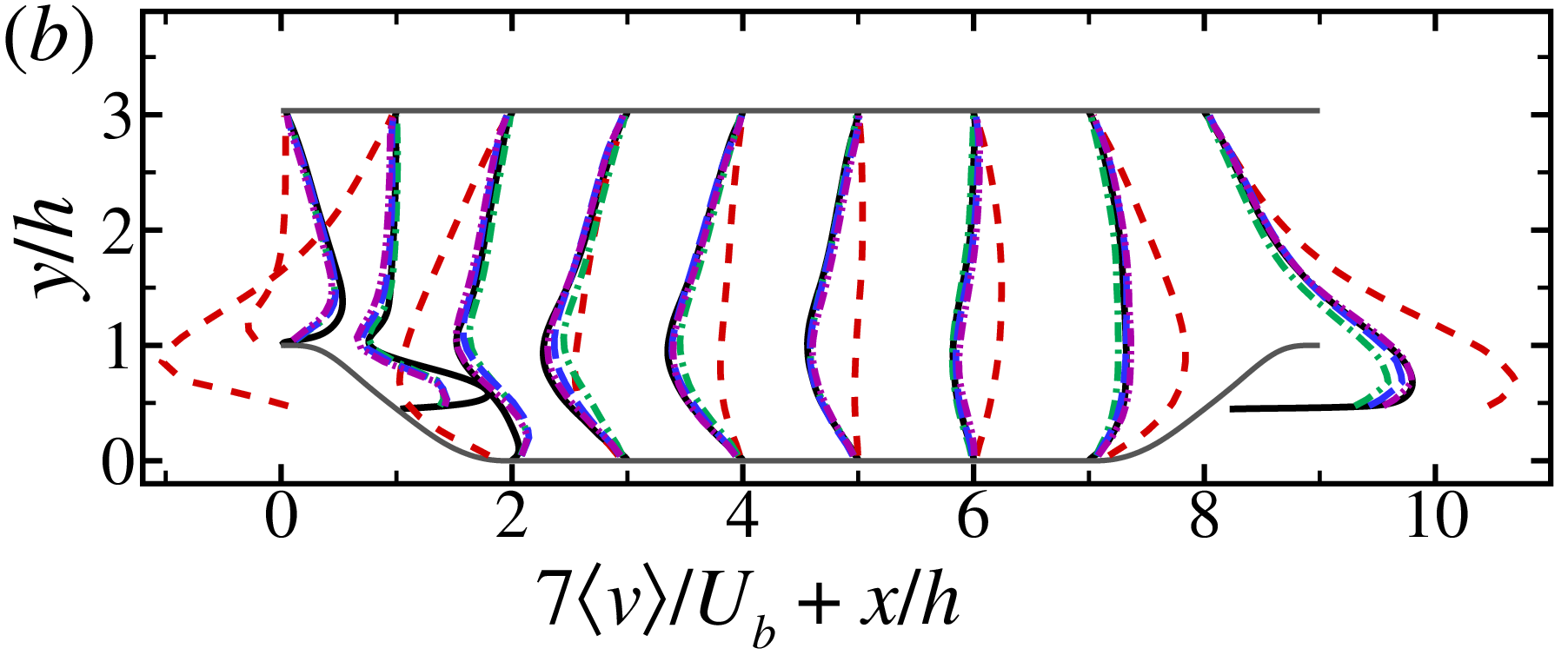}}
\centering{\includegraphics[width=0.495\textwidth]{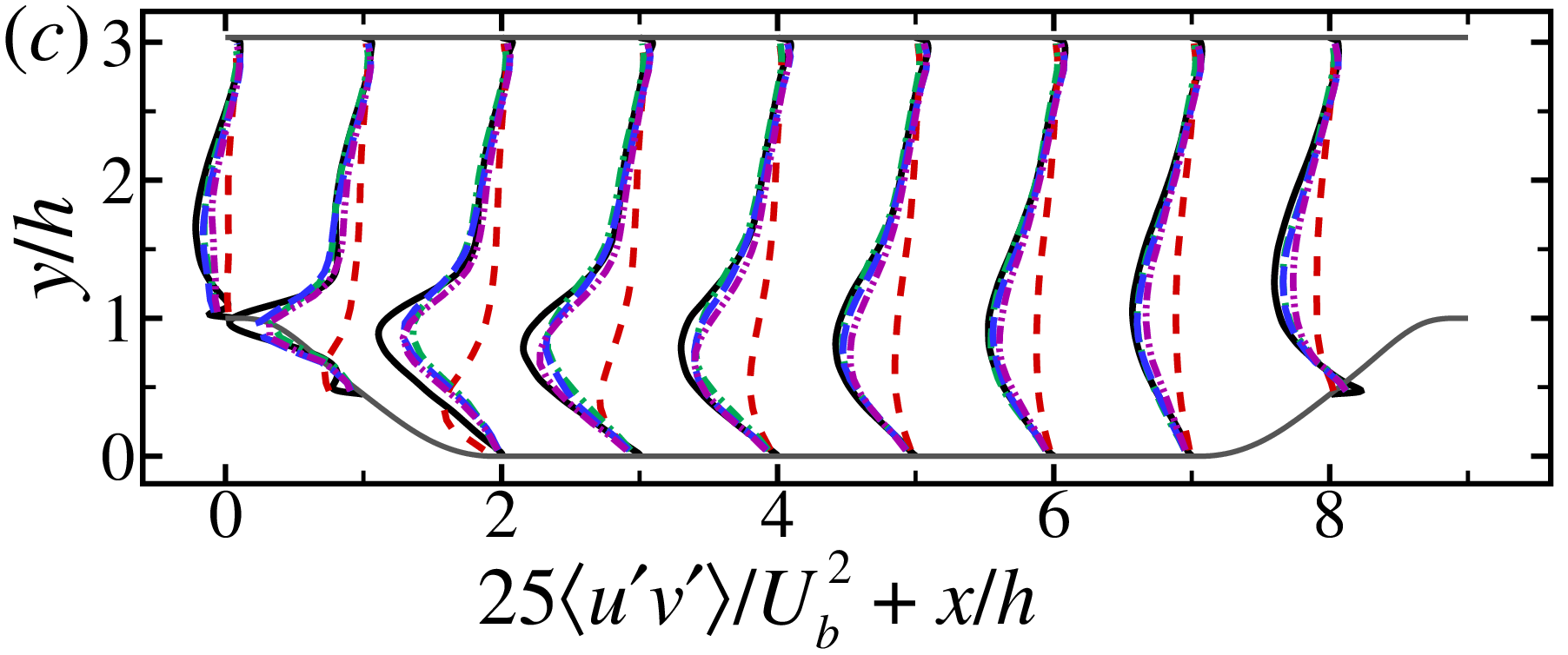}}
\centering{\includegraphics[width=0.495\textwidth]{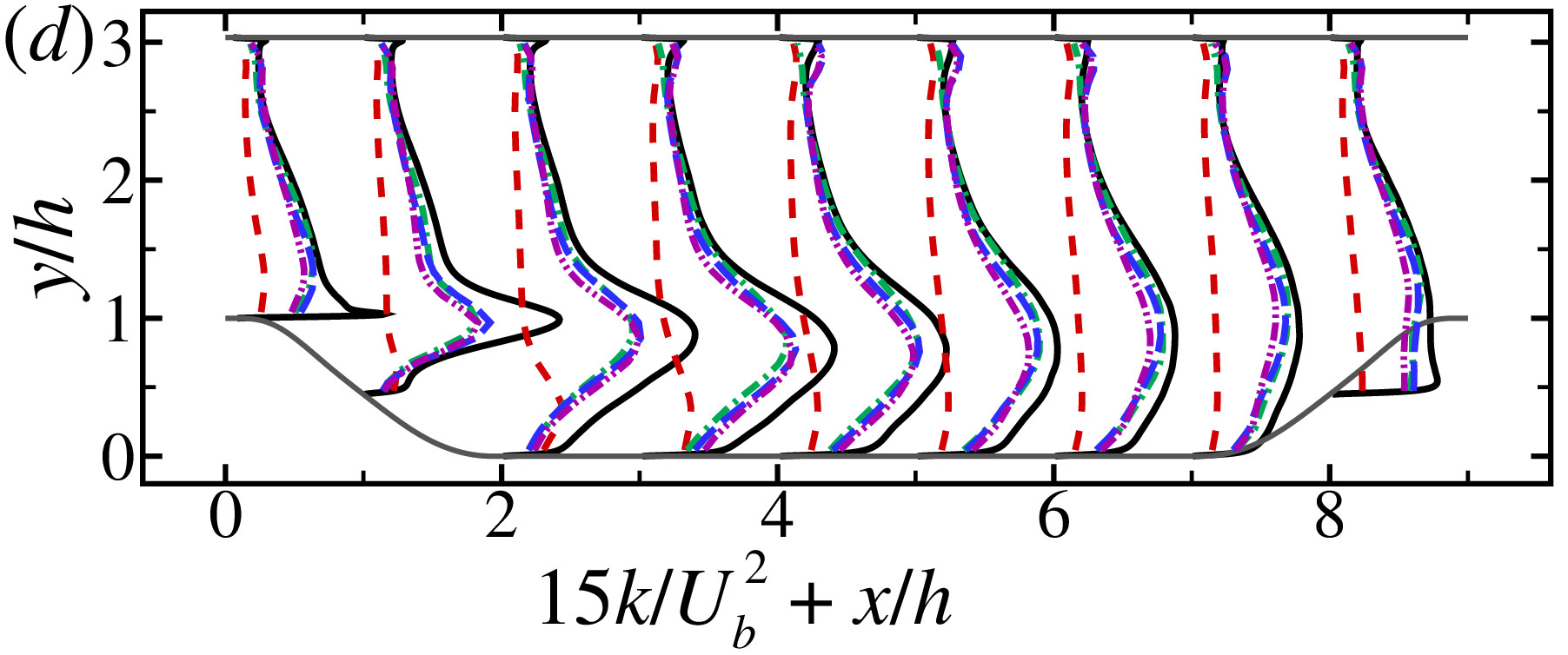}}
  \caption{{\color{black}Vertical profiles of (a) time-averaged streamwise velocity $\left\langle u \right\rangle$ and (b) vertical velocity $\left\langle v \right\rangle$, (c) primary Reynolds shear stress $\left\langle u'v' \right\rangle$, and (d) turbulence kinetic energy $k$ from the H1.0-WR case and the H1.0-WM-0.09 cases with different model combinations. The periodic hill case with $Re_h = 10595$ is employed for testing.}}
\label{fig:profile_nut_HB}
\end{figure}
%
%\begin{figure}
%\centering{\includegraphics[width=0.95\textwidth]{Fig_nut_profile_err_legend.eps}}
%\centering{\includegraphics[width=0.4\textwidth]{Fig_nut_profile_err_u_HB.eps}} \quad\quad
%\centering{\includegraphics[width=0.4\textwidth]{Fig_nut_profile_err_TKE_HB.eps}}
%  \caption{Relative errors of (a) time-averaged streamwise velocity $\left\langle u \right\rangle$ and (b) turbulence kinetic energy $k$ between the H1.0-WR case and the H1.0-WM-0.09 cases with different models at $Re_h = 10595$.}
%\label{fig:profile_nut_HB_err}
%\end{figure}
%
\begin{figure}
\centering{\includegraphics[width=0.48\textwidth]{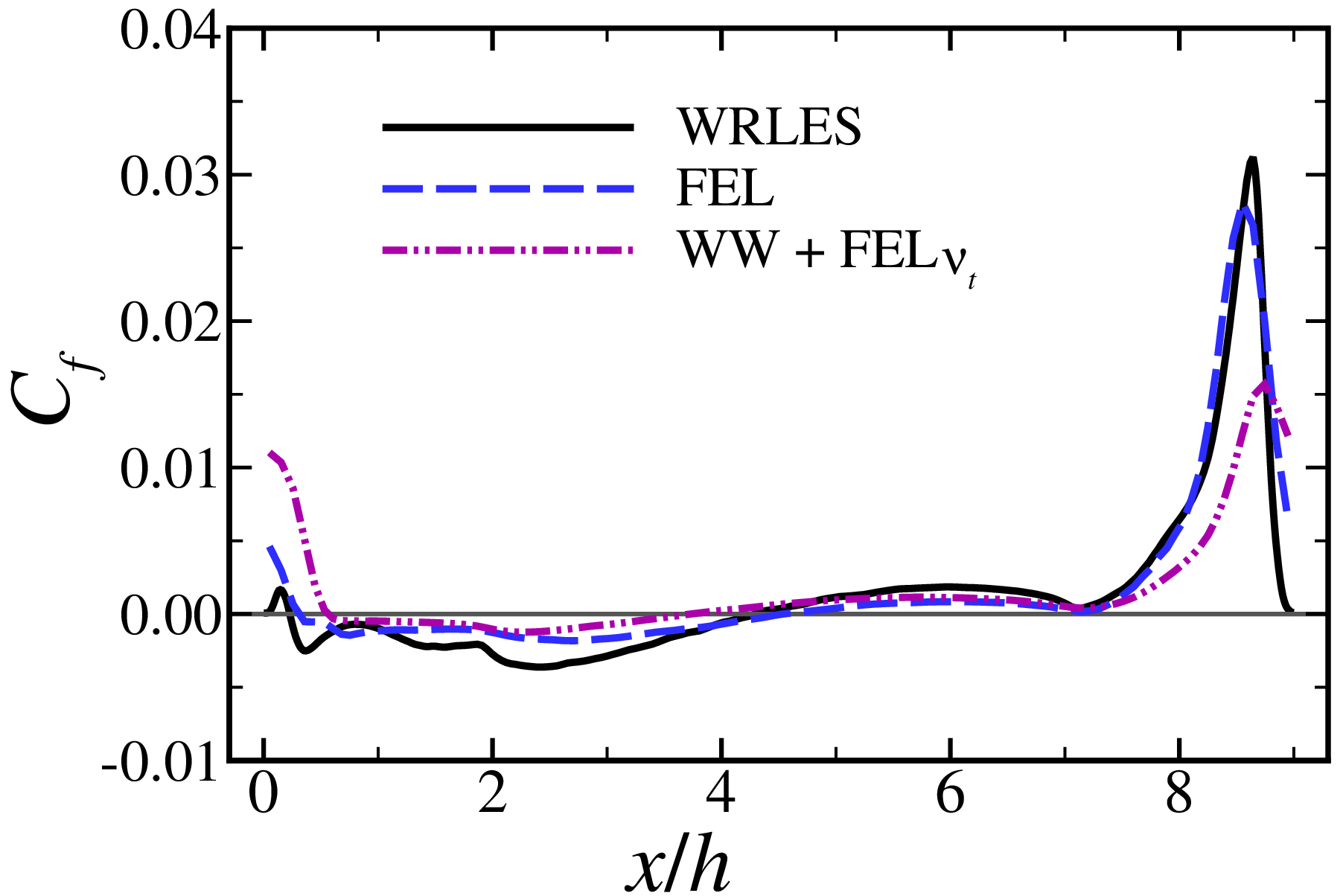}} \;
  \caption{{\color{black}Comparison of the time-averaged skin friction coefficients between the WRLES, the WMLES with the FEL and WW + $\text{FEL}_{\nu_t}$ models for the H1.0 case at $Re_h = 10595$.}}
\label{fig:Cf_HB_WW}
\end{figure}
\begin{figure}
\centering{\includegraphics[width=0.495\textwidth]{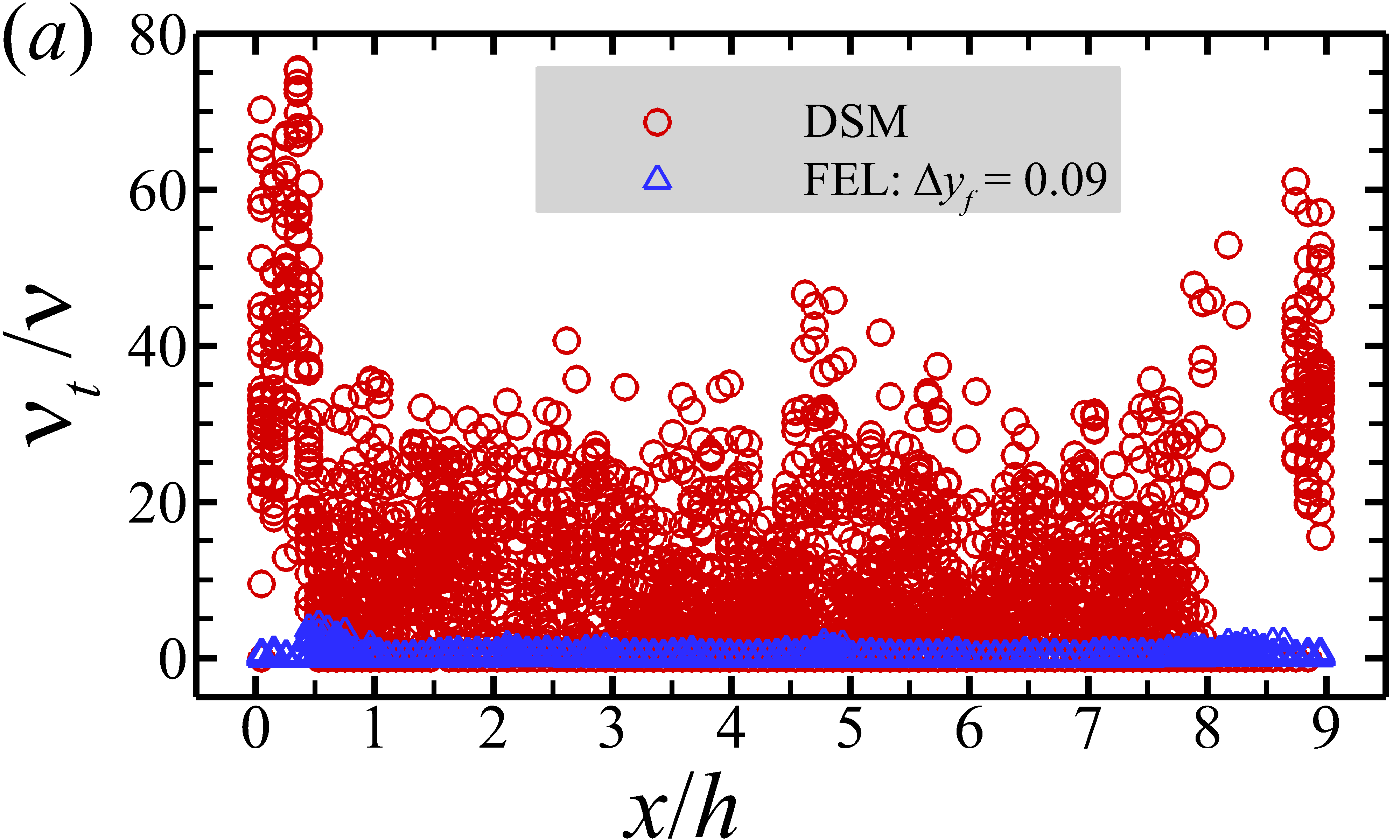}}
\centering{\includegraphics[width=0.495\textwidth]{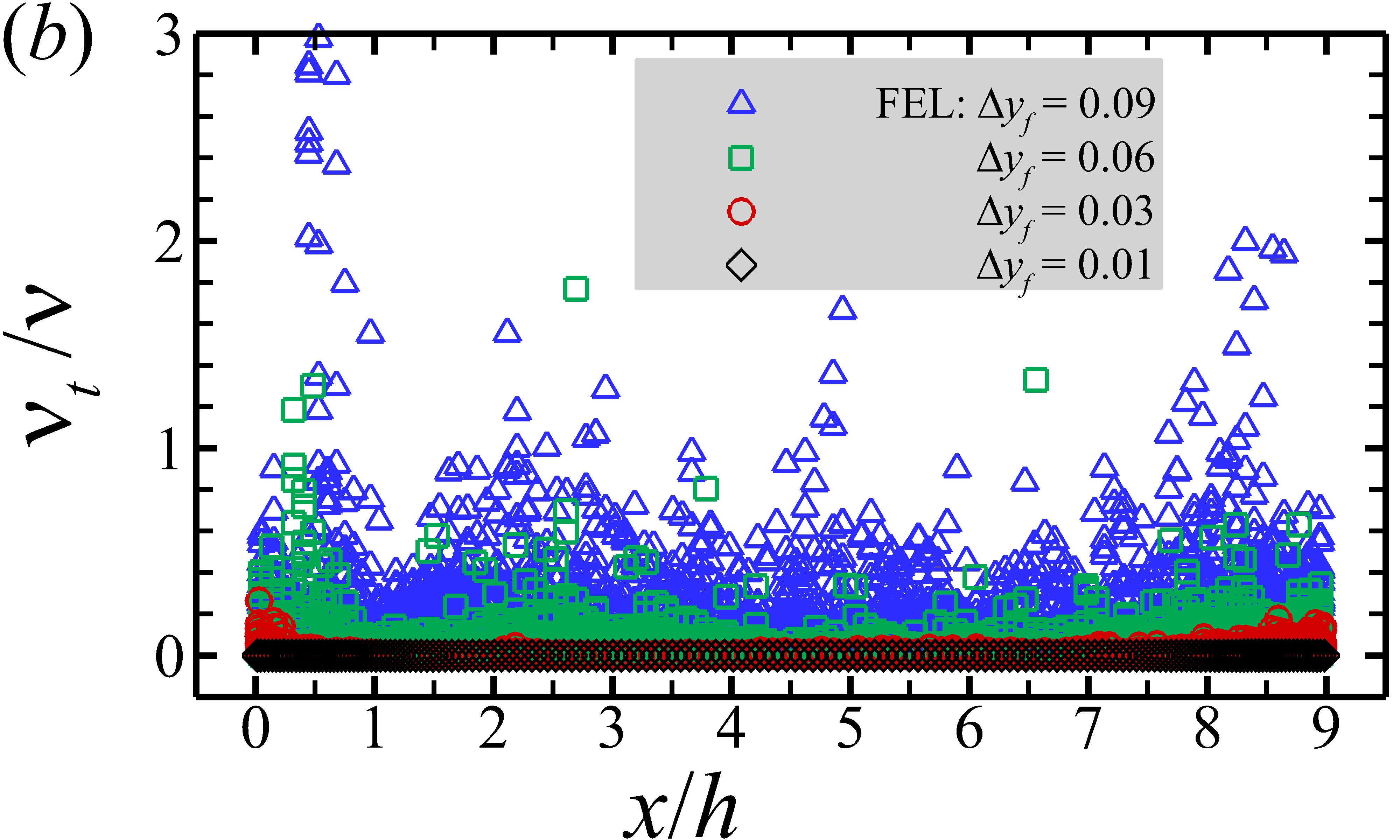}}
  \caption{{\color{black}Comparison of the eddy viscosity $\nu_t$ at the first off-wall grid nodes of an instantaneous flow field: (a) the DSM and the FEL model (eq.~(\ref{eq_nut})) simultaneously calculated from the H1.0-WM-0.09 case with the FEL model, (b) the FEL model calculated from the H1.0-WM-0.01/0.03/0.06/0.09 cases at $Re_h = 10595$.}}
\label{fig:scatter_nut_HB}
\end{figure}

The proposed eddy viscosity model, i.e., the $\text{FEL}_{\nu_t}$ submodel, is further assessed using the results from various combinations of the wall shear stress model and the eddy viscosity model for the first off-wall grid nodes, which include the FEL model, the $\text{FEL}_{\tau_w} + \text{DSM}$ model using the $\text{FEL}_{\tau_w}$ model with only the $\tau_w$ submodel and the DSM~\citep{Germano_etal_PoF_1991}, the $\text{FEL}_{\tau_w} + \text{zero} \ \nu_t$ model with the zero eddy viscosity ($\nu_t=0$) at the first off-wall grid nodes, the WW model with the DSM, and the WW model with the $\text{FEL}_{\nu_t}$ submodel.
Figure~\ref{fig:profile_nut_HB} shows the vertical profiles of the flow statistics obtained from the H1.0-WR case and H1.0-WM-0.09 cases at $Re_h = 10595$. It is seen that the $\text{FEL}_{\tau_w} + \text{DSM}$ model fails to capture the separation bubble and flow statistics. With the learned eddy viscosity submodel ($\text{FEL}_{\nu_t}$), the predictions of flow statistics at various streamwise locations are significantly improved for the WW model, although the $\text{FEL}_{\nu_t}$ submodel is not trained for the WW model. As for the comparison between the zero $\nu_t$ and the $\text{FEL}_{\nu_t}$, the predicted mean velocities in the lower region with the separation bubble are somewhat better for the FEL model using the $\text{FEL}_{\nu_t}$ submodel.
%
%To probe into the role of the wall shear stress submodel and the eddy viscosity submodel in the proposed FEL model, results from different type of models, i.e.,
%, the predictions of flow statistics are also improved for the WW model, only with a little larger of TKE than that from the FEL model, as shown in figure~\ref{fig:profile_nut_HB_err}.
%On the other hand, the predictions from the $\text{FEL}_{\tau_w}$ model with zero eddy viscosity are a little worse than those from the FEL model at the lower half of vertical profiles of flow statistics, but more accurate near the top wall, which results in the smaller errors of time-averaged streamwise velocity and TKE. 

To further examine the difference caused by the wall shear stress condition, the comparison of the time-averaged skin friction coefficients predicted by the FEL model and the WW + $\text{FEL}_{\nu_t}$ model is shown in figure~\ref{fig:Cf_HB_WW}. As seen, the skin friction coefficient predicted by the FEL model generally agrees with that from the WRLES, especially on the peak value at the windward of the hill ($x/h \in [7.0, 9.0]$). As for the WW + $\text{FEL}_{\nu_t}$ model, the peak value is overestimated at the leeward hill face while underestimated at the windward hill face. As for the flow separation, it is delayed by the WW + $\text{FEL}_{\nu_t}$ model ($x_{\text{sep}}/h = 0.57$ (0.22 for WRLES)), resulting an early occurrence of the reattachment ($x_{\text{ret}}/h = 3.78$ (4.35 for WRLES)).
%The discrepancy of skin friction coefficient can affect the flow separation.
%The streamwise locations of the separation point and reattachment point are [0.54, 4.15] and [0.57, 3.78] for the FEL model and WW model, respectively, demonstrating that a small change in the separation location could be accompanied with a big change in the reattachment location~\citep{Frohlich_etal_JFM_2005}.

The scatter distributions of the eddy viscosity at the first off-wall grid nodes at one time instant, which are computed using the DSM and the $\text{FEL}_{\nu_t}$ submodel are compared in figure~\ref{fig:scatter_nut_HB}. It is observed in figure~\ref{fig:scatter_nut_HB}(a) that the eddy viscosity predicted by the $\text{FEL}_{\nu_t}$ submodel is much smaller than that from the DSM, explaining the not bad performance of the zero $\nu_t$ model. Figure~\ref{fig:scatter_nut_HB}(b) compares the scatter distributions of the eddy viscosity calculated from the H1.0-WM-0.01/0.03/0.06/0.09 cases with the $\text{FEL}_{\nu_t}$ submodel. As seen, the eddy viscosity monotonously decreases with the grid resolution, which demonstrates the grid convergence property of the $\text{FEL}_{\nu_t}$ submodel.

Overall, the assessment has shown a better performance of the $\text{FEL}_{\nu_t}$ submodel, which gives a sound predictions of the SGS stresses in terms of both the amplitude and PDF distributions. Using the $\text{FEL}_{\nu_t}$ submodel with the conventional WW model can improve the predictions of the overall flow patterns, but cannot accurately predict the flow separation and reattachment.
%Using a zero $\nu_t$ at the first off-wall grid nodes can increase the predictions of the overall flow patterns, but cannot accurately predict the flow separation and reattachment, the critical statistics of the flow.
}

{\color{black}
\section{Applications to other flow configurations}\label{sec:Application_flow_config}
In this section, the FEL model is further tested using other three flow configurations including flows over a two-dimensional (2D) wavy wall, a three-dimensional (3D) wavy wall, and a 2D Gaussian bump. Schematics of the corresponding geometries are shown in figure~\ref{fig:case_validation}.
\begin{figure}
\centering
	\begin{subfigure}[b]{1.0\textwidth}
	\centering
	\includegraphics[width = 0.44\textwidth]{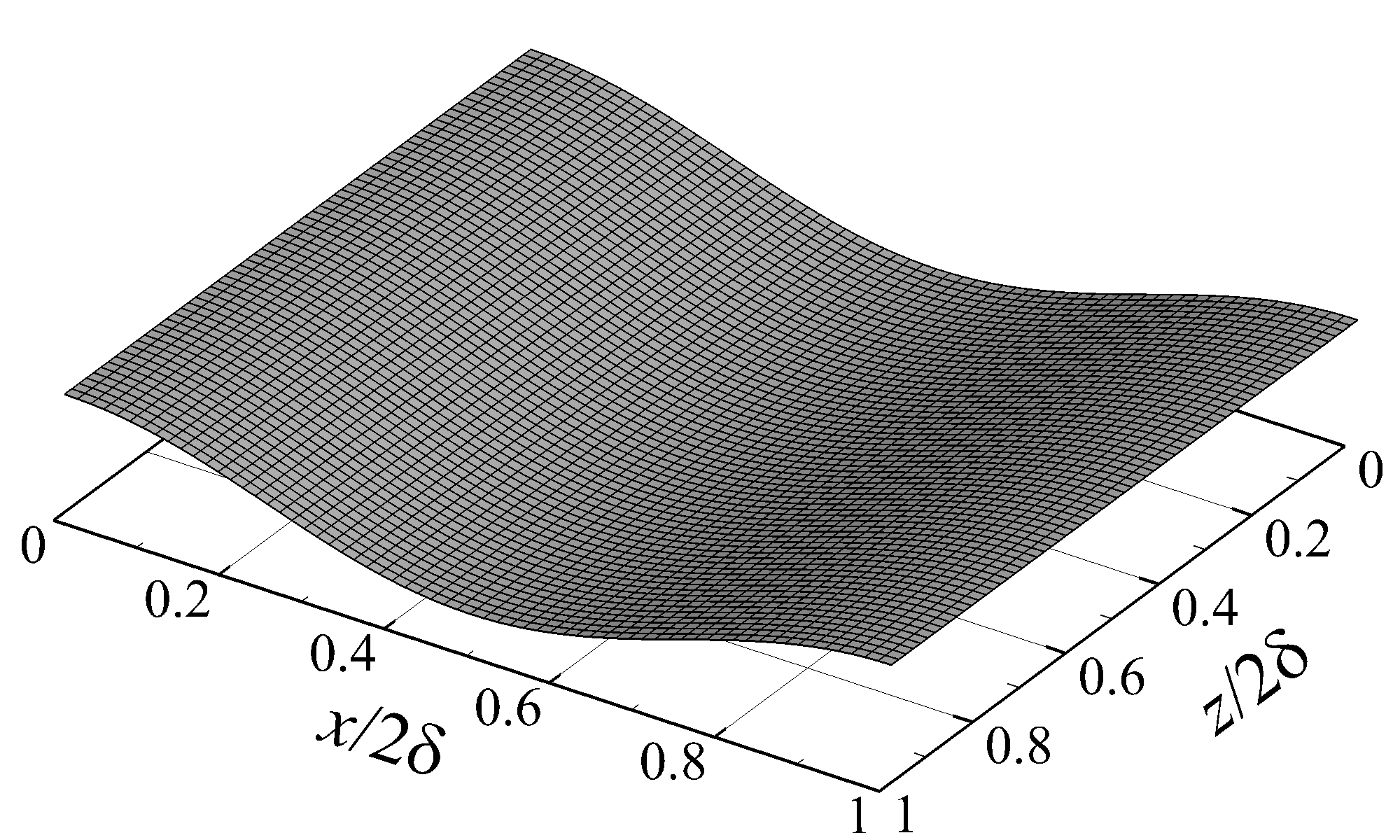}\quad
	\includegraphics[width = 0.44\textwidth]{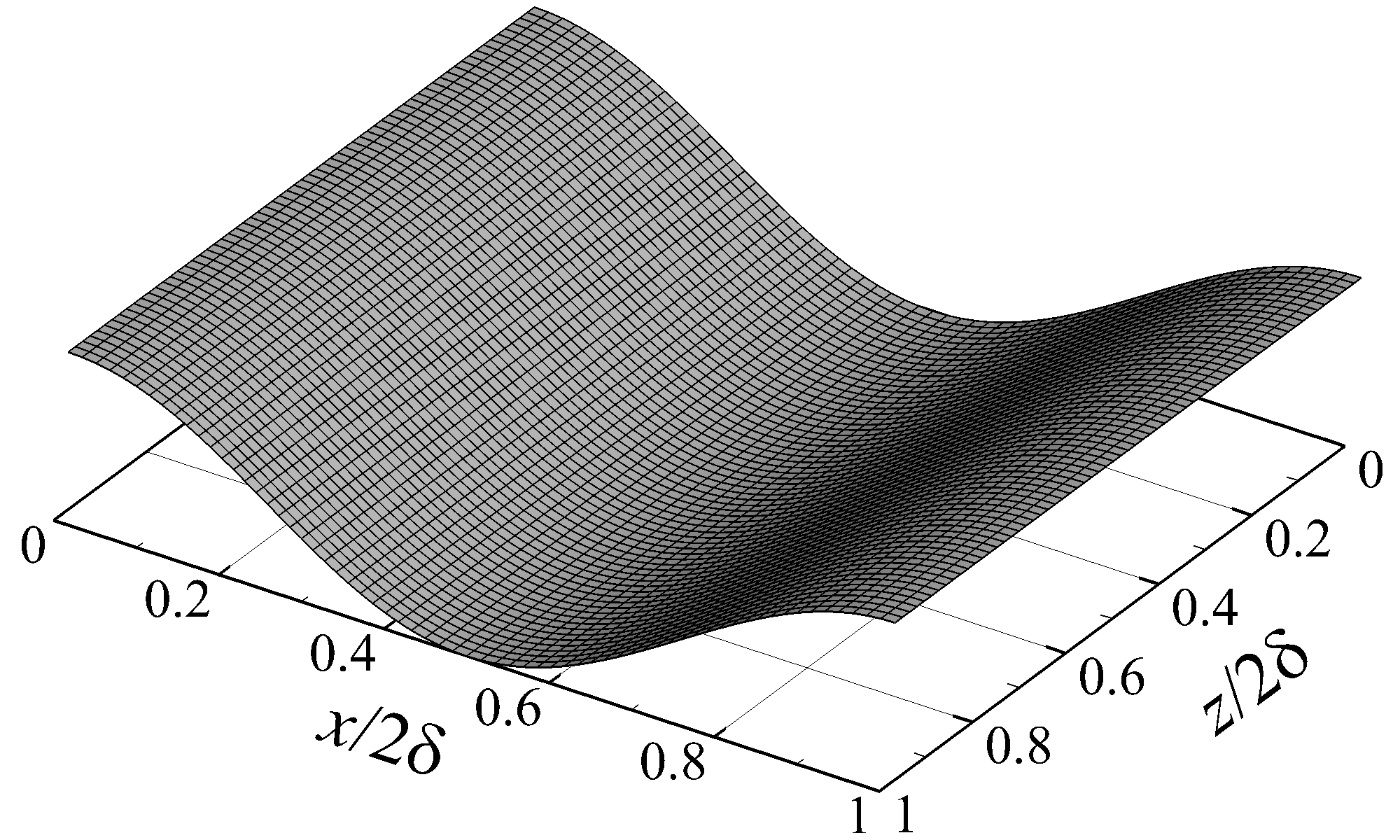}
	\subcaption{2D Wavy wall: $\alpha = 0.05$, $\chi = 0.1$ (left) and $\alpha = 0.1$, $\chi = 0.2$ (right)}
	\end{subfigure}
	\begin{subfigure}[b]{1.0\textwidth}
	\centering
	\includegraphics[width = 0.44\textwidth]{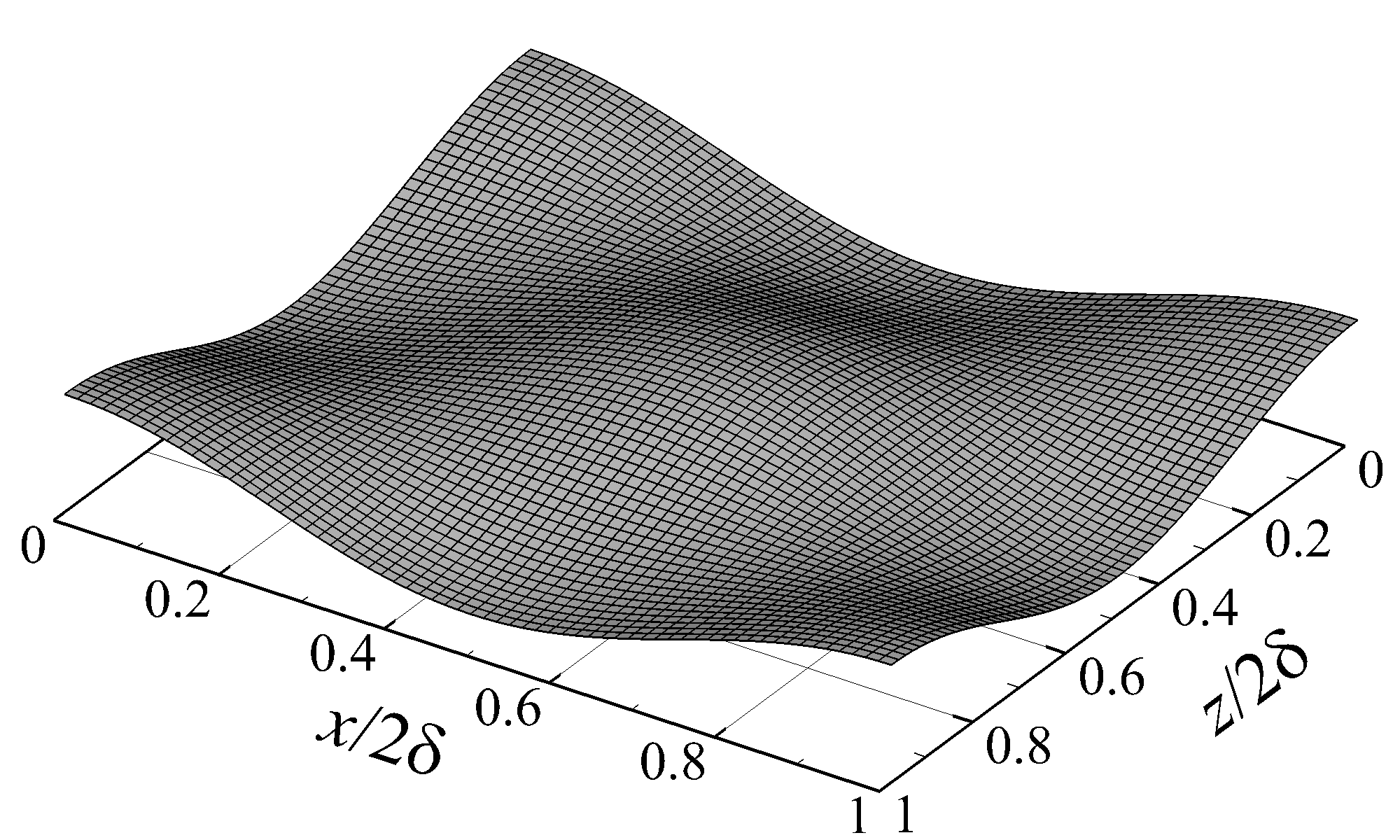}\quad
	\includegraphics[width = 0.44\textwidth]{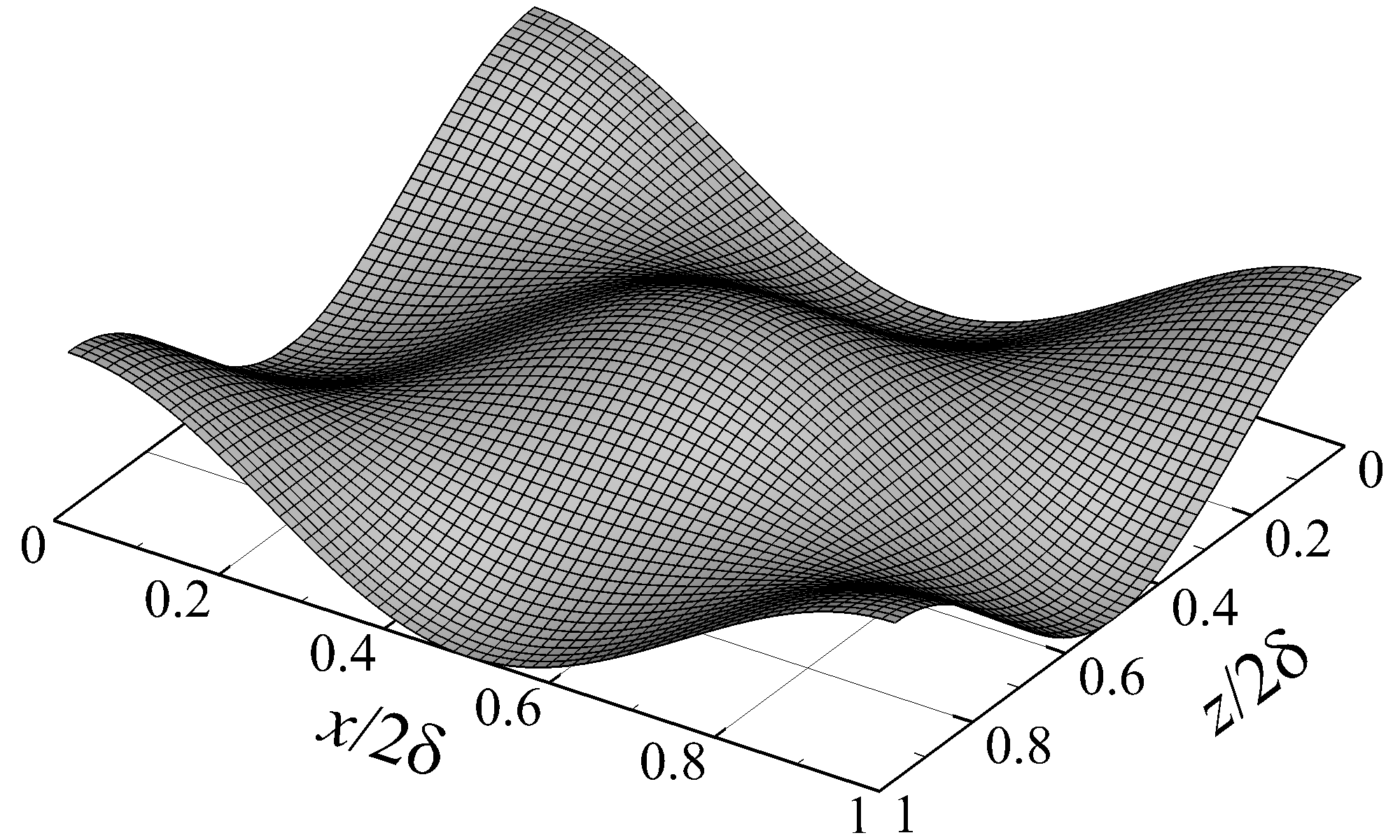}
	\subcaption{3D Wavy wall: $\alpha = 0.05$, $\chi = 0.1$ (left) and $\alpha = 0.1$, $\chi = 0.2$ (right)}
	\end{subfigure}
	\begin{subfigure}[b]{1.0\textwidth}
	\centering
	\includegraphics[width = 0.9\textwidth]{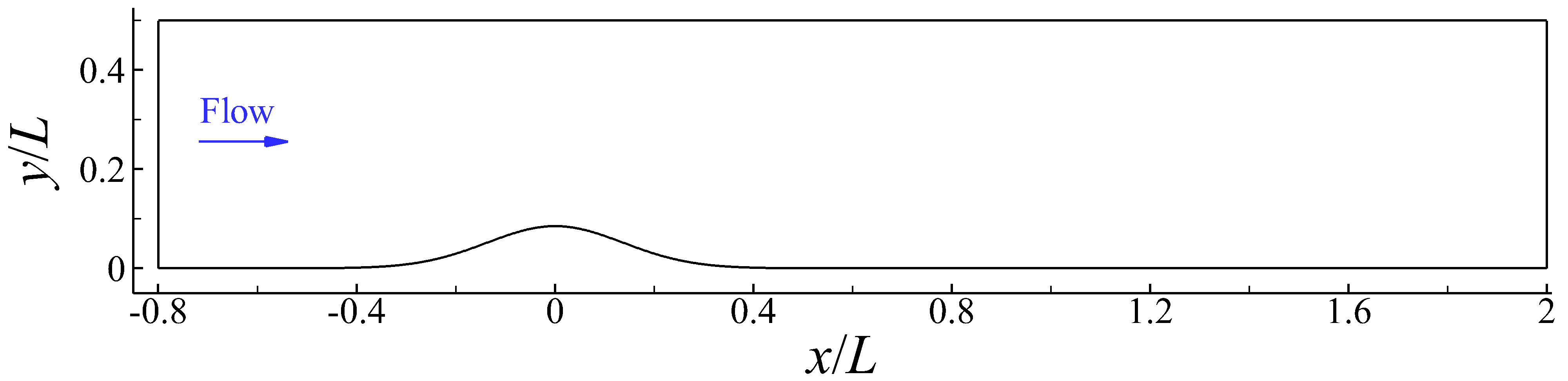}
	\subcaption{2D Gaussian bump}
	\end{subfigure}
  \caption{{\color{black}Geometries of four flow configurations tested in section~\ref{sec:Application_flow_config}.}}
\label{fig:case_validation}
\end{figure}
\begin{table}
  \begin{center}
\def~{\hphantom{0}}
  \begin{tabular}{c|c|c|cc}
  \cline{1-5}
  Case       &  Reynolds number 			   &  $N_x \times N_y \times N_z$ &  $\Delta y_f$  &  $\Delta y_c^+$ \\ \cline{1-5}\rule{0pt}{8pt}
  2D-Wavy-WR & \multirow{2}{*}{$Re_b = 11200$} & $128 \times 192 \times 128$  & $\Delta y_f/2\delta=0.001$ & 0.5 \\ %\cline{0-0}\cline{3-6}
  2D-Wavy-WM &          					   & $64 \times 48 \times 64$     & $\Delta y_f/2\delta=0.02$  & 10  \\ \cline{1-5}
  3D-Wavy-WR & \multirow{2}{*}{$Re_b = 11200$} & $128 \times 192 \times 128$  & $\Delta y_f/2\delta=0.001$ & 0.5 \\ %\cline{0-0}\cline{3-6}
  3D-Wavy-WM &          					   & $64 \times 48 \times 64$     & $\Delta y_f/2\delta=0.02$  & 10  \\ \cline{1-5}
  2D-bump-WM & $Re_L=2 \times 10^6$  		   & $920 \times 90 \times 50$    & $\Delta y_f/h = 0.03$  	   & 198$\sim$318 \\ \cline{1-5}
  \end{tabular}
  \caption{{\color{black}Parameters of WRLES and WMLES cases for the simulated flow configurations in section 5.}}
  \label{tab:case_validation}
  \end{center}
\end{table}
\begin{figure}
\centering
	\begin{subfigure}[b]{1.0\textwidth}
	\centering
	\includegraphics[width = 0.32\textwidth]{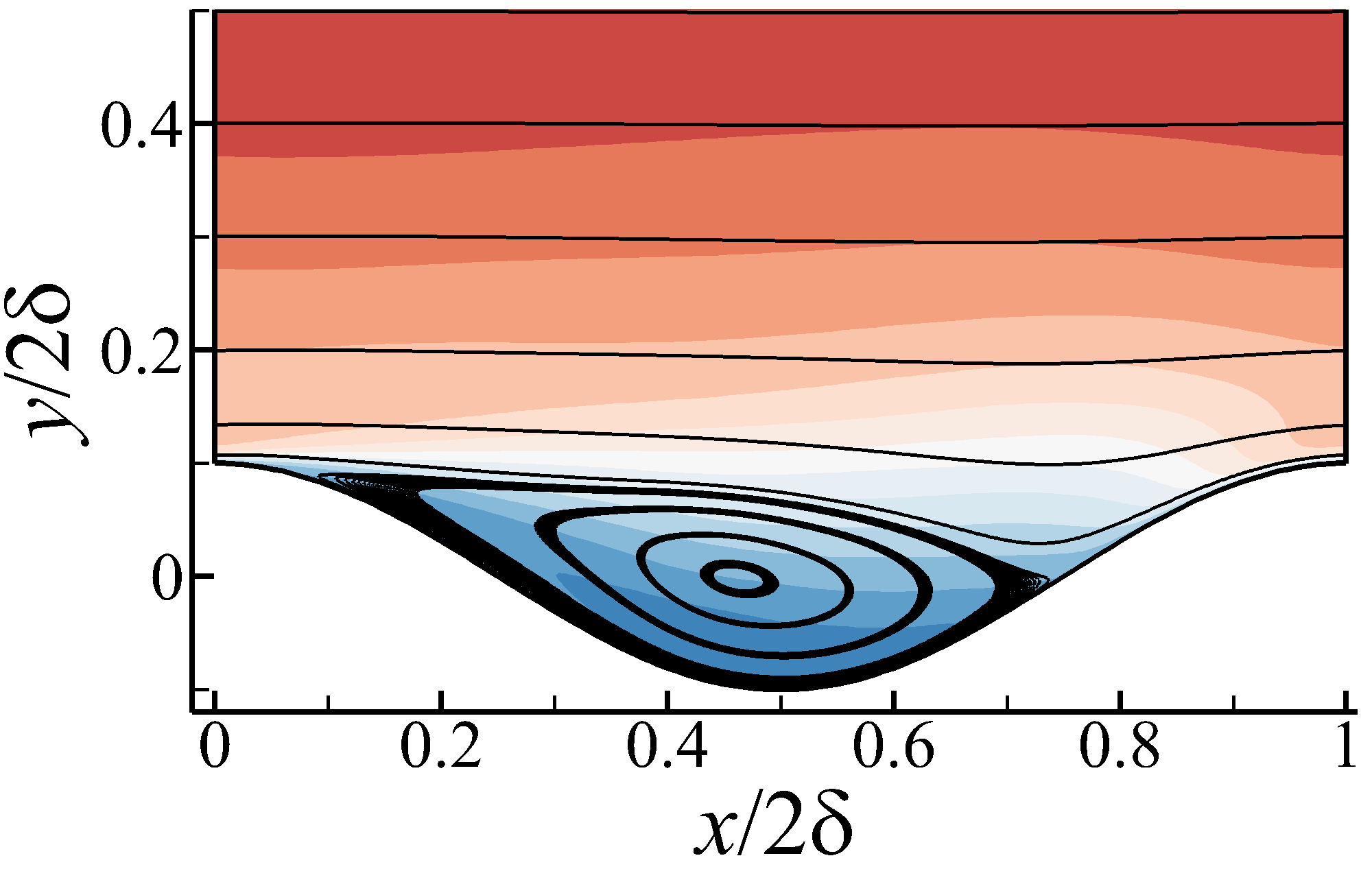}\;
	\includegraphics[width = 0.32\textwidth]{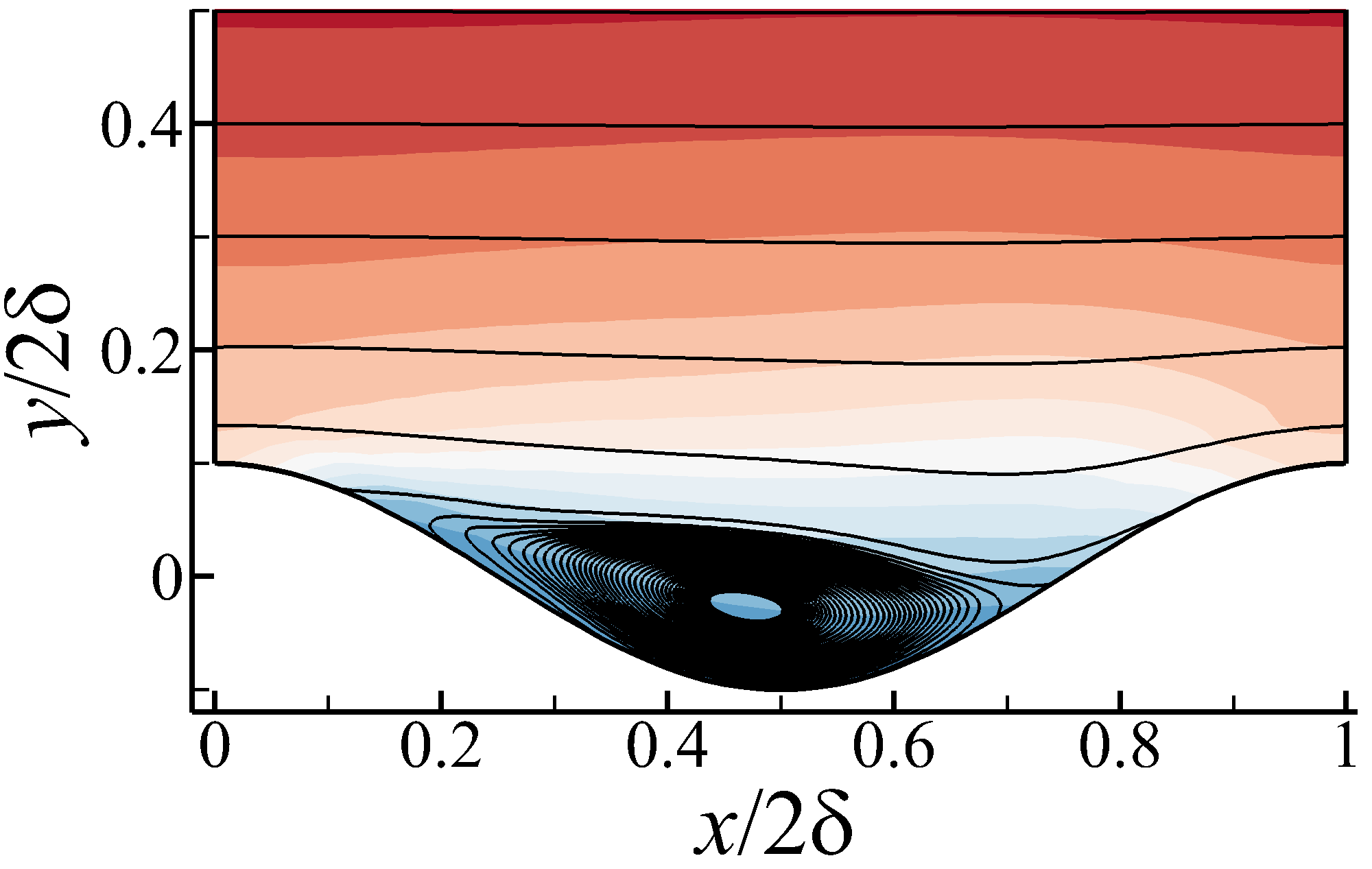}\;
	\includegraphics[width = 0.32\textwidth]{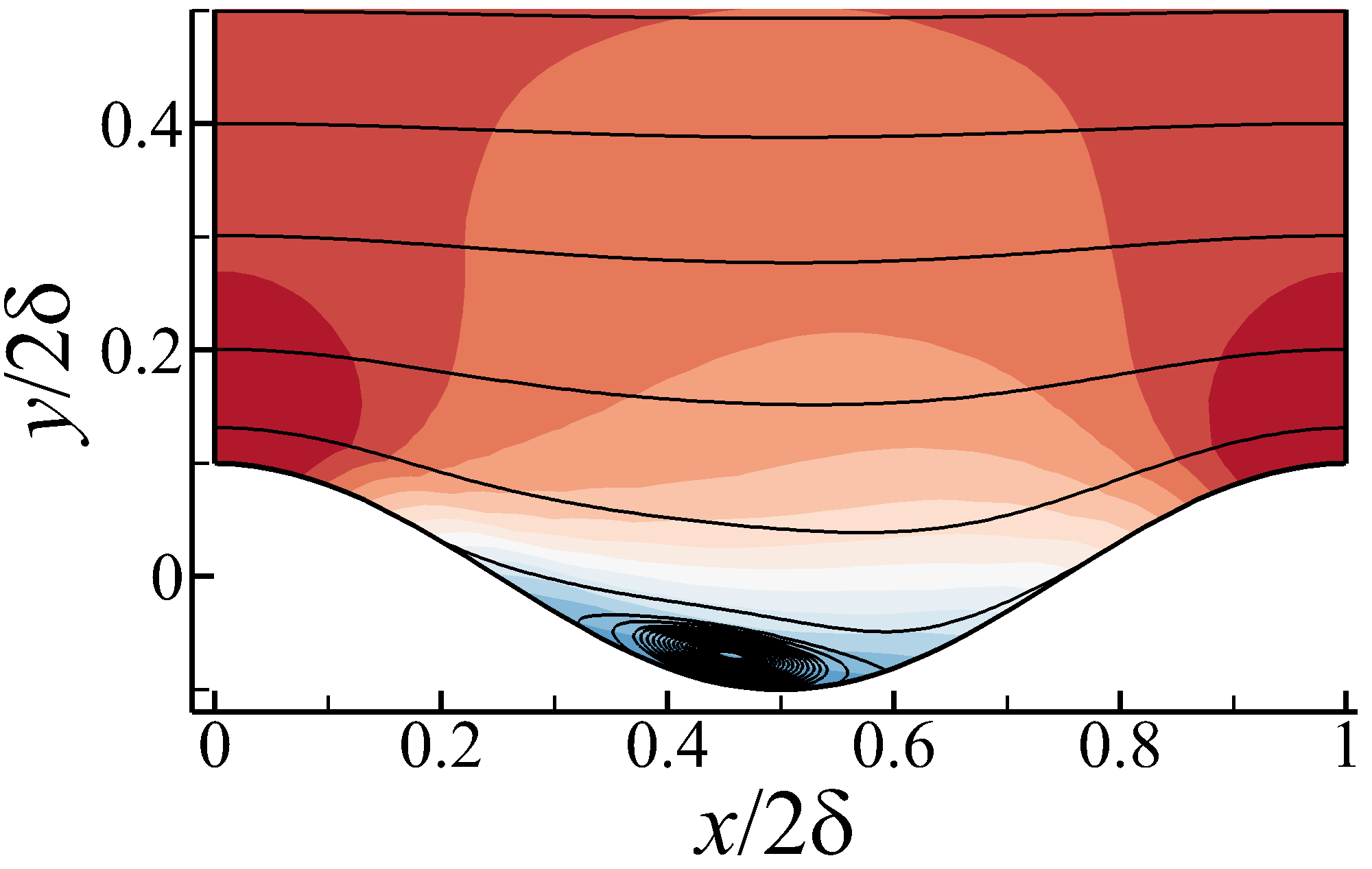}
	\subcaption{2D wavy wall ($\alpha = 0.1$, $\chi = 0.2$): WR vs. FEL vs. WW}
	\end{subfigure}
	\begin{subfigure}[b]{1.0\textwidth}
	\centering
	\includegraphics[width = 0.3\textwidth]{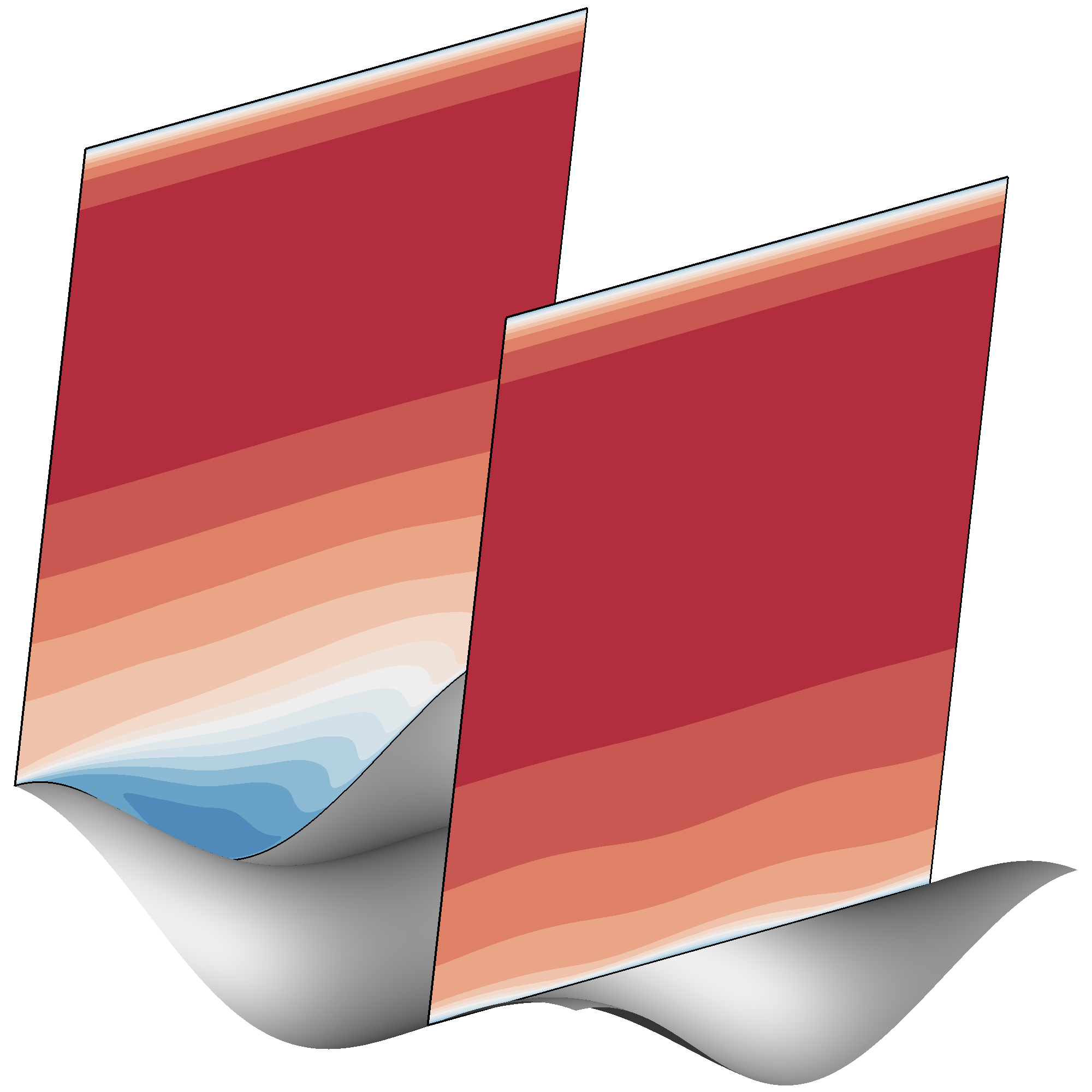}\;
	\includegraphics[width = 0.3\textwidth]{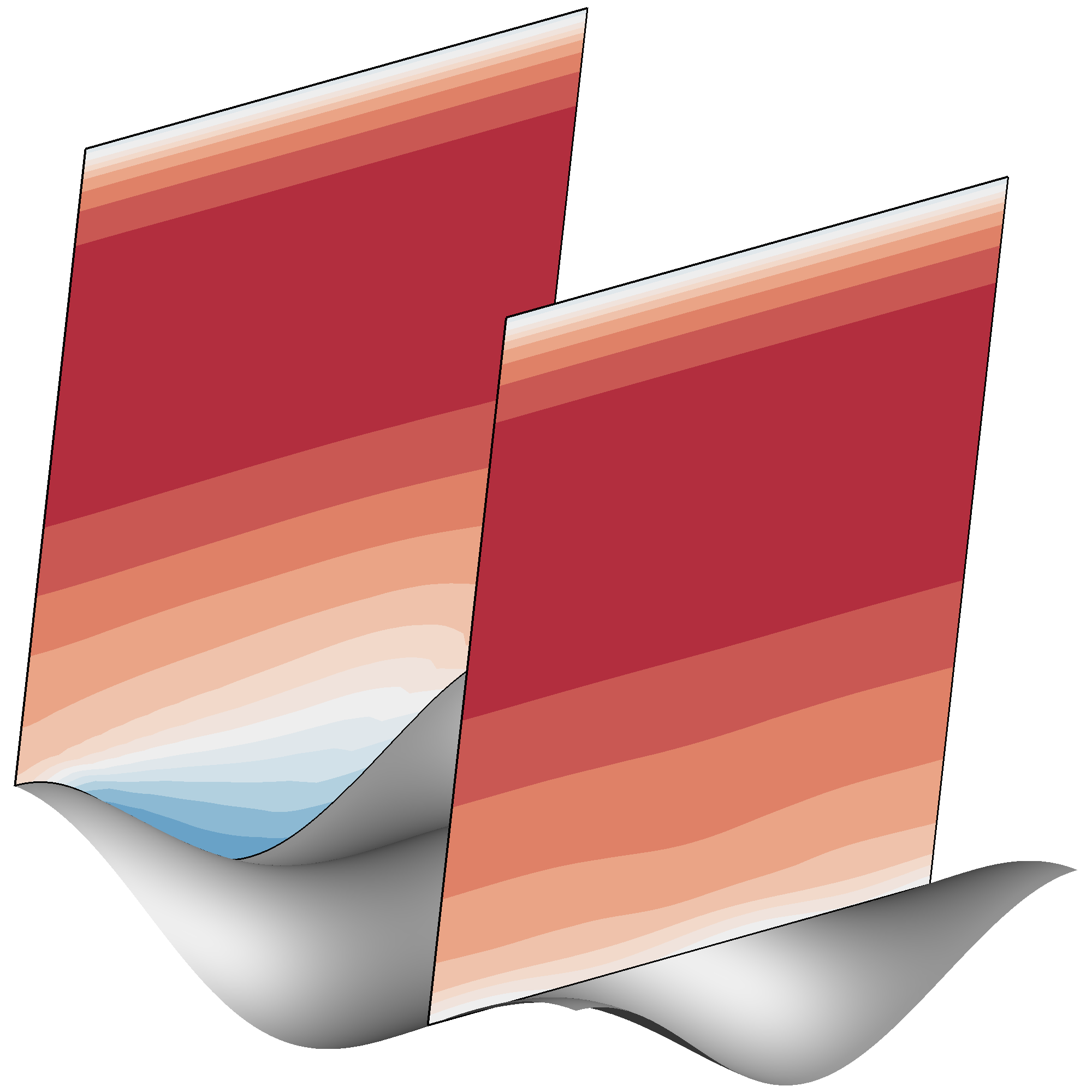}\;
	\includegraphics[width = 0.3\textwidth]{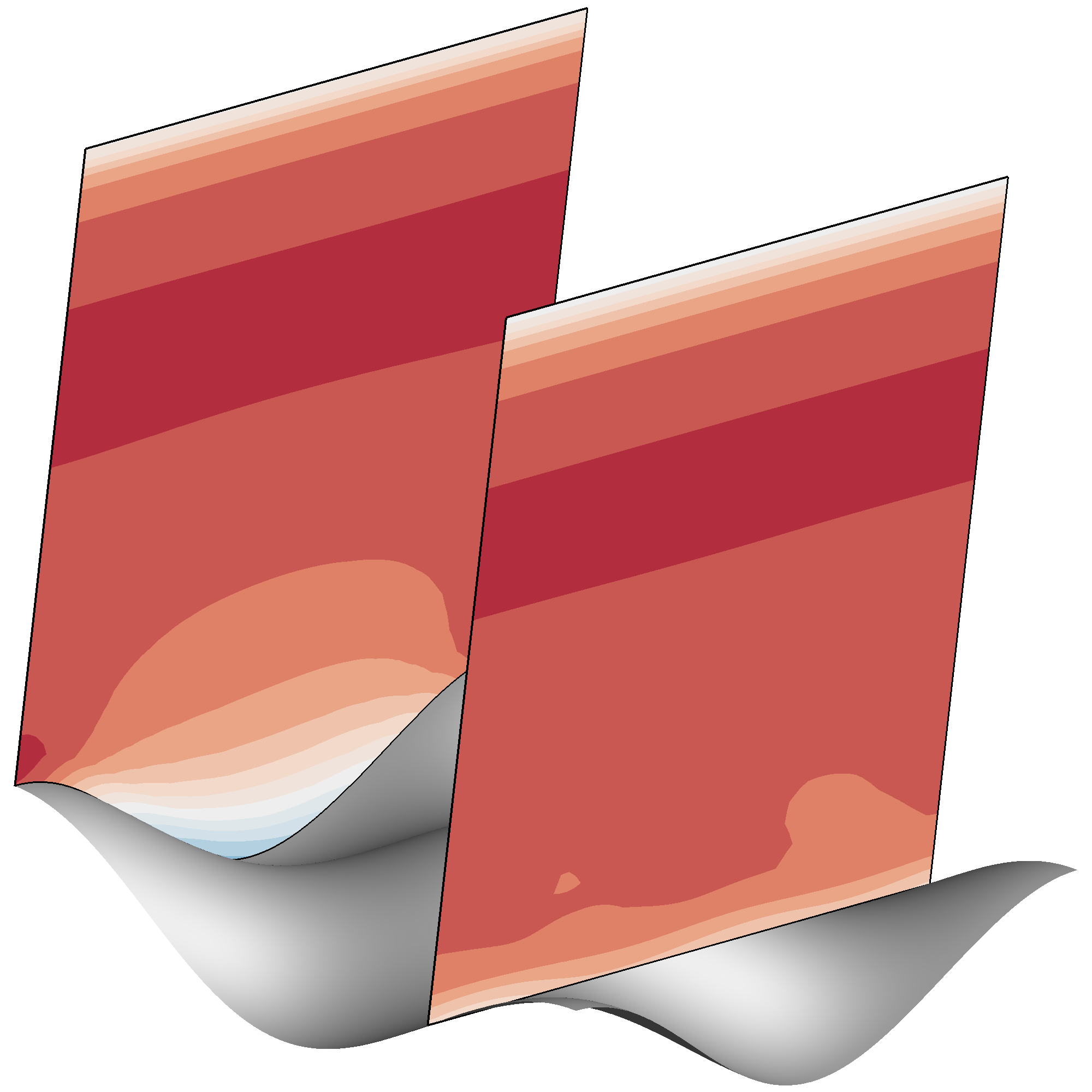}
	\subcaption{3D wavy wall ($\alpha = 0.1$, $\chi = 0.2$): WR vs. FEL vs. WW}
	\end{subfigure}
	\begin{subfigure}[b]{1.0\textwidth}
	\centering
	\includegraphics[width = 0.495\textwidth]{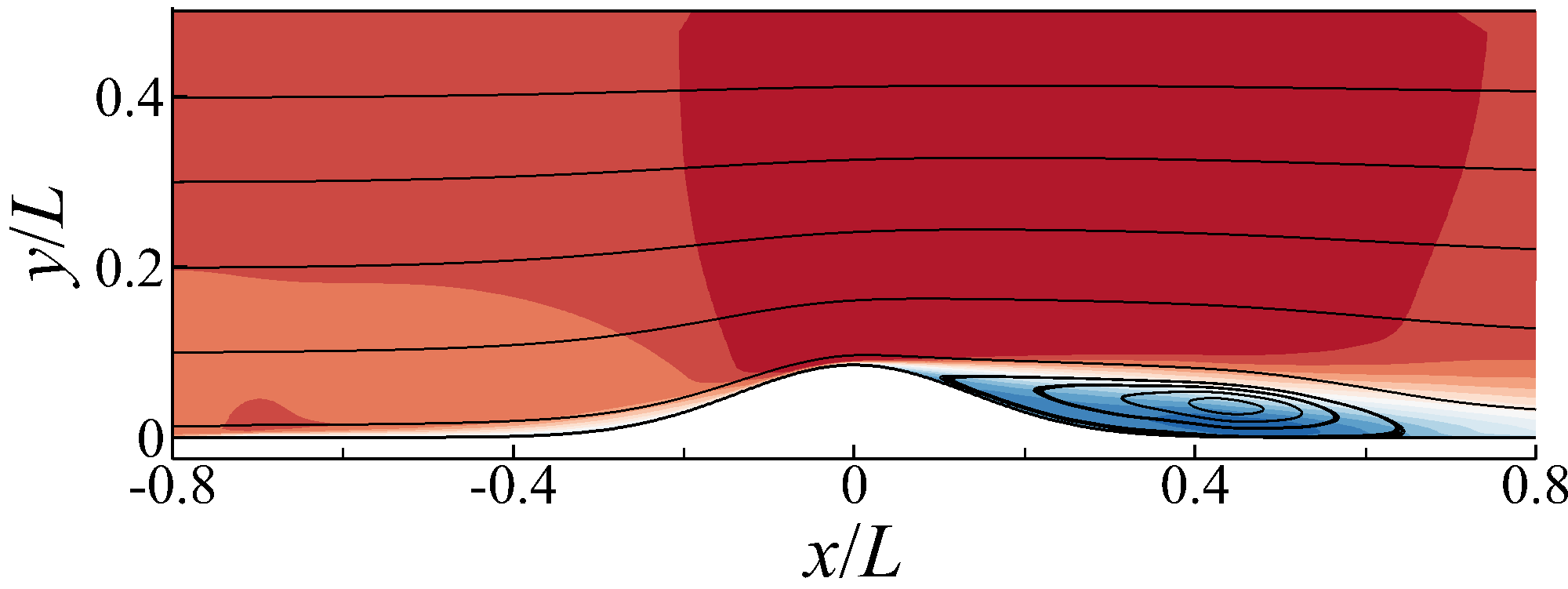}
	\includegraphics[width = 0.495\textwidth]{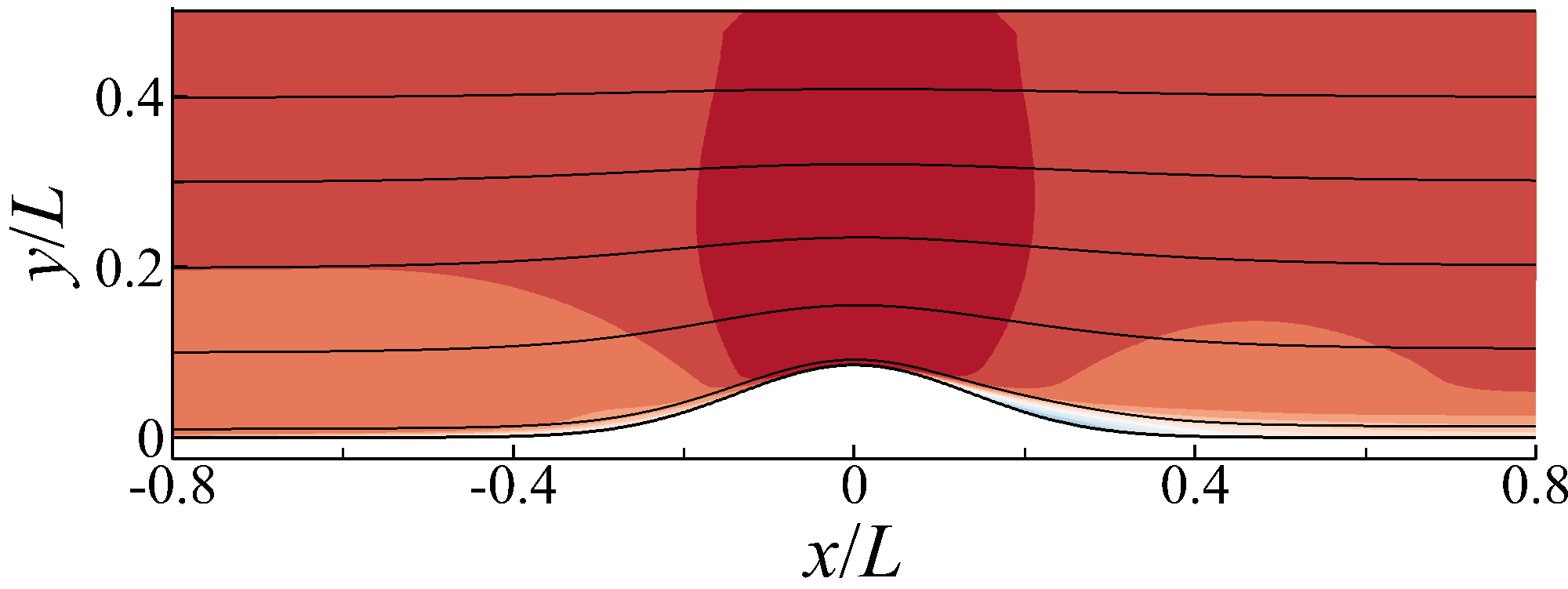}
	\subcaption{2D Gaussian bump: FEL vs. WW}
	\end{subfigure}
\caption{{\color{black}Contours of time-averaged streamwise velocity with streamlines obtained from WRLES and WMLES cases for the simulated flow configurations (figure~\ref{fig:case_validation} and table~\ref{tab:case_validation}).}}
\label{fig:case_validation_contour}
\end{figure}
\begin{figure}
\centering{\includegraphics[width=0.66\textwidth]{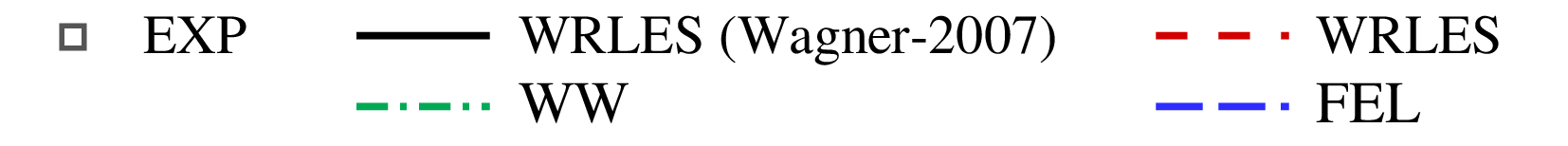}}
\centering{\includegraphics[width=0.42\textwidth]{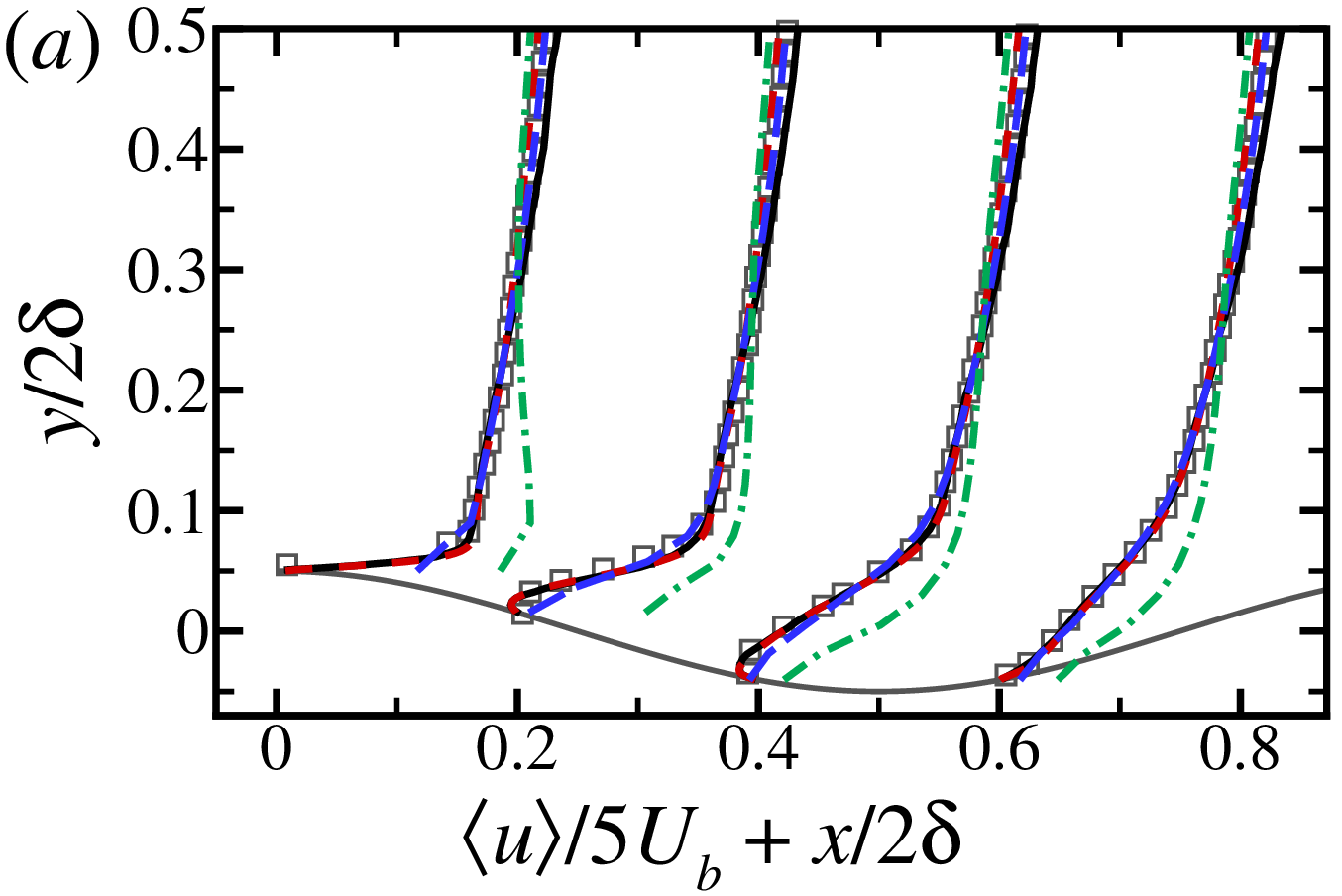}}\quad
\centering{\includegraphics[width=0.42\textwidth]{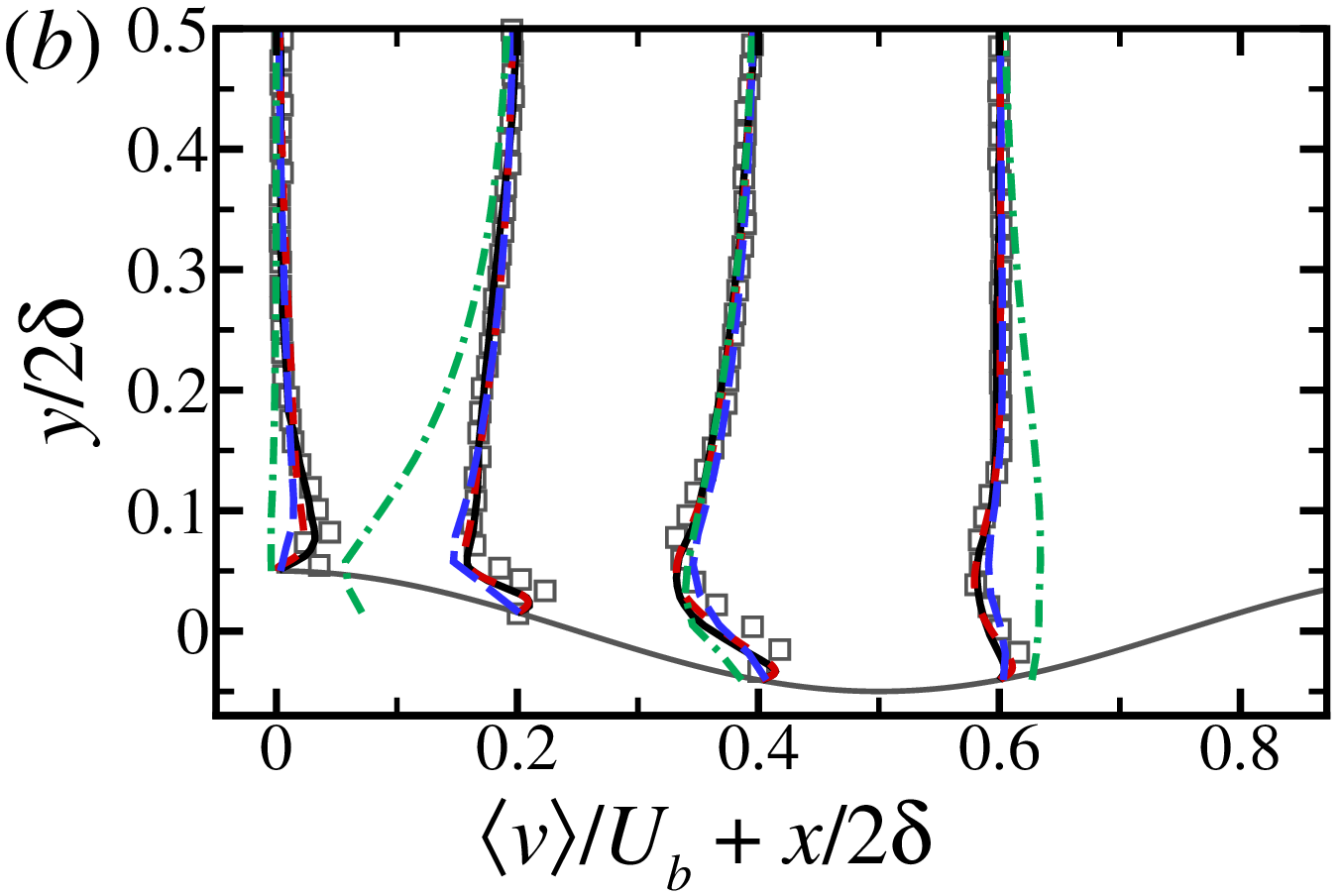}}
\centering{\includegraphics[width=0.42\textwidth]{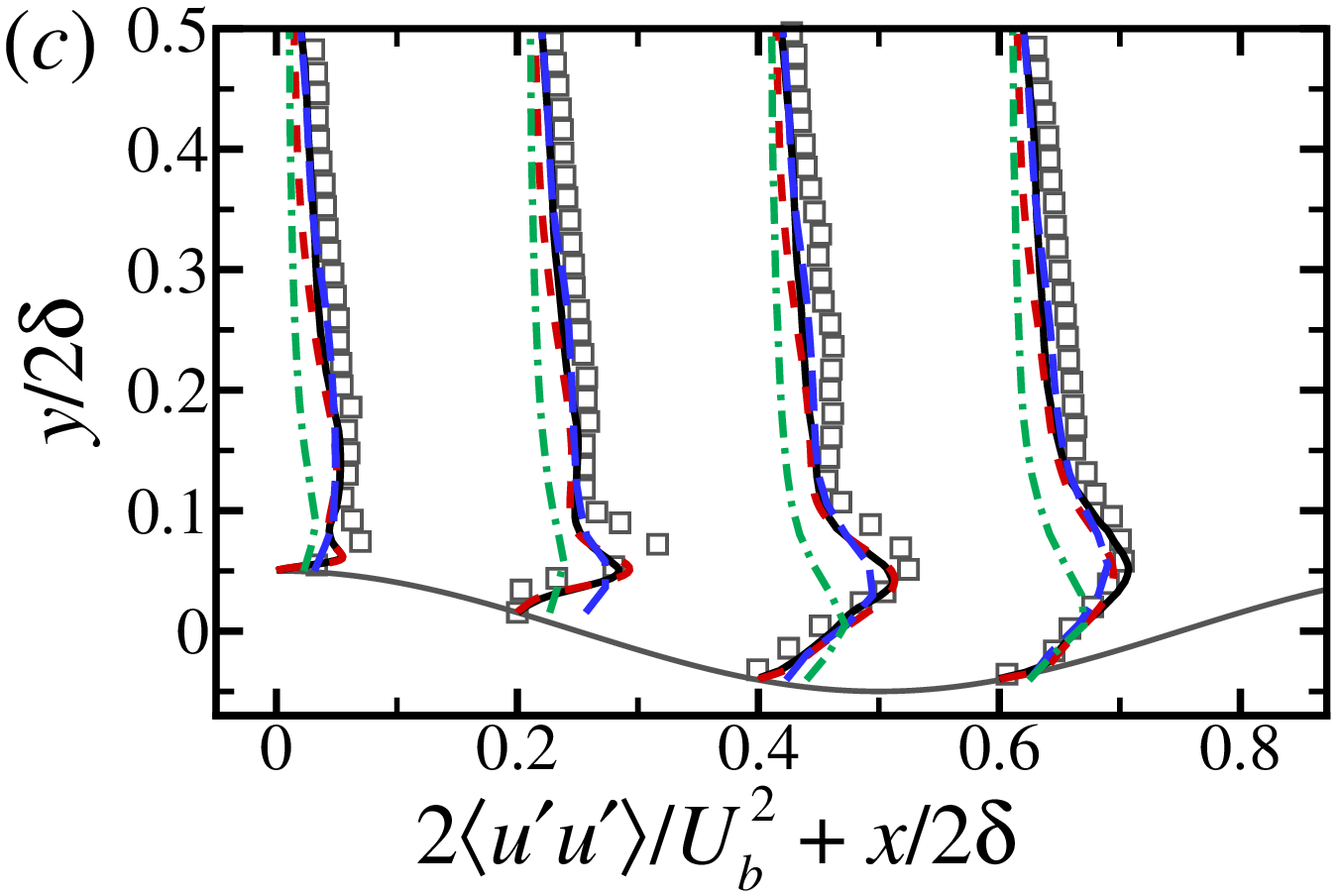}}\quad
\centering{\includegraphics[width=0.42\textwidth]{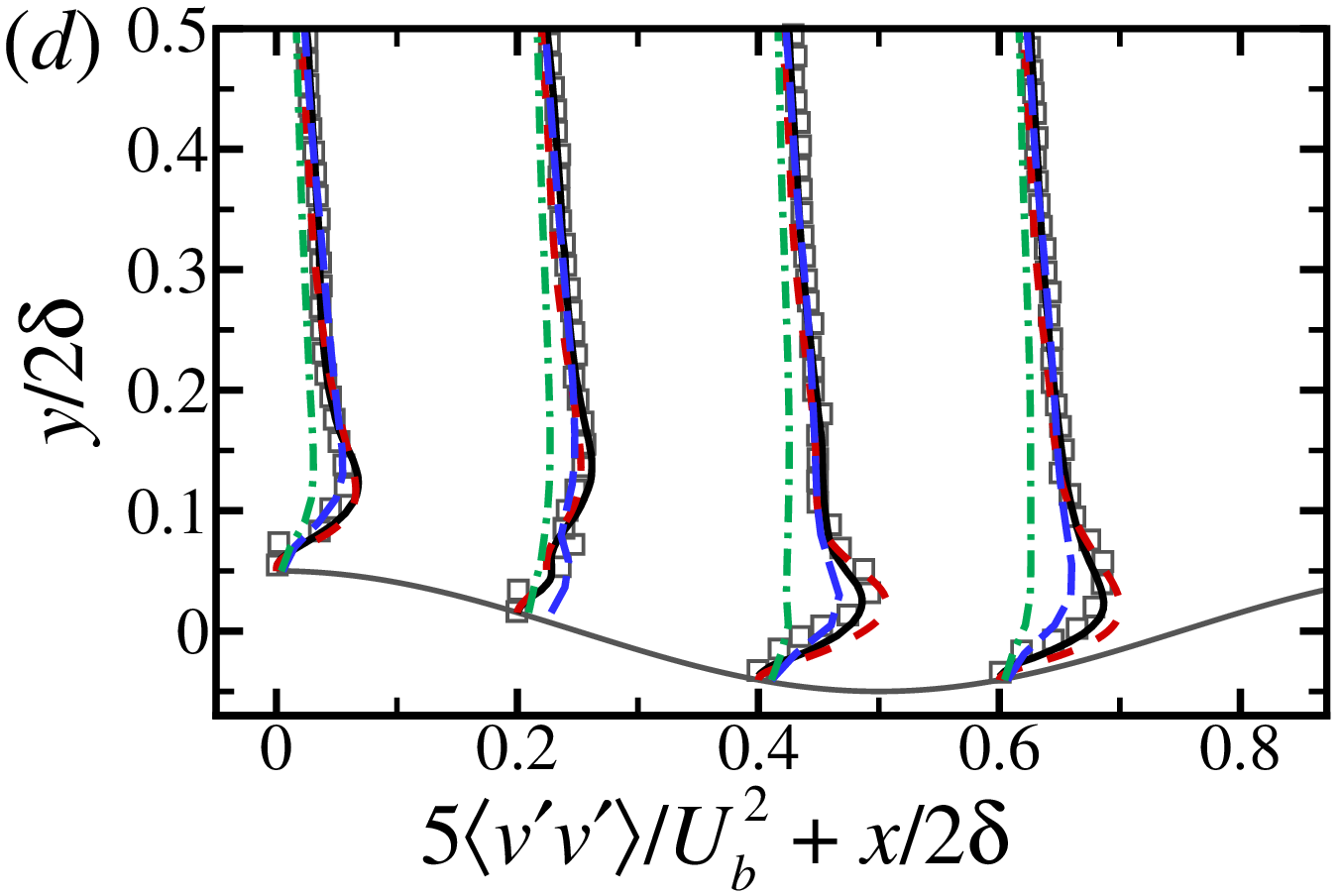}}
\centering{\includegraphics[width=0.42\textwidth]{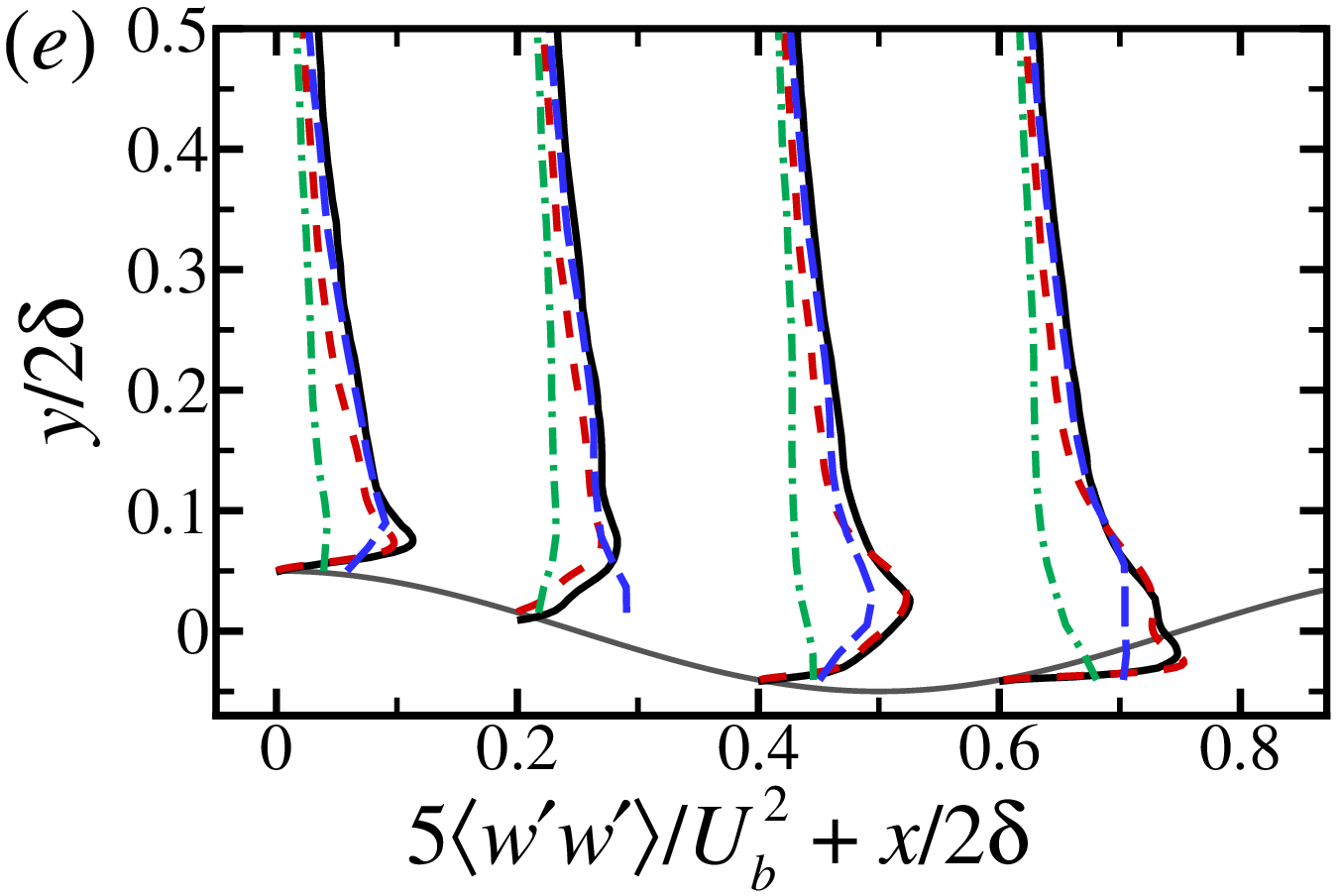}}\quad
\centering{\includegraphics[width=0.42\textwidth]{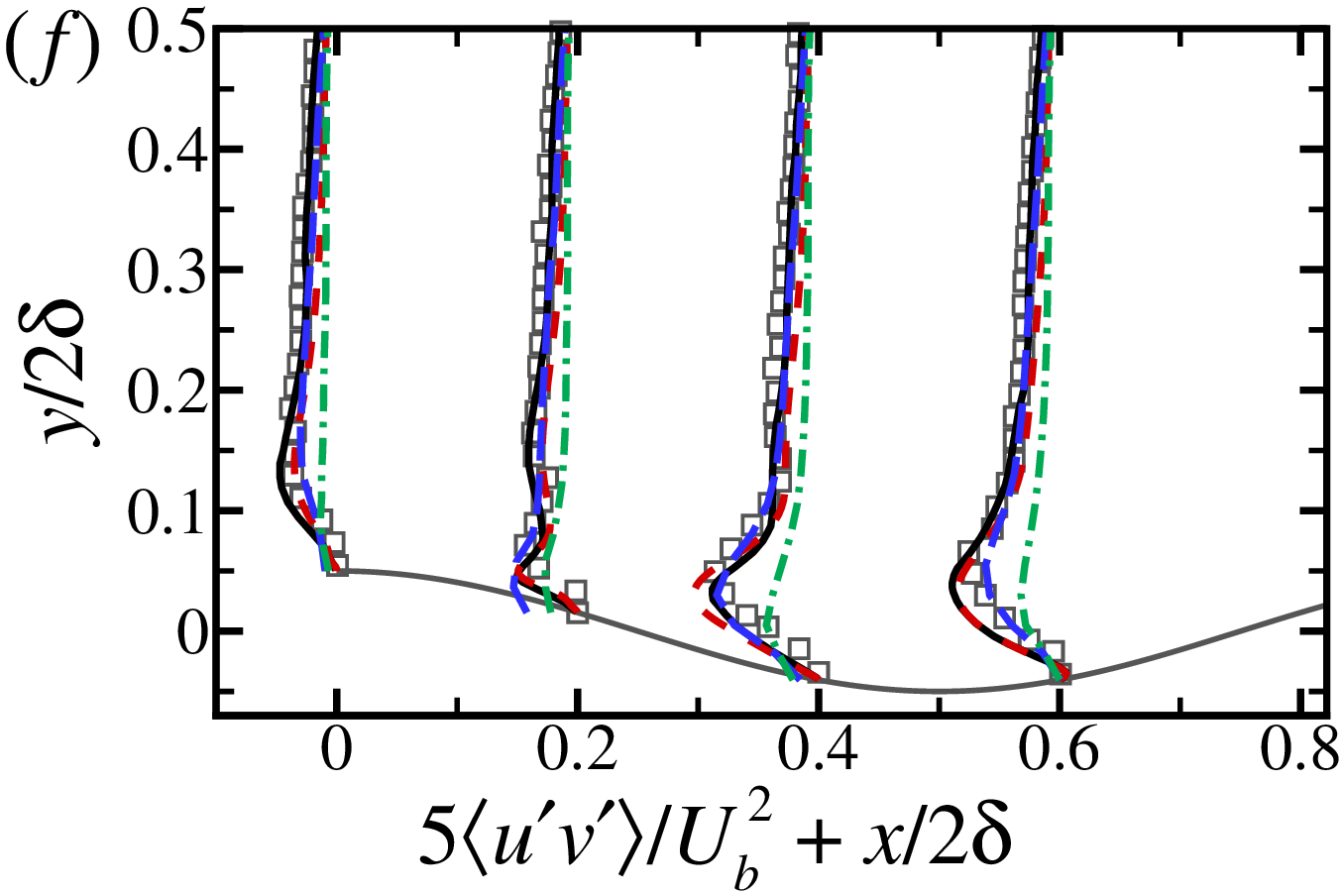}}
  \caption{{\color{black}Vertical profiles of (a)$\sim$(f) $\left\langle u \right\rangle$, $\left\langle v \right\rangle$, $\left\langle u'u' \right\rangle$, $\left\langle v'v' \right\rangle$, $\left\langle w'w' \right\rangle$ and $\left\langle u'v' \right\rangle$ from the 2D-wavy-WR and WM cases with $\alpha = 0.05$, $\chi = 0.1$ (figure~\ref{fig:case_validation}(a) and table~\ref{tab:case_validation}). The experimental data (``EXP'') of \citet{Wagner_etal_EF_2007} and WRLES data of \citet{Wagner_etal_JoT_2010} are included as comparison.}}
\label{fig:profile_case_wavy2D_0p05}
\end{figure}
\begin{figure}
\centering{\includegraphics[width=0.68\textwidth]{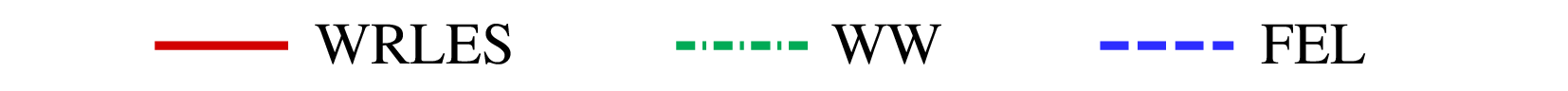}}
\centering{\includegraphics[width=0.42\textwidth]{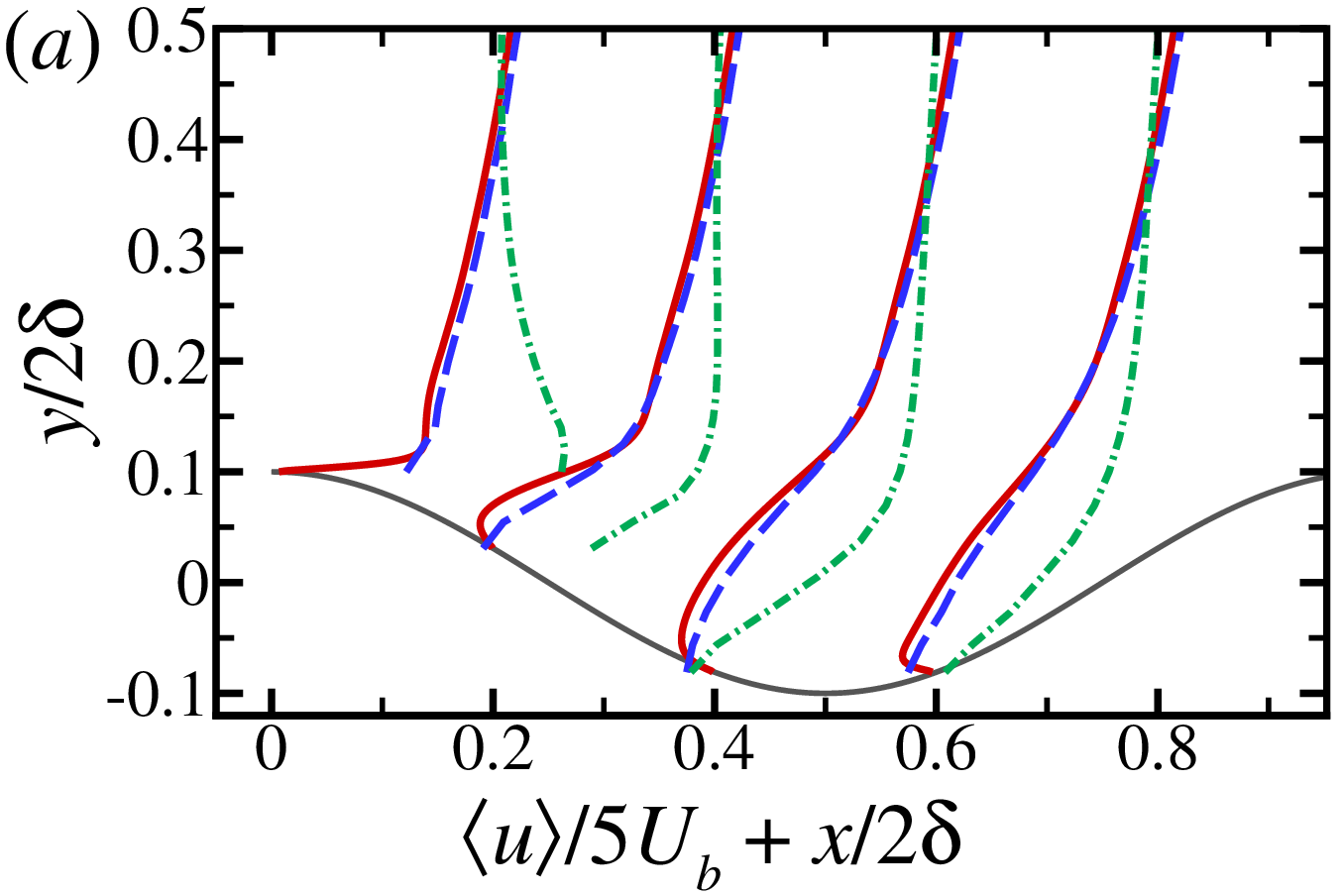}}\quad
\centering{\includegraphics[width=0.42\textwidth]{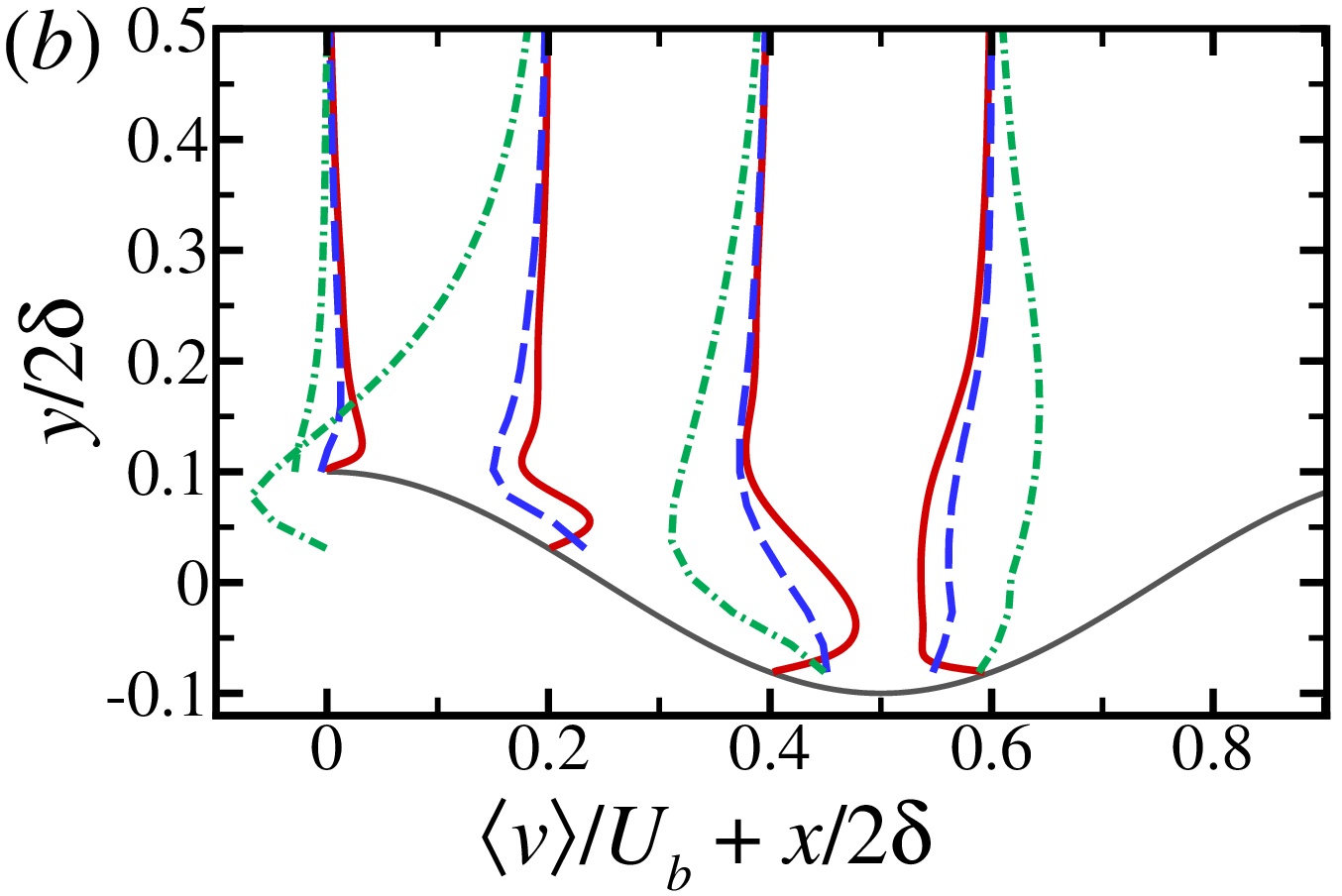}}
\centering{\includegraphics[width=0.42\textwidth]{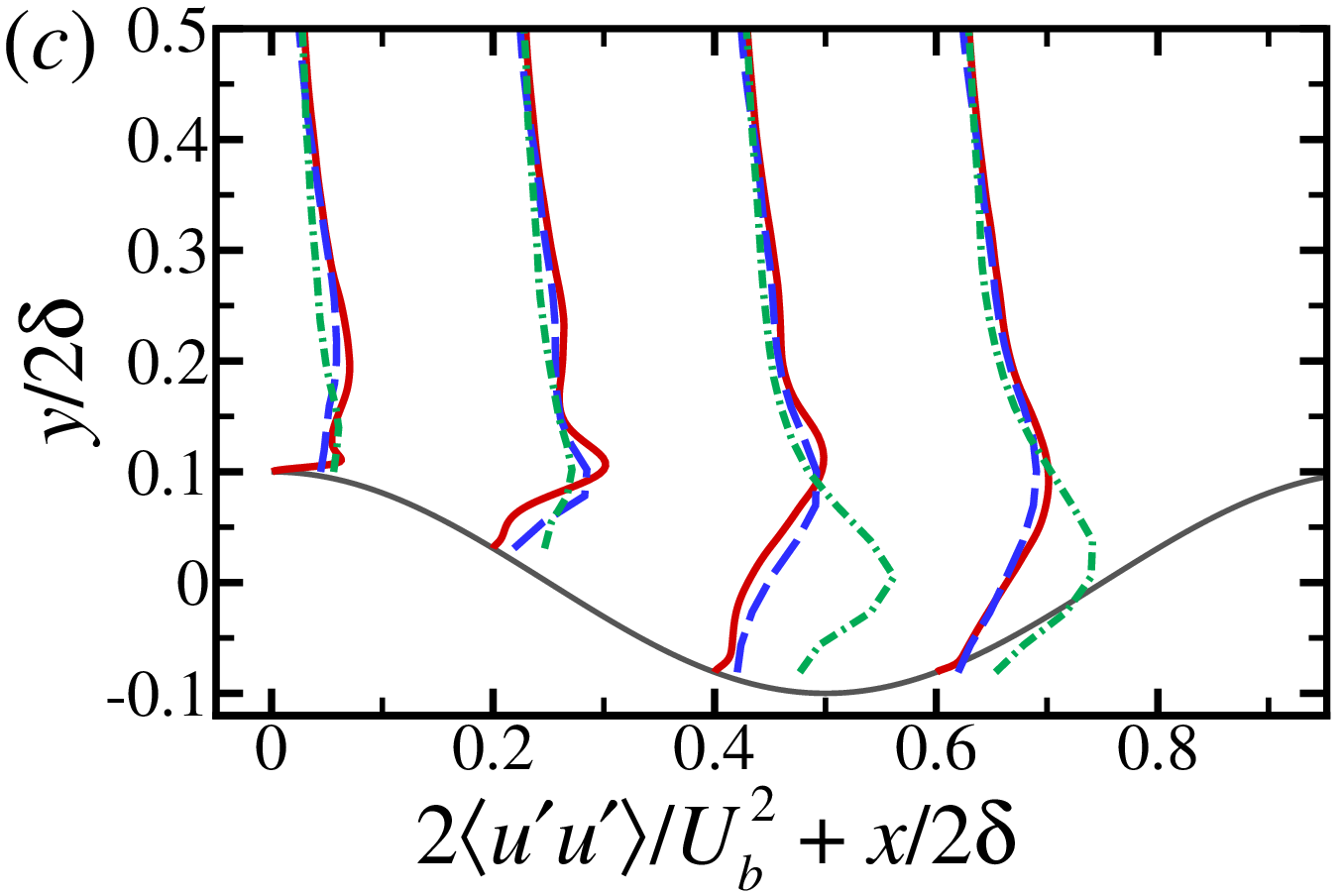}}\quad
\centering{\includegraphics[width=0.42\textwidth]{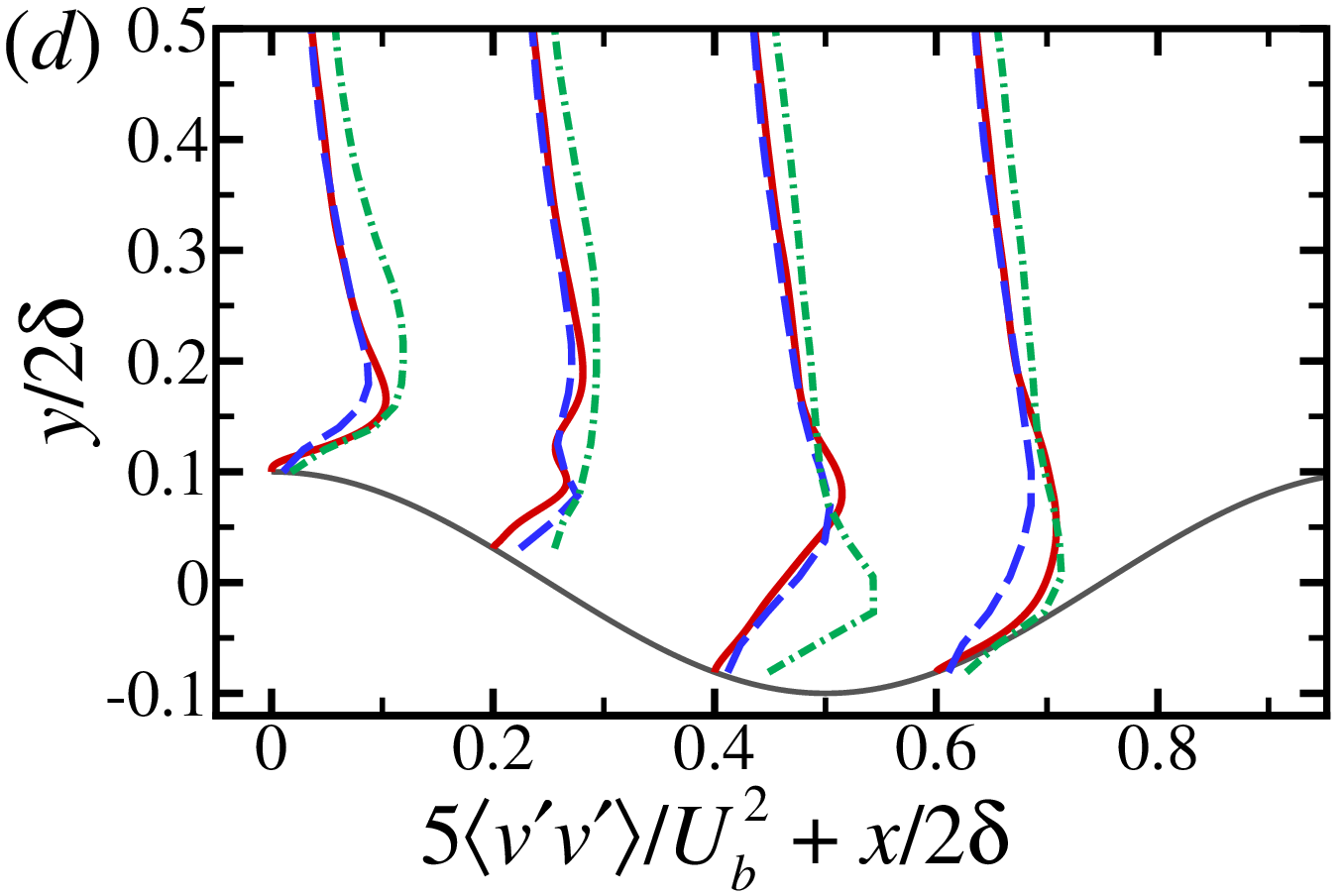}}
\centering{\includegraphics[width=0.42\textwidth]{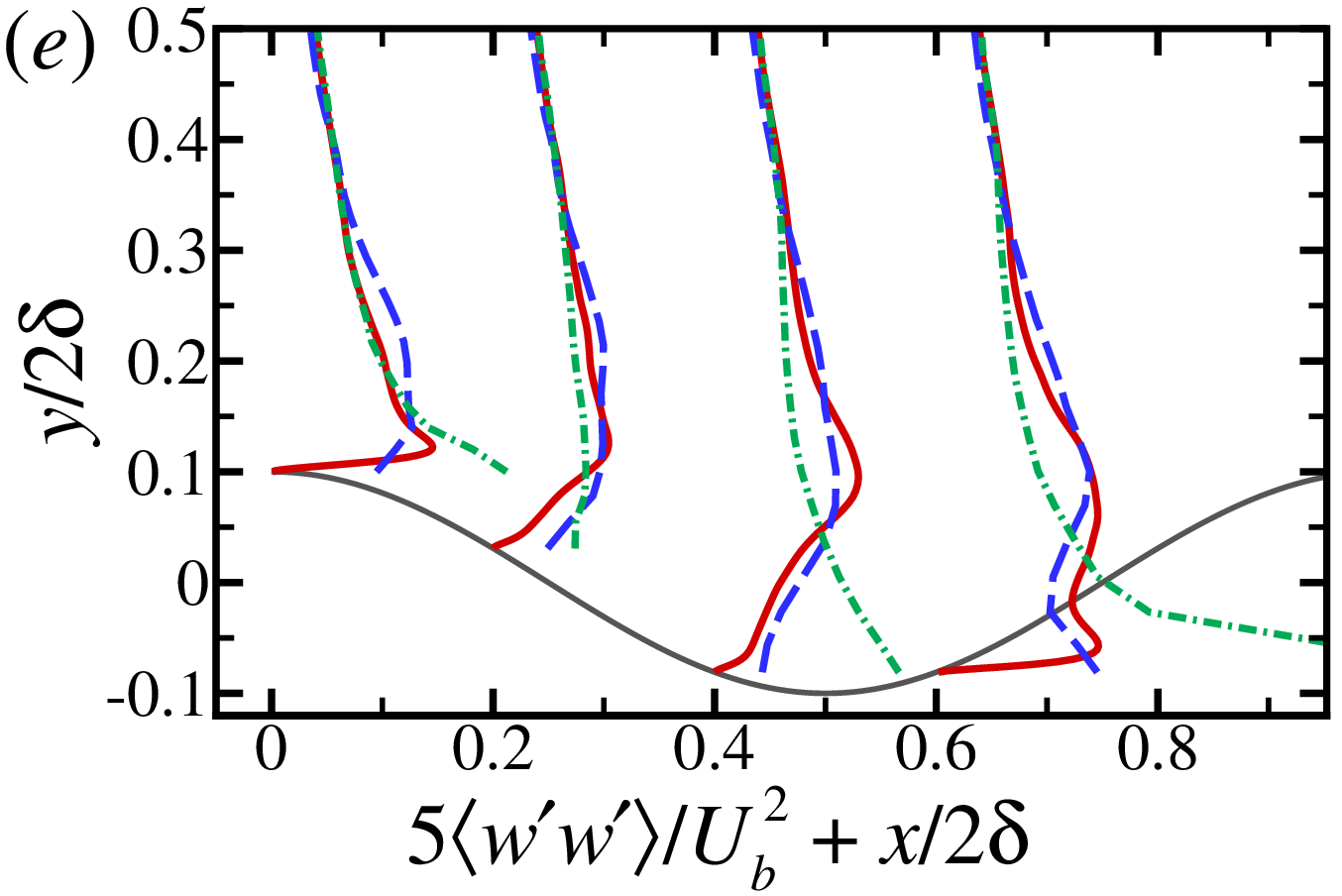}}\quad
\centering{\includegraphics[width=0.42\textwidth]{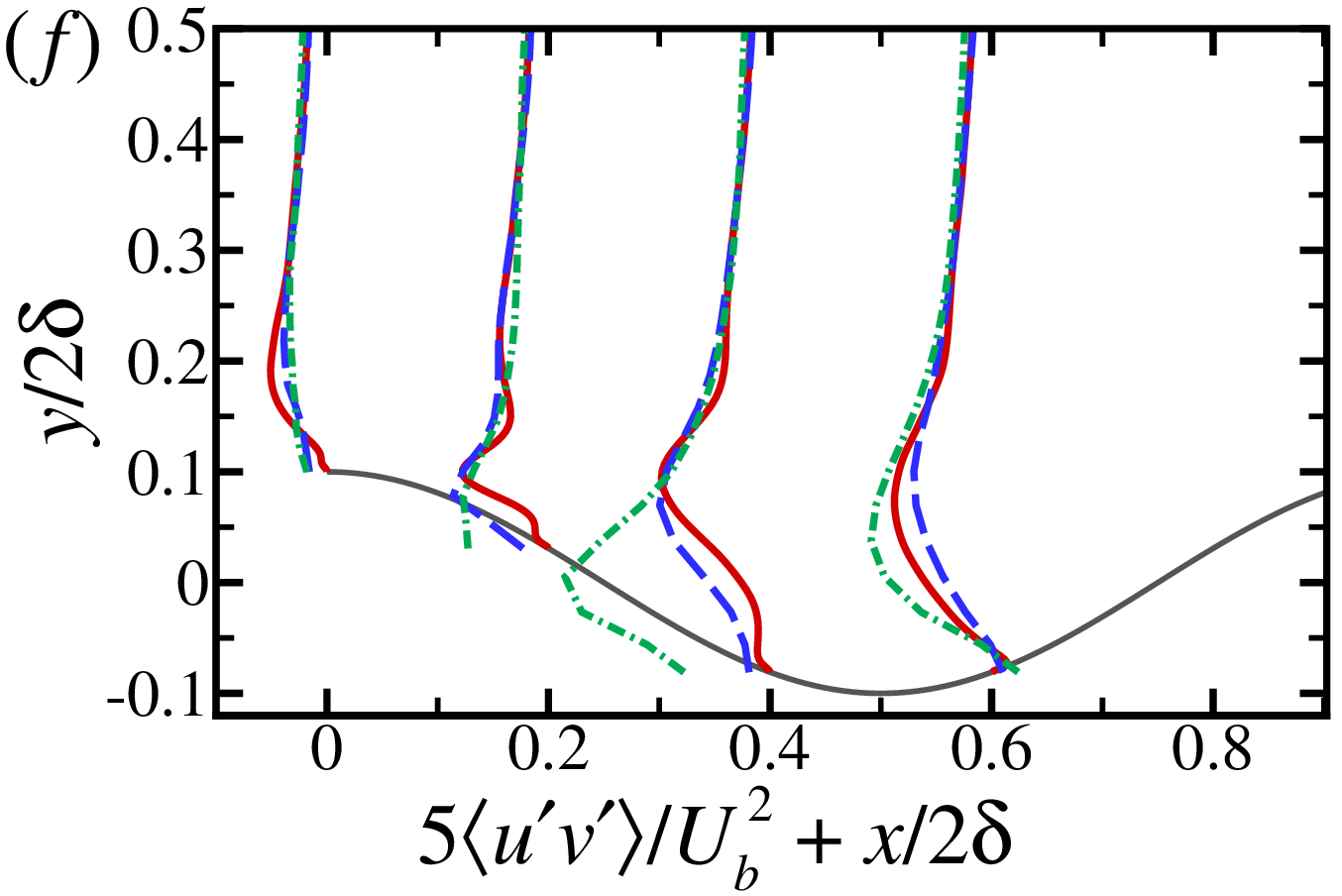}}
  \caption{{\color{black}Vertical profiles of (a)$\sim$(f) $\left\langle u \right\rangle$, $\left\langle v \right\rangle$, $\left\langle u'u' \right\rangle$, $\left\langle v'v' \right\rangle$, $\left\langle w'w' \right\rangle$ and $\left\langle u'v' \right\rangle$ from the 2D-wavy-WR and WM cases with $\alpha = 0.1$, $\chi = 0.2$ (figure~\ref{fig:case_validation}(a) and table~\ref{tab:case_validation}).}}
\label{fig:profile_case_wavy2D_0p1}
\end{figure}
\begin{figure}
\centering{\includegraphics[width=0.88\textwidth]{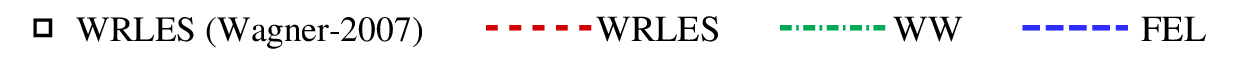}}
\centering{\includegraphics[width=0.42\textwidth]{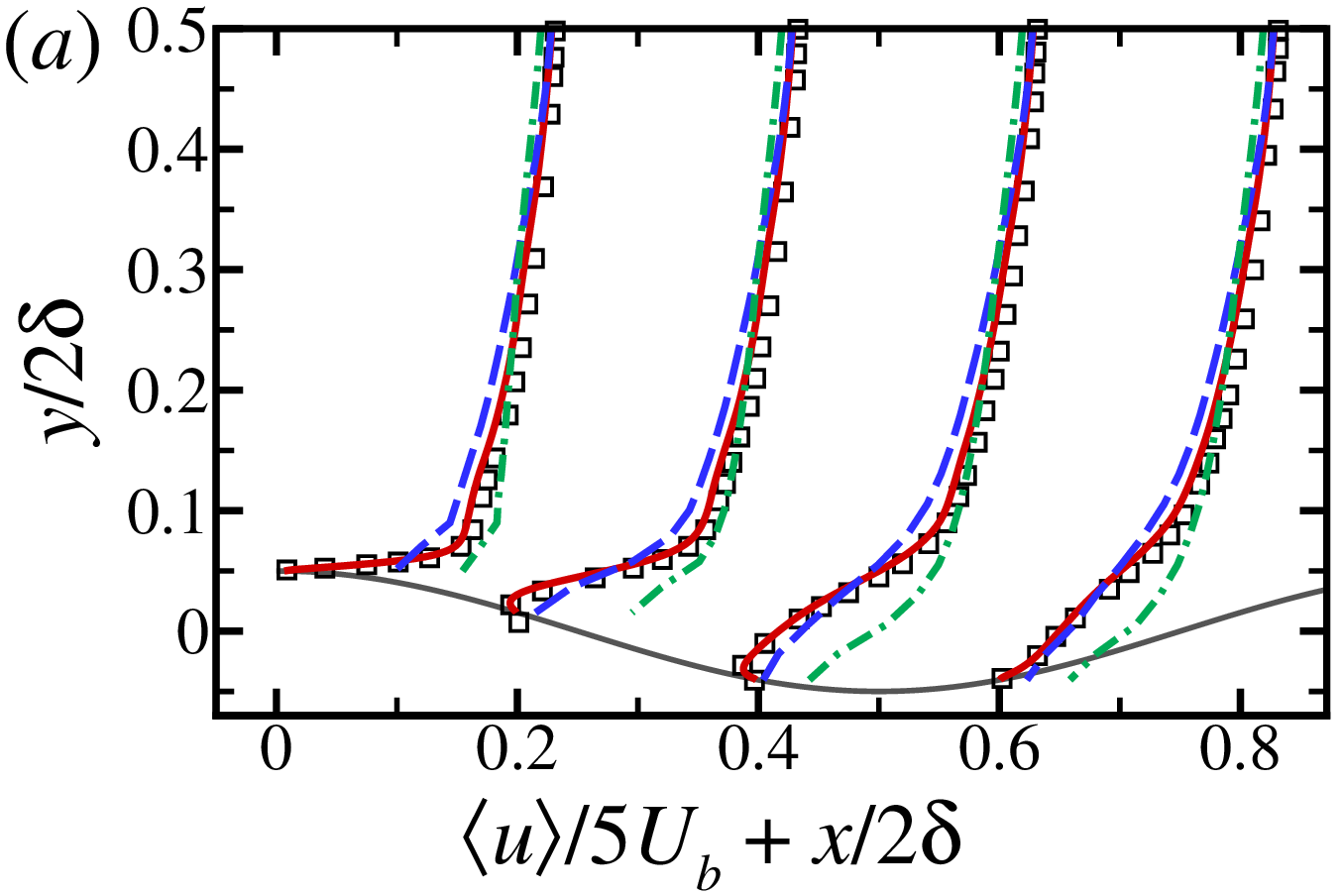}}\quad
\centering{\includegraphics[width=0.42\textwidth]{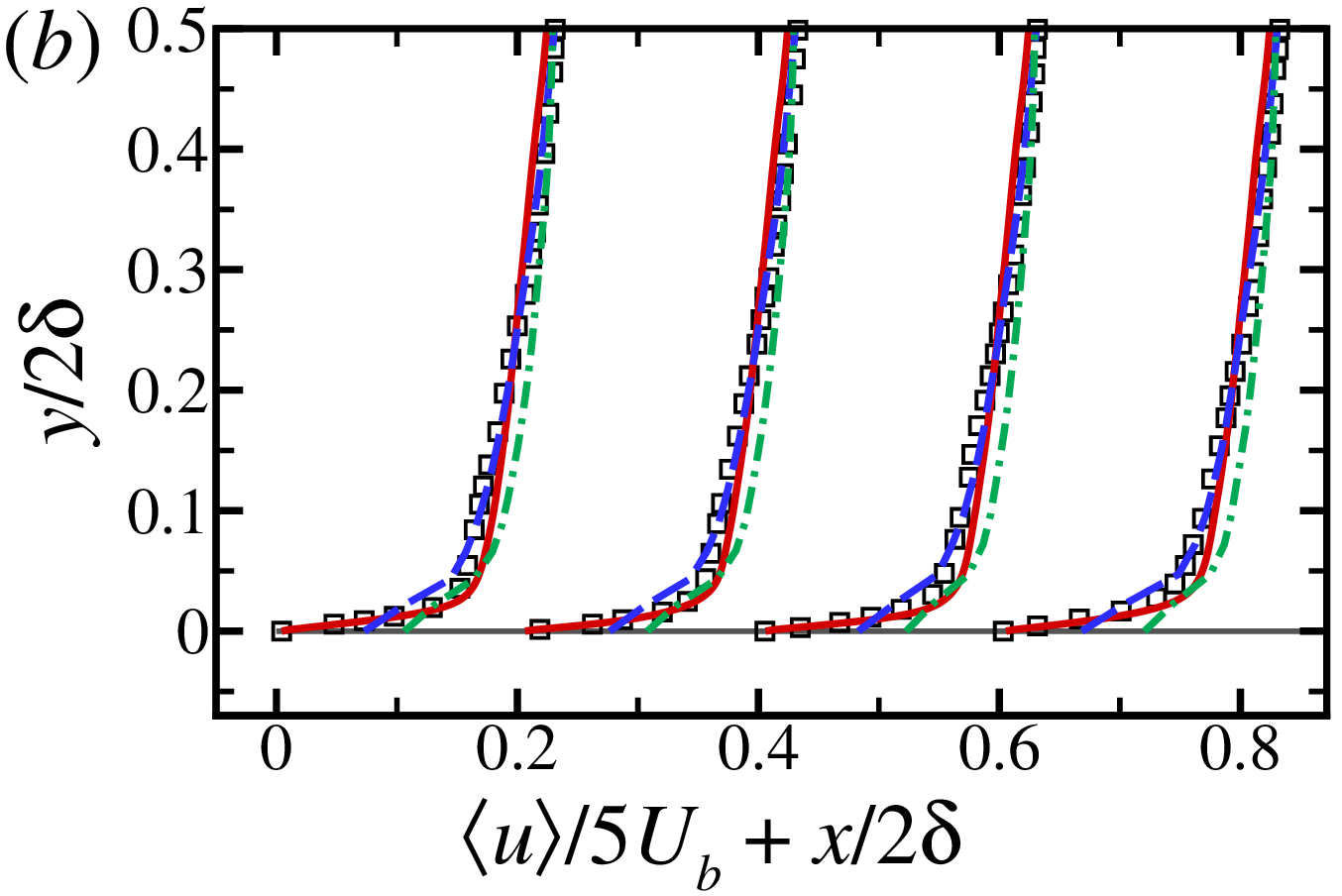}}
\centering{\includegraphics[width=0.42\textwidth]{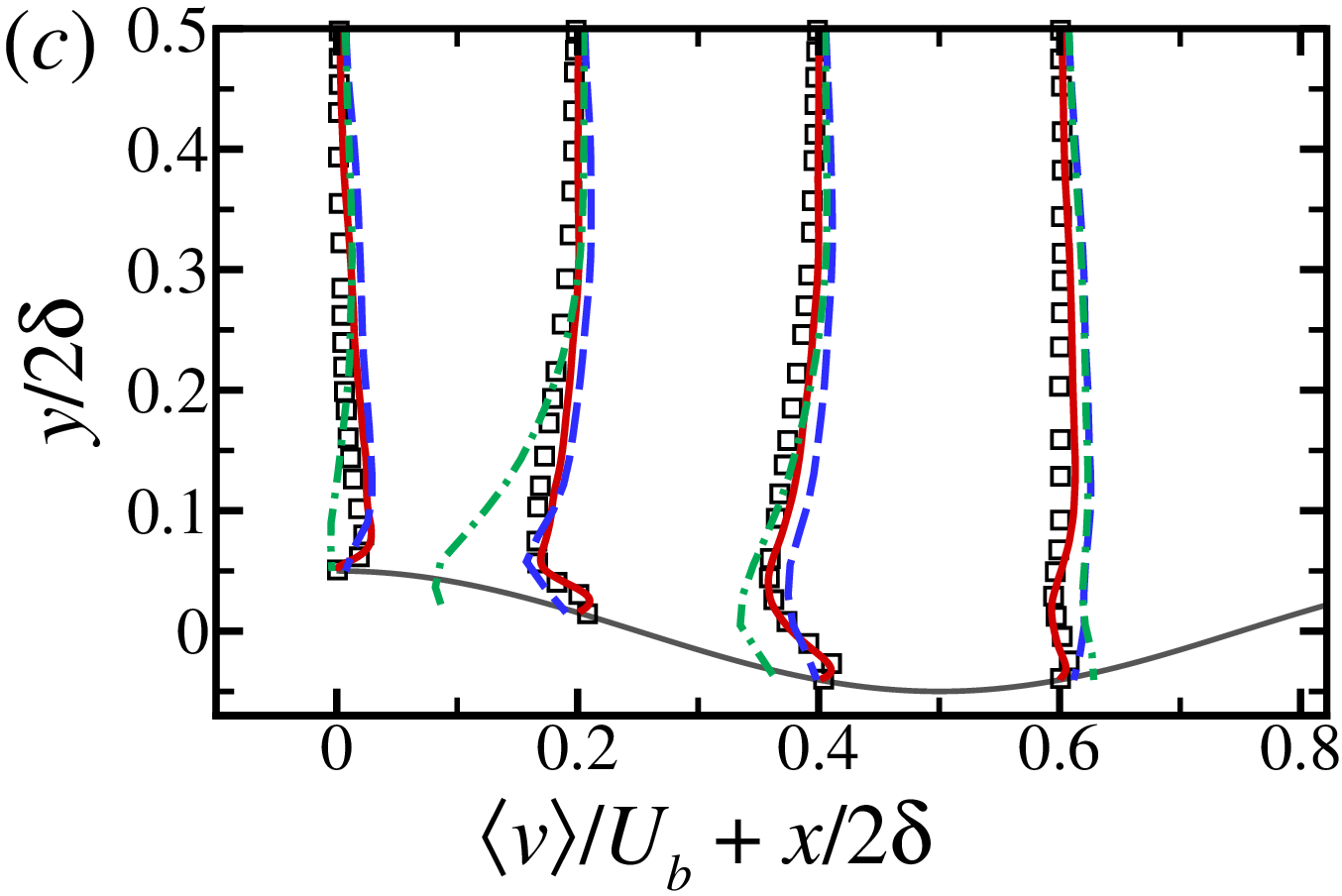}}\quad
\centering{\includegraphics[width=0.42\textwidth]{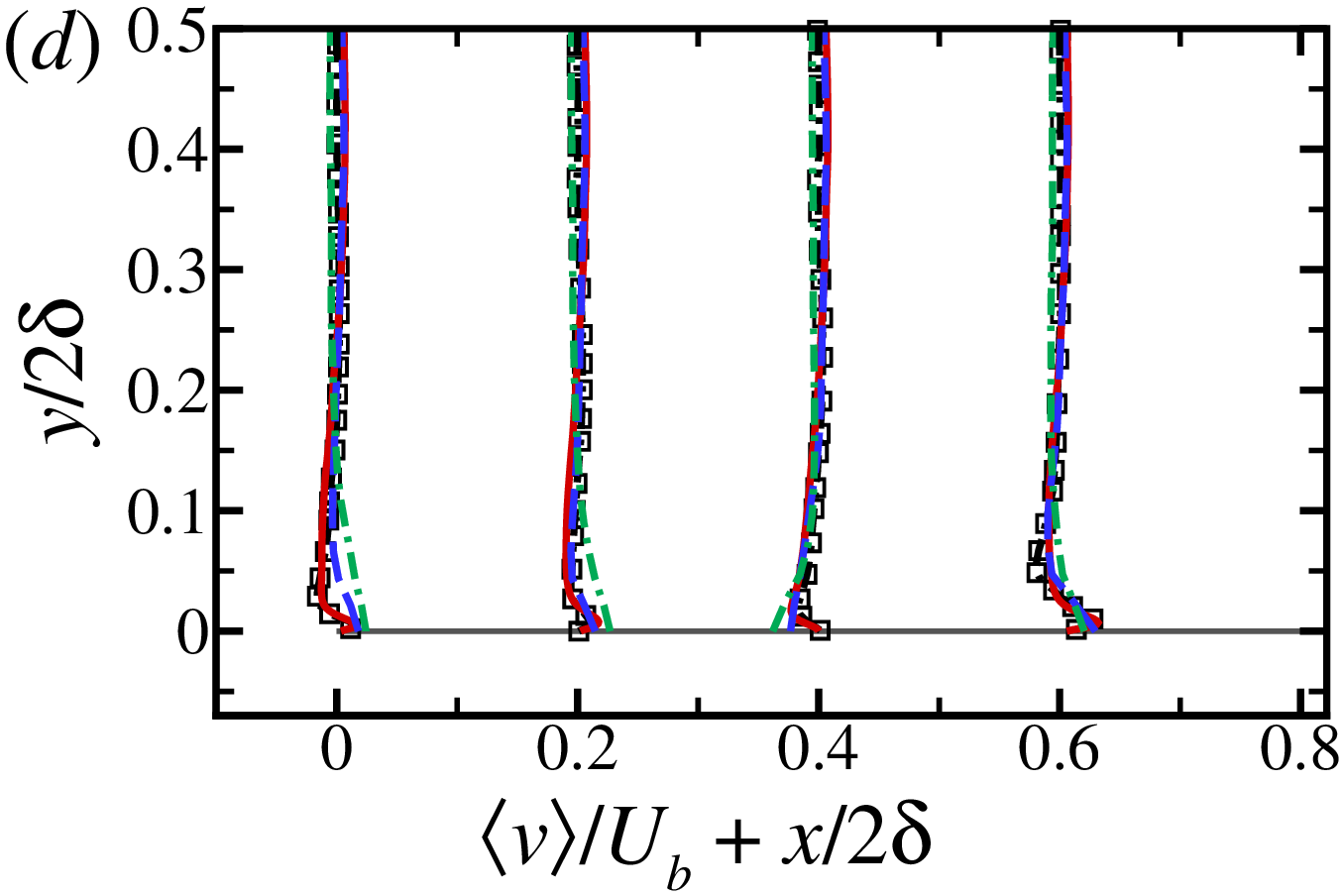}}
  \caption{{\color{black}Vertical profiles of (a, b) time-averaged streamwise velocity $\left\langle u \right\rangle$ and (c, d) vertical velocity $\left\langle v \right\rangle$ at $z/2\delta = 0.0$ (a, c) and $z/2\delta = 0.25$ (b, d) from the 3D-wavy-WR and WM cases with $\alpha = 0.05$, $\chi = 0.1$ (figure~\ref{fig:case_validation}(b) and table~\ref{tab:case_validation}). The WRLES data of \citet{Wagner_etal_JoT_2010} are included as comparison.}}
\label{fig:profile_case_wavy3D_0p05}
\end{figure}
\begin{figure}
\centering{\includegraphics[width=0.68\textwidth]{Fig_wavy0p1_profile_legend.eps}}
\centering{\includegraphics[width=0.42\textwidth]{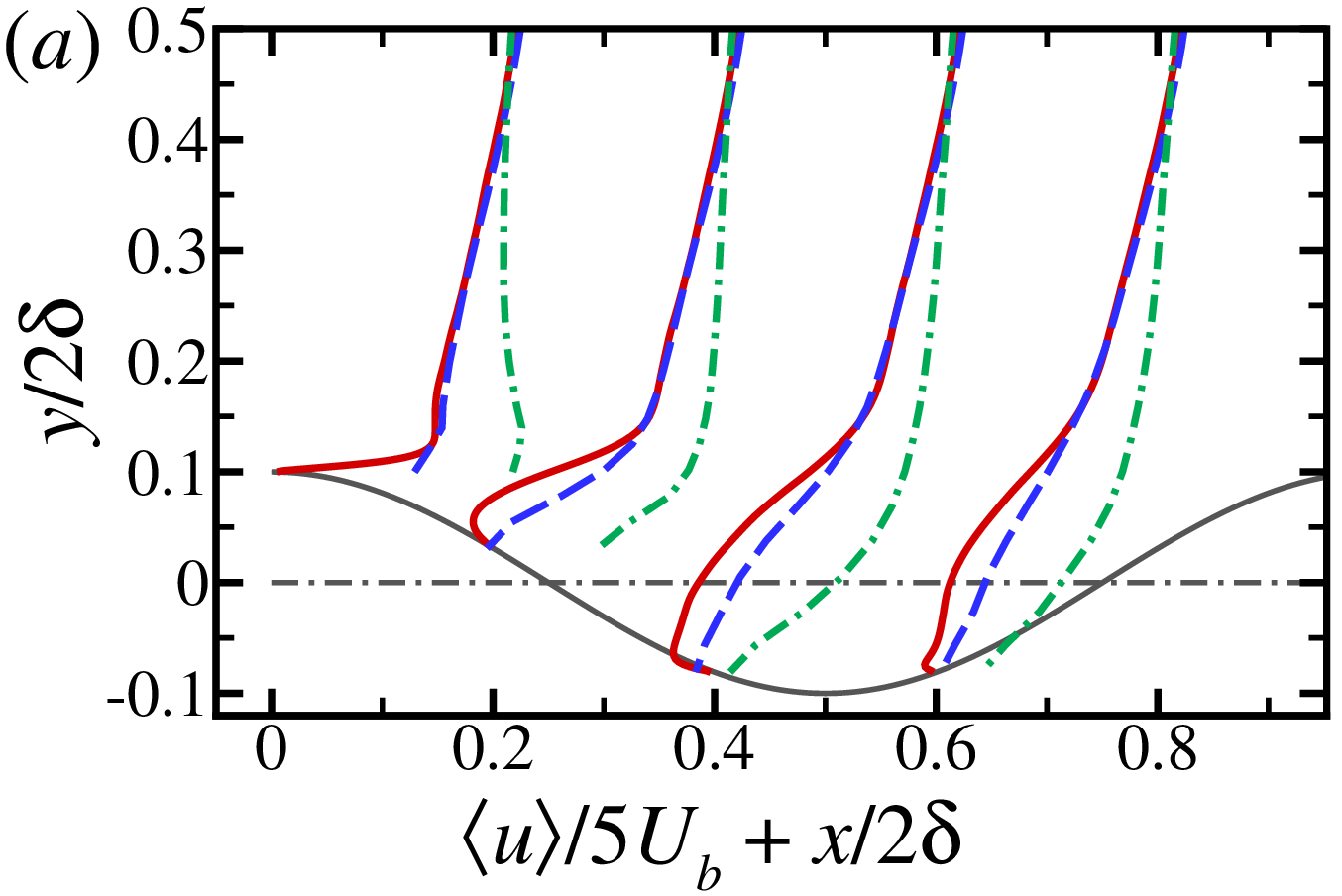}}\quad
\centering{\includegraphics[width=0.42\textwidth]{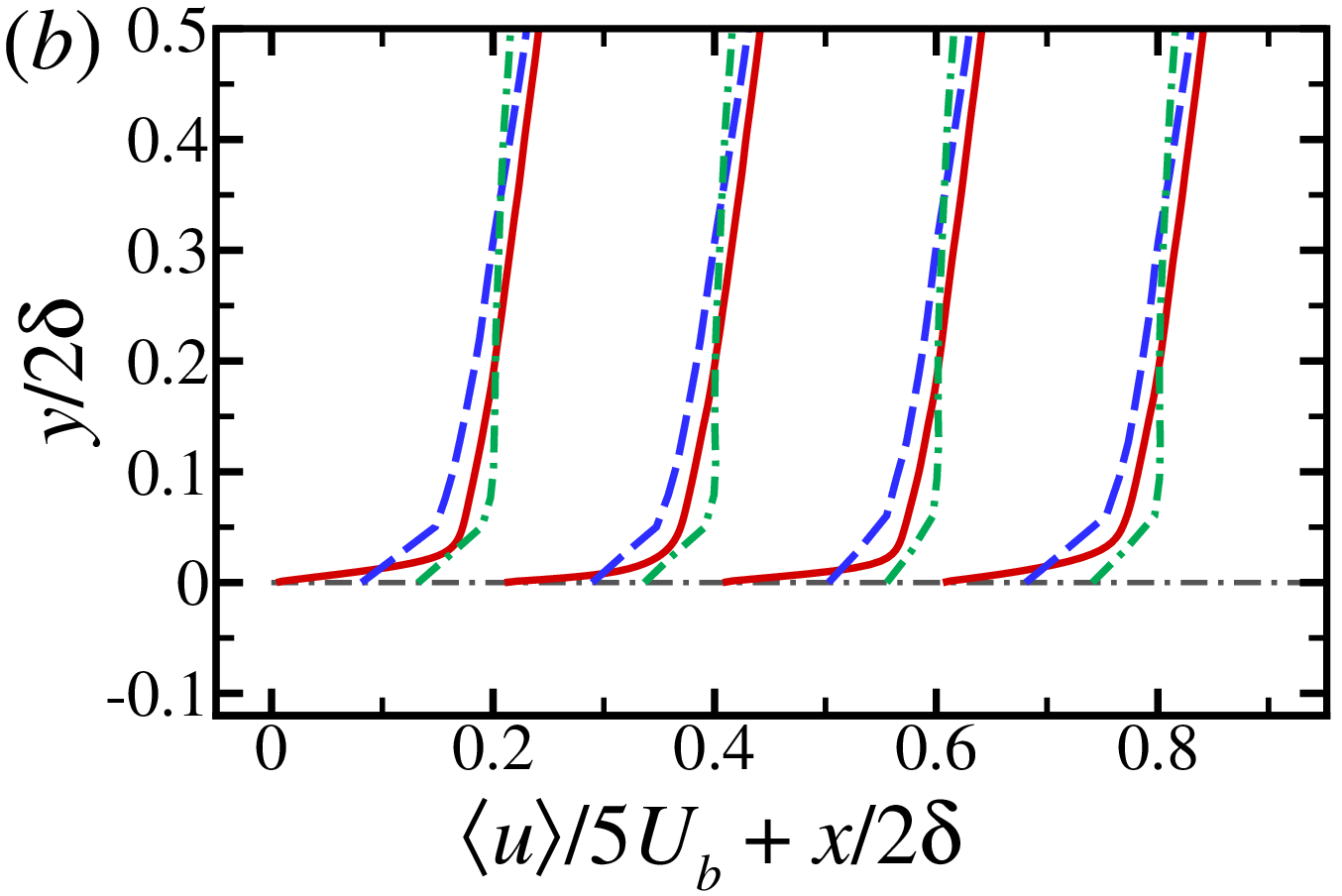}}
\centering{\includegraphics[width=0.42\textwidth]{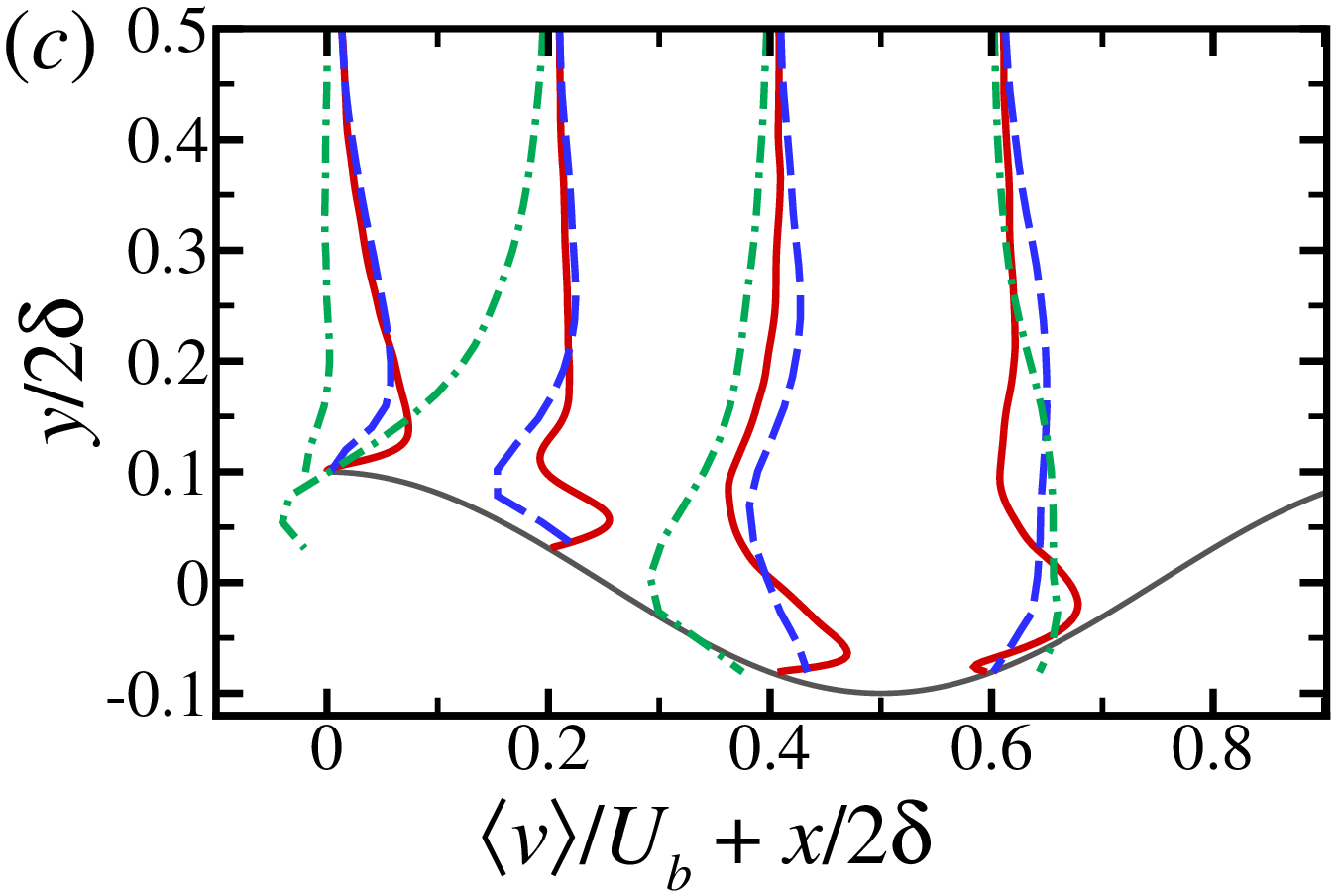}}\quad
\centering{\includegraphics[width=0.42\textwidth]{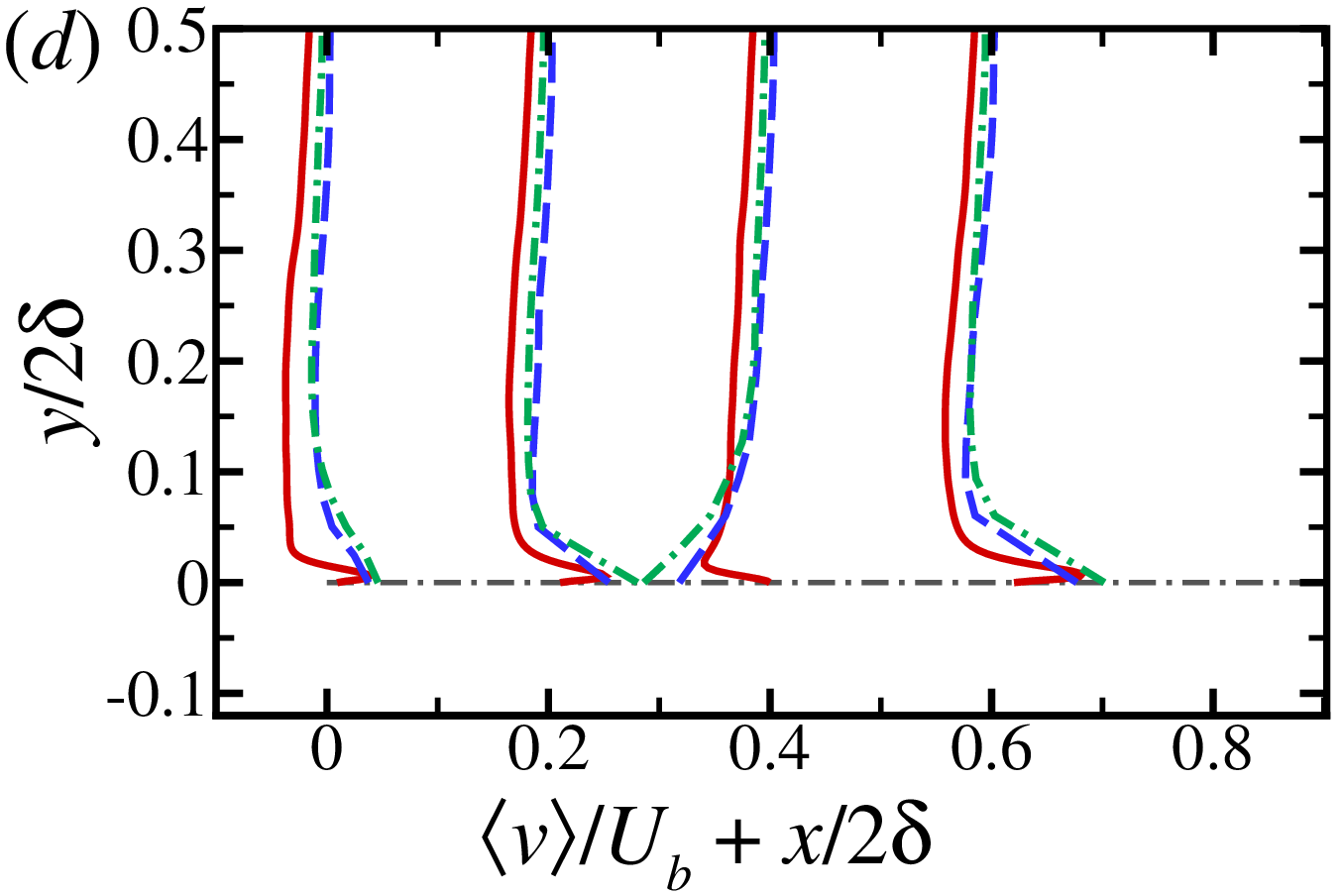}}
  \caption{{\color{black}Vertical profiles of (a, b) time-averaged streamwise velocity $\left\langle u \right\rangle$ and (c, d) vertical velocity $\left\langle v \right\rangle$ at $z/2\delta = 0.0$ (a, c)and $z/2\delta = 0.25$ (b, d) from the 3D-wavy-WR and WM cases with $\alpha = 0.1$, $\chi = 0.2$ (figure~\ref{fig:case_validation}(b) and table~\ref{tab:case_validation}).}}
\label{fig:profile_case_wavy3D_0p1}
\end{figure}
\begin{figure}
\centering{\includegraphics[width=0.68\textwidth]{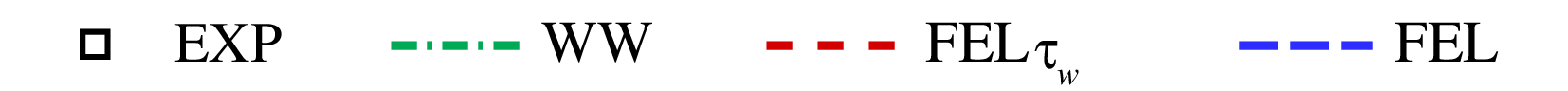}}\\
\centering
	\begin{subfigure}[b]{0.244\textwidth}
	\centering
	\includegraphics[width = 1.0\textwidth]{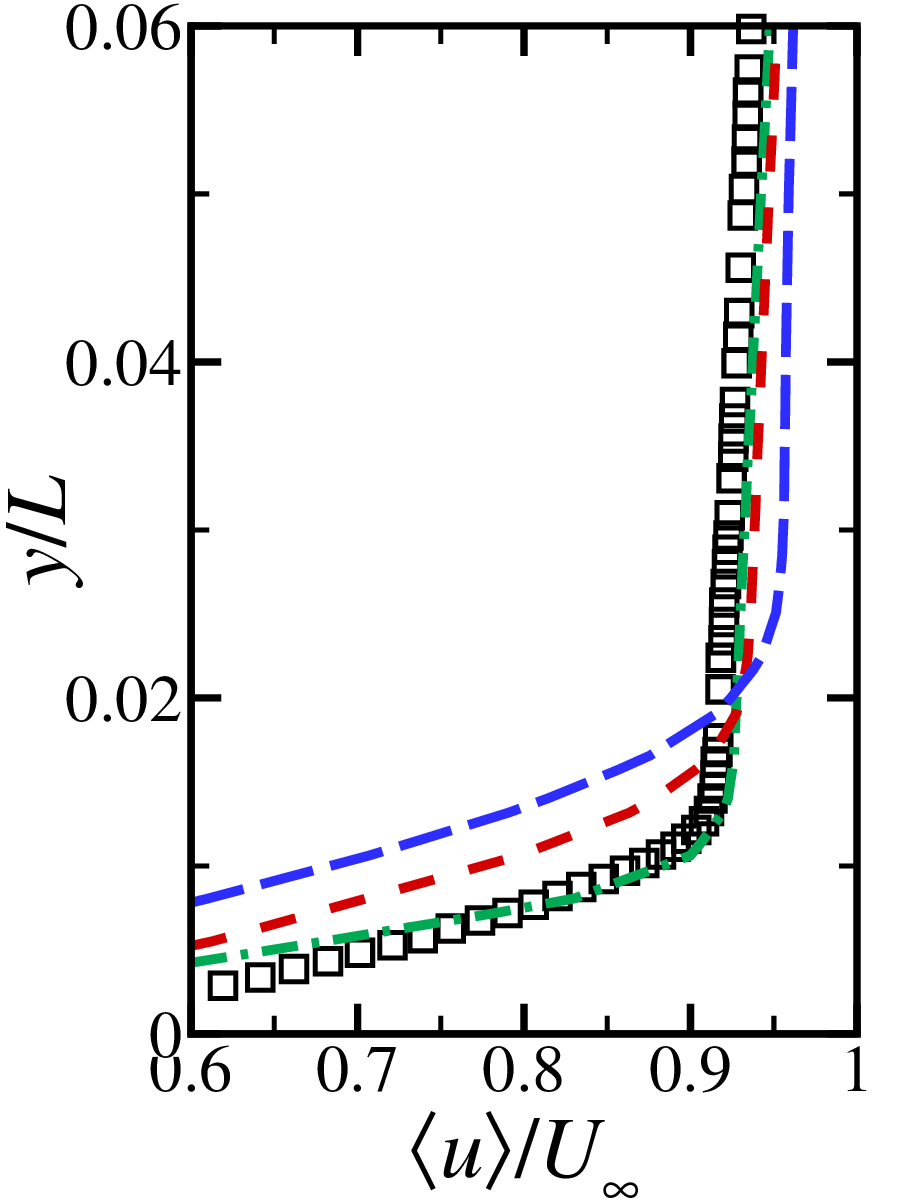}
	\subcaption{$x/L=-0.4$}
	\end{subfigure}
	\begin{subfigure}[b]{0.244\textwidth}
	\centering
	\includegraphics[width = 1.0\textwidth]{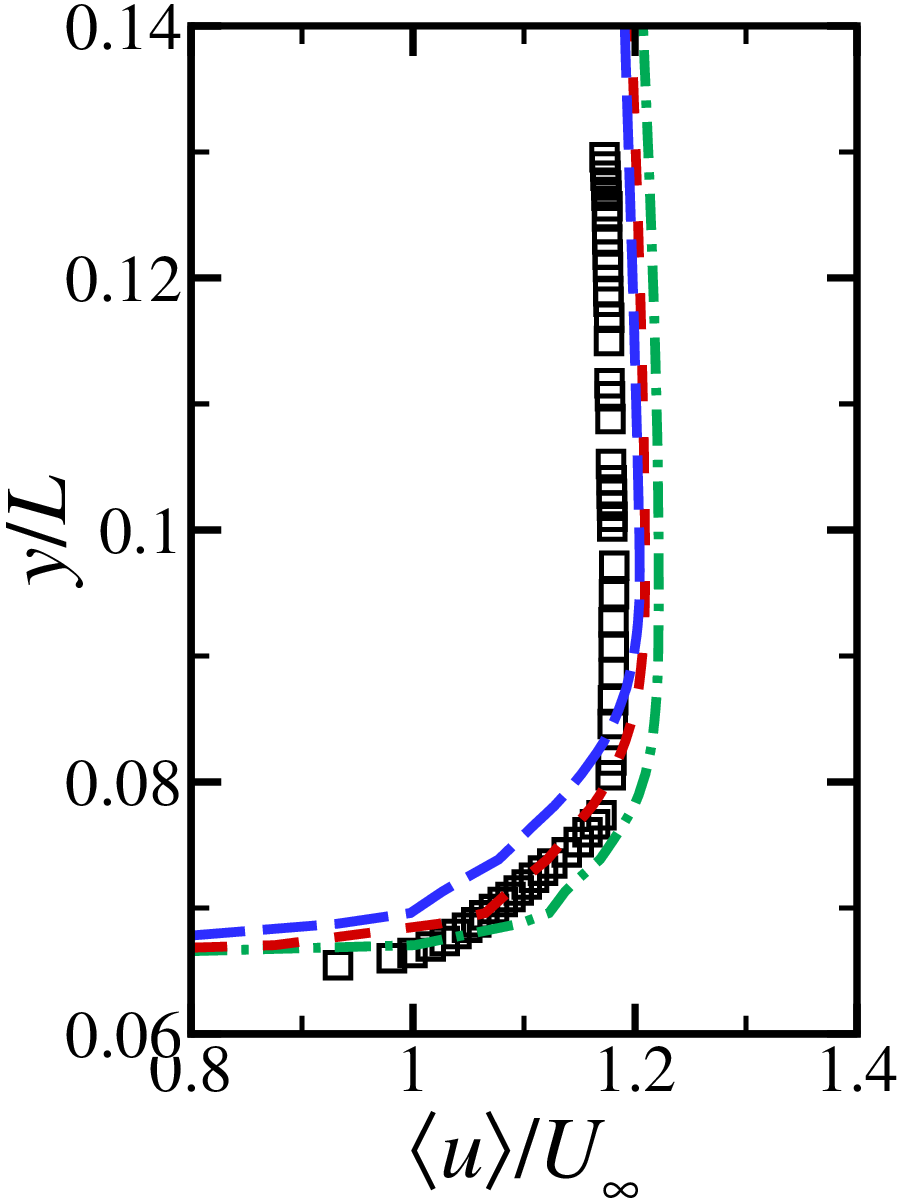}
	\subcaption{$x/L=-0.1$}
	\end{subfigure}
	\begin{subfigure}[b]{0.244\textwidth}
	\centering
	\includegraphics[width = 1.0\textwidth]{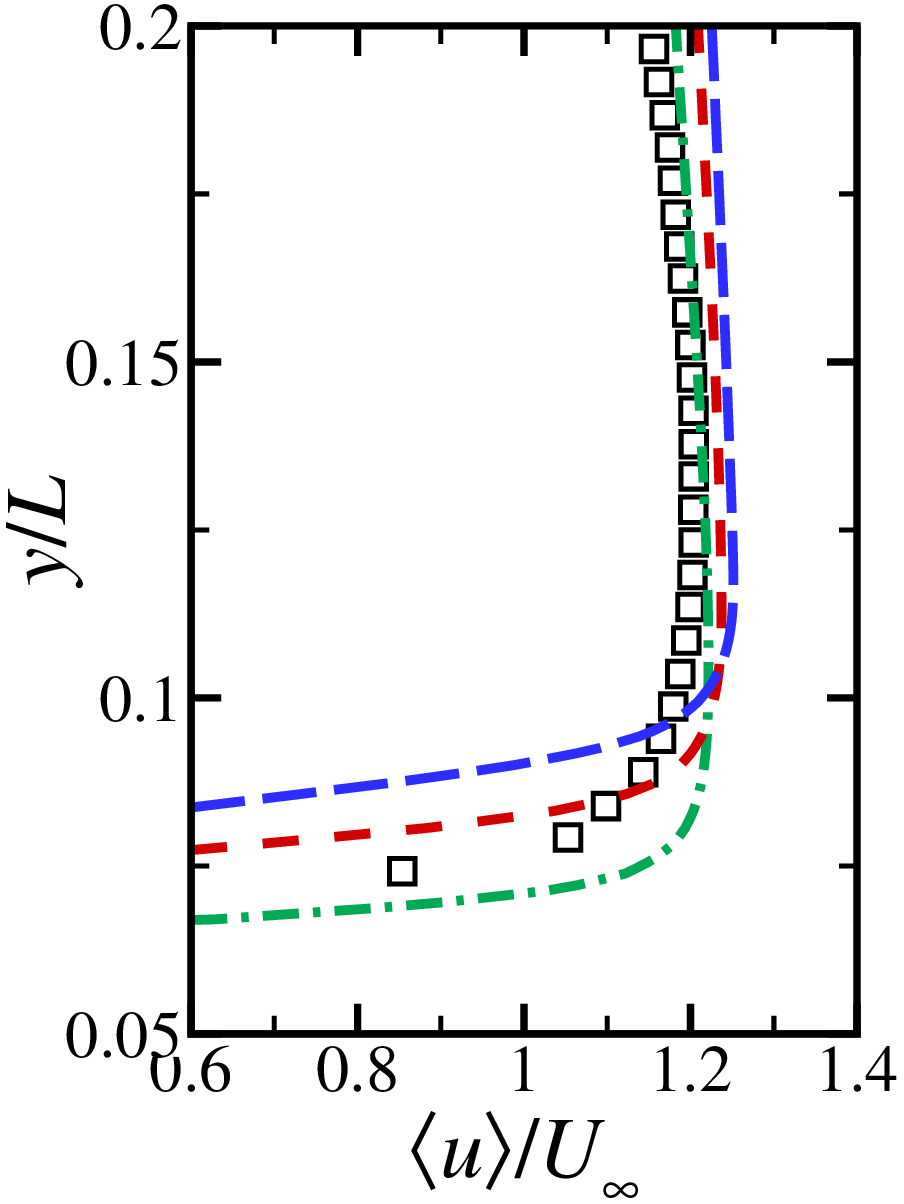}
	\subcaption{$x/L=0.1$}
	\end{subfigure}
	\begin{subfigure}[b]{0.244\textwidth}
	\centering
	\includegraphics[width = 1.0\textwidth]{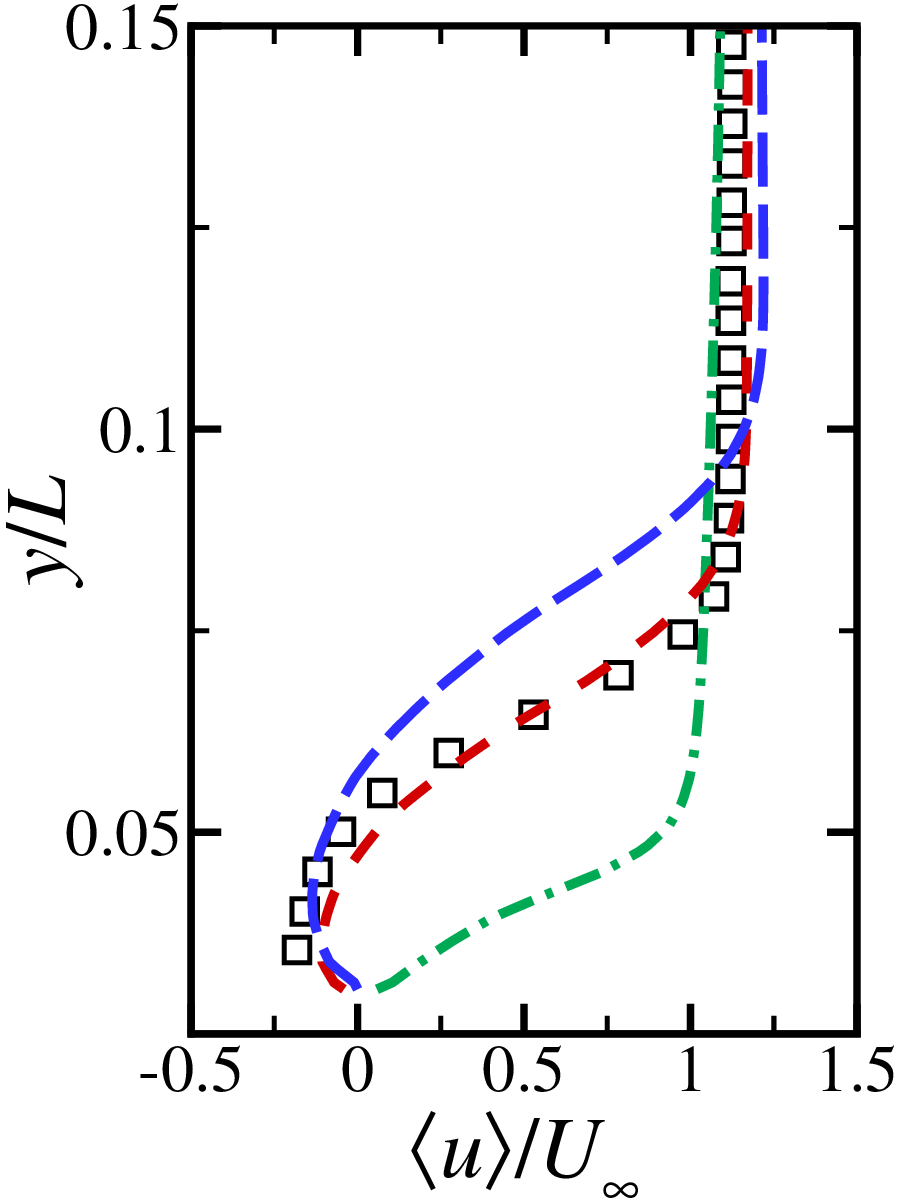}
	\subcaption{$x/L=0.2$}
	\end{subfigure}
\caption{{\color{black}Vertical profiles of time-averaged streamwise velocity $\left\langle u \right\rangle$ from the 2D-bump-WM cases with the WW model (figure~\ref{fig:case_validation}(c) and table~\ref{tab:case_validation}), the FNN model for only $\tau_w$ and the FEL model, while the experimental data (``EXP'') of~\citet{Gray_etal_AIAA_2022} is included as comparison.}}
\label{fig:profile_case_bump}
\end{figure}

The case setups for the simulated flow configurations are described as follows:
\begin{enumerate}
    \item {\bf Flow over a 2D wavy wall.} The geometry is given by the sine function
    \begin{equation}
        y_w(x) = a\sin\left( 2\pi x / \lambda_w \right),
        \label{eq_wavy_2D}
    \end{equation}
    where $a$ is the amplitude and $\lambda_w$ is the wave length of the wavy wall. Two ratios, i.e., $\alpha = a / \lambda_w$ and $\chi = a / \delta$ (where $\delta$ is the channel half height) are employed for characterizing the flow. Two cases with $\alpha = 0.05$, $\chi = 0.1$ and $\alpha = 0.1$, $\chi = 0.2$ are simulated as shown in figure~\ref{fig:case_validation}(a). The Reynolds number based on $2\delta$ and $U_b$ is $Re_b = 11200$. The computational domain is $2.0\delta \times 2.0\delta \times 2.0\delta$ and discretized using a curvilinear grid with the resolution of $128 \times 192 \times 128$ and $64 \times 48 \times 64$ for WRLES and WMLES cases, respectively, as shown in table~\ref{tab:case_validation}. Periodic boundary condition is applied in the streamwise and spanwise directions. At the bottom wavy wall and the top flat wall, the no-slip boundary condition is applied in WRLES, the wall model is employed in WMLES.
    \item {\bf Flow over the three-dimensional (3D) wavy wall.} The geometry is given by 
    \begin{equation}
        y_w(x, z) = a\sin\left( 2\pi x / \lambda_{w_x} \right) \sin\left( 2\pi z / \lambda_{w_z} \right) 
        \label{eq_wavy_3D}
    \end{equation}
    as shown in figure~\ref{fig:case_validation}(b). The geometrical parameters $\alpha$ and $\chi$ are the same in the $x$ and $z$ directions. The Reynolds number based on $2\delta$ and $U_b$ is also $Re_b = 11200$. Two cases are simulated with the values of $\alpha$ and $\chi$, the computational domain, the employed grid resolution and the boundary condition the same as the 2D wavy wall cases.
    \item {\bf Flow over a 2D Gaussian-shaped bump.} The geometry is given by 
    \begin{equation}
        y(x) = h\exp\left[ -(x/x_0)^2 \right], 
    \end{equation}
    where the bump height $h = 0.085L$, the constant $x_0 = 0.195L$, $L$ is the spanwise width of the 3D bump configuration~\citep{Slotnick_exp_2019}. Since the bump height is more than one order of magnitude smaller than its width, the flow away from two ends can be considered as two-dimensional. In this study, the bump is considered as 2D with the spanwise computational domain reduced to $L_z = 0.04L$, half of that in the high-fidelity simulation~\citep{Uzun_Malik_AIAA_2022}. The streamwise and vertical sizes of the computational domain are $L_x = 2.8L$ and $L_z = 0.5L$ as shown in figure~\ref{fig:case_validation}(c), with the grid resolution of $920 \times 90 \times 50$ in the streamwise, vertical, and spanwise directions, respectively. The free-slip boundary condition is applied at the top wall. At the bottom wall, the wall model is employed. In the spanwise direction, the periodic boundary condition is applied. At the inlet positioned at $x/L = -0.8$, the turbulent boundary layer profile (which is computed using the solver of~\citet{Qin_Dong_AMM_2016}) with the superimposed synthetic turbulence generated using a digital filtering technique~\citep{Klein_etal_JCP_2003} is employed. The recycling-rescaling technique~\citep{Morgan_etal_AIAA_2011} is employed to recycle the turbulent fluctuations while keeping the mean inflow profile fixed. The Reynolds number based on $L$ and the upstream reference velocity $U_\infty$ is $Re_L = 2\times 10^6$.
\end{enumerate}

The results from the simulated cases are shown in the following. Figure~\ref{fig:case_validation_contour} compares the contours of time-averaged streamwise velocity with streamlines obtained from the WRLES and WMLES cases for the three flow configurations. For the 2D and 3D wavy wall cases (figure~\ref{fig:case_validation_contour}(a, b)), it is seen that the FEL model accurately predict the separation bubble and the global flow field, while the WW model can hardly predict the flow separation. For the 2D Gaussian bump, which has a Reynolds number significantly larger than the training case, a reasonable prediction of the flow separation is also observed for the FEL model.

We then quantitatively assess the performance of the FEL model on predicting the first- and second-order flow statistics. Figures~\ref{fig:profile_case_wavy2D_0p05}$\sim$\ref{fig:profile_case_wavy2D_0p1} compare the vertical profiles of the flow statistics obtained from the 2D-wavy-WR and WM cases with different wave amplitudes. In figure~\ref{fig:profile_case_wavy2D_0p05}, compared with experimental data of \citet{Wagner_etal_EF_2007} and WRLES data of \citet{Wagner_etal_JoT_2010}, the present WRLES is verified. As for the proposed FEL wall model, it is seen that it can accurately predict the time-averaged velocities $\left\langle u \right\rangle$ and $\left\langle v \right\rangle$, the primary Reynolds shear stress $\left\langle u'v' \right\rangle$, and streamwise velocity fluctuation $\left\langle u'u' \right\rangle$, while slightly underpredicts the vertical and spanwise velocity fluctuations $\left\langle v'v' \right\rangle$, $\left\langle w'w' \right\rangle$ at the peak region. The WW model, on the other hand, fails to accurately predict the flow statistics. As for the steeper wavy wall with $\alpha = 0.1$, $\chi = 0.2$, the separation bubble grows larger. Good agreements with the reference data are still observed for the proposed FEL wall model, while large discrepancies are observed for the WW model as shown in figure~\ref{fig:profile_case_wavy2D_0p1}.

In the 3D wavy wall case, the flow is influenced by the curvatures in both the streamwise and spanwise directions. It is well beyond the training data of the FEL wall model, which are from cases with 2D configuration, and is a challenge for model validation. In figure~\ref{fig:case_validation_contour}(b), we have seen that the flow patterns in the spanwise slices $z/2\delta = 0.0$ and 0.5 are similar to those from the 2D case, but the bubbles exhibit phase difference due to the wave crest. In the spanwise slice $z/2\delta = 0.25$, on the other hand, the flow pattern is similar to the turbulent channel flow. Figures~\ref{fig:profile_case_wavy3D_0p05}$\sim$\ref{fig:profile_case_wavy3D_0p1} compares the vertical profiles of time-averaged streamwise and vertical velocities in the two spanwise slices $z/2\delta = 0.0$ and 0.25 obtained from the 3D-wavy-WR and WM cases with different wave amplitudes. Using the gentle wavy surface case (figure~\ref{fig:profile_case_wavy3D_0p05}), the present WRLES cases are verified using the data from \citet{Wagner_etal_JoT_2010}. Compared with the WRLES results, it is seen that the FEL model preforms well on predicting the velocity profiles in both the gentle and steep wavy surfaces, better than the WW model especially in the slice $z/2\delta = 0.0$.

Test results from the 2D-bump-WM cases are shown in figure~\ref{fig:profile_case_bump}, where the vertical profiles of time-averaged streamwise velocity $\left\langle u \right\rangle$ obtained from the three wall models are compared. Compared with the experimental data (``EXP'') of~\citet{Gray_etal_AIAA_2022}, the WW model simulates well the developing region of the turbulent boundary layer ($x/L=-0.4$ and -0.1), but fails to predict the flow separation downstream of the Gaussian bump ($x/L=0.2$). The FEL model, on the other hand, underestimates the near-wall velocity at $x/L=-0.4$ and 0.1, while well predicts the separation bubble at $x/L=0.2$. Surprisingly, the FEL model with only the $\tau_w$ sub-model ($\text{FEL}_{\tau_w}$) gives an overall good prediction of $\left\langle u \right\rangle$ at various streamwise locations. It is noted that we do not attempt to draw the conclusion that the $\text{FEL}_{\tau_w}$ model outperforms the FEL model for such cases, as neither of them was trained for simulating a developing boundary layer.
}

Overall, the test cases have shown that the FEL model has a strong generalization ability in predicting {\color{black}separated flow with different configurations}, grid resolutions, and Reynolds numbers. The learned eddy-viscosity coefficient of the FEL model results in a smaller energy dissipation when compared to the dynamic approach, which is beneficial for flow separation near the hill crest {\color{black}while does not favour the developing boundary layer}.

\section{Conclusions}\label{sec:Conclusion}
%
%Lack of generalizability and suboptimal \emph{a posteriori} performance are two challenges for data-driven LES wall models. 
%Both submodels are embedded into the WMLES solver. The former directly uses the FNN\_PH-LoW model, which was learned from the WRLES data of PH case and the law of the wall. For the latter, a new neural-network-enabled mixing length model is proposed and trained
%of turbulent flow over periodic hills, considering different geometries, at
%
In this study, we proposed a features-embedded-learning (FEL) wall model for improving the \emph{a posteriori} performance in wall-modelled large-eddy simulation (WMLES) of separated flows. The FEL wall model comprises two submodels: a wall shear stress model and an eddy viscosity model for the first off-wall grid. The former was trained using the wall-resolved simulation data and the law of the wall. The latter was modelled via a modified mixing length model with the model coefficient learned in an embedded way in the WMLES environment using the ensemble Kalman method.

The embedded training of the FEL model was conducted in the WMLES environment with two cases, i.e., the periodic hill flow with two different hill slopes (H0.5 and H1.5) for one grid resolution with the height of the first off-wall grid cell $\Delta y_f/h=0.06$ (where $h$ is the hill's height). The learned model was systematically assessed using the periodic hill cases with grid resolutions, hill slopes, and Reynolds numbers different from the training cases, {\color{black} the two-dimensional wavy wall cases, the three-dimensional wavy wall cases, and the two-dimensional Gaussian bump case.}
The key results include: 
\begin{enumerate}
    \item Good \emph{a posteriori} performance was achieved for the FEL model in predicting key flow statistics, including the pattern of the separation bubble, the skin friction, and the mean velocity and second-order turbulence statistics. 
    \item Good generalizability was observed for the proposed model for cases with different {\color{black} flow configurations,} grid resolutions, and Reynolds numbers. 
    \item The relative errors of the FEL model are less than 10\% for mean streamwise velocity $\left\langle u \right\rangle$ and 20\% for turbulence kinetic energy (TKE) $k$, respectively, being significantly lower than the Werner-Wengle (WW) model {\color{black}for most cases}.
    \item The underestimation of the TKE is caused by the fact that a considerable amount of TKE is not resolved by the coarse grid, which is confirmed by the good agreement between the $k$ predicted by the FEL model and the $k$ computed from the filtered WRLES flow fields.
    \item The FEL model improves the predictions of the SGS stresses and energy transfer rate at the first off-wall grid, which is the key reason for the good \emph{a posteriori} performance of the FEL model {\color{black} in simulating separated flows}.
    \item {\color{black}Using the $\text{FEL}_{\nu_t}$ submodel with the conventional WW model can improve the predictions of the overall flow patterns, but cannot accurately predict the flow separation and reattachment.}
\end{enumerate}

%This work has been focused on the periodic hill flow cases. More complicated cases, for instance, three-dimensional boundary layer flows, will be carried out to further test the FEL model. Moreover, 
%{\color{black}It should be noted that there are still some discrepancies on the prediction of separated flows between the FEL model and the WRLES, e.g., the H1.5-WM-0.06 case used for training, the inconsistency of prediction accuracy with grid refinement.}
%The prediction accuracy of WMLES is affected by many factors, such as grid resolution, discretization errors, errors from the SGS models and wall models, which should be further identified.

The model assessments have been focused on the basic turbulence statistics. Further evaluation of the models on predicting spatiotemporal flow structures ~\citep{He_Jin_Yang_ARFM_2017} needs to be performed. 
{\color{black}
Analyzing the underlying physics causing the discrepancies is important but challenging because of the many factors involved, which include the discretization error, the subgrid scale model error, and the wall model error. It is beyond the scope of this paper, and will be carried out in the future work. 
}
%Moreover, the proposed framework can be extended to accommodate any kind of flow configuration, as long as sufficient observation data is available. 
%To address more complex flow configurations, the embedded learning of wall model can be enhanced by incorporating progressive machine learning techniques~\citep{Bin_etal_PRF_2022} to account for more intricate flow physics. 
%Evaluation of the models for capturing spatiotemporal flow structures is certainly needed~\cite{He_Jin_Yang_ARFM_2017}. 

%\citet{Bin_etal_PRF_2022} proposed a general framework of progressive machine learning for near-wall turbulence modeling to control the network’s behavior when extrapolating a simple model to more complex ones. The framework was tested in four examples, i.e., log layer, channel, boundary layer, and rotating channel, and the eddy-viscosity model trained with a neural network preserved the law of the wall and was open to corrections that account for other physics.
%The third strategy is the online learning, in which the model training interacts with the prediction environment.

\section*{Acknowledgements}
This work was supported by NSFC Basic Science Center Program for ``Multiscale Problems in Nonlinear Mechanics'' (NO. 11988102), the Strategic Priority Research Program of Chinese Academy of Sciences (CAS, NO. XDB0620102), National Natural Science Foundation of China (NOs. 12002345, 12172360), and CAS Project for Young Scientists in Basic Research (YSBR-087).

\section*{Declaration of Interests}
The authors report that they have no conflict of interest.

\appendix
{\color{black}
\section{Data preparation for the training of the wall shear stress model}\label{appendix:data}
The data preparation for the training of the wall shear stress model is described in this appendix. The training data consist of the periodic hill flow data and the law of the wall data.
%and the The data from the periodic hill case is generated by applying the triangulation with linear interpolation approach to
The Reynolds numbers of the periodic hill cases are $Re_h = 5600$ and 10595. For each case, nine snapshots per one flow-through time ($T=L_x/U_b$) on four spanwise ($x-y$) slices are saved for a total simulation time $50T$. On each snapshot, the input-output pairs at 95 nodes, which are uniformly distributed in $y_n/h \in [0.006, 0.1]$ are extracted for interpolation. As for the logarithmic law, the data are generated in the following three steps: (i) Sample $N_R = 701$ cases for $Re_\tau = u_\tau \delta/\nu \in [10^2, 10^9]$, (ii) For each $Re_{\tau}$, sample the velocity and wall shear stress data at $y_n = 10^{\lg \frac{30}{Re_\tau} + j\cdot dh}$ in the range of $y_n \in [ \frac{30}{Re_\tau}, 0.1 ]$, where $j$ is the index of grid point, (iii) Normalize the velocity and wall shear stress using the bulk velocity ($U_b = \int_{0}^{Re_\tau} \frac{U^+}{Re_\tau}dy^+ +0.5$) calculated by integrating on the logarithmic law. The number of input-output samples from the periodic hill flow and the logarithmic law are both $1.1 \times 10^6$, of which $90\%$ are used for model training and the rest $10\%$ are employed for validation. More details on the data preparation can be found in our previous papers~\citep{Zhou_He_Yang_PRF_2021, Zhou_etal_PoF_2023}.
}

\section{Comparison with the filtered WRLES results}\label{appendix:filtering}
\begin{figure}
\centering{\includegraphics[width=0.85\textwidth]{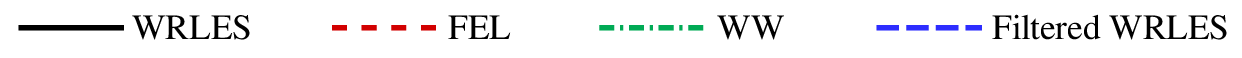}}
\centering{\includegraphics[width=0.495\textwidth]{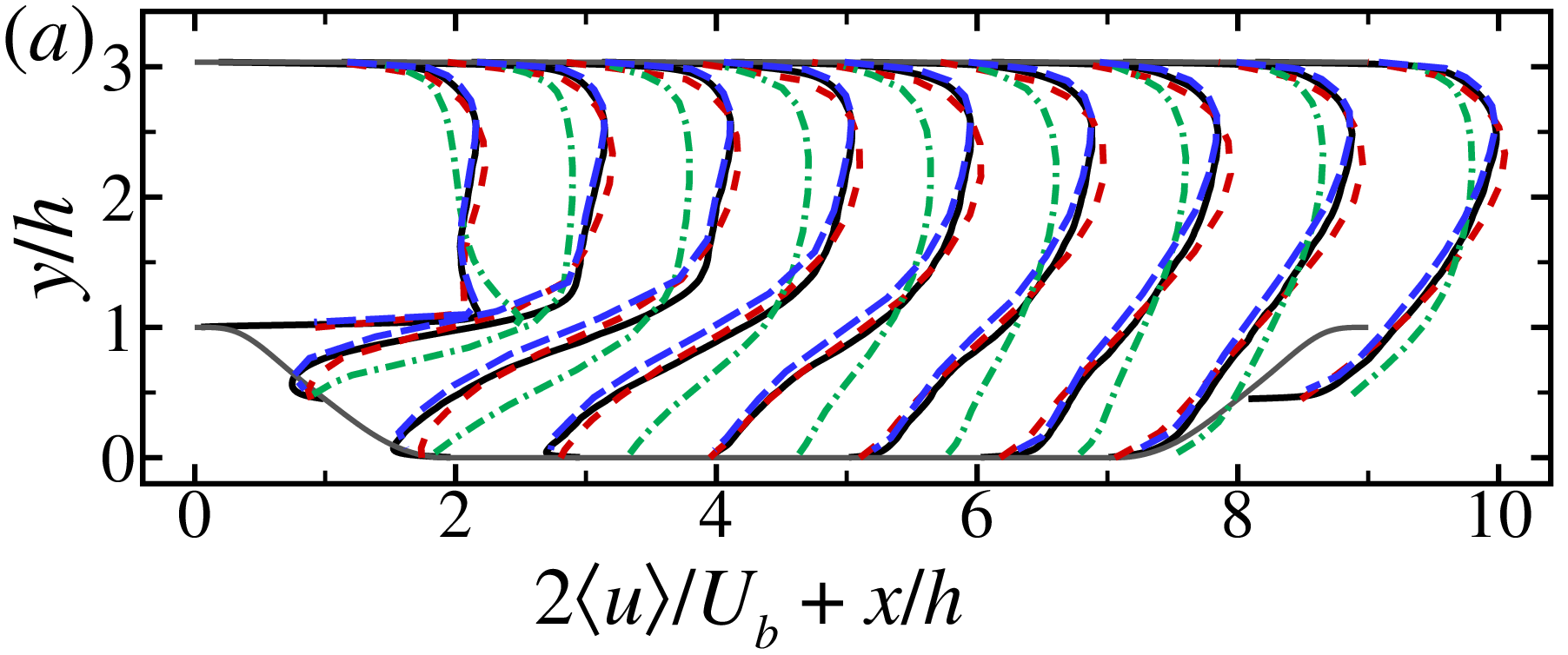}}
\centering{\includegraphics[width=0.495\textwidth]{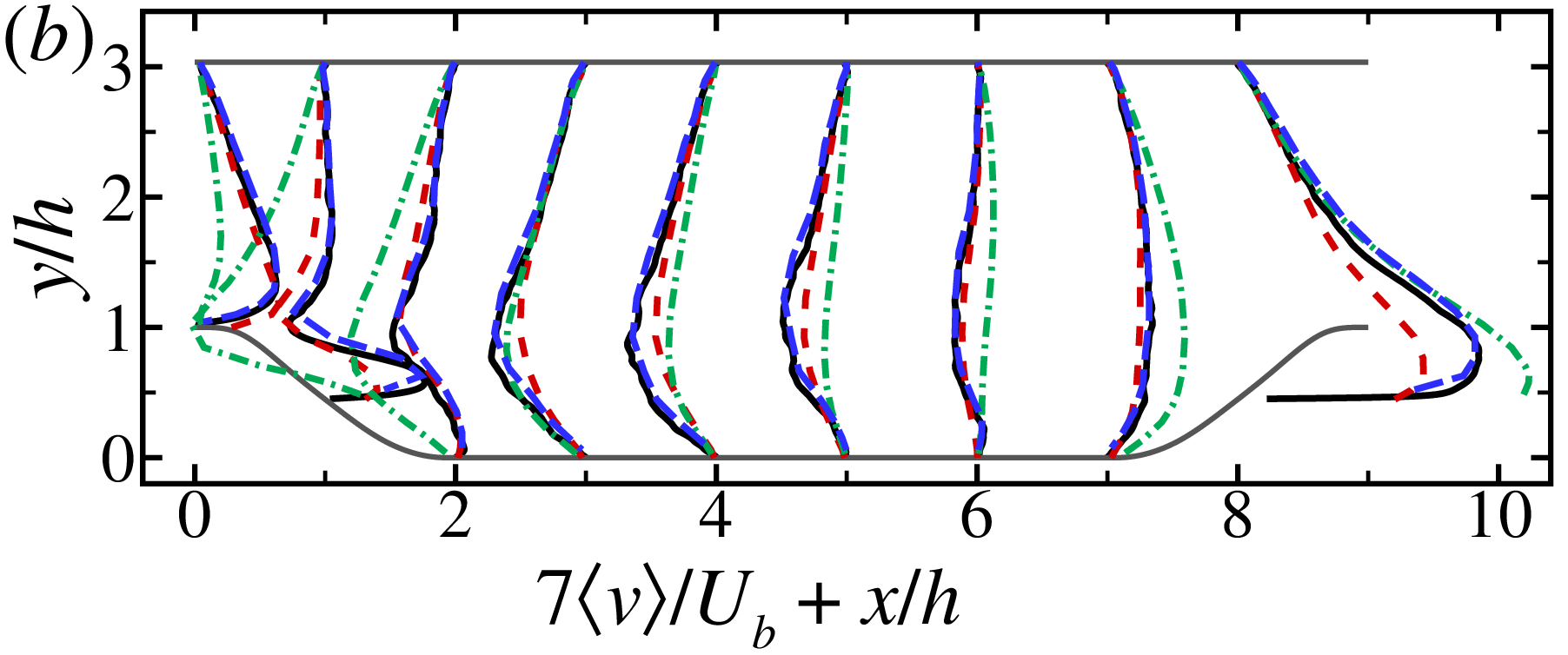}}
\centering{\includegraphics[width=0.495\textwidth]{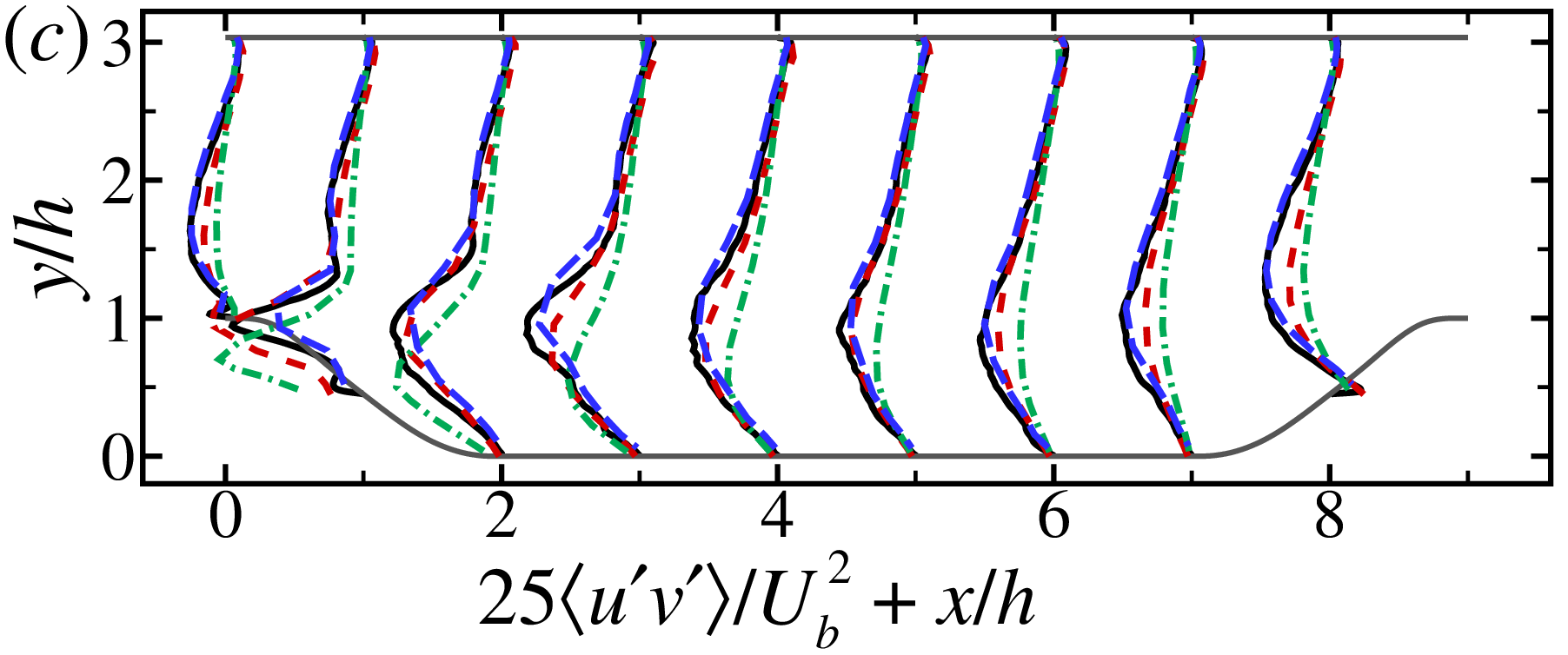}}
\centering{\includegraphics[width=0.495\textwidth]{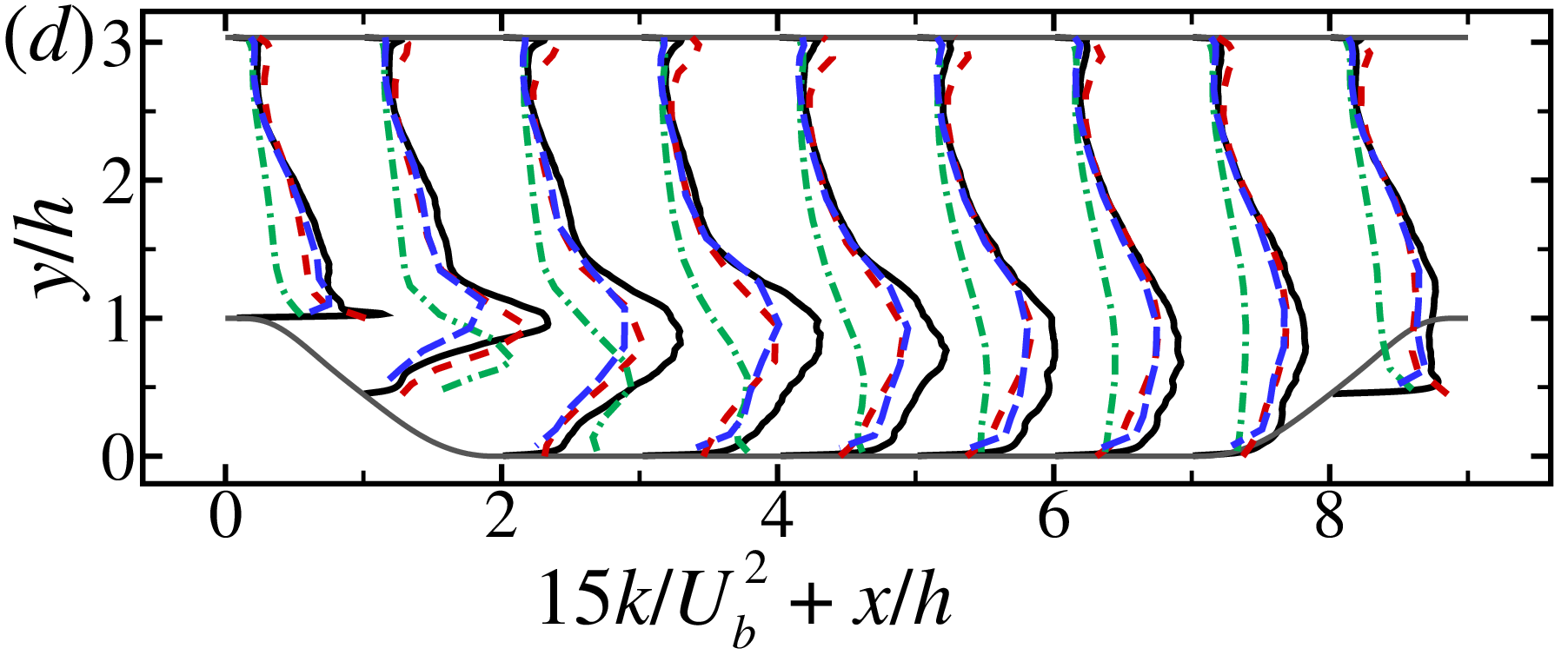}}
  \caption{Vertical profiles of (a) time-averaged streamwise velocity $\left\langle u \right\rangle$ and (b) vertical velocity $\left\langle v \right\rangle$, (c) primary Reynolds shear stress $\left\langle u'v' \right\rangle$, and (d) turbulence kinetic energy $k$ from the WRLES, the WMLES with the FEL or WW model, and the spatial filtering of WRLES for the H1.0 case {\color{black}with $\Delta y_f/h \approx 0.06$} at $Re_h = 10595$. {\color{black}Note that it is not proper to consider the difference between the filtered WRLES results and the WRLES results as error. The way it plotted in the figure is just for comparison.}}
\label{fig:filter_profile}
\end{figure}
\begin{figure}
\centering{\includegraphics[width=0.82\textwidth]{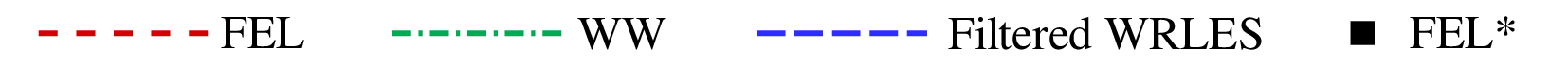}}
\centering{\includegraphics[width=0.4\textwidth]{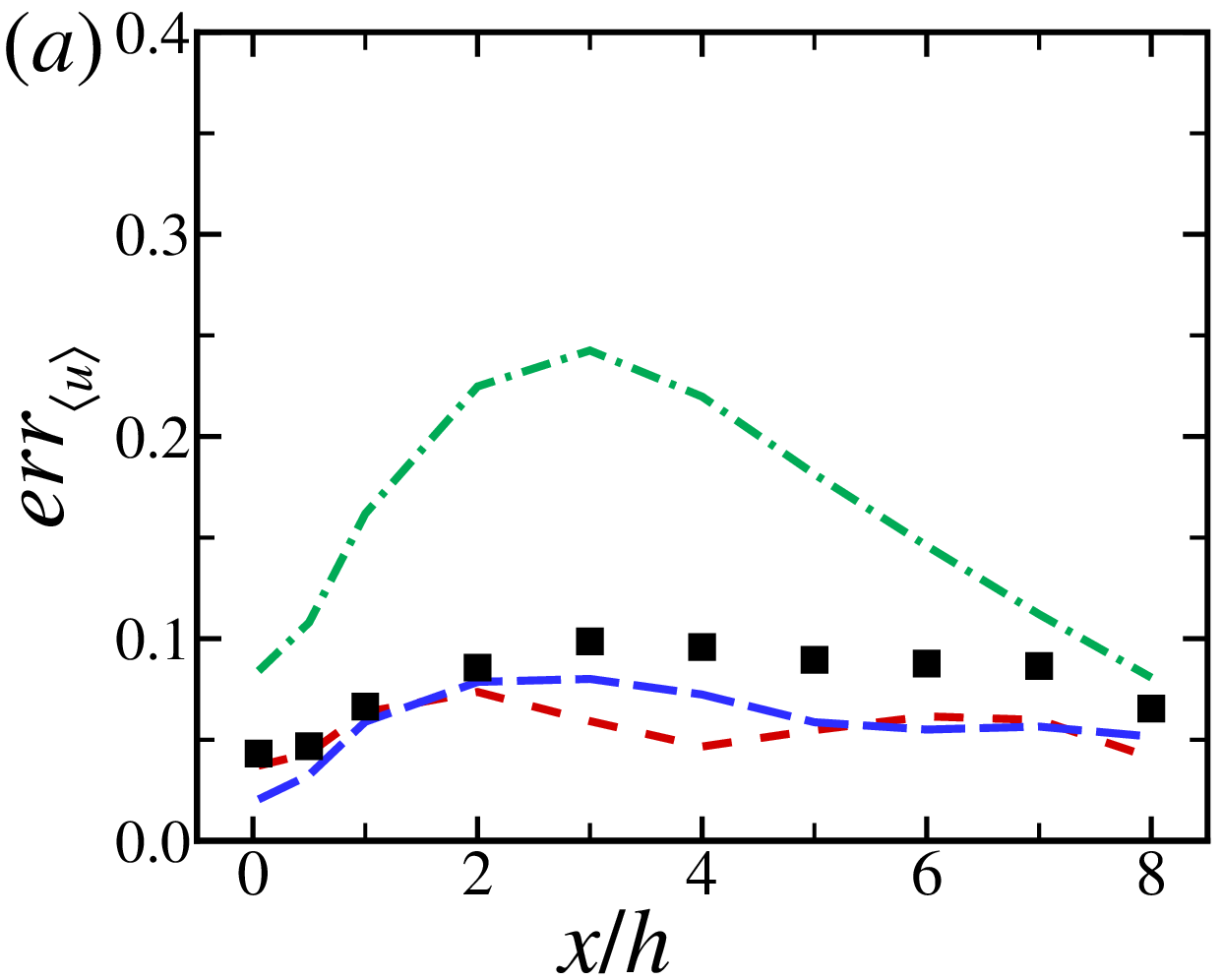}}\quad
\centering{\includegraphics[width=0.4\textwidth]{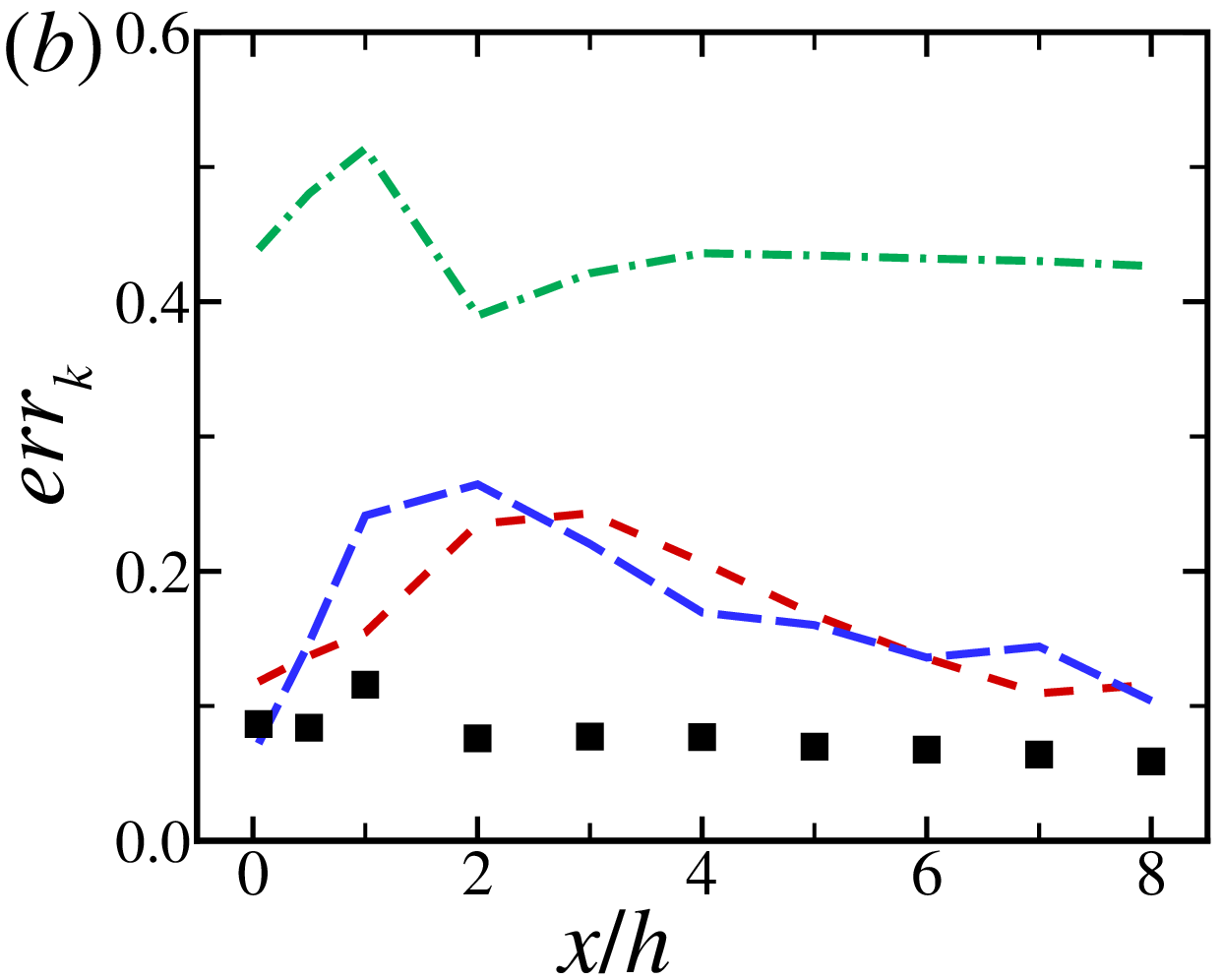}}
  \caption{Relative errors of (a) time-averaged streamwise velocity $\left\langle u \right\rangle$ and (b) turbulence kinetic energy $k$ between the WMLES, the spatial filtering and the WRLES for the H1.0 case {\color{black}with $\Delta y_f/h \approx 0.06$} at $Re_h = 10595$.}
\label{fig:filter_profile_err}
\end{figure}

Because of the employed coarse grid, a considerable amount of TKE is not resolved in WMLES. It is fair to compare the TKE predicted by WMLES with the filtered WRLES predictions. According to section~\ref{subsec:viscosity}, the flow fields from the H1.0-WR case are spatially filtered onto a coarse grid with $\Delta y_f/h \approx 0.06$, and the WMLES cases with the FEL and WW models are then carried out.

Figure~\ref{fig:filter_profile} compare the flow statistics obtained from the WMLES with the filtered WRLES results. As seen, the TKE profiles predicted by the FEL model compare well with the filtered WRLES predictions. The errors in WMLES predictions are shown in figure~\ref{fig:filter_profile_err}. {\color{black}It should be noted that the symbols in ``FEL*'' represents the error between the FEL model and the filterd WRLES, while the three lines represent the errors between the respective cases and the WRLES results.} Compared with the WRLES results, the errors from the FEL model are almost the same as those from the filtered WRLES. Significant errors, on the other hand, are still observed for the WW model. {\color{black}When the FEL model is directly compared with the filter WRLES, the error for TKE is less than 8\% at most streamwise locations.}

%These results demonstrate that the WMLES prediction can be significantly improved by suppressing the SGS eddy viscosity at the first off-wall grid nodes.

%{\color{black}
%\section{\emph{A priori} test of the skin friction coefficient} \label{appendix:Cf}
%
%\begin{figure}
%\centering{\includegraphics[width=0.48\textwidth]{Fig_Cf_HB_WW_priori.eps}}
%  \caption{{\color{black}Comparison of the time-averaged skin friction coefficients between the WRLES, the FEL and WW models in the \emph{a priori} test. The instantaneous flow fields of H1.0-WR case at $Re_h = 10595$ are used for calculation.}}
%\label{fig:Cf_HB_priori}
%\end{figure}
%
%As a \emph{a priori} test of the FEL and WW models, the instantaneous flow fields of H1.0-WR case at $Re_h = 10595$, which consists of 450 snapshots covering 50 flow-through times, are used as the input. The wall-normal distance is set to $\Delta y_c/h=0.045$, corresponding to the H1.0-WM case with $\Delta y_f/h=0.09$. The instantaneous outputs of skin friction coefficient are then averaged and compared with the WRLES results, as shown in figure~\ref{fig:Cf_HB_priori}. It is observed that the time-averaged skin friction coefficient predicted by the FEL model coincides well with the WRLES result at all streamwise locations. As for the WW model, the time-averaged skin friction coefficient is underestimated at the windward of the hill ($x/h \in [7.0, 9.0]$) and the recirculation zone ($x/h \in [0.3, 4.0]$), and overestimated at the leeward hill face ($x/h \in [0, 0.3]$).
%}

\bibliographystyle{jfm}
% Note the spaces between the initials
\bibliography{main}

\end{document}